\documentclass[12pt,oneside,a4paper,semicolon]{book} 

\usepackage{amsmath, amssymb, amsthm}
\usepackage{graphicx,color}
\usepackage{multirow}
\usepackage[left=1.5in, right=1in, top=1in, bottom=1in]{geometry}
\usepackage{epsfig}
\usepackage{color}
\usepackage{setspace}
\usepackage{latexsym}
\usepackage{mathrsfs}
\usepackage{graphicx}
\usepackage{tikz}
\usepackage[titletoc]{appendix}
\usepackage{alltt}
\graphicspath{ {images/} }
\usepackage{longtable}
\usepackage{tikz}
\usepackage{acronym}
\usepackage{pgfplots}
\usepackage{csquotes}
\usepackage[hidelinks]{hyperref}
\usepackage{titletoc,tocloft}
\usepackage{times}

\usepackage{sectsty}
\usepackage{titlesec}
\usepackage{float}  

\newtheorem{thm}{Theorem}[section]
 \newtheorem{cor}[thm]{Corollary}
 \newtheorem{lem}[thm]{Lemma}
 
 \theoremstyle{definition}
 
 \theoremstyle{remark}

\newcounter{subchapter}[chapter]
\renewcommand{\thesubchapter}{\Roman{subchapter}}

\newcommand{\subchapter}[1]{%
    \refstepcounter{subchapter}
    \addcontentsline{toc}{section}{PART \thesubchapter\quad #1}
    \begin{center}
        \vspace{1cm}
        {\Large\bfseries PART \thesubchapter}\\[0.5cm]
        {\Large\bfseries #1}
        \vspace{1cm}
    \end{center}
} 

\chapterfont{\centering }
\titleformat{\chapter}[display]
{\normalfont%
	\huge
	\bfseries}{\chaptertitlename\ \thechapter}{16pt}{%
	\Huge 
}

\usepackage{textcase}

\sectionfont{\normalfont}
\titleformat{\section}
{\normalfont\fontsize{14pt}{16pt}  \selectfont}{ \thesection}{1em}{}
\titleformat{\subsection}
{\normalfont\fontsize{12pt}{14pt} \selectfont}{\thesubsection}{1em}{}
\titleformat{\subsubsection}
{\normalfont\fontsize{12pt}{14pt} \selectfont}{\thesubsubsection}{1em}{}
\usepackage{caption}
\newtheorem{definition}{Definition}
\newcommand{\brdefinition}{\begin{definition}}
	\newcommand{\erdefinition}{\end{definition}}

\newtheorem{corollary}{Corollary}
\newcommand{\bcorollary}{\begin{corollary}}
	\newcommand{\ecorollary}{\end{corollary}}
\newtheorem{example}{Example}
\newcommand{\bexample}{\begin{example}}
	\newcommand{\eexample}{\end{example}}
\newtheorem{remark}{Remark}
\newcommand{\bremark}{\begin{remark}}
	\newcommand{\eremark}{\end{remark}}

\newtheorem{theorem}{Theorem}
\newcommand{\btheorem}{\begin{theorem}}
	\newcommand{\etheorem}{\end{theorem}}
\newtheorem{lemma}{Lemma}
\newcommand{\blemma}{\begin{lemma}}
	\newcommand{\elemma}{\end{lemma}}

\newcommand{\brdef}{\begin{defi}}
	\newcommand{\erdef}{\end{defi}}
\newcommand{\bcor}{\begin{cor}}
	\newcommand{\ecor}{\end{cor}}
\newcommand{\bth}{\begin{thm}}
	\newcommand{\ble}{\begin{lem}}
		\newcommand{\ele}{\end{lem}}
	\newcommand{\bcha}{\end{cha}}

\renewcommand{\thechapter}{\arabic{chapter}}
\renewcommand{\thesection}{\thechapter.\arabic{section}}

\theoremstyle{definition}

\usepackage{sectsty}
\usepackage{titlesec}
\chapterfont{\centering }
\titleformat{\chapter}[display]
{\bf\centering}
{\chaptertitlename\ \thechapter}{16pt}{\large}
\sectionfont{\normalfont}
\cftsetindents{chapter}{0em}{3em}      
\cftsetindents{section}{0em}{3em}

\usepackage{natbib}

\usepackage[acronym,toc=false,section=section, nogroupskip]{glossaries}
\makeglossaries

\definecolor{butter1}{rgb}{0.988,0.914,0.310}
\definecolor{chocolate1}{rgb}{0.914,0.725,0.431}
\definecolor{chameleon1}{rgb}{0.541,0.886,0.204}
\definecolor{skyblue1}{rgb}{0.447,0.624,0.812}
\definecolor{plum1}{rgb}{0.678,0.498,0.659}
\definecolor{scarletred1}{rgb}{0.937,0.161,0.161}

\usepackage{array}
\newcolumntype{L}[1]{>{\raggedright\let\newline\\\arraybackslash\hspace{0pt}}m{#1}}
\newcolumntype{C}[1]{>{\centering\let\newline\\\arraybackslash\hspace{0pt}}m{#1}}
\newcolumntype{R}[1]{>{\raggedleft\let\newline\\\arraybackslash\hspace{0pt}}m{#1}}

\usepackage{etoolbox}

\makeatletter
\patchcmd{\ttlh@hang}{\parindent\z@}{\parindent\z@\leavevmode}{}{}
\patchcmd{\ttlh@hang}{\noindent}{}{}{}
\makeatother

\begin{document}

\onehalfspacing
\begin{center}
\thispagestyle{empty}

\vskip .7cm
~\\~\\~\\~
{\Large \textbf{Dynamical and Cosmological Perspectives on Symmetric Teleparallel Gravity and its Extensions}}
\vskip 4 cm
{\it \large Submitted in partial fulfillment of the requirements for the degree of}
\vskip 0.5cm
{\bfseries \huge Doctor of Philosophy}
\vskip 2 cm
~\\~\\~
\textit \large {by} \\
\vskip .5cm
\textbf{\Large Pooja Vishwakarma}
\vskip 6.5 cm
\centerline{\includegraphics[scale=0.3]{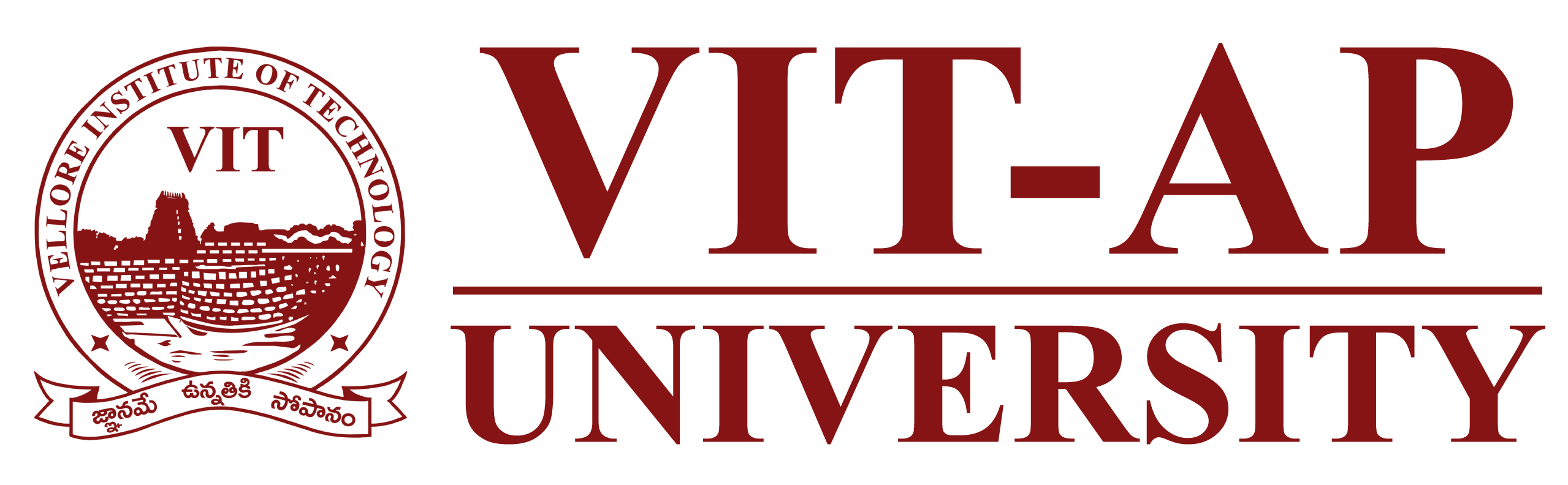}}
\vskip 0.7cm
{
\textbf{September, 2026}}
\end{center}

\newpage
\begin{spacing}{1.5}
\begin{center}
\thispagestyle{empty}
\vskip .5cm
{\centerline {{\large \textbf{DECLARATION}}}}
\end{center}

\par \quad \quad I hereby declare that the thesis entitled  \textbf{\enquote {Dynamical and Cosmological Perspectives on Symmetric Teleparallel Gravity and its Extensions}} submitted to VIT-AP University for the award of the degree of {\it Doctor of Philosophy}  is a record of bonafide work carried out by me under the supervision of \textbf{Dr. Aswathy~R~K}, Assistant Professor, Department of Mathematics, School of Advanced Sciences, VIT-AP University, Andhra Pradesh, India.
\vskip 0.5cm
\noindent
\par \qquad
I further declare that the work reported in this thesis has not been submitted and will not be submitted, either in part or in full, for the award of any other degree or diploma in this institute or any other institute or university.
~\\
~\\
~\\
~\\
~\\
~\\
~\\
\noindent
Place: VIT-AP University \qquad \qquad \qquad \qquad \qquad \quad \quad          			\textbf{Signature of the Candidate}
\vskip 0.3cm
\noindent
Date: 15-09-2026 \qquad  \quad \qquad  \qquad \qquad \qquad \qquad \qquad \quad \textbf{(Pooja Vishwakarma)}

\newpage
\begin{center}
\thispagestyle{empty}
\vskip .5cm

{\centerline{ {\large \textbf{CERTIFICATE}}}}
\end{center}

\par \quad \quad This is to certify that the thesis entitled\; \textbf{\enquote{Dynamical and Cosmological Perspectives on Symmetrical Teleparallel Gravity and its Extensions}} submitted by \textbf{Ms. POOJA VISHWAKARMA}, Department of Mathematics, School of Advanced Sciences, VIT-AP University, Andhra Pradesh, India for the award of the degree of {\it Doctor of Philosophy}, is a record of bonafide work carried out by him under my supervision, as per the VIT-AP University code of academic and research ethics.
\vskip 0.2cm
\noindent
\par
\qquad  The contents of this report have not been submitted and will not be submitted either in part or in full, for the award of any other degree or diploma in this institute or any other institute or university. The thesis fulfills the requirements and regulations of the University and in my opinion meets the necessary standards for submission.

~\\
~\\
~\\
~\\
~\\
\noindent
Place: VIT-AP University \qquad \qquad \qquad \qquad \qquad \quad \quad \quad         						 \textbf{Signature of the Guide}
\vskip 0.3cm
\noindent
Date: 15-09-2026  \qquad \quad  \qquad  \qquad \qquad \qquad \qquad \qquad \quad \quad \textbf{(Dr. Aswathy~R~K)}

\newpage
\pagenumbering{roman}
\addcontentsline{toc}{section}{\bf ABSTRACT}
{\centerline { {\large \textbf{ABSTRACT}}}}
~\\
\newacronym{gr}{GR}{General Relativity}
\newacronym{tegr}{TEGR}{Teleparallel Equivalent of General Relativity}
\newacronym{stegr}{STEGR}{Symmetric Teleparallel Equivalent of General Relativity}
\newacronym{de}{DE}{Dark Energy}
\newacronym{dm}{DM}{Dark Matter}
\newacronym{eos}{EoS}{Equation of State}
\newacronym{lcdm}{$\Lambda$CDM}{Lambda-Cold-Dark-Matter}
\newacronym{flrw}{FLRW}{Friedmann–Lemaître–Robertson–Walker}
\newacronym{cmb}{CMB}{Cosmic Microwave Background}
\newacronym{cc}{CC}{Cosmic Chronometers}
\newacronym{snia}{SNIa}{Supernovae Type Ia}
\newacronym{bao}{BAO}{Baryon Acoustic Oscillation}
\newacronym{trgb}{TRGB}{Tip of the Red Giant Branch}
\newacronym{ode}{ODE}{Ordinary Differential Equation}
\newacronym{cmt}{CMT}{Center Manifold Theory}
\newacronym{bbn}{BBN}{Big Bang Nucleosynthesis}
\newacronym{pnplus}{PN+}{Pantheon+}
\newacronym{da}{$D_A$}{Angular Diameter Distance}
\newacronym{dl}{$D_L$}{Luminosity Distance}
\newacronym{dmdist}{$D_M$}{Comoving Distance}

\noindent This thesis explores the phenomenon of late-time cosmic acceleration through the lens of modified gravity theories, with a particular emphasis on $f(Q)$ gravity and scalar-tensor extensions involving the trace of the energy-momentum tensor. Unlike the standard \gls{lcdm} model, which relies on a cosmological constant to account for the accelerated expansion of the Universe, these alternative frameworks aim to address persistent issues such as the unexplained nature of dark energy and dark matter, as well as observational inconsistencies. In particular, symmetric teleparallel gravity, where gravitational effects are described by non-metricity rather than curvature, offers a compelling platform for formulating alternative cosmological models. By incorporating a dynamical scalar field and modifications to the gravitational interaction, such models can reproduce a broad spectrum of cosmological behaviors. This work employs dynamical system techniques to analyze the evolution of the Universe within these modified teleparallel frameworks, highlighting their potential to account for various cosmic phases and the observed late-time acceleration without invoking a constant cosmological term.

\textit{Chapter 1} presents a comprehensive overview of the foundational formulations and mathematical structures underlying various modified gravity theories, including $f(Q)$, $f(Q, B)$, and $f(Q, \phi)$. It also introduces the theory of dynamical systems and outlines its relevance to cosmological modeling. Furthermore, the chapter discusses key cosmological observations and datasets that form the basis for the analyses conducted in subsequent chapters.

\textit{Chapter 2} presents a comprehensive literature review on $f(Q)$ gravity, highlighting its theoretical development, key cosmological implications, and recent advancements in the context of late-time cosmic acceleration and modified gravity frameworks.

\textit{Chapter 3} investigates cosmological models within the framework of $f(Q)$ gravity, considering both the background evolution and linear perturbations. The study focuses on three specific functional forms of $f(Q)$, from which relevant cosmological parameters are expressed in terms of dynamical variables. For each model, the corresponding critical points are determined along with the conditions required for their existence. A detailed dynamical systems analysis is carried out for each case individually, assessing stability through eigenvalue analysis and phase space diagrams. The results consistently indicate the presence of at least one stable node per model. Furthermore, the time evolution of key cosmological parameters demonstrates that the models are capable of describing accelerated expansion of the Universe.

\textit{Chapter 4} extends the analysis of cosmological models in $f(Q)$ gravity by incorporating observational constraints, with a particular focus on \gls{bbn}. The chapter explores two specific functional forms of $f(Q)$, examining their behavior at both the background and perturbation levels. Cosmological parameters are derived using suitable dynamical variables, and the associated critical points are identified along with their existence conditions. Each model is analyzed separately, with stability evaluated through eigenvalue analysis and phase portraits, revealing at least one stable node in every case. The influence of \gls{bbn} constraints is considered to ensure consistency with early-universe physics, providing bounds on model parameters to demonstrate the capacity of these models to account for the observed late-time acceleration of the Universe while remaining consistent with the early-universe constraints imposed by nucleosynthesis.

\textit{Chapter 5} investigates the application of the Noether symmetry approach within the framework of extended symmetric teleparallel gravity, specifically the $f(Q,\phi)$ formulation, where the non-metricity scalar $ Q $ is coupled to a scalar field $\phi$ in a Friedmann–Lemaître–Robertson–Walker (FLRW) background. By implementing the Noether symmetry method, we identify a viable functional form of $f(Q,\phi)$ that admits a nontrivial Noether vector. The existence of this symmetry enables the construction of the corresponding point-like Lagrangian and facilitates the derivation of exact cosmological solutions. The resulting solutions are fully compatible with the underlying symmetry structure and naturally lead to an accelerated expansion of the universe. In addition, we compute the relevant cosmological parameters-namely, the Hubble parameter, the deceleration parameter, and the effective equation-of-state parameter and show that their behavior remains consistent with current observational constraints.

\textit{Chapter 6} presents a dynamical systems analysis within the framework of $f(Q, \phi)$ gravity. The study investigates a specific models characterized by various functional forms of the non-metricity scalar, identifying the associated critical points and assessing their stability to understand the resulting cosmological dynamics. The chapter includes graphical illustrations of the evolution of the equation of state parameter, along with the density parameters corresponding to radiation, matter, and dark energy. Based on the trajectories of these cosmological parameters, it is concluded that the model successfully reproduces a phase of late-time accelerated expansion, consistent with current observational data.

\textit{Chapter 7} concludes the thesis by summarizing the main findings and discussions from the preceding chapters. It reflects on the overall outcomes of the research, emphasizing the key contributions made to the field, and outlines possible directions for future work and applications within the context of modified gravity and cosmology.


\newpage
\addcontentsline{toc}{section}{\bf ACKNOWLEDGMENT}
{\centerline { {\large \textbf{ACKNOWLEDGMENT}}}}

\vspace*{1cm}
\indent\indent First and foremost, I would like to express my sincere thanks to my Ph.D. supervisor, \textbf{Dr. Aswathy R~K}, Assistant Professor, Department of Mathematics, School of Advanced Sciences, VIT-AP University, Andhra Pradesh, India. Her constant support, and enthusiastic guidance helped shape my research and encouraged me to grow into an independent researcher.

\indent\indent I am truly grateful to my doctoral committee members for their time, support, and valuable feedback throughout this journey. My special thanks to {\bf Dr. Saroj~R.~Yadav}, Assistant Professor, Department of Mathematics and Humanities, SVNIT Surat, for being the external expert and for his thoughtful suggestions that improved the quality of this work. I also express my heartfelt thanks to {\bf Dr. Pratik Premadarshi Ray} and {\bf Dr. Prashant Patel}, Assistant Professors, School of Advanced Sciences, VIT-AP University, for their academic input and encouragement during the evaluation process.

\indent\indent I would like to thank {\bf Dr. G. Viswanathan}, Founder and Chancellor of VIT, for fostering a research-friendly environment, and {\bf Dr. P. Arulmozhivarman}, Vice Chancellor of VIT-AP University, for his constant encouragement. I am also grateful to {\bf Dr. Jagadish Chandra Mudiganti}, Registrar, for his administrative support. My sincere thanks to {\bf Dr. Ravindra Dhuli}, Dean Academic Research, for promoting a strong research culture, {\bf Dr. Srinivas S}, Dean, School of Advanced Sciences, for his continued support, and {\bf Dr. Vemula Ramakrishna Reddy}, Head of the Department of Mathematics, for his guidance and motivation.

\indent\indent I am truly thankful to all my {\bf colleagues} and {\bf friends} for their support, thoughtful discussions, and encouragement during difficult times. I also express my deepest gratitude to {\bf all my family members}, especially my mother {\bf Smt. Poonam Vishwakarma} and my husband {\bf Anil Mayani}, for their unconditional love, strength, and constant support. I also warmly thank {\bf Dr. Shah Parth Mukeshbhai} for his kindness, encouragement, and moral support. Their belief in me has been a powerful source of motivation throughout this journey.
\vskip 0.1cm
\noindent
Place: VIT-AP University
\vskip 0.1cm
\noindent
Date: 15-09-2026  \qquad \qquad \qquad \qquad \qquad  \qquad \qquad \qquad  \textbf{POOJA VISHWAKARMA}

\end{spacing}
\newpage
\begin{spacing}{1.5}
	\tableofcontents	
\end{spacing}
\newpage
\let\origaddvspace\addvspace
\renewcommand{\addvspace}[1]{}
\begin{spacing}{1.5}
	\addcontentsline{toc}{section}{\bf LIST OF FIGURES}
	\setcounter{lofdepth}{2} \listoffigures
\end{spacing}
\newpage
\begin{spacing}{1.5}
	\addcontentsline{toc}{section}{\bf LIST OF TABLES}
	\setcounter{lotdepth}{2} \listoftables 	
\end{spacing}

\vspace*{0.5cm}
\begin{spacing}{1.5}
\addcontentsline{toc}{section}{\bf LIST OF TERMS AND ABBREVIATIONS}
\chapter*{LIST OF TERMS AND ABBREVIATIONS}
\begin{acronym}[TDMA]
\acro{LCDM}[$\Lambda$CDM]{Lambda-Cold-Dark-Matter}
\acro{GR}{General Relativity}
\acro{TEGR}{Teleparallel Equivalent of General Relativity}
\acro{STEGR}{Symmetric Teleparallel Equivalent of General Relativity}
\acro{DE}{Dark Energy}
\acro{DM}{Dark Matter}
\acro{EoS}{Equation of State}
\acro{FLRW}{Friedmann-Lemaître-Robertson-Walker}
\acro{CMB}{Cosmic Microwave Background}
\acro{CC}{Cosmic Chronometers}
\acro{SNIa}{Supernovae Type Ia}
\acro{BAO}{Baryon Acoustic Oscillation}
\acro{TRGB}{Tip of the Red Giant Branch}
\acro{ODE}{Ordinary Differential Equation}
\acro{CMT}{Center Manifold Theory}
\acro{BBN}{Big Bang Nucleosynthesis}
\acro{PN+}{Pantheon+}
\acro{$D_A$}{Angular Diameter Distance}
\acro{$D_L$}{Luminosity Distance}
\acro{$D_M$}{Comoving Distance}
\acro{CGR}{Coincident General Relativity}
\acro{GTEGR}{General Teleparallel Equivalent of General Relativity}
\acro{TG}{Teleparallel Gravity}
\acro{STG}{Symmetric Teleparallel Gravity}
\acro{GHY}{Gibbons-Hawking-York}
\end{acronym}
\end{spacing}

\newpage
\let\origaddvspace\addvspace
\renewcommand{\addvspace}[1]{}
\vspace*{0.5cm}
\begin{spacing}{1.5}
\addcontentsline{toc}{section}{\bf LIST OF SYMBOLS}
\begin{center}{{\Large \bfseries LIST OF SYMBOLS}} \end{center}
\vspace*{1.5cm}
\begin{tabular}{cp{0.6\textwidth}}
$H$ & : Hubble parameter\\
$q$ & : Deceleration parameter\\
$s$ & : Snap parameter\\
$\mathrm{j}$ & : Jerk parameter\\
$l$ & : Lerk parameter\\
$L_{m}$ & : Matter Lagrangian\\
$g_{ij}$ &: Lorentzian metric\\
 $g$ &: Determinant of $g_{ij}$  \\ 
$G_{ij}$ & : Einstein tensor\\
$\Lambda(T)$ & : Effective time variable cosmological constant\\
 $\Gamma^{k}_{ij}$  &: General affine connection (Christoffel symbols) \\ 
 $\{^{k} _{ij}\}$ &: Levi-Civita connection\\  
$\nabla_{i}$ &: Co-variant derivative \\ 
$R^{k}_{\sigma ij}$ &: Riemann tensor \\
$R_{ij}$ &: Ricci tensor \\
$R$ &: Ricci scalar \\
$S_{M}$ & : Matter action\\
$T_{ij}$ & : Energy-momentum tensor\\
$T$ & : Trace of the energy-momentum tensor \\
$C_{s}^{2}$ & : Adiabatic speed of sound\\
$\delta(t)$ & : Deviation of Hubble parameter\\
$\delta_{m}(t)$ & : Deviation of energy density\\
$\rho_{c}$ & : Critical density\\
$Q$ & : Non-metricity scalar\\
$L^{k}_{ij}$  &: Disformation tensor \\ 
$P^{k}_{ij}$  &: Super potential \\
$n_{s}$ & : Spectral index\\
$r$ & : Tensor to scalar ratio\\ 
\end{tabular}

\end{spacing}
\renewcommand{\addvspace}[1]{\origaddvspace{#1}}

\mainmatter

\pagenumbering{arabic}

\chapter{Introduction} \label{c1}
\newacronym{gr}{GR}{General Relativity}
\newacronym{tegr}{TEGR}{Teleparallel Equivalent of General Relativity}
\newacronym{stegr}{STEGR}{Symmetric Teleparallel Equivalent of General Relativity}
\newacronym{de}{DE}{Dark Energy}
\newacronym{dm}{DM}{Dark Matter}
\newacronym{eos}{EoS}{Equation of State}
\newacronym{lcdm}{$\Lambda$CDM}{Lambda-Cold-Dark-Matte r}
\newacronym{flrw}{FLRW}{Friedmann–Lemaître–Robertson–Walker}
\newacronym{cmb}{CMB}{Cosmic Microwave Background}
\newacronym{cc}{CC}{Cosmic Chronometers}
\newacronym{snia}{SNIa}{Supernovae Type Ia}
\newacronym{bao}{BAO}{Baryon Acoustic Oscillation}
\newacronym{trgb}{TRGB}{Tip of the Red Giant Branch}
\newacronym{ode}{ODE}{Ordinary Differential Equation}
\newacronym{cmt}{CMT}{Center Manifold Theory}
\newacronym{bbn}{BBN}{Big Bang Nucleosynthesis}
\newacronym{pnplus}{PN+}{Pantheon+}
\newacronym{da}{$D_A$}{Angular Diameter Distance}
\newacronym{dl}{$D_L$}{Luminosity Distance}
\newacronym{dmdist}{$D_M$}{Comoving Distance}
\newacronym{cgr}{CGR}{Coincident General Relativity}
\newacronym{gtegr}{GTEGR}{General Teleparallel Equivalent of General Relativity}
\newacronym{tg}{TG}{Teleparallel Gravity}
\newacronym{stg}{STG}{Symmetric Teleparallel Gravity}
\newacronym{ghy}{GHY}{Gibbons-Hawking-York}

Since ancient times, humanity has been fascinated by profound questions concerning the Universe: What initiated its existence? What fundamental laws govern its evolution? Is the Universe finite or infinite? Will it persist indefinitely, or will it meet an ultimate fate? These timeless inquiries have given rise to \textit{cosmology}-the scientific discipline devoted to exploring the origin, structure, evolution, and destiny of the Universe. By reconstructing the cosmic history from its earliest moments to its current accelerated expansion, cosmology provides a framework for understanding the underlying mechanisms of cosmic evolution and predicting future developments.

The proposed thesis, titled \textit{``Dynamical and Cosmological Perspectives on Symmetric Teleparallel Gravity and Its Extension''}, focuses on exploring the cosmological dynamics of \(f(Q)\) gravity through a dynamical systems approach. It aims to reconstruct viable cosmological models, analyze their stability, and investigate their compatibility with observed cosmic evolution. The subsequent chapters provide a comprehensive review of the theoretical foundations, followed by mathematical formulations, phase space analysis, and discussion of physical implications within the context of symmetric teleparallel gravity.

This chapter is divided into two parts. In the first part, we discuss the theoretical aspects of cosmology, while the second part is devoted to the dynamical-system and stability analysis.

\subchapter{Theoretical Aspects of Cosmology}

\vspace{1em}
\section{Emergence of Modern Cosmology}
The study of cosmology seeks to explain the evolution of the Universe from its birth to its ultimate fate. The cornerstone of modern cosmology is the \textit{cosmological principle}, which postulates that the Universe is homogeneous and isotropic on sufficiently large scales. The theoretical foundation for describing the gravitational interaction within this framework was laid by Albert Einstein in 1915 through the formulation of the \textit{General Theory of Relativity (GR)}. In this geometric theory, gravity arises as a manifestation of the curvature of spacetime induced by matter and energy \citep{einstein2005riemann, aldrovandi2012teleparallel, bahamonde2023teleparallel}, described by the Einstein field equations:
\begin{equation}
    G_{\mu\nu} = 8\pi G T_{\mu\nu},
\end{equation}
where \(G_{\mu\nu}\) is the Einstein tensor, \(G\) is the gravitational constant, and \(T_{\mu\nu}\) denotes the energy–momentum tensor. \gls{gr} successfully explains a broad range of astrophysical and cosmological phenomena, such as gravitational lensing, perihelion precession, and the large-scale dynamics of the Universe.

In 1917, Einstein applied his theory to the entire Universe, proposing a static cosmological model stabilized by a cosmological constant (\(\Lambda\)). However, this model was soon replaced after observations revealed that the Universe is not static but expanding. The pioneering works of Alexander Friedmann\citep{friedman1922krummung, friedmann1924moglichkeit}, Georges Lemaître\citep{lemaitre4univers}, Howard P. Robertson\citep{robertson1935kinematics, robertson1936kinematics}, and Arthur G. Walker\citep{walker1937milne} provided dynamic solutions to Einstein’s field equations, leading to the formulation of the \gls{flrw} metric. This metric describes an expanding, homogeneous, and isotropic Universe and forms the foundation of the standard cosmological model.

The observational confirmation of cosmic expansion came in 1929 \citep{bahamonde2023teleparallel} when Edwin Hubble discovered a linear relationship between the recessional velocity of galaxies and their distance-a result now known as \textit{Hubble’s law} (\(v = H_{0} d\)) \citep{Hubble:1931zz}. This discovery gave strong support to the \textit{Big Bang} model, which posits that the Universe originated from an extremely hot and dense state approximately $13.8$ billion years ago. Further evidence emerged in 1965 with the discovery of the \gls{cmb} radiation by Penzias and Wilson, which provided a remnant signature of the early Universe and solidified the Big Bang paradigm.

Despite large-scale homogeneity, the Universe exhibits significant inhomogeneities on smaller scales, manifested through the clustering of galaxies, galaxy clusters, and cosmic voids\citep{Peebles:1980yev, Peebles:1994xt}. Minute anisotropies in the \gls{cmb}, at the level of one part in \(10^{5}\), indicate primordial fluctuations that later evolved into the observed large-scale structures\citep{Smoot1992, Planck:2018vyg}. To explain these features and resolve issues such as the horizon and flatness problems, \textit{cosmic inflation}-a brief epoch of exponential expansion-was introduced in the late 1970s by Starobinsky, Guth, and Linde\citep{Starobinsky1980,Guth1981,Linde1982}. Quantum fluctuations during inflation are believed to have seeded the large-scale structure of the Universe observed today\citep{Mukhanov1981,Mukhanov1992}.

By the late 20th century, significant observational advances reshaped our understanding of cosmic evolution. In 1998, two independent research teams-the \textit{Supernova Cosmology Project} and the \textit{High-Z Supernova Search Team}-discovered that distant \gls{snia} appeared dimmer than expected, implying that the Universe’s expansion is accelerating rather than decelerating\citep{Riess1998,Perlmutter1999}. This unexpected result suggested the presence of a mysterious repulsive component, termed \gls{de}, which dominates the current cosmic energy density\citep{Peebles2003,Weinberg2013}.

Subsequent observations from the \textit{Planck} satellite and other cosmological probes confirmed that ordinary (baryonic) matter constitutes only about \(4\text{--}5\%\) of the total energy content of the Universe, while \gls{dm} and \gls{de} contribute approximately \(25\%\) and \(70\%\), respectively\citep{Planck:2018vyg}. The current concordance model, known as the \gls{lcdm} model, attributes cosmic acceleration to a cosmological constant (\(\Lambda\)) and describes the Universe as composed of cold \gls{dm} and \gls{de}\citep{Spergel2003, Planck:2018vyg}. Although \gls{lcdm} successfully accounts for most observational data, it suffers from fine-tuning and coincidence problems\citep{Weinberg1989,Carroll2001,Padmanabhan2003}, motivating the exploration of \textit{modified theories of gravity} as potential alternatives\citep{Clifton2012,Nojiri2011,Capozziello2011}.

The development of \gls{de} in 1915 marked a paradigm shift by describing gravity as spacetime curvature governed by Riemannian geometry\citep{Einstein1915,Einstein1916}. However, Einstein himself later explored generalizations of his theory in search of a unified description of gravity and electromagnetism\citep{Einstein1925,Einstein1928}. These attempts introduced new geometric structures such as \textit{torsion} and \textit{non-metricity}, leading to alternative formulations of gravity within the broader framework of \textit{metric-affine geometry} \citep{Hehl1995, Obukhov2006, BeltranJimenez:2019esp}.

Postulating that gravitational effects can be attributed not only to curvature but also to torsion or non-metricity gives rise to the \textit{geometric trinity of gravity} \citep{Hehl1995, BeltranJimenez:2017tkd, BeltranJimenez:2019esp}. This trinity comprises three dynamically equivalent formulations of GR:
\begin{enumerate}
    \item \gls{gr} -- based on spacetime curvature,
    \item \gls{tegr} -- formulated via torsion in a curvature-free manifold, and
    \item \gls{stegr} -- described through non-metricity in a curvature- and torsion-free geometry.
\end{enumerate}

These formulations offer different geometrical perspectives while yielding the same field equations as \gls{gr}. Moreover, they provide alternative pathways for extending Einstein’s theory to address outstanding cosmological puzzles\citep{BeltranJimenez:2019tme, Capozziello2022}.

In recent decades, the increasing tension between \gls{gr} predictions and cosmological observations-such as cosmic acceleration, \gls{dm}, and early-Universe inflation-has stimulated the study of modified theories of gravity\citep{Clifton2012,Nojiri2017}. Rather than introducing exotic matter components, these models seek to explain the observed phenomena through extensions of spacetime geometry. Within this context, the \textit{symmetric teleparallel framework} has emerged as a compelling alternative, where gravity is attributed to non-metricity rather than curvature or torsion\citep{BeltranJimenez:2017tkd, BeltranJimenez:2019esp, Capozziello2022}.

A particularly successful generalization is the \(f(Q)\) gravity model, in which the gravitational Lagrangian is generalized from the non-metricity scalar \(Q\) to an arbitrary function \(f(Q)\) \citep{BeltranJimenez:2019tme}. This framework has garnered considerable attention for its rich cosmological implications. Studies have explored its role in large-scale structure formation, inflationary and bouncing cosmologies, black hole and wormhole solutions, modified stellar structures, and even quantum cosmology \citep{Xu2019, Barros2020, Khyllep2021, Mandal2020, Bajardi2020, BeltranJimenez:2019tme}. Furthermore, \(f(Q)\) gravity offers new insights into the nature of gravitational energy–momentum and provides potential explanations for the late-time acceleration of the Universe without invoking a cosmological constant \citep{Lazkoz2019,Anagnostopoulos2021}.

However, despite its successes, \(f(Q)\) gravity also presents theoretical challenges, such as issues in Hamiltonian analysis and the identification of propagating degrees of freedom \citep{Hohmann2021}. Ongoing research aims to refine its mathematical foundations and confront its predictions with observational data \citep{Ayuso2021}.

\vspace{1em}
\section{Fundamental Principles of Cosmology}
In this section, we introduce the fundamental concepts of cosmology and outline the physical motivations that guide our exploration of theoretical dark energy models in the subsequent chapters of this thesis. The focus will be placed on the underlying physical insights and phenomenological observations rather than on detailed mathematical formulations. Furthermore, the discussion will primarily concern late-time cosmology, emphasizing the universe’s evolution in its recent epochs. Consequently, topics related to the physics of the very early universe will not be covered here.

\vspace{1em}
\subsection{Cosmological Principal}
Modern cosmology is founded on the cosmological principle (also referred to as the Copernican principle), which posits that on sufficiently large scales-approximately ($10^{8}$) light-years-the universe is both homogeneous and isotropic. In essence, this principle implies that the Earth does not occupy a privileged or central position in the cosmos. Although it serves as a fundamental assumption in cosmology, the cosmological principle cannot be directly verified through experiments with current observational capabilities.

From our vantage point on Earth, the universe appears highly isotropic; however, verifying homogeneity requires observations from multiple locations separated by cosmological distances-something presently unfeasible. Despite this limitation, the cosmological principle is generally accepted as a valid working hypothesis, enabling the universe to be investigated within a scientific framework. Ultimately, its credibility rests on the extent to which cosmological models based on this principle agree with observational evidence.

\vspace{1em}
\subsection{Hubble's Observations}
The discovery of the expanding Universe marks one of the most transformative milestones in modern cosmology, largely credited to the meticulous observations of Edwin Hubble in the late 1920s, with key assistance from Milton Humason. Their work provided the first compelling empirical evidence that the Universe is dynamic rather than static, challenging the long-held belief in an unchanging cosmos. Hubble determined the distances to galaxies by using Cepheid variable stars as standard candles, capitalizing on the well-established Leavitt Law, which relates a Cepheid’s luminosity to its pulsation period. By measuring their brightness variations and apparent magnitudes, he was able to estimate distances far beyond the Milky Way, extending the cosmic distance scale. Meanwhile, Humason’s precise spectroscopic studies revealed that the spectral lines of distant galaxies were systematically shifted toward longer wavelengths, indicating that these galaxies were receding from the observer--a manifestation of the Doppler effect.

By correlating the galaxies’ distances with their recessional velocities, Hubble discovered a linear relationship, now known as Hubble’s Law, expressed as
\begin{equation}
    v = H_{0} D,
\end{equation}
where $v$ denotes the recessional velocity, $D$ the distance, and $H_{0}$ the Hubble constant, representing the current rate of cosmic expansion. This empirical law demonstrated that galaxies farther from Earth recede faster, implying that space itself is stretching uniformly. The implications of this finding were profound: the Universe was revealed to be expanding, in alignment with the earlier theoretical models proposed by Alexander Friedmann and Georges Lemaître, who had derived expanding solutions to Einstein’s field equations. Hubble’s discovery provided the observational foundation for the Big Bang theory, suggesting that the Universe began from an extremely hot and dense initial state and has been expanding ever since. This framework has since been reinforced by observations such as the \gls{cmb} radiation and primordial nucleosynthesis. Beyond its theoretical significance, Hubble’s work revolutionized cosmology, establishing it as an observational science grounded in empirical evidence and inaugurating a new era in astrophysics that continues to refine our understanding of the Universe’s origin, structure, and evolution.

\vspace{1em}
\subsection{Distances in an Expanding Universe}

In the context of an expanding Universe, the concept of ``distance'' becomes nuanced and requires careful definitions depending on the observational context. Several distance measures are commonly used in cosmology to characterize separations between objects, such as galaxies. In this section, we summarize three frequently employed cosmological distance definitions.

\vspace{1em}
\subsubsection{Comoving Distance}

The \textit{\gls{dmdist}}, denoted by \gls{dmdist}, represents the separation between two objects measured along a coordinate grid that expands along with the Universe. The actual physical distance \( r_p(t) \) at cosmic time \( t \) is related to the comoving distance via:
\begin{equation}
r_p(t) = a(t) \, D_{M}(t), \label{eq:proper_comoving}
\end{equation}
where \( a(t) \) is the scale factor.
To derive an expression for \gls{dmdist}, consider a photon traveling at the speed of light \( c \). The infinitesimal interval of light travel time is related to the comoving distance by:
\begin{equation}
dt = \frac{dr_p(t)}{c} = \frac{a(t)}{c} dD_{M}.
\end{equation}
Rearranging and integrating from the time of emission \( t \) to the present time \( t_0 \), one obtains:
\begin{equation}
D_{M}(t) = c \int_{t}^{t_0} \frac{dt'}{a(t')}.
\end{equation}
Changing variables from cosmic time to redshift, using
\begin{equation}
\frac{dz}{dt} = -(1+z) H(z),
\end{equation}
the comoving distance can be expressed as:
\begin{equation}
D_{M}(z) = \int_{0}^{z} \frac{c \, dz'}{H(z')}. \label{eq:comoving_distance}
\end{equation}
For small redshifts (\( z \ll 1 \)), the above reduces to the linear approximation:
\begin{equation}
D_{M} \approx \frac{cz}{H_0}, \label{eq:hubble_law}
\end{equation}
which corresponds to the familiar Hubble-Lemaître law, where redshift is approximately proportional to distance at low redshifts.

\vspace{1em}
\subsubsection{Angular Diameter Distance}

The \textit{\gls{da}} denoted by \gls{da} is defined by relating the physical size \( s \) of an object to its observed angular size \( \theta \) via:
\begin{equation}
\theta = \frac{s}{D_A}. \label{eq:angular_diameter_definition}
\end{equation}
Assuming small angles (which is usually valid in astronomical observations), this relation provides a direct geometric definition of distance.

The physical size of an object can also be written as:
\begin{equation}
s = a(t) \, D_{M}(t) \, \theta,
\end{equation}
which leads to the expression for the angular diameter distance:
\begin{equation}
D_A(z) = \frac{a(t)}{1} D_{M}(z) = \frac{1}{1+z} \int_{0}^{z} \frac{c \, dz'}{H(z')}. \label{eq:angular_diameter_distance}
\end{equation}

\vspace{1em}
\subsubsection{Luminosity Distance}

The \textit{\gls{dl}} denoted by \gls{dl} is defined through the relation between the observed flux \( F \) and the intrinsic luminosity \( L \) of a source:
\begin{equation}
F = \frac{L}{4 \pi D_{L}^{2}}. \label{eq:luminosity_definition}
\end{equation}

In an expanding Universe, two effects contribute to the reduction in observed flux: (i) the redshifting of photon energies, and (ii) the stretching of the interval between photon arrivals due to cosmic expansion. Together, these effects lead to an additional factor of \( (1+z)^2 \) in the flux dilution compared to a static Universe.

Consequently, the luminosity distance is related to the comoving distance by:
\begin{equation}
D_{L}(z) = (1+z) \, D_{M}(z) = (1+z) \int_{0}^{z} \frac{c \, dz'}{H(z')}. \label{eq:luminosity_distance}
\end{equation}

\vspace{1em}
\subsection{Friedmann–Lemaître–Robertson–Walker (FLRW)}
Modern cosmology rests on the cosmological principle, which asserts that on sufficiently large scales, the Universe is both homogeneous and isotropic. Homogeneity implies that the large-scale properties of the Universe are uniform and independent of the observer’s location, while isotropy indicates that the Universe appears the same in every direction. These assumptions provide the foundation for constructing cosmological models and allow the use of a common coordinate system known as comoving coordinates, which expand along with the Universe. In this framework, the relation between the physical position vector ( $\vec{r}$ ) and the comoving coordinate ( $\vec{x}$ ) is expressed as
\begin{equation*}
\vec{r} = a(t)\vec{x},
\end{equation*}
where $a(t)$  is the scale factor, describing how distances evolve with cosmic expansion.

The geometry of a homogeneous and isotropic Universe is described by the \gls{flrw} metric, derived independently by Friedmann \citep{friedman1922krummung, friedmann1924moglichkeit}, Lemaître \citep{lemaitre4univers}, Robertson \citep{robertson1935kinematics, robertson1936kinematics}, and Walker \citep{walker1937milne}. Incorporating cosmic expansion, the metric takes the general form
\begin{equation}
ds^2 = -N(t)^2 dt^2 + a^2(t) \left[ \frac{dr^2}{1 - \kappa r^2} + r^2 d\theta^2 + r^2 \sin^2\theta, d\phi^2 \right],
\end{equation}
where $N(t)$ is the lapse function, $a(t)$ is the scale factor, and $\kappa$ represents the spatial curvature of the Universe. The parameter $\kappa$ determines the large-scale geometry:
\begin{itemize}
\item $\kappa = 0$: Flat Universe, consistent with Euclidean geometry where parallel lines never meet or diverge.
\item $\kappa > 0 $: Closed Universe, corresponding to a spherical geometry with finite size but no boundaries—analogous to the surface of a sphere where a traveler eventually returns to the starting point.
\item $\kappa < 0 $: Open Universe, exhibiting hyperbolic geometry, where parallel lines diverge, resembling a saddle-shaped surface.
\end{itemize}

For computational convenience, the speed of light is often set to ( $c = 1$ ). Under the assumption of spatial flatness $\kappa = 0$, the \gls{flrw} metric simplifies to
\begin{equation}
ds^2 = -N(t)^2 dt^2 + a^2(t)(dx^2 + dy^2 + dz^2),
\end{equation}
which is frequently adopted in cosmological analyses.

The dynamics of the Universe are governed by the Einstein field equations, obtained from the variational principle applied to the Einstein–Hilbert action. By applying the symmetries of homogeneity and isotropy encapsulated in the \gls{flrw} metric, one derives the Friedmann equations, which describe the temporal evolution of the scale factor $a(t)$ in relation to the matter-energy content of the Universe. These equations form the cornerstone of modern cosmology, enabling the study of cosmic expansion, spatial curvature, and the influence of dark matter and dark energy on the Universe’s evolution.

\vspace{1em}
\subsection{The Friedmann Equations}

Substituting the \gls{flrw} metric into Einstein's field equations (1.1), and assuming the Universe is filled with a perfect fluid characterized by the energy-momentum tensor
\begin{equation}
T^{\mu\nu} = (\rho + p) u^\mu u^\nu + p g^{\mu\nu},
\end{equation}
where \( \rho \) and \( p \) are the energy density and pressure of the fluid, respectively, and \( u^\mu \) is the 4-velocity, we obtain the Friedmann equations \citep{friedman1922krummung}:
\begin{align}
\left( \frac{\dot{a}}{a} \right)^2 + \frac{k}{a^2} &= \frac{8\pi G_N}{3} \rho + \frac{\Lambda}{3}, \label{friedmann1} \\
\frac{\ddot{a}}{a} &= -\frac{4\pi G_N}{3} (\rho + 3p) + \frac{\Lambda}{3}. \label{friedmann2}
\end{align}

Cosmologists examine the dynamics of the Universe using the deceleration parameter, defined as \( q = -\frac{\ddot{a}a}{\dot{a}^2} \). A negative value of \( q \) signifies that the Universe is currently experiencing accelerated expansion, whereas a positive value corresponds to a phase of decelerated expansion. As discussed in the following section, recent observational evidence provides strong support for the ongoing accelerated expansion of the Universe. Differentiating equation~(1.17) and using equation~(1.18), one obtains the continuity equation, expressing conservation of energy:
\begin{equation}
\dot{\rho} = -3 (\rho + p) \frac{\dot{a}}{a}. \label{continuity}
\end{equation}

Together, equations~(1.17--1.19) govern the evolution of the Universe, given its matter-energy content.

\vspace{1em}
\subsection{Dynamics of the Scale Factor in a Flat Universe}
The astronomical evidence indicates a flat universe \citep{verde2003first}. Therefore, it is beneficial to conduct a more thorough examination of the change in the scale factor for a flat geometry. With $k = 0$ for a flat universe, the initial Friedmann equation equation (1.17) becomes simplified to:
\begin{equation}
H^2 = \left( \frac{\dot{a}}{a} \right)^2 = \frac{8\pi G_{N}}{3} \rho, 
\end{equation}
which implies
\begin{equation}
\left( \frac{\dot{a}}{a} \right)^2 \propto a^{-3(1+\omega)} \implies \dot{a} \propto a^{-\frac{1}{2}(1+3\omega)} \implies a \propto t^{\frac{2}{3(1+\omega)}}. 
\end{equation}

Now, as we are aware that $H = \frac{\dot{a}}{a}$, i.e., $H dt = \frac{1}{a} da$, integrating both sides generates:
\begin{equation}
a \propto t^{\frac{2}{3(1+\omega)}}. 
\end{equation}
We can infer that the Universe governed by a cosmological constant would result in rapid expansion. This phenomenon is commonly referred to as de-Sitter expansion. Another scenario to consider is an empty Universe with $\rho = 0$. In this situation, the Friedmann equation simplifies to $\dot{a} = 0$, which means that the scale factor $a$ remains constant. To summarize our findings, the scale factor for a flat universe changes in the following manner:
\begin{equation}
a(t) \propto
\begin{cases} 
t^{\frac{2}{3(1+\omega)}} & \omega \neq -1, \\
e^{Ht} & \omega = -1, \\
1 & \rho = 0.
\end{cases} 
\end{equation}
We have gathered the results for various dominance scenarios in Table 1.1. From this, it is evident that only a universe dominated by a cosmological constant results in accelerated expansion.This observation has led to the inference that the ``\gls{de}'' thought to drive this acceleration is attributed to the cosmological constant $\Lambda$ \citep{Riess1998, Perlmutter1999}.

\begin{table}[htp!]
\centering
\begin{tabular}{|c|c|c|c|}
\hline
 $ a(t)$ & Name of an Era & $\rho(a)$ & $\omega$ \\ 
\hline
$t^{\frac{1}{2}}$ & Radiation Dominated & $a^{-4}$ & $\frac{1}{3}$ \\ 
\hline
$t^{\frac{2}{3}}$ & Matter Dominated & $a^{-3}$ & $0$ \\ 
\hline
$e^{Ht}$ & Cosmological Constant $\Lambda$ & $a^0$ & $-1$ \\ 
\hline
$t$ & Vacuum & $0$ & Undefined \\ 
\hline
\end{tabular}
\caption{Evolution of the scale factor representing different important eras of the Universe's evolution.}
\end{table}

\vspace{1em}
\subsection{The Hubble Parameter and \gls{eos} Parameter}

The rate of expansion of the Universe is described by the Hubble parameter:
\begin{equation}
H(t) = \frac{\dot{a}}{a}, 
\end{equation}
and its present-day value is referred to as the Hubble constant:
\begin{equation}
H_0 = H(t = t_0).
\end{equation}
Assuming a linear equation of state of the form \( p = w \rho \), where \( w \) is the \gls{eos} parameter, the continuity equation (\ref{continuity}) integrates to give:
\begin{equation}
\rho(a) = \rho_0 \left( \frac{a}{a_0} \right)^{-3(1+w)},
\end{equation}
where \( \rho_0 \) and \( a_0 \) denote present-day values.

For the different components of the Universe, the values of \( w \) are typically:
\begin{itemize}
    \item \( w = 0 \) for non-relativistic matter (baryons and cold \gls{dm}),
    \item \( w = \frac{1}{3} \) for radiation,
    \item \( w = -1 \) for \gls{de} in the form of a cosmological constant.
\end{itemize}

\vspace{1em}
\subsection{The Critical Density and Density Parameters}

The critical density \( \rho_{\text{crit}} \), which corresponds to a spatially flat Universe (\( k = 0 \)), is defined as:
\begin{equation}
\rho_{\text{crit}} = \frac{3 H^2}{8\pi G_N}.
\end{equation}
The density parameters can be written as,
\begin{equation}\label{ch1_densityparametesr}
\Omega_{m}=\frac{8 \pi G_N \rho_{m} }{3 H^{2}}\,, \hspace{0.8cm} \Omega_{r}=\frac{8 \pi G_N \rho_{r} }{3 H^{2}}\,, \hspace{0.8cm} \Omega_{DE}=\frac{8 \pi G_N \rho_{DE} }{3 H^{2}}\,,  
\end{equation}
satisfying
\begin{equation}\label{ch1_constraints}
\Omega_{m}+\Omega_{r}+\Omega_{DE}=1.    
\end{equation}

\vspace{1em}
\subsection{The Friedmann Equation in Terms of Redshift}

The Friedmann equation (\ref{friedmann1}) can be expressed in terms of these fractional energy densities as:
\begin{equation}
H^2(a) = H_0^2 \left[ \Omega_r \left( \frac{a_0}{a} \right)^4 + \Omega_m \left( \frac{a_0}{a} \right)^3 + \Omega_k \left( \frac{a_0}{a} \right)^2 + \Omega_\Lambda \right],
\end{equation}
where \( \Omega_k \) is defined by:
\begin{equation}
\Omega_k = -\frac{k}{H_0^2 a_0^2}.
\end{equation}
Defining the redshift \( z \) as:
\begin{equation}
1 + z = \frac{a_0}{a},
\end{equation}
and setting \( a_0 = 1 \), the Friedmann equation becomes:
\begin{equation}\label{LCDMmodel}
H^2(z) = H_0^2 \left[ \Omega_r (1+z)^4 + \Omega_m (1+z)^3 + \Omega_k (1+z)^2 + \Omega_\Lambda \right].
\end{equation}

The expression of the equation (1.33), called as the standard \gls{lcdm} model, where $H_{0}$ is the Hubble constant. The \gls{lcdm} model, widely regarded as the standard or Benchmark model of cosmology, aligns well with observational data and successfully explains the Universe's evolution \citep{Planck:2018vyg}. This framework provides a robust description of cosmic dynamics across both early and late times, and at both background and perturbative levels. However, in recent years, certain observations have raised potential concerns, particularly the well-known \(H_0\) tension \citep{DiValentino:2021izs, Brout:2022vxf, Bernal:2016gxb, Benisty:2021cmq}. The \(H_0\) tension refers to the persistent discrepancy between different methods of measuring the Hubble constant, which challenges the completeness of the standard \(\Lambda\)CDM scenario \citep{vagnozzi2020new, Abdalla:2022yfr, moresco2022unveiling, Planck:2018vyg, DiValentino:2021izs}.

Measurements based on early Universe observations, such as the \gls{cmb} data from the Planck satellite \citep{ade2014planck}, yield values of the Hubble constant that are significantly lower than those obtained from late-time, local distance indicators like Cepheid variables and \gls{snia} \citep{Riess:2021jrx}. For instance, the SH0ES collaboration led by Riess et al. \citep{Riess:2021jrx} reported \(H_0 = 73.30 \pm 1.04 \, \text{km s}^{-1} \, \text{Mpc}^{-1}\) using \gls{snia} observations. Similarly, the H0LiCOW project \citep{wong2017h0licow} obtained \(H_0 = 73.3^{+1.7}_{-1.8} \, \text{km s}^{-1} \, \text{Mpc}^{-1}\) through strong gravitational lensing of quasars. In contrast, Freedman et al. \citep{freedman2019carnegie} estimated a lower value of \(H_0 = 69.8 \pm 1.9 \, \text{km s}^{-1} \, \text{Mpc}^{-1}\) using the \gls{trgb} method.

Regarding early Universe constraints, the Planck Collaboration \citep{Planck:2018vyg} reported \(H_0 = 67.4 \pm 0.5 \, \text{km s}^{-1} \, \text{Mpc}^{-1}\), while Abbott et al. \citep{ligo2017gravitational} presented a similar value of \(67.2^{+1.2}_{-1.0} \, \text{km s}^{-1} \, \text{Mpc}^{-1}\). This ongoing discrepancy, known as the \(H_0\) tension, presents a significant challenge to the standard cosmological model and continues to inspire new approaches and theories in modern cosmology.

\vspace{1em}
\section{The History of the Universe Evolution:}
The Big Bang theory is a foundational concept in cosmology, providing a comprehensive framework to explain the origin and evolution of the Universe. Following the initial singularity, the Universe experienced a period of rapid exponential expansion, known as inflation. This was succeeded by distinct evolutionary phases, including a radiation-dominated era, followed by a matter-dominated phase, and, at present, a phase characterized by the dominance of dark energy.

\begin{figure}[htp!]
\begin{center}
\includegraphics[width=0.85\textwidth]{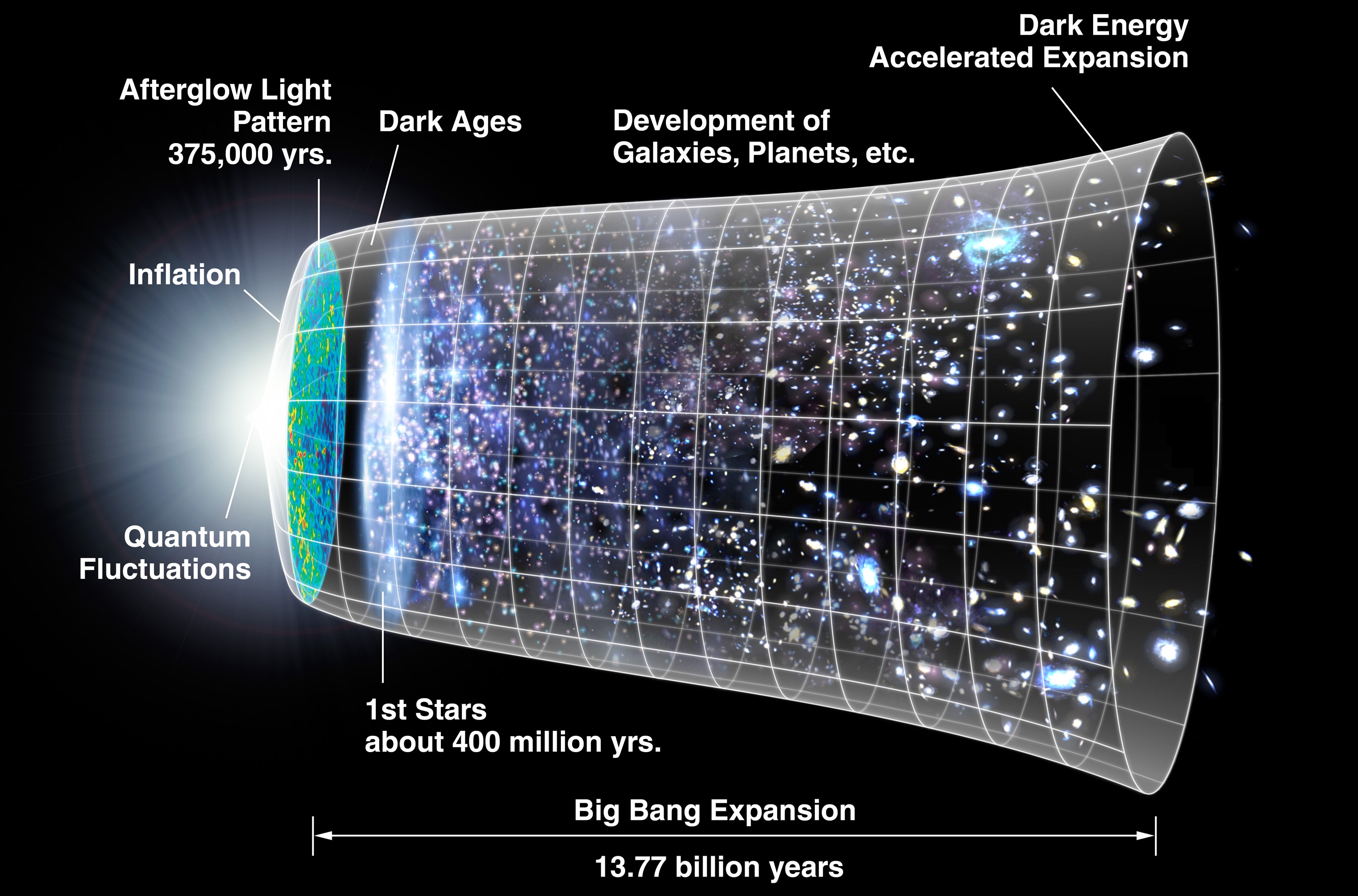}
\caption{$\Lambda$CDM Model of Cosmology [Credit: NASA Goddard Space Flight Center]}
\end{center}
\end{figure}

\vspace{1em}
\subsection{Inflation:}
In 1981, Alan Guth \citep{Guth1981} proposed the concept of the inflationary phase, which was later supported by John D. Barrow, Michael S. Turner, and Andrei D. Linde \citep{hakim2005inflationary, linde1983chaotic, barrow1981inflation}. During the inflationary phase, the Universe undergoes a rapid and exponential expansion, increasing its volume by a factor of $e^{60}$ to $e^{70}$ compared to its initial size before inflation. This expansion takes place over a very short period, estimated to last from $10^{-36}$ seconds to approximately $10^{-33}$ to $10^{-32}$ seconds following the Big Bang. The inflation phenomenon describes the early Universe epoch during which the scale factor exponentially expanded in just a fraction of a second ($\ddot{a} > 0$). When this condition is applied to the acceleration equation, it violates the strong energy condition $(\rho + 3p) < 0$. In cosmology, it is always found that the energy density $\rho > 0$, and therefore, to fulfill the above condition, the pressure must be negative.

The inflationary model provides a graceful solution to the standard Big Bang issues such as:

\begin{enumerate}
    \item \textit{The Flatness Problem:} This problem originates from the extraordinary balance between the matter density and expansion rate of the Universe over its evolution. Minor variations in these factors at the Universe’s inception should have led to substantial discrepancies. However, it remains enigmatic why these slight variations did not magnify with the Universe’s expansion \citep{Guth1981, Linde1982, kolb1990early, mukhanov2005physical}.
    
    \item \textit{The Horizon Problem:} This problem was introduced by Austrian physicist Wolfgang Rindler in 1956 \citep{rindler1956visual}. It emerges from the challenges in explaining the observed uniformity in causally disconnected regions of space in the absence of a mechanism that establishes identical initial conditions everywhere. This issue is associated with the uniformity of the Universe, commonly referred to as the homogeneity problem was later recognized as one of the key motivations for the inflationary scenario \citep{Guth1981, Linde1982, kolb1990early, mukhanov2005physical}.
    
    \item \textit{The Magnetic Monopole Problem:} This is an intriguing puzzle at the intersection of cosmology and particle physics. It arises from the predictions of grand unified theories (GUTs) and theories concerning the early Universe \citep{Georgi:1974sy}. As per these theories, a significant number of magnetic monopoles-hypothetical particles with only one magnetic pole-should have been produced during the early Universe, particularly during phase transitions as the Universe cooled and the fundamental forces differentiated. Despite extensive searches, no magnetic monopoles have been detected in nature or in cosmic rays \citep{Guth1981}. This scarcity of evidence suggests that magnetic monopoles are either extremely rare or non-existent, challenging our understanding of the fundamental forces governing the Universe \citep{goddard1978magnetic}.
\end{enumerate}

The inflation period was exhausted as the inflation field disintegrated into particles and radiation, effectively warming up the Universe and creating the hot, dense state that marks the conventional start of the Big Bang theory \citep{Guth1981}.

\vspace{1em}
\subsection{Radiation Dominated Era:}
The Universe in its early stages went through a phase called the \textit{Radiation-dominated epoch}. This phase began at the Planck epoch, which was around $10^{-43}$ seconds after the Big Bang. The Radiation-dominated epoch continued until approximately 47,000 years after the Big Bang, when the Universe shifted to being matter-dominated after cooling enough for matter to become dominant.

During the early stages of the universe, the dominant energy density consisted mainly of photons and other relativistic particles. The scale factor $a(t)$, which indicates the universe’s size over time, followed the behaviour of $a(t) \propto t^{1/2}$. This behaviour suggested rapid expansion that gradually decelerated. The universe experienced a cooling process due to its expansion, as it had extremely high temperatures at the outset of the radiation-dominated era, around $10^{10}$ K. 
This era established the initial conditions for the formation of large-scale structures by introducing minor irregularities in the \gls{cmb}, which acted as the starting points for future galaxy and structure formation \citep{Planck:2018vyg}. The transition from the radiation-dominated epoch to matter domination left marks in the \gls{cmb}, providing valuable insights into the early universe. 

In essence, the radiation-dominated epoch played a crucial role in cosmology, laying the groundwork for the universe’s present-day observed state \citep{kolb1990early}.

\vspace{1em}
\subsection{Matter Dominated Era:}
The era dominated by matter began approximately 47,000 years after the Big Bang and lasted about 9.8 billion years until \gls{de} took over. During this period, the universe’s energy density was mainly driven by matter, including baryonic matter and dark matter. The scale factor $a(t)$ continued to follow $a(t) \propto t^{2/3}$, indicating a slowing down of the expansion rate compared to the radiation-dominated era \citep{mukhanov2005physical, Peebles:1994xt}. 

The universe continued to cool as it expanded, leading to a significant drop in temperatures. Additionally, this era facilitated the gravitational collapse of residual density fluctuations from the radiation-dominated era, ultimately leading to the formation of cosmic structures such as galaxies and galaxy clusters. Notably, the large-scale structure of the universe transitioned from a uniform distribution of matter to the complex, web-like structure observed today \citep{Peebles:1980yev}.
The redshift $z$ was related to the scale factor $a(t)$ through the equation:  
\[
1 + z = \frac{1}{a(t)}.
\]  
Distant objects displayed an increasing redshift as the universe expanded, indicating their recession. The matter-dominated era significantly influenced the distribution of matter and the \gls{cmb} radiation, playing a crucial role in shaping the large-scale structure of the universe. 

Essentially, this era represented a substantial shift in the universe’s evolution from a radiation-dominated, hot and dense state to a cooler, more organized cosmos governed by the gravitational influence of matter, laying the groundwork for the complex structure and dynamics observed in the universe today.

\vspace{1em}
\subsection{Dark energy dominated Era:}
In the current epoch of cosmic evolution, the Universe is believed to be dominated by \gls{de}-a mysterious component with negative pressure that drives the accelerated expansion of space. Observational evidence, particularly from \gls{snia}, the \gls{cmb}, and \gls{bao}, indicates that approximately 70\% of the total energy density of the Universe today is attributed to dark energy. This phase, known as the dark energy dominated era, began after the matter-dominated era and marks a transition in the Universe's expansion history. The cosmological constant \(\Lambda\), introduced in the \gls{lcdm} model, is the simplest form of \gls{de}, though various dynamical models such as quintessence and phantom fields have also been proposed. The onset of this era significantly influences the large-scale structure formation, the fate of the Universe, and poses deep theoretical challenges regarding the nature and origin of dark energy.

\vspace{1em}
\section{Identification of Late-Time Cosmic Acceleration}
The groundbreaking discoveries of 1998 \citep{Riess1998, Perlmutter1999} demonstrated that the expansion of our Universe is accelerating, fundamentally transforming our understanding of cosmology \citep{WMAP:2012nax}. This accelerated expansion has been attributed to an enigmatic form of energy referred to as \gls{de} \citep{DiValentino:2016hlg}. While \gls{gr} is widely accepted as an accurate description of gravity for large masses, particularly at the scale of the solar system \citep{Peebles2003, Padmanabhan2003, Weinberg1989}, its application to the accelerating Universe requires the introduction of a cosmological constant, $\Lambda$. This cosmological constant behaves as a fluid with negative pressure, enabling the violation of certain energy conditions and the emergence of a repulsive gravitational force.
Despite its success, the cosmological constant $\Lambda$ lacks a comprehensive theoretical explanation \citep{misner1973gravitation}, leading researchers to explore alternative frameworks to understand the late-time acceleration of the Universe. A prominent avenue of exploration is the development of modified theories of gravity, which involve altering or replacing \gls{gr} to better align with cosmological observations. Additionally, the quest to explain other unresolved phenomena, such as the nature of \gls{dm} and the dynamics of inflation, provides further motivation for modifying \gls{gr} \citep{Capozziello2011, faraoni2008f}.

It is well known that \gls{gr} employs the Levi-Civita connection, which is symmetric and torsionless. An alternative approach is to replace this connection with the Weitzenböck connection \citep{Weitzenbock1923}, which is skew-symmetric, curvatureless, and characterized by a non-zero torsion tensor \citep{Maluf:1994ji, Aldrovandi2013, Andrade2000MG9}. The field equations of Symmetric Teleparallel Gravity are analogous to those of \gls{gr}, leading to this formalism being referred to as the \gls{stegr}. Both \gls{gr} and \gls{stegr} provide equivalent explanations of gravity. However, while \gls{stegr} produces the same observational predictions as \gls{gr}, it offers distinct physical and mathematical interpretations \citep{Pereira:2019woq, bahamonde2023teleparallel, Chervon:2023gio}.
It is crucial to note that modifications to \gls{stegr} differ fundamentally from those applied to \gls{gr}. This thesis investigates modifications to \gls{stegr} by incorporating various scalars, such as the Symmetric Teleparallel boundary term. Additionally, this framework emphasizes the role of scalar field coupling functions, which enable the exploration of a broader range of viable cosmic models to study the evolution of the Universe. The primary focus of this work is to analyze the Universe's late-time accelerating behavior within the framework of modified symmetric teleparallel gravity.

Although modifications of \gls{gr} are often employed to address theoretical challenges, the proposition of modifications to \gls{stg} is a relatively recent development. This raises key questions regarding the potential of modified \gls{stg} theories to effectively resolve or mitigate cosmological challenges such as \gls{de} and \gls{dm}. 
Are there specific advantages of modified \gls{stg} in cosmology compared to standard modifications? How does higher-order \gls{stg} contribute to understanding late-time cosmic acceleration in the evolution of the Universe? Furthermore, do these theories provide a framework for examining the history of the Universe's evolution?
This thesis aims to explore these questions by presenting and analyzing several modified \gls{stg} theories.

\vspace{1em}
\section{Frameworks for Understanding the Accelerating Universe}
This section explores the underlying motivations and core principles that drive the study of modified gravity theories as alternative approaches to address limitations inherent in \gls{gr} or the \gls{stegr}. In the context of cosmology, theoretical models are generally categorized into two main classes:

\begin{enumerate}

\item \textit{Dark Energy Models:} These models operate within the framework of Einstein's field equations by introducing an energy-momentum tensor that includes a component of exotic matter possessing negative pressure. Such a formulation effectively accounts for the observed evolution of the Universe.

\item \textit{Modified Gravity Models:} These models provide an alternative perspective on the nature of gravitation, distinct from \gls{gr}. One of their notable features is the capability to explain the evolution of the Universe without the need to invoke exotic matter components.

\end{enumerate}

\vspace{1em}
\section{Dark Energy Models}

Dynamical \gls{de} models have been developed to account for the late-time acceleration of the Universe without invoking the cosmological constant $\Lambda$. These models typically introduce an additional energy component on the matter side of Einstein’s field equations, either with a constant or evolving \gls{eos}. The dynamical nature of these models allows the \gls{eos} parameter to vary with time or redshift, offering a broader framework for cosmological evolution than a static $\Lambda$.

\begin{itemize}

\item Several phenomenological parameterizations of the \gls{eos} as functions of redshift $z$ have been proposed. Among the most notable are the Chevallier–Polarski–Linder (CPL) parameterization \citep{Chevallier:2000qy, Linder:2002et}, the Jassal–Bagla–Padmanabhan (JBP) model \citep{jassal2005wmap}, the Barboza–Alcaniz formulation \citep{Barboza:2008rh}, and the Wetterich parameterization \citep{Wetterich:1987fm}, among others. These models allow for tracking the evolution of the \gls{de} component more flexibly across cosmic time.

\item One of the most widely studied classes of dynamical \gls{de} is based on scalar field theories, particularly the \textit{quintessence} model. In this framework, a scalar field $\phi$, minimally coupled to gravity, evolves with time under the influence of a self-interacting potential $V(\phi)$. The nature of cosmic acceleration is determined by the relative magnitudes of the kinetic and potential energy terms. In the regime where $V(\phi) \ll \dot{\phi}^2$, the \gls{eos} parameter approximates $\omega \approx 1$, which behaves like stiff matter and does not significantly contribute to \gls{de}. In contrast, when $V(\phi) \gg \dot{\phi}^2$, the \gls{eos} approaches $\omega \approx -1$, resembling a cosmological constant. Intermediate cases with $-1 < \omega < 1$ lead to energy densities evolving as $\rho \propto a^{-m}$, which supports accelerated expansion when $0 \leq m < 2$. A wide range of quintessence potentials have been investigated in the literature to model various cosmic acceleration scenarios (e.g., exponential, power-law, tracker, thawing, etc.) \citep{Tsujikawa:2013fta, Mao:2008ug, muromachi2015search}.

\item An alternative scalar field model, known as the \textit{phantom field}, was introduced by Caldwell \citep{Caldwell:1999ew}. In this model, the kinetic term has a negative sign, leading to negative kinetic energy and an \gls{eos} $\omega < -1$. This results in a super-accelerated expansion and may culminate in a finite-time singularity known as the \textit{Big Rip}. Models that incorporate elements of both quintessence and phantom fields are termed \textit{quintom} models \citep{Feng:2004ad, Guo:2004fq}. These frameworks provide a richer dynamical behavior by allowing the \gls{eos} to transition smoothly between $\omega > -1$ and $\omega < -1$, thereby accommodating various phases of cosmic acceleration.

\item Another scalar field approach, known as \textit{K-essence} \citep{Armendariz-Picon:1999hyi}, originates from early Universe inflationary models and is also applicable to late-time acceleration \citep{Armendariz-Picon:2000ulo, Chiba:1999ka}. Unlike quintessence, K-essence models are dominated by non-canonical kinetic terms in the scalar field Lagrangian, and the acceleration is primarily driven by these kinetic contributions. These models have been employed to describe both early inflation and the recent accelerated expansion within a unified framework.
\end{itemize}

\vspace{1em}
\section{ Modified gravity model}\label{modifiedgravitymodel} 
To gain deeper insights into the nature of \gls{dm} and \gls{de}, numerous modified gravity theories have been developed that extend beyond the traditional framework of \gls{gr}. These alternative theories aim to resolve both observational discrepancies and theoretical challenges. By unifying early-universe inflation with late-time cosmic acceleration, modified gravity presents a compelling avenue for addressing these issues. Additionally, it holds potential for bridging the gap between gravity and the fundamental forces described by quantum field theory, including electromagnetic, weak, and strong nuclear interactions.
\gls{gr} can be formulated within different but equivalent geometric frameworks-namely, the curvature-based and teleparallel representations. In the teleparallel formulation, curvature and non-metricity are absent, and torsion governs the gravitational interaction. In contrast, the curvature-based approach eliminates torsion and non-metricity, relying solely on curvature. A further alternative is symmetric teleparallel gravity, where gravity is described through the non-metricity scalar $Q$, which captures how the length of a vector changes during parallel transport along a closed path.

A wide range of modified gravity theories have been proposed in the literature. Notable examples include \( f(Q) \) gravity \citep{BeltranJimenez:2017tkd,, Lazkoz:2019sjl, Barros2020, Anagnostopoulos2021, Frusciante:2021sio}, \( f(Q, \mathcal{T}) \) gravity \citep{Xu2019, Xu:2020yeg}, \( f(Q, \phi) \) gravity \citep{Gonzalez-Espinoza:2021mwr}, \( f(Q, B) \) gravity, and various scalar-tensor theories \citep{Amendola:1999er, Copeland:2006wr, Tsujikawa:2013fta}, among others.

\vspace{1em}
\subchapter{Dynamical Systems and Stability}

This part provides a concise overview of the aspects of dynamical systems theory that are most relevant to cosmological applications. The discussion highlights the principal tools commonly employed in the literature, without attempting to cover the wide range of mathematical approaches used in other branches of dynamical systems theory. Readers seeking a more comprehensive mathematical treatment may consult standard textbooks such as those by Arrowsmith and Place \citep{arrowsmith1990introduction}, Wiggins \citep{wiggins2003introduction}, or Perko \citep{perko2013differential}. For an introduction tailored specifically to cosmology, see Böhmer and Chan \citep{bohmer2017dynamical}.

\vspace{1em}
\section{Basic Concepts of Dynamical Systems}

To begin, let us address a fundamental question: \emph{What constitutes a dynamical system?} Broadly speaking, a dynamical system provides a mathematical framework capable of describing processes ranging from simple mechanical setups, such as the motion of a pendulum, to highly complex phenomena such as the evolution of the Universe. In abstract terms, any dynamical system consists of:
\begin{enumerate}
    \item a \emph{state space} (or phase space), and
    \item a \emph{rule} that governs how points in this state space evolve with respect to a chosen time parameter.
\end{enumerate}

The second component is particularly significant: for most complex physical phenomena, deriving an exact evolution law directly from first principles is often difficult or impossible. Thus, we typically introduce the governing equations as a model input. To characterise the system, we specify a set of quantities that fully describe its state. The state space then corresponds to the set of all possible values these quantities may take. 

In ecological systems, for example, populations of different species often serve as the state variables in predator--prey models. In cosmology, one might initially consider using the energy densities of various cosmic components. However, these are not always the most suitable variables for recasting the cosmological equations into an autonomous dynamical system; alternative variables often lead to a clearer and more tractable formulation.

Let $x = (x_1, x_2, \dots, x_n) \in X$ be a point in the state space $X \subseteq \mathbb{R}^n$. A general autonomous system can be written in the form \citep{wiggins2003introduction}
\begin{equation}
    \dot{x} = f(x), \label{eq:dynsys_general}
\end{equation}
where $f : X \to X$ is a smooth vector field, and the overdot denotes differentiation with respect to some time parameter $t \in \mathbb{R}$, which need not coincide with physical cosmic time. The vector field has the form
\begin{equation}
    f(x) = \big(f_1(x), \dots, f_n(x)\big).
\end{equation}
A solution of equation (1.34) is referred to as a \emph{trajectory} or \emph{orbit} in the phase space. In this work we restrict attention to finite-dimensional, continuous systems with sufficiently smooth vector fields. Although cosmological dynamical systems often involve elementary functions and are smooth in large regions of the phase space, some models may nevertheless exhibit singularities. In such cases, standard analytical tools apply only in those regions where $f$ remains continuous.
\vspace{1em}
\begin{definition}[Critical point / fixed point / equilibrium point]
A point $x_0 \in X$ is a \emph{critical point} of the autonomous system $\dot{x} = f(x)$ if
\begin{equation}
    f(x_0) = 0.
\end{equation}
\end{definition}

A critical point corresponds to a state in which the system experiences no evolution. Whether such a state is physically attainable or stable under perturbations is a key question when analysing cosmological dynamics.

\vspace{1em}
\begin{definition}[Stable fixed point]
A fixed point $x_0$ is said to be \emph{stable} if, for every $\epsilon > 0$, there exists a $\delta > 0$ such that any solution $\phi(t)$ of equation (1.34) satisfying $\|\phi(t_0) - x_0\| < \delta$ remains within $\epsilon$ of $x_0$ for all $t \geq t_0$.
\end{definition}

\vspace{1em}
\begin{definition}[Asymptotically stable fixed point]
A stable fixed point $x_0$ is \emph{asymptotically stable} if there exists a $\delta > 0$ such that $\|\phi(t_0) - x_0\| < \delta$ implies
\[
\lim_{t \to +\infty} \phi(t) = x_0.
\]
\end{definition}

The distinction between stability and asymptotic stability lies in whether nearby trajectories merely remain close to the fixed point or actually converge to it. In cosmological models, stable points are typically also asymptotically stable, except in special cases.

A fixed point that is not stable is termed \emph{unstable}. Understanding the stability of fixed points is central in cosmological dynamical systems, as these points typically correspond to cosmological epochs such as matter domination, radiation domination, or late-time accelerated expansion.

Two further concepts frequently encountered in phase-space analyses are:

\begin{itemize}
    \item \textbf{Invariant set:} A subset $S \subseteq X$ is \emph{invariant} if any trajectory that begins in $S$ remains in $S$ for all time.
    \item \textbf{Heteroclinic orbit:} A trajectory connecting two distinct critical points, i.e., a solution $\phi(t)$ such that $\lim_{t \to -\infty} \phi(t) = x_i$ and $\lim_{t \to +\infty} \phi(t) = x_f$ for critical points $x_i \neq x_f$.
\end{itemize}

To investigate the stability properties of critical points, one typically applies \emph{linear stability theory}, which is generally sufficient for most cosmological models. Additional tools, such as Lyapunov functions and centre manifold theory, become relevant in the presence of \emph{non-hyperbolic points}. Other frameworks for analysing stability also exist, including the Kosambi--Cartan--Chern approach, which has been applied to cosmology in several works \citep{bohmer2017dynamical}.

\vspace{1em}
\subsection{Linear Stability Theory}

Linear stability theory provides a systematic method for analysing the behaviour of a dynamical system in the vicinity of a fixed point. Consider the autonomous system
\begin{equation}
    \dot{x} = f(x),
\end{equation}
with a fixed point located at $x_0$. Assuming that the vector field $f$ is sufficiently smooth, one may expand each function $f_i(x)$ in a Taylor series about $x_0$:
\begin{equation}
    f_i(x) = f_i(x_0) + \sum_{j=1}^{n} \left.\frac{\partial f_i}{\partial x_j}\right|_{x_0} y_j 
    + \frac{1}{2} \sum_{j,k=1}^{n} 
    \left.\frac{\partial^2 f_i}{\partial x_j \partial x_k}\right|_{x_0} y_j y_k + \cdots ,
    \label{eq:taylor_expansion}
\end{equation}
where $y = x - x_0$. Linear stability theory retains only the first-order terms of this expansion. The collection of first derivatives defines the Jacobian matrix
\begin{equation}
    J = \left( \frac{\partial f_i}{\partial x_j} \right)
    =
    \begin{pmatrix}
        \frac{\partial f_1}{\partial x_1} & \cdots & \frac{\partial f_1}{\partial x_n} \\
        \vdots & \ddots & \vdots \\
        \frac{\partial f_n}{\partial x_1} & \cdots & \frac{\partial f_n}{\partial x_n}
    \end{pmatrix},
    \label{eq:jacobian}
\end{equation}
which is also referred to as the \emph{stability matrix}. The eigenvalues of $J$, evaluated at the fixed point $x_0$, determine the qualitative behaviour of trajectories near that point.

Since the Jacobian is an $n \times n$ matrix, it possesses $n$ eigenvalues, which may in general be complex. The applicability of linear stability theory depends critically on the nature of these eigenvalues, which motivates the following definition \citep{wiggins2003introduction}.

\vspace{1em}
\begin{definition}[Hyperbolic point]
A fixed point $x_0$ of the system $\dot{x} = f(x)$ is called \emph{hyperbolic} if none of the eigenvalues of $J(x_0)$ has zero real part. A fixed point for which at least one eigenvalue has real part equal to zero is termed \emph{non-hyperbolic}.
\end{definition}

Linear stability theory is inconclusive for non-hyperbolic points, in which case one must employ more advanced tools such as centre manifold theory or Lyapunov methods. For cosmological applications, it is often sufficient to classify fixed points into three broad categories:

\begin{itemize}
    \item \textbf{Unstable (repelling) point:} all eigenvalues of $J(x_0)$ have positive real parts.
    \item \textbf{Stable (attracting) point:} all eigenvalues have negative real parts.
    \item \textbf{Saddle point:} the eigenvalues have real parts of mixed signs.
\end{itemize}

While a complete classification is straightforward in two- or three-dimensional systems, higher-dimensional cosmological models may become significantly more involved. Nevertheless, the above categories typically suffice for most cosmological dynamical analyses.

\vspace{1em}
\subsubsection{Example: Two-Dimensional Systems}

Consider the planar system
\begin{equation}
    \dot{x} = f(x,y), \qquad
    \dot{y} = g(x,y),
    \label{eq:2d_system}
\end{equation}
with a hyperbolic fixed point $(x_0, y_0)$ such that $f(x_0,y_0)=0$ and $g(x_0,y_0)=0$. The Jacobian of this system is
\begin{equation}
    J =
    \begin{pmatrix}
        f_{,x} & f_{,y} \\
        g_{,x} & g_{,y}
    \end{pmatrix},
    \label{eq:2d_jacobian}
\end{equation}
where subscripts denote partial derivatives. The eigenvalues $\lambda_{1,2}$ of $J$ are
\begin{align}
    \lambda_{1} &= \frac{1}{2}\left( f_{,x} + g_{,y} + \sqrt{(f_{,x}-g_{,y})^{2} + 4 f_{,y} g_{,x}} \right), \label{eq:lambda1}\\
    \lambda_{2} &= \frac{1}{2}\left( f_{,x} + g_{,y} - \sqrt{(f_{,x}-g_{,y})^{2} + 4 f_{,y} g_{,x}} \right), \label{eq:lambda2}
\end{align}
evaluated at $(x_0, y_0)$. In two dimensions, the eigenvalues are easy to compute explicitly, enabling a complete stability classification. 

\vspace{1em}
\subsection{Lyapunov's Method}

Beyond linear stability analysis, more general techniques exist that can determine the stability of fixed points even when linearisation fails. One of the most powerful of these is the method introduced by Lyapunov, which can establish both local and global stability and is applicable even in the presence of non-hyperbolic points. The principal challenge lies in identifying a suitable Lyapunov function, as no universal procedure exists for constructing one.

\vspace{1em}
\begin{definition}[Lyapunov function]
Let $\dot{x} = f(x)$ be a smooth dynamical system with a fixed point $x_0$. A continuous function $V : \mathbb{R}^n \to \mathbb{R}$ is a Lyapunov function in a neighbourhood $U$ of $x_0$ if:
\begin{enumerate}
    \item $V$ is differentiable on $U \setminus \{x_0\}$,
    \item $V(x) > V(x_0)$ for all $x \in U \setminus \{x_0\}$,
    \item $\dot{V}(x) \le 0$ for all $x \in U \setminus \{x_0\}$.
\end{enumerate}
\end{definition}

The third condition is decisive. Using the chain rule,
\begin{equation}
    \frac{dV}{dt}
    = \sum_{i=1}^{n} \frac{\partial V}{\partial x_i} \dot{x}_i
    = \nabla V \cdot f(x),
    \label{eq:lyapunov_derivative}
\end{equation}
which must be non-positive throughout the neighbourhood.

The connection between Lyapunov functions and fixed-point stability is expressed through the following classical theorem (see, e.g., \citep{brauer1989qualitative}, with cosmological applications discussed in \citep{Charters:2001hi}).

\vspace{1em}
\begin{theorem}[Lyapunov stability]
Let $x_0$ be a fixed point of $\dot{x} = f(x)$ and let $U$ be a domain containing $x_0$.  
\begin{itemize}
    \item If there exists a Lyapunov function $V(x)$ such that $\dot{V}(x) \le 0$ in $U$, then $x_0$ is stable.
    \item If $\dot{V}(x) < 0$ in $U$, then $x_0$ is asymptotically stable.
\end{itemize}
Moreover, if $V(x) \to \infty$ as $\|x\| \to \infty$, then $x_0$ is globally (asymptotically) stable.
\end{theorem}

Thus, once an appropriate Lyapunov function is found, the stability of a fixed point may be established without solving the dynamical system explicitly. However, the converse is not true: failure to find such a function does not imply instability—it merely reflects the absence of a suitable choice.

\vspace{1em}
\subsection{Center Manifold Theory}
\gls{cmt} constitutes a significant tool in the qualitative analysis of dynamical systems, particularly near non-hyperbolic fixed points. It provides a framework for examining the local behavior of such systems by reducing their dimensionality. While linear stability theory becomes inconclusive in the presence of eigenvalues with zero real parts, \gls{cmt} allows one to determine the stability characteristics by investigating the behavior confined to the center manifold, denoted by $W_c$.

A detailed exposition of the mathematical foundation of \gls{cmt} can be found \citep{wiggins2003introduction}, and additional applications are presented in Refs.\citep{carr2012applications}. In the vicinity of a non-hyperbolic critical point, the system's dynamics are governed by the evolution along $W_c$, which is tangent to the eigenspace associated with the eigenvalues whose real parts are zero.

Consider a dynamical system of the form:
\begin{equation}
\tau' = F(\tau),
\label{eq:cmt1}
\end{equation}
where $\tau = (\mu, \nu) \in \mathbb{R} \times \mathbb{R}$. The center manifold for this system can be locally expressed as:
\begin{equation}
W_c = \{ (\mu, \nu) \in \mathbb{R} \times \mathbb{R} \, | \, \nu = h(\mu),\, |\mu| < \delta,\, h(0) = 0,\, \nabla h(0) = 0 \},
\label{eq:cmt2}
\end{equation}
for a sufficiently small $\delta > 0$. Here, $h(\mu)$ is a smooth function representing the center manifold.

The steps involved in applying \gls{cmt} are outlined below:

\begin{enumerate}
    \item \textbf{Coordinate Transformation:} Shift the non-hyperbolic critical point to the origin by introducing new coordinates. This results in a revised autonomous system centered at the origin.

    \item \textbf{System Reformulation:} The transformed system is recast in the form:
    \begin{equation}
    \tau' = 
    \begin{pmatrix}
        \mu' \\
        \nu'
    \end{pmatrix}
    =
    \begin{pmatrix}
        A\mu \\
        B\nu
    \end{pmatrix}
    +
    \begin{pmatrix}
        \varphi(\mu, \nu) \\
        \psi(\mu, \nu)
    \end{pmatrix},
    \label{eq:cmt3}
    \end{equation}
    where the functions $\varphi$ and $\psi$ satisfy the conditions:
    \begin{equation}
    \varphi(0, 0) = 0,\quad \nabla \varphi(0, 0) = 0,\quad \psi(0, 0) = 0,\quad \nabla \psi(0, 0) = 0.
    \end{equation}
    The matrices $A$ and $B$ are square matrices such that the eigenvalues of $A$ have zero real parts, and those of $B$ have strictly negative real parts.

    \item \textbf{Determining the Center Manifold Function:} A function $h(\mu)$ is sought, typically via a series expansion involving powers of $\mu$. This function satisfies the quasilinear partial differential equation:
    \begin{equation}
    \mathcal{N}(h(\mu)) = \nabla h(\mu) \left[ A\mu + \varphi(\mu, h(\mu)) \right] - B h(\mu) - \psi(\mu, h(\mu)) = 0,
    \label{eq:cmt4}
    \end{equation}
    under the constraints $h(0) = 0$ and $\nabla h(0) = 0$.

    \item \textbf{Reduced Dynamics on the Center Manifold:} Substituting the approximate solution $h(\mu)$ from Eq.~\eqref{eq:cmt4} into the system yields the reduced system:
    \begin{equation}
    \mu' = A\mu + \varphi(\mu, h(\mu)).
    \label{eq:cmt5}
    \end{equation}

    \item \textbf{Analysis of the Reduced System:} The system in equation (1.49) can often be simplified further to a form such as:
    \begin{equation}
    \mu' = k \mu^n,
    \label{eq:cmt6}
    \end{equation}
    where $k$ is a constant and $n$ is a positive integer. The stability of the origin depends on the sign of $k$ and the parity of $n$:
    \begin{itemize}
        \item If $k < 0$ and $n$ is odd, the critical point is locally stable.
        \item In all other cases, the system exhibits local instability.
    \end{itemize}
\end{enumerate}

\gls{cmt} thus offers a rigorous method to analyze the stability of equilibrium points when traditional linearization fails due to zero eigenvalues.

\vspace{1em}
\subsection{Further Advanced Methods in Dynamical Systems}
\vspace{1em}
\subsubsection{Poincar\'e Sphere and Behaviour at Infinity}

In this subsection, we investigate the behaviour of the flow as $\|x\| = \sqrt{x^{2}+y^{2}} \rightarrow +\infty$ for two–dimensional autonomous systems. For planar systems, one can use suitable projection techniques to study the asymptotic flow at large distances. When applied correctly, these techniques allow us to construct the global phase portrait, including the behaviour of trajectories at infinity. Standard references for these methods include \citep{lefschetz1957, lynch2007, perko2013differential}.

To compactify the phase space, we introduce the \textit{Poincar\'e sphere}, which maps all points at infinity to its equator.

\vspace{1em}
\begin{definition}[Poincar\'e sphere]
The Poincar\'e sphere is the unit sphere
\begin{equation}
S^{2} = \{(X,Y,Z)\in\mathbb{R}^{3} \mid X^{2}+Y^{2}+Z^{2}=1\},
\end{equation}
with its north (or south) pole tangent to the $(x,y)$–plane at the origin. Points of the $(x,y)$–plane are mapped to the upper hemisphere via central projection. The transformation is given by
\begin{equation}
X = xZ, \qquad Y = yZ, \qquad Z = \frac{1}{\sqrt{1+x^{2}+y^{2}}}.
\end{equation}
\end{definition}

Consider the polynomial system
\begin{align}
\dot{x} &= P(x,y), \\
\dot{y} &= Q(x,y),
\end{align}
where the maximum degree among all terms in $P$ and $Q$ is $m$, and let $P_{m}$ and $Q_{m}$ denote the homogeneous components of degree $m$. The behaviour at infinity is summarised in the following result from \citep{lefschetz1957, perko2013differential}.

\vspace{1em}
\begin{theorem}[Critical points at infinity]
The critical points at infinity correspond to points on the equator of the Poincar\'e sphere, $(X,Y,0)$ with $X^{2}+Y^{2}=1$, satisfying
\begin{equation}
X Q_{m}(X,Y) - Y P_{m}(X,Y) = 0.
\end{equation}
Equivalently, in polar coordinates, the angles $\theta_{j}$ and $\theta_{j}+\pi$ are determined by the solutions of
\begin{equation}
G_{m+1}(\theta) = \cos\theta\, Q_{m}(\cos\theta,\sin\theta) - \sin\theta\, P_{m}(\cos\theta,\sin\theta) = 0,
\end{equation}
which, if not identically zero, has at most $m+1$ distinct pairs of solutions. Moreover, the orientation of the flow on the equator is clockwise when $G_{m+1}(\theta)<0$ and counterclockwise when $G_{m+1}(\theta)>0$.
\end{theorem}

Since each projective line intersects the equator twice, critical points at infinity always appear in antipodal pairs.

To determine the stability of these points, the flow on the sphere is projected onto the planes tangent to the equator at $Y=1$ and $X=1$. This leads to the result below \citep{lefschetz1957, perko2013differential}.

\vspace{1em}
\begin{theorem}[Stability at infinity]
Near any point of the equator other than $(0,\pm1,0)$, the flow is topologically equivalent to the system
\begin{align}
\pm \dot{y} &= y z^{m} P\!\left(\frac{1}{z},\frac{y}{z}\right) - z^{m} Q\!\left(\frac{1}{z},\frac{y}{z}\right), \\
\pm \dot{z} &= z^{m+1} P\!\left(\frac{1}{z},\frac{y}{z}\right),
\end{align}
where the sign is determined by the direction of the flow on the equator.  
Similarly, near any equatorial point other than $(\pm 1,0,0)$, the system is equivalent to
\begin{align}
\pm \dot{x} &= x z^{m} Q\!\left(\frac{x}{z},\frac{1}{z}\right) - z^{m} P\!\left(\frac{x}{z},\frac{1}{z}\right), \\
\pm \dot{z} &= z^{m+1} Q\!\left(\frac{x}{z},\frac{1}{z}\right).
\end{align}
Thus, if $(0,\pm1,0)$ are not critical points at infinity, one studies stability using the first pair of equations; if $(\pm1,0,0)$ are not critical points, the second pair suffices. If all four appear, both systems must be analysed. Applications in cosmology appear in section~5.1.
\end{theorem}

\vspace{1em}
\subsubsection{Invariant Submanifolds}

Earlier we introduced invariant sets, regions of phase space that trajectories cannot leave. Fixed points are the simplest examples, but higher–dimensional invariant manifolds, such as periodic orbits, also play a key role. The term \emph{invariant submanifold} is often used to emphasise that such sets have lower dimension than the full phase space but still extend throughout it.

If an orbit begins on an invariant submanifold, it remains there for all time. Since a $C^{1}$ vector field cannot generate trajectories that abruptly change direction, no orbit may cross such submanifolds. Consequently, invariant submanifolds partition the phase space into dynamically disconnected regions. An attractor cannot be global unless it lies in the intersection of all invariant submanifolds.

A simple criterion for detecting invariant submanifolds relies on factorisation of the dynamical equations. If the right–hand side of $\dot{y}$ contains a constant root $y=\beta$, then $y=\beta$ defines an invariant submanifold. For points satisfying $y=\beta$, one has $\dot{y}=0$, so this condition persists along the entire orbit.

Consider the example
\begin{align}
\dot{x} &= 6x^{3}+y^{2}, \\
\dot{y} &= 3y+4y^{2}+2yx,
\end{align}
for which the second equation can be written as
\begin{equation}
\dot{y}=y(3+4y+2x).
\end{equation}
Thus $y=0$ is an invariant submanifold dividing the plane into two disconnected regions $y>0$ and $y<0$. A global attractor, if it exists, must lie on this line.

\vspace{1em}
\subsubsection{Global Phase–Space Properties and Monotonic Functions}

Various tools may assist in determining global phase–space properties such as periodic orbits, homoclinic connections, separatrices, and strange attractors. Only a limited number of works attempt a complete global analysis due to the high technical complexity involved; see \citep{perko2013differential, wiggins2003introduction}. Here we focus on two relatively simple methods useful for excluding periodic orbits or fixed points.

We begin with \textit{Dulac's criterion}, which provides a sufficient condition for the absence of periodic solutions in planar systems \citep{perko2013differential}. Consider the system $\dot{x}=f(x)$ and suppose that in a simply connected region there exists a scalar function $H(x)$ such that
\begin{equation}
\nabla\cdot (Hf) \neq 0.
\end{equation}
Then the phase space contains no periodic orbits.

For a two–dimensional system
\begin{align}
\dot{x} &= P(x,y), \\
\dot{y} &= Q(x,y),
\end{align}
assume hypothetically that a closed orbit $\Gamma$ exists. Because $\Gamma$ is an orbit of the system, the integral
\begin{equation}
\oint_{\Gamma} H(P\,dy - Q\,dx)
\end{equation}
must vanish. Using Stokes' theorem, this becomes
\begin{equation}
\iint_{S} \left( \frac{\partial(HP)}{\partial x} + \frac{\partial(HQ)}{\partial y} \right) dx\,dy,
\end{equation}
where $S$ is the interior of $\Gamma$. However, by assumption this integral cannot be zero, producing a contradiction. Thus no closed orbits may exist. Dulac's criterion is only sufficient, and other results—such as the Poincar\'e–Bendixson theorem \citep{arrowsmith1990introduction, wiggins2003introduction, perko2013differential}—are often used to complement it.

Another tool for ruling out fixed points involves constructing a positive semidefinite function $W$ \citep{wainwright1989}. Since
\begin{equation}
\dot{W} = \nabla W \cdot f
\end{equation}
is a linear combination of the derivatives of the phase–space variables, $\dot{W}$ can vanish only at fixed points. Therefore, if $\dot{W}$ is strictly positive away from the origin, no other fixed points exist.

As an illustration, consider
\begin{align}
\dot{x} &= \beta x + \gamma^{2} y, \\
\dot{y} &= -\beta x + \delta^{2} y,
\end{align}
which has an evident fixed point at the origin. Let $W=x^{2}+y^{2}$. Then
\begin{equation}
\dot{W} = 2\gamma^{2} x^{2} + 2\delta^{2} y^{2},
\end{equation}
which is positive definite except at $(0,0)$. Hence no other fixed points can exist.

\vspace{1em}
\section{Application of Dynamical Systems Techniques in Cosmological Studies}
The development of modern cosmological models often relies on systems of non-linear differential equations, which are inherently challenging to solve analytically. One of the effective methodologies for gaining insights into the qualitative behavior of such systems is the application of dynamical systems theory. This approach transforms the complex cosmological equations into a system of autonomous differential equations by introducing dimensionless, normalized variables and a redefined time parameter.

These normalized variables are carefully chosen so that they correspond to physically observable quantities and exhibit well-behaved dynamics. The analysis typically focuses on identifying the fixed (or critical) points of the autonomous system and studying their stability through techniques such as linearization and eigenvalue analysis. This method allows researchers to probe into the possible past and future states of the Universe. For instance, unstable fixed points can be interpreted as representing early cosmological epochs, while stable attractors may signify the long-term asymptotic behavior of the Universe, including late-time acceleration.

Furthermore, the presence of heteroclinic orbits-trajectories connecting unstable and stable critical points-illustrates transitional cosmic dynamics, enriching our understanding of the evolutionary pathways of the Universe. The dynamical systems framework has been extensively utilized within the context of general relativity and cosmological models. Prominent examples include analyses of modified gravity theories, scalar-tensor models, Bianchi-type cosmologies, and scalar fields with non-minimal coupling.

This approach has proven indispensable for exploring and classifying a wide range of cosmological behaviors across various gravitational theories. A comprehensive survey of such applications in modified gravity frameworks can be found in References \citep{Rave-Franco:2021yvu, Kadam:2022lgq, Khyllep:2022spx, Zonunmawia:2018xvf, Vishwakarma:2024bnd, Vishwakarma:2024qvw}.

\vspace{1em}
\section{Current Study}
This thesis is organized as follows: a brief discussion of the standard \gls{flrw} universe and the prerequisite fundamental formulas for formulating STEGR are presented. A basic overview of late-time cosmic acceleration is summarized in the remaining section of \textit{Chapter 1}, which also elaborates on the cosmological implications of dynamical system analysis, a powerful tool for selecting and analyzing cosmologically viable models. 

\textit{Chapter 2} provides a detailed literature review on $f(Q)$ gravity, outlining its theoretical foundation, cosmological relevance, and recent developments. The focus is on its role in explaining late-time cosmic acceleration within the framework of modified gravity.

\textit{Chapter 3} analyzes three cosmological models in $f(Q)$ gravity using both background evolution and perturbation theory. Critical points are identified and studied via dynamical systems methods, showing at least one stable node in each model. The models exhibit behavior consistent with accelerated cosmic expansion.

\textit{Chapter 4} extends the cosmological analysis by incorporating \gls{bbn} constraints. Two forms of $f(Q)$ are studied, with stability examined through dynamical systems tools. BBN bounds are used to constrain model parameters, ensuring compatibility with both early and late-time cosmology.

\textit{Chapter 5} explores the Noether symmetry approach to extended symmetric teleparallel gravity in the \( f(Q,\phi) \) framework, where the non-metricity scalar \( Q \) is coupled to a scalar field \( \phi \) in an FLRW background. By identifying a nontrivial Noether symmetry, we determine the corresponding functional form of \( f(Q,\phi) \) and construct the point-like Lagrangian, which allows us to obtain exact cosmological solutions. These symmetry-admitted solutions naturally describe an accelerated phase of the universe. We also compute the Hubble parameter, deceleration parameter, and effective equation-of-state parameter, showing that they align with present observational constraints.

\textit{Chapter 6} investigates a model of $f(Q, \phi)$ gravity through dynamical systems analysis. Critical points and their stability are examined, and graphical results illustrate the evolution of the equation of state and energy densities. The model supports late-time acceleration consistent with observations.

\textit{Chapter 7} summarizes the research findings across all chapters, highlighting key results and their cosmological implications. It reflects on the contribution to modified gravity studies and suggests future directions for theoretical and observational investigations.

\chapter{Review of Modified Theories of Gravity and their Mathematical Foundations}\label{Chapter_2}

\section{Introduction}
The exploration of gravitational theories has long been driven by the desire to comprehend the dynamics and evolution of the Universe. General Relativity (GR), formulated by Einstein in 1915, has stood as the cornerstone of modern gravitational physics, successfully describing the large-scale structure of spacetime, black holes, and gravitational waves. However, the discovery of the late-time cosmic acceleration \citep{Riess1998, Perlmutter1999}, the horizon and flatness problems, and the nature of dark matter and dark energy have motivated researchers to look beyond GR. The observed cosmic acceleration cannot be explained by standard matter and radiation within GR without invoking a cosmological constant, which in turn leads to the well-known fine-tuning and coincidence problems \citep{Padmanabhan2003, Copeland:2006wr}. 

To address these challenges, several modified and extended theories of gravity have been developed. These include $f(R)$ gravity \citep{Nojiri:2006be, Sotiriou:2008rp, DeFelice:2010aj}, $f(T)$ gravity \cite{Cai:2015emx, Bahamonde:2015zma}, and more recently, the symmetric teleparallel formulation of gravity based on the non-metricity scalar $Q$ \citep{BeltranJimenez:2019esp, BeltranJimenez:2018vdo}. Among these, $f(Q)$ gravity, which modifies the action of symmetric teleparallel gravity, provides a compelling alternative to curvature-based and torsion-based formulations. The theory maintains second-order field equations and local Lorentz invariance, making it theoretically elegant and mathematically consistent.

This chapter provides a comprehensive review of the literature related to $f(Q)$ gravity and its extensions. The discussion begins with a brief overview of modified gravity theories, proceeds to the geometric foundations of symmetric teleparallel gravity, and then examines the developments, cosmological implications, and dynamical analyses of $f(Q)$ gravity and its generalizations.

\vspace{1em}
\section{Modified Theories of Gravity: An Overview}

The motivation for modifying GR arises from both theoretical and observational grounds. On cosmological scales, the accelerated expansion of the Universe suggests the existence of a repulsive component or a modification of the gravitational interaction. On theoretical grounds, GR fails to be renormalizable at high energies, motivating the search for a more fundamental description of gravity. 

A natural way to generalize GR is to replace the Ricci scalar $R$ in the Einstein–Hilbert action by a general function $f(R)$:
\begin{equation}
S = \frac{1}{2\kappa^2} \int f(R) \sqrt{-g}\, d^4x + S_m,
\end{equation}
where $\kappa^2 = 8\pi G$ and $S_m$ denotes the matter action. The $f(R)$ gravity models have been successful in describing inflationary behavior and late-time acceleration \citep{Nojiri:2006be, DeFelice:2010aj}. However, they often lead to fourth-order differential field equations, which complicate their mathematical structure and can introduce instabilities.

In contrast, teleparallel gravity (TEGR) is formulated using the torsion scalar $T$ rather than the curvature scalar. Its modified version, $f(T)$ gravity, maintains second-order equations but breaks local Lorentz invariance \citep{Cai:2015emx}. To overcome these limitations, a new geometric approach, known as symmetric teleparallel gravity (STG), was introduced, where gravity is attributed to the non-metricity of spacetime rather than curvature or torsion. The corresponding modification, $f(Q)$ gravity, offers second-order equations and preserves local Lorentz invariance, making it a strong candidate for exploring cosmological dynamics.

\vspace{1em}
\section{Geometric Foundations of Symmetric Teleparallel Gravity}

In differential geometry, three fundamental quantities describe gravitational interactions: \emph{curvature, torsion, and non-metricity}. These quantities are defined in terms of the affine connection $\Gamma^{\alpha}{}_{\mu\nu}$:
\begin{align}
R^{\alpha}{}_{\beta\mu\nu} &= \partial_{\mu}\Gamma^{\alpha}{}_{\beta\nu} - \partial_{\nu}\Gamma^{\alpha}{}_{\beta\mu} + \Gamma^{\alpha}{}_{\sigma\mu}\Gamma^{\sigma}{}_{\beta\nu} - \Gamma^{\alpha}{}_{\sigma\nu}\Gamma^{\sigma}{}_{\beta\mu}, \\
T^{\alpha}{}_{\mu\nu} &= \Gamma^{\alpha}{}_{\nu\mu} - \Gamma^{\alpha}{}_{\mu\nu}, \\
Q_{\alpha\mu\nu} &= \nabla_{\alpha} g_{\mu\nu}.
\end{align}
Each of these corresponds to a specific geometric framework:
\begin{itemize}
    \item \emph{Curvature-based geometry (Riemannian):} General Relativity.
    \item \emph{Torsion-based geometry (Teleparallel):} $f(T)$ gravity.
    \item \emph{Non-metricity-based geometry (Symmetric Teleparallel):} $f(Q)$ gravity.
\end{itemize}

In symmetric teleparallel gravity, both curvature and torsion vanish ($R = 0$, $T = 0$), and the gravitational interaction is fully described by the non-metricity tensor $Q_{\alpha\mu\nu}$. The non-metricity scalar $Q$ is constructed as
\begin{equation}
Q = -\frac{1}{4}Q_{\alpha\mu\nu}Q^{\alpha\mu\nu} + \frac{1}{2}Q_{\alpha\mu\nu}Q^{\mu\nu\alpha} + \frac{1}{4}Q_{\alpha}Q^{\alpha} - \frac{1}{2}\tilde{Q}_{\alpha}Q^{\alpha},
\end{equation}
where $Q_{\alpha} = Q_{\alpha\mu}{}^{\mu}$ and $\tilde{Q}_{\alpha} = Q_{\mu\alpha}{}^{\mu}$ are the two independent traces of the non-metricity tensor.

The action for $f(Q)$ gravity is given by
\begin{equation}
S = \int \left[\frac{1}{2}f(Q) + \mathcal{L}_m \right] \sqrt{-g}\, d^4x,
\end{equation}
where $\mathcal{L}_m$ denotes the matter Lagrangian density. For $f(Q) = -Q/2$, the theory reduces to the symmetric teleparallel equivalent of GR, known as \gls{stegr} \cite{BeltranJimenez:2018vdo}. The resulting field equations are of second order, offering an advantage over the fourth-order equations of $f(R)$ gravity.

\vspace{1em}
\section{Curvature, Torsion, and Non-Metricity: The Fundamental Objects of Metric-Affine Geometry}

Metric-affine geometry is characterized by three fundamental geometric quantities: curvature, torsion, and non-metricity. These quantities arise naturally from the connection $\Gamma$, and in the case of non-metricity, they depend on both $\Gamma$ and the metric tensor $g_{\mu\nu}$. In this section, we begin by introducing the mathematical definitions of these geometric quantities through the notion of \emph{parallel transport}, which offers an intuitive understanding of their underlying structure \citep{Heisenberg:2023lru, DAmbrosio:2023asf}. Building on this foundation, we then present the formal definitions of \emph{curvature, torsion, and non-metricity}, highlighting how each of these tensorial objects characterises distinct aspects of the affine connection. This discussion will subsequently provide the groundwork for analysing teleparallel and symmetric teleparallel formulations of gravity in the following sections.

\vspace{1em}
\subsection{Parallel Transport}

There is no canonical procedure for comparing vectors-or more generally, tensors-located at different points on a differentiable manifold, as such comparisons require additional geometric structure. This limitation poses a fundamental challenge when attempting to define directional derivatives of tensor fields. The introduction of a connection $\Gamma$ allows us to overcome this issue. Alternatively, the concept of \emph{parallel transport} provides a different yet equivalent approach that helps us better understand the geometry $(M, g, \Gamma)$ of a metric-affine manifold.

When defining the covariant derivative, two major difficulties arise. First, expressions such as ``$p + \epsilon u$'' are not meaningful because $p$ belongs to the manifold $M$, while $u$ lies in the tangent space $T_pM$-two distinct mathematical objects. Second, in generalizing the difference quotient, one needs to compute $v_q - v_p$, i.e., the difference between vectors in $T_qM$ and $T_pM$, which is also undefined.

Parallel transport resolves both of these issues. The first is handled by introducing a smooth curve $\gamma(s)$ passing through the point $p$, where $u = \dot{\gamma}(s)$ serves as its tangent vector. The second is addressed by defining a rule that prescribes how a vector $v$ at point $p$ is transported to another point $q$ along $\gamma$. However, there is no unique prescription for this process-various choices are possible-reflecting the freedom inherent in defining a connection $\Gamma$. 

To formalize this, we define the \textit{parallel transport map}
\begin{equation}
    P^{(\gamma)}_{t s} : T_{\gamma(s)}M \rightarrow T_{\gamma(t)}M, \quad 
    v_{\gamma(s)} \mapsto P^{(\gamma)}_{t s}v_{\gamma(s)},
    \label{eq:3.1}
\end{equation}
where $\gamma(s) = p$ and $\gamma(t) = q$. The map satisfies the following axioms:
\begin{enumerate}
    \item $P^{(\gamma)}_{t t} = \text{id}$,
    \item $P^{(\gamma)}_{t u} \circ P^{(\gamma)}_{u s} = P^{(\gamma)}_{t s}$,
    \item $P^{(\gamma)}_{t s}$ is smooth in its arguments $(s, t, \gamma)$.
\end{enumerate}

Given a vector $v_p \in T_pM$, we can now transport it along $\gamma$ to obtain the vector
\begin{equation}
    P^{(\gamma)}_{t s} v_p \in T_qM.
    \label{eq:3.2}
\end{equation}
The particular form of $P^{(\gamma)}_{t s}$ is not unique-any choice satisfying the above axioms suffices. This construction allows vectors at different points to be compared, which is essential for defining derivatives on manifolds. Consequently, we define the covariant derivative as
\begin{equation}
    \nabla_w v := \lim_{s \to 0} \frac{P^{(\gamma)}_{0 s} v_{\gamma(s)} - v_{\gamma(0)}}{s}
    = \frac{d}{ds}\bigg( P^{(\gamma)}_{0 s} v_{\gamma(s)} \bigg) \bigg|_{s=0},
    \label{eq:3.3}
\end{equation}
where $w = \dot{\gamma}$ denotes the tangent vector to $\gamma$.

Since the connection $\Gamma$ determines the covariant derivative, there must exist a correspondence between $\Gamma$ and the parallel transport map $P^{(\gamma)}_{t s}$. Working in a local coordinate chart $\{x^\mu\}$ and expanding equation (2.8), we have
\begin{equation}
    [P^{(\gamma)}_{0 s}]^{\alpha}_{\ \mu} v^\mu_{\gamma(s)} .
    \label{eq:3.4}
\end{equation}
Expanding this expression around $s=0$ using Taylor's theorem yields
\begin{equation}
    [P^{(\gamma)}_{0 s}]^{\alpha}_{\ \mu} v^\mu_{\gamma(s)} 
    = v^\alpha_p + s \left( w^\nu_p \partial_\nu v^\alpha_p + v^\mu_p \frac{d [P^{(\gamma)}_{0 s}]^{\alpha}_{\ \mu}}{d s}\Big|_{s=0} \right) + \mathcal{O}(s^2).
    \label{eq:3.6}
\end{equation}

Since $[P^{(\gamma)}_{0 0}]^{\alpha}_{\ \mu} = \delta^{\alpha}_{\mu}$ (from Axiom A1) and the differentiability of $P^{(\gamma)}_{0 s}$ is guaranteed by Axiom A3, we may write
\begin{equation}
    \frac{d [P^{(\gamma)}_{0 s}]^{\alpha}_{\ \mu}}{d s}\Big|_{s=0}
    = \frac{d [P^{(\gamma)}_{0 s}]^{\alpha}_{\ \mu}}{d \gamma^\nu}\Big|_{s=0} w^\nu_p.
    \label{eq:3.7}
\end{equation}
Substituting equation (2.12) into equation (2.11), we obtain
\begin{equation}
    (\nabla_w v)^\alpha = w^\nu_p \left( \partial_\nu v^\alpha_p + \frac{d [P^{(\gamma)}_{0 s}]^{\alpha}_{\ \mu}}{d \gamma^\nu}\Big|_{s=0} v^\mu_p \right).
    \label{eq:3.8}
\end{equation}
Comparing this with the standard coordinate expression of the covariant derivative,
\begin{equation}
    w^\nu \nabla_\nu v^\alpha = w^\nu (\partial_\nu v^\alpha + \Gamma^{\alpha}_{\ \nu \mu} v^\mu),
    \label{eq:3.9}
\end{equation}
we find the equivalence relation
\begin{equation}
    \frac{d [P^{(\gamma)}_{0 s}]^{\alpha}_{\ \mu}}{d \gamma^\nu}\Big|_{s=0} = \Gamma^{\alpha}_{\ \nu \mu}.
    \label{eq:3.10}
\end{equation}
Hence, a given connection $\Gamma$ uniquely determines a notion of parallel transport $P^{(\gamma)}_{t s}$ via equation (2.15), and conversely, a parallel transport law defines the connection coefficients.

Furthermore, from Axioms A1--A3, one immediately obtains the inverse property
\begin{equation}
    P^{(\gamma)}_{0 s} \circ P^{(\gamma)}_{s 0} = \text{id},
    \label{eq:3.11}
\end{equation}
implying that $P^{(\gamma)}_{0 s}$ is an element of the general linear group $\mathrm{GL}(n, \mathbb{R})$.

\vspace{1em}
\subsubsection{Parallel Transport Equation}

A vector $v$ is said to be \textit{parallel transported} along $\gamma$ if
\begin{equation}
    P^{(\gamma)}_{t s} v_{\gamma(s)} = v_{\gamma(t)}.
    \label{eq:3.12}
\end{equation}
For infinitesimally close points $t = s + \epsilon$, this condition leads to the \textit{parallel transport equation}
\begin{equation}
    \nabla_{\dot{\gamma}(s)} v_{\gamma(s)} = 0,
    \label{eq:3.13}
\end{equation}
which, in coordinate form, becomes
\begin{equation}
    \dot{\gamma}^\mu \left( \partial_\mu v^\nu + \Gamma^{\nu}_{\ \mu \lambda} v^\lambda \right) = 0.
    \label{eq:3.14}
\end{equation}
This equation describes how a vector evolves under infinitesimal parallel transport. For nearby points $p = \gamma(s)$ and $q = \gamma(s + \epsilon)$, the transported vector satisfies
\begin{equation}
    v^\nu_q = v^\nu_p - \epsilon \Gamma^{\nu}_{\ \mu \lambda}(p) \dot{\gamma}^\mu v^\lambda_p,
    \label{eq:3.16}
\end{equation}
indicating that parallel transport modifies a vector by an infinitesimal term depending on the connection, the original vector, and the tangent direction of the curve.

\vspace{1em}
\subsubsection{Geometric Interpretation}

Having defined parallel transport, we can now interpret how it affects the intrinsic properties of vectors in metric-affine spaces $(M, g, \Gamma)$. The essential features of vectors are:
\begin{enumerate}
    \item Each vector possesses a specific direction.
    \item Vectors can be combined via head-to-tail addition, producing a resultant vector.
    \item When the manifold is equipped with a metric $g_{\mu\nu}$, each vector also has a well-defined magnitude.
\end{enumerate}
In Euclidean geometry, parallel transport preserves both the direction and magnitude of vectors, and vector addition follows the familiar parallelogram law. However, in a general metric-affine geometry, where the connection $\Gamma$ may include curvature, torsion, or non-metricity, these properties no longer necessarily hold. The deviation from Euclidean behavior reflects how the connection modifies direction, addition, and length during parallel transport.

\vspace{1em}
\subsection{Curvature}

We now examine how the \emph{direction} of a vector changes under parallel transport. To quantify this change, one must compare a vector with itself after being transported along a closed curve. The variation in direction resulting from this closed-path transport provides a natural measure of curvature.

To make this concept precise, consider an infinitesimal closed loop composed of four curve segments, as illustrated schematically in Figure (2.1). The loop connects four nearby points $p$, $q$, $r$, and $s$ on the manifold. Let the curve $\gamma_u$ connect $p$ and $q$ with tangent vector $u$, and let $v_p$ denote the vector to be transported around the loop.

\begin{figure}[htp!]
\begin{center}
\includegraphics[width=0.98\textwidth]{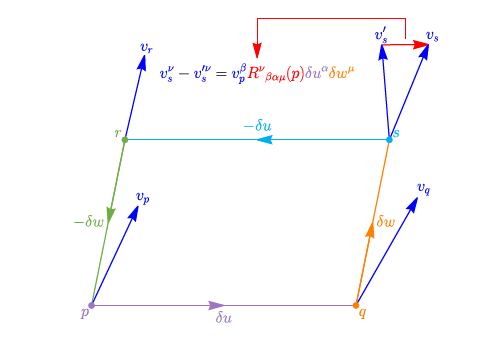}
\caption{Parallel transport of a vector around an infinitesimal closed loop results in a change of direction. This deviation is encoded in the curvature tensor $R^{\alpha}{}_{\mu\nu\rho}$.}
\end{center}
\end{figure}

We begin by transporting $v_p$ from $p$ to $q$ along $\gamma_u$. For an infinitesimal displacement, equation (2.2) applies, yielding
\begin{equation}
    v^{\nu}_{q} = v^{\nu}_{p} - \Gamma^{\nu}{}_{\mu\lambda}(p)\,\delta u^{\mu}\, v^{\lambda}_{p},
    \label{eq:3.17}
\end{equation}
where $\delta u^{\mu} := \epsilon u^{\mu}$ with $\epsilon \ll 1$. The subscript $p$ indicates that all quantities on the right-hand side are evaluated at $p$.

Next, the vector $v_q$ is transported from $q$ to $s$ along the curve $\gamma_w$ with tangent vector $w$. Since the displacement between $q$ and $s$ is also infinitesimal, we denote it by $\delta w^{\mu} := \lambda w^{\mu}$, where $\lambda \ll 1$. Applying equation (2.2) once more, we obtain
\begin{align}
    v^{\nu}_{s} &= v^{\nu}_{q} - \Gamma^{\nu}{}_{\mu\lambda}(q)\, \delta w^{\mu}\, v^{\lambda}_{q}, \nonumber \\
    &= v^{\nu}_{p} - \Gamma^{\nu}{}_{\mu\lambda}(p)\, \delta u^{\mu}\, v^{\lambda}_{p} 
    - \Big[ \Gamma^{\nu}{}_{\mu\lambda}(p) + \partial_{\rho}\Gamma^{\nu}{}_{\mu\lambda}(p)\, \delta u^{\rho} \Big]
    \Big[ v^{\lambda}_{p} - \Gamma^{\lambda}{}_{\alpha\beta}(p)\, \delta u^{\alpha}\, v^{\beta}_{p} \Big] \delta w^{\mu}.
    \label{eq:3.18}
\end{align}

In the second step, the quantities at $q$ were expanded about $p$ to first order in $\delta u^{\mu}$. Retaining only terms up to second order in the infinitesimal displacements $\delta u$ and $\delta w$, we find
\begin{equation}
    v^{\nu}_{s} = v^{\nu}_{p} - \Gamma^{\nu}{}_{\mu\lambda}(p)\, \delta u^{\mu}\, v^{\lambda}_{p}
    - \Gamma^{\nu}{}_{\mu\lambda}(p)\, \delta w^{\mu}\, v^{\lambda}_{p}
    - v^{\beta}_{p} \Big[ \partial_{\alpha}\Gamma^{\nu}{}_{\mu\beta}(p)
    - \Gamma^{\nu}{}_{\mu\lambda}(p)\Gamma^{\lambda}{}_{\alpha\beta}(p) \Big] 
    \delta u^{\alpha}\, \delta w^{\mu}.
    \label{eq:3.19}
\end{equation}

Instead of continuing this process through all four segments, we adopt a more efficient route. We can equivalently transport $v_p$ first from $p$ to $r$ along $\delta w$ and then from $r$ to $s$ along $\delta u$, obtaining a new vector $v'_{s}$. Repeating the same reasoning as above, one finds
\begin{equation}
\begin{split}
    v'^{\nu}_{s} & = v^{\nu}_{p} - \Gamma^{\nu}{}_{\mu\lambda}(p)\, \delta w^{\mu}\, v^{\lambda}_{p}
    - \Gamma^{\nu}{}_{\mu\lambda}(p)\, \delta u^{\mu}\, v^{\lambda}_{p}
     \\  & - v^{\beta}_{p} \Big[ \partial_{\mu}\Gamma^{\nu}{}_{\alpha\beta}(p)
    - \Gamma^{\nu}{}_{\alpha\lambda}(p)\Gamma^{\lambda}{}_{\mu\beta}(p) \Big] 
    \delta u^{\alpha}\, \delta w^{\mu}.
    \label{eq:3.20}
    \end{split}
\end{equation}

The difference between the two transported vectors $v_s$ and $v'_s$ isolates the effect of curvature. Since their zeroth- and first-order terms coincide, the difference arises only at second order:
\begin{align}
\begin{split}
    v^{\nu}_{s} - v'^{\nu}_{s} 
    =\; v^{\beta}_{p} \Big[
        \partial_{\mu}\Gamma^{\nu}{}_{\alpha\beta}(p)
        - \partial_{\alpha}\Gamma^{\nu}{}_{\mu\beta}(p)
        + \Gamma^{\nu}{}_{\mu\lambda}(p)\Gamma^{\lambda}{}_{\alpha\beta}(p)
        \\[4pt]
        -\, \Gamma^{\nu}{}_{\alpha\lambda}(p)\Gamma^{\lambda}{}_{\mu\beta}(p)
    \Big] \delta u^{\alpha}\, \delta w^{\mu}.
\end{split}
\label{eq:3.21}
\end{align}

Defining the \textit{curvature tensor} $R^{\alpha}{}_{\mu\nu\rho}$ as
\begin{equation}
\begin{split}
    R^{\alpha}{}_{\mu\nu\rho}
    & :=\; 2\,\partial_{[\nu}\Gamma^{\alpha}{}_{\rho]\mu}
    + 2\,\Gamma^{\alpha}{}_{[\nu|\lambda|}\Gamma^{\lambda}{}_{\rho]\mu}
    \\[4pt]
   &  =\; \partial_{\nu}\Gamma^{\alpha}{}_{\rho\mu}
    - \partial_{\rho}\Gamma^{\alpha}{}_{\nu\mu}
    + \Gamma^{\alpha}{}_{\nu\lambda}\Gamma^{\lambda}{}_{\rho\mu}
    - \Gamma^{\alpha}{}_{\rho\lambda}\Gamma^{\lambda}{}_{\nu\mu}.
\end{split}
\label{eq:3.22}
\end{equation}
we can rewrite equation \eqref{eq:3.21} succinctly as
\begin{equation}
    v^{\nu}_{s} - v'^{\nu}_{s} = v^{\beta}_{p}\, R^{\nu}{}_{\beta\alpha\mu}\, \delta u^{\alpha}\, \delta w^{\mu}.
\end{equation}

Thus, the curvature tensor quantifies the infinitesimal rotation (or change in direction) experienced by a vector parallel transported around a closed loop. By construction, it satisfies the antisymmetry property
\begin{equation}
    R^{\alpha}{}_{\mu(\nu\rho)} = 0,
    \label{eq:3.23}
\end{equation}
and it depends solely on the connection $\Gamma$, independent of any metric structure. A connection for which the curvature tensor vanishes is referred to as \textit{flat}.

\vspace{1em}
\subsubsection{Contractions of the Curvature Tensor}

Given the four-index tensor $R^{\alpha}{}_{\mu\nu\rho}$, several traces (or contractions) can be constructed. Owing to its antisymmetry in the last two indices, two distinct contractions can be formed even in the absence of a metric:
\begin{align}
    R_{\mu\nu} &:= R^{\lambda}{}_{\mu\lambda\nu}, \label{eq:3.24} \\
    H_{\mu\nu} &:= R^{\lambda}{}_{\lambda\mu\nu}. \label{eq:3.25}
\end{align}
The tensor $R_{\mu\nu}$ is known as the \textit{Ricci tensor}, while $H_{\mu\nu}$ is the \textit{homothetic tensor}. As will be discussed later in Section, the homothetic tensor can be related to the Ricci tensor and the torsion tensor.

If a metric $g_{\mu\nu}$ is present, two additional traces can be formed. By raising one of the indices, we define the \textit{co-Ricci tensor} as
\begin{equation}
    P_{\mu}{}^{\nu} := g^{\rho\lambda} R_{\mu\rho}{}^{\nu}{}_{\lambda}
    = R_{\mu\lambda}{}^{\nu\lambda}.
    \label{eq:3.26}
\end{equation}
The co-Ricci tensor can be expressed in terms of the Ricci tensor and the non-metricity tensor. Finally, contracting the Ricci tensor with the metric yields the \textit{Ricci scalar},
\begin{equation}
    R := g^{\mu\nu} R_{\mu\nu}
    = R^{\mu}{}_{\mu}
    = R^{\lambda\mu}{}_{\lambda\mu}.
    \label{eq:3.27}
\end{equation}

These various contractions provide distinct scalar and tensorial measures of curvature and play central roles in constructing gravitational actions and field equations. This concludes our discussion of the curvature tensor and its associated traces.

\vspace{1em}
\subsection{Torsion}

We now consider property (b) outlined previously. In Euclidean geometry, the addition of two vectors can be visualized through the construction of a parallelogram. Let two vectors at a point $p \in M$ be denoted as $u_p \in T_pM$ and $v_p \in T_pM$. Translating $v_p$ along $u_p$ such that their tails coincide is equivalent to translating $u_p$ along $v_p$. In this case, both operations lead to the same resultant vector $w_p := u_p + v_p$, representing the total displacement from $p$ to $s$ (see Figure 2.2).

In curved or non-Euclidean geometries, this equivalence no longer necessarily holds. To examine this behavior, we restrict our attention to infinitesimal vectors $\delta u_p := \epsilon u_p$ and $\delta v_p := \lambda v_p$, where $\epsilon, \lambda \ll 1$. These infinitesimal displacements are assumed to be linearly independent; otherwise, no parallelogram can be defined.

According to equation (2.2), the infinitesimal parallel transport of $\delta v_p$ to the tip of $\delta u_p$ (i.e., to point $q$) is
\begin{equation}
    \delta v^{\alpha}_{q} = \delta v^{\alpha}_{p} - \Gamma^{\alpha}{}_{\mu\nu}(p)\, \delta u^{\mu}_{p}\, \delta v^{\nu}_{p}.
    \label{eq:3.28}
\end{equation}
The total displacement from $p$ to $s_1$ then reads
\begin{equation}
    \delta w^{\alpha}_{p} := \delta u^{\alpha}_{p} + \delta v^{\alpha}_{p} - \Gamma^{\alpha}{}_{\mu\nu}(p)\, \delta u^{\mu}_{p}\, \delta v^{\nu}_{p}.
    \label{eq:3.29}
\end{equation}
Alternatively, transporting $\delta u_p$ first to $r$ (the tip of $\delta v_p$) and then combining yields
\begin{equation}
    \delta w'^{\alpha}_{p} := \delta v^{\alpha}_{p} + \delta u^{\alpha}_{p} - \Gamma^{\alpha}{}_{\nu\mu}(p)\, \delta u^{\mu}_{p}\, \delta v^{\nu}_{p}.
    \label{eq:3.30}
\end{equation}

\begin{figure}[htp!]
\begin{center}
\includegraphics[width=1.1\textwidth]{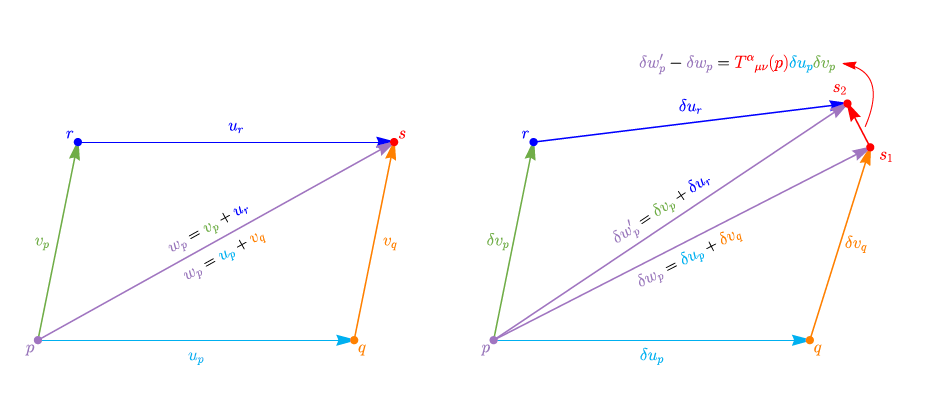}
    \caption{Left: In Euclidean geometry, transporting $v_p$ along $u_p$ or $u_p$ along $v_p$ results in the same endpoint $s$, forming a closed parallelogram. 
    Right: In a general affine connection, the parallelogram fails to close, producing two distinct endpoints $s_{1}$ and $s_{2}$. The discrepancy is described by the torsion tensor $T^{\alpha}_{\mu\nu}$.}
    \end{center}
\end{figure}

The failure of the parallelogram to close is quantified by comparing $\delta w_p$ and $\delta w'_p$:
\begin{equation}
    \delta w'^{\alpha}_{p} - \delta w^{\alpha}_{p} = \left( \Gamma^{\alpha}{}_{\mu\nu}(p) - \Gamma^{\alpha}{}_{\nu\mu}(p) \right) 
    \delta u^{\mu}_{p}\, \delta v^{\nu}_{p} =: T^{\alpha}{}_{\mu\nu}\, \delta u^{\mu}_{p}\, \delta v^{\nu}_{p}.
    \label{eq:3.31}
\end{equation}
This motivates the definition of the \emph{torsion tensor}:
\begin{equation}
    T^{\alpha}{}_{\mu\nu} := 2\,\Gamma^{\alpha}{}_{[\mu\nu]} 
    = \Gamma^{\alpha}{}_{\mu\nu} - \Gamma^{\alpha}{}_{\nu\mu}.
    \label{eq:3.32}
\end{equation}
The parallelogram closes if and only if $T^{\alpha}{}_{\mu\nu} = 0$, meaning the connection is torsion-free. Nonzero torsion thus measures the failure of infinitesimal parallelograms to close under parallel transport.

The antisymmetry of $T^{\alpha}{}_{\mu\nu}$ in its lower indices implies that if $\delta u_p$ and $\delta v_p$ are linearly dependent, e.g., $\delta u_p = f\, \delta v_p$ for a scalar $f$, then
\[
T^{\alpha}{}_{\mu\nu}\, \delta u^{\mu}_{p}\, \delta v^{\nu}_{p}
= f\,T^{\alpha}{}_{\mu\nu}\, \delta v^{\mu}_{p}\, \delta v^{\nu}_{p} = 0,
\]
since the contraction of an antisymmetric tensor with a symmetric product vanishes. Thus, no failure of closure occurs in this degenerate case.

From a geometric perspective, torsion represents the antisymmetric part of the affine connection. While $\Gamma^{\alpha}{}_{\mu\nu}$ does not transform tensorially due to its inhomogeneous term, the antisymmetric difference $\Gamma^{\alpha}{}_{\mu\nu} - \Gamma^{\alpha}{}_{\nu\mu}$ eliminates the non-tensorial component, rendering $T^{\alpha}{}_{\mu\nu}$ a true tensor.

A connection satisfying $T^{\alpha}{}_{\mu\nu} = 0$ is called \textit{torsion-free}. Even without a metric, one can define a trace of the torsion tensor:
\begin{equation}
    T_{\mu} := T^{\alpha}{}_{\mu\alpha}.
    \label{eq:3.33}
\end{equation}
The order of contracted indices is crucial:
\begin{equation}
    T^{\alpha}{}_{\alpha\mu} = -T^{\alpha}{}_{\mu\alpha} = -T_{\mu}.
    \label{eq:3.34}
\end{equation}
This trace is unique. In the presence of a metric, one might consider $g^{\mu\nu}T^{\alpha}{}_{\mu\nu}$, but this vanishes identically due to the symmetry of $g^{\mu\nu}$ and antisymmetry of $T^{\alpha}{}_{\mu\nu}$.

\vspace{1em}
\subsection{Non-Metricity}
We now address the third property of vectors-their magnitude. Let $v \in TM$ with components $v^{\mu}$ in a coordinate chart, and let $g_{\mu\nu}$ denote the components of the metric tensor. The squared magnitude of $v$ is defined as
\begin{equation}
    \|v\|^{2} := g(v, v) = g_{\mu\nu} v^{\mu} v^{\nu}.
    \label{eq:3.35}
\end{equation}
We examine how this magnitude changes when $v$ is parallel transported along a curve $\gamma$ with tangent vector $u = u^{\mu}\partial_{\mu}$, as shown in Figure 2.3. The condition for parallel transport is
\begin{equation}
    u^{\alpha} \nabla_{\alpha} v^{\mu} = 0,
    \label{eq:3.36}
\end{equation}
where $\nabla$ denotes the covariant derivative associated with $\Gamma^{\alpha}{}_{\mu\nu}$.

\begin{figure}[htp!]
\begin{center}
\includegraphics[width=1.1\textwidth]{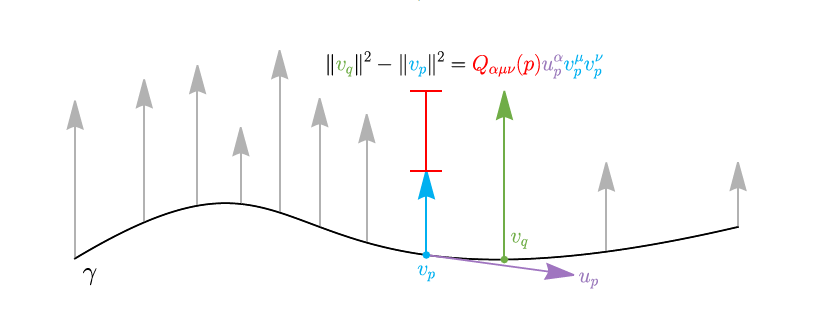}
    \caption{A vector field $v$ parallel transported along a curve $\gamma$. The change in its squared magnitude between nearby points $p$ and $q$ is governed by the non-metricity tensor $Q_{\alpha\mu\nu}$.}
    \end{center}
\end{figure}

Using Leibniz's rule, the rate of change of the squared magnitude along $\gamma$ is
\begin{equation}
    \frac{d}{dt}\bigg|_{\gamma(t)} \|v\|^{2}
    = u^{\alpha}\nabla_{\alpha}(g_{\mu\nu}v^{\mu}v^{\nu})
    = (u^{\alpha}\nabla_{\alpha} g_{\mu\nu}) v^{\mu}v^{\nu}
    + 2g_{\mu\nu}(u^{\alpha}\nabla_{\alpha}v^{\mu})v^{\nu}.
    \label{eq:3.37a}
\end{equation}
The second term vanishes by equation \eqref{eq:3.36}, yielding
\begin{equation}
    \frac{d}{dt}\|v\|^{2} = Q_{\alpha\mu\nu}\, u^{\alpha}v^{\mu}v^{\nu},
    \label{eq:3.37}
\end{equation}
where
\begin{equation}
    Q_{\alpha\mu\nu} := \nabla_{\alpha} g_{\mu\nu}
    \label{eq:3.38}
\end{equation}
is the \textit{non-metricity tensor}. Nonzero $Q_{\alpha\mu\nu}$ measures the change in the metric (and thus vector magnitudes) under parallel transport. The condition
\begin{equation}
    \nabla_{\alpha} g_{\mu\nu} = 0
    \label{eq:3.39}
\end{equation}
is known as the \textit{metricity condition}, and a connection satisfying it is termed \textit{metric-compatible}.

Raising the last two indices gives
\begin{equation}
    Q^{\alpha}{}_{\mu\nu} \neq \nabla_{\alpha} g^{\mu\nu}.
    \label{eq:3.40}
\end{equation}
From the identity $g_{\mu\lambda} g^{\lambda\nu} = \delta_{\mu}^{\nu}$, one finds
\begin{equation}
    Q^{\alpha}{}_{\mu\nu} = -\nabla_{\alpha} g^{\mu\nu}.
    \label{eq:3.41}
\end{equation}

Beyond measuring length variation, $Q_{\alpha\mu\nu}$ quantifies the change in all metric-dependent quantities under parallel transport. For instance, it affects the angle between two vectors and the $n$-dimensional volume of a region $\Omega \subset M$, given by
\begin{equation}
    \mathrm{Vol}(\Omega) := \int_{\Omega} \sqrt{|g|}\, d^{n}x.
    \label{eq:3.42}
\end{equation}
Transporting $\mathrm{Vol}(\Omega)$ along $\gamma$ yields
\begin{align}
    \frac{d}{dt}\mathrm{Vol}(\Omega)
    &= \int_{\Omega} u^{\alpha} \nabla_{\alpha}\sqrt{|g|}\, d^{n}x
    = \frac{1}{2} \int_{\Omega} \sqrt{|g|}\, u^{\alpha} g^{\mu\nu} \nabla_{\alpha} g_{\mu\nu}\, d^{n}x \nonumber \\
    &= \frac{1}{2} \int_{\Omega} \sqrt{|g|}\, u^{\alpha} g^{\mu\nu} Q_{\alpha\mu\nu}\, d^{n}x.
    \label{eq:3.43}
\end{align}
It is therefore convenient to define two independent traces of the non-metricity tensor:
\begin{equation}
    Q_{\alpha} := g^{\mu\nu} Q_{\alpha\mu\nu} = Q_{\alpha\mu}{}^{\mu},
    \qquad
    \bar{Q}_{\alpha} := g^{\mu\nu} Q_{\mu\nu\alpha} = Q_{\mu}{}^{\mu}{}_{\alpha}.
    \label{eq:3.44}
\end{equation}
These traces play important roles in the formulation of gravitational actions based on non-metricity.

\vspace{1em}
\subsection{Classification of Metric-Affine Geometries and Decomposition of the Connection}

Since the affine connection $\Gamma^{\alpha}{}_{\mu\nu}$ is not a tensor, it does not, by itself, possess an intrinsic geometric interpretation that remains invariant under coordinate transformations. However, the connection together with the metric $g_{\mu\nu}$ gives rise to tensorial quantities--namely the curvature, torsion, and non-metricity tensors, which fully encode the geometric properties of a given metric-affine space $(M, g, \Gamma)$. Thus, while $\Gamma^{\alpha}{}_{\mu\nu}$ alone lacks direct geometric meaning, the pair $(g_{\mu\nu}, \Gamma^{\alpha}{}_{\mu\nu})$ does contain all necessary information to characterize the geometry of spacetime.

Based on the structures introduced, one can identify different levels of geometrical complexity as follows:
\begin{enumerate}
    \item \emph{Bare Manifold:} The simplest possible geometric setup consists solely of a differentiable manifold $M$ without a metric or connection. This structure allows the definition of smooth functions, vector fields, tensor fields, and curves. However, no notions of distance, angle, or covariant differentiation are available. From a physical standpoint, such a manifold provides minimal utility for describing dynamical or gravitational phenomena.
    
    \item \emph{Affine Geometry:} An \emph{affine geometry} is defined by the pair $(M, \Gamma)$. In addition to the properties of a bare manifold, the presence of a connection allows the definition of a covariant derivative. This, in turn, enables one to compute the curvature and torsion tensors. Nonetheless, the absence of a metric means that concepts such as length, angle, or geodesics remain undefined.

    \item \emph{Riemannian Geometry:} A \emph{Riemannian geometry} consists of the pair $(M, g)$, which introduces a metric structure to the manifold. This allows one to define lengths, distances, areas, and volumes. Moreover, a natural connection, the Levi-Civita connection, can be constructed solely from the metric, leading to a well-defined covariant derivative $D$. The Levi-Civita connection is both torsion-free and metric-compatible, implying that Riemannian geometries possess curvature but neither torsion nor non-metricity. \gls{gr} is formulated within this geometric framework.

    \item \emph{Metric-Affine Geometry:} The most general framework is the \emph{metric-affine geometry}, characterized by the triplet $(M, g, \Gamma)$. In this setup, all geometric quantities: curvature, torsion, and non-metricity, are, in general, nonzero. Depending on which of these quantities vanish, several well-known subcases can be identified:

\begin{itemize}
    \item $R^{\alpha}{}_{\mu\nu\rho}=0$, $T^{\alpha}{}_{\mu\nu}=0$, $Q_{\alpha\mu\nu}=0$: This corresponds to a flat geometry, which reduces to Euclidean or Minkowskian spacetime depending on the metric signature.
    
    \item $R^{\alpha}{}_{\mu\nu\rho}\neq0$, $T^{\alpha}{}_{\mu\nu}=0$, $Q_{\alpha\mu\nu}=0$: A purely Riemannian geometry where only curvature is nonzero. This is the setting of Einstein’s General Relativity.
    
    \item $R^{\alpha}{}_{\mu\nu\rho}=0$, $T^{\alpha}{}_{\mu\nu}\neq0$, $Q_{\alpha\mu\nu}=0$: The geometry where torsion alone is nonvanishing, forming the foundation of the \gls{tegr} and its extensions.
    
    \item $R^{\alpha}{}_{\mu\nu\rho}=0$, $T^{\alpha}{}_{\mu\nu}=0$, $Q_{\alpha\mu\nu}\neq0$: This geometry possesses only non-metricity, providing the basis of the \gls{stegr} and its generalizations such as $f(Q)$ gravity.
    
    \item $R^{\alpha}{}_{\mu\nu\rho}=0$, $T^{\alpha}{}_{\mu\nu}\neq0$, $Q_{\alpha\mu\nu}\neq0$: Both torsion and non-metricity are present. This setting defines the General Teleparallel Equivalent of GR (GTEGR), which extends \gls{tegr} and \gls{stegr} simultaneously.
    
    \item $R^{\alpha}{}_{\mu\nu\rho}\neq0$, $T^{\alpha}{}_{\mu\nu}\neq0$, $Q_{\alpha\mu\nu}=0$: A geometry featuring both curvature and torsion, known as Riemann--Cartan geometry.
    
    \item $R^{\alpha}{}_{\mu\nu\rho}\neq0$, $T^{\alpha}{}_{\mu\nu}=0$, $Q_{\alpha\mu\nu}\neq0$: A geometry exhibiting curvature and non-metricity but no torsion.
    
    \item $R^{\alpha}{}_{\mu\nu\rho}\neq0$, $T^{\alpha}{}_{\mu\nu}\neq0$, $Q_{\alpha\mu\nu}\neq0$: The most general case, where all geometric quantities are nonzero, representing the full metric--affine framework.
\end{itemize}
\end{enumerate}

\begin{figure}[htp!]
\begin{center}
\includegraphics[width=1.00\textwidth]{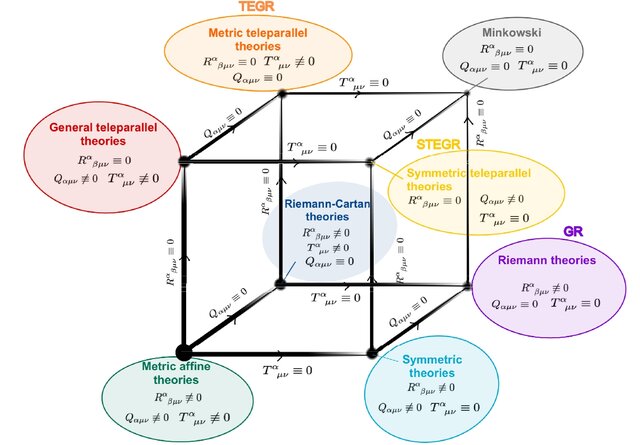}
    \caption{A schematic representation can be constructed to illustrate all possible forms of metric-affine geometries denoted by \( (M, g, \Gamma) \). This classification is organized according to the presence or absence of the fundamental geometric quantities: the curvature tensor \( R^{\alpha}{}_{\mu \nu \rho} \), the torsion tensor \( T^{\alpha}{}_{\mu \nu} \), and the non-metricity tensor \( Q_{\alpha \mu \nu} \).}
    \end{center}
\end{figure}

Although the connection $\Gamma^{\alpha}{}_{\mu\nu}$ itself does not constitute an invariant geometric object, it still encapsulates the structure leading to curvature, torsion, and non-metricity. This motivates a decomposition of the connection that explicitly displays its geometric content.

\vspace{1em}
\subsubsection{Decomposition of the Connection}

Let us consider a general metric-affine space $(M, g, \Gamma)$. The goal is to express the connection in a form that isolates contributions from the Levi-Civita part, torsion, and non-metricity. We begin with the covariant derivative of the metric and its cyclic permutations:
\begin{align}
\partial_{\alpha} g_{\mu\nu} - \Gamma^{\beta}{}_{\alpha\nu} g_{\mu\beta} - \Gamma^{\beta}{}_{\alpha\mu} g_{\beta\nu} &= Q_{\alpha\mu\nu}, \nonumber\\
\partial_{\mu} g_{\nu\alpha} - \Gamma^{\beta}{}_{\mu\nu} g_{\alpha\beta} - \Gamma^{\beta}{}_{\mu\alpha} g_{\beta\nu} &= Q_{\mu\nu\alpha}, \nonumber\\
\partial_{\nu} g_{\alpha\mu} - \Gamma^{\beta}{}_{\nu\alpha} g_{\beta\mu} - \Gamma^{\beta}{}_{\nu\mu} g_{\alpha\beta} &= Q_{\nu\alpha\mu}. \label{eq:cyclic}
\end{align}
Adding the first two equations and subtracting the third yields
\begin{equation}
T^{\beta}{}_{\nu\mu} g_{\alpha\beta} + T^{\beta}{}_{\nu\alpha} g_{\mu\beta} + T^{\beta}{}_{\alpha\mu} g_{\nu\beta}
+ \partial_{\alpha} g_{\mu\nu} + \partial_{\mu} g_{\nu\alpha} - \partial_{\nu} g_{\alpha\mu}
- 2 \Gamma^{\beta}{}_{\alpha\mu} g_{\nu\beta}
= Q_{\alpha\mu\nu} + Q_{\mu\nu\alpha} - Q_{\nu\alpha\mu}. \label{eq:combine}
\end{equation}
Solving for the connection $\Gamma^{\alpha}{}_{\mu\nu}$, we find
\begin{equation}
\Gamma^{\alpha}{}_{\mu\nu} = \left\{^{\alpha}{}_{\mu\nu}\right\} + K^{\alpha}{}_{\mu\nu} + L^{\alpha}{}_{\mu\nu}, \label{eq:connection_decomp}
\end{equation}
where $\left\{^{\alpha}{}_{\mu\nu}\right\}$ denotes the Levi-Civita connection, $K^{\alpha}{}_{\mu\nu}$ is the \textit{contortion tensor}, and $L^{\alpha}{}_{\mu\nu}$ is the \emph{disformation tensor}, defined respectively as
\begin{align}
K^{\alpha}{}_{\mu\nu} &= \tfrac{1}{2} \left( T^{\alpha}{}_{\mu\nu} + T_{\mu}{}^{\alpha}{}_{\nu} + T_{\nu}{}^{\alpha}{}_{\mu} \right), \label{eq:contortion}\\
L^{\alpha}{}_{\mu\nu} &= \tfrac{1}{2} \left( - Q^{\alpha}{}_{\mu\nu} + Q_{\mu}{}^{\alpha}{}_{\nu} + Q_{\nu}{}^{\alpha}{}_{\mu} \right). \label{eq:disformation}
\end{align}

From this decomposition, one immediately recovers that if both torsion and non-metricity vanish, the connection reduces uniquely to the Levi-Civita connection:
\begin{equation}
T^{\alpha}{}_{\mu\nu} = 0, \quad Q_{\alpha\mu\nu} = 0 \quad \Rightarrow \quad \Gamma^{\alpha}{}_{\mu\nu} = \left\{^{\alpha}{}_{\mu\nu}\right\}. \label{eq:levi-civita}
\end{equation}
This result confirms that Riemannian geometry, governed by the Levi-Civita connection, is a special case of the broader metric-affine framework where both torsion and non-metricity are constrained to vanish.

\vspace{1em}
\subsection{The Lie Derivative Revisited: Symmetries of Metric-Affine Geometries}
The concept of the Lie derivative is inherently independent of any notion of connection or parallel transport. It is defined purely in terms of the flow generated by vector fields on the manifold \( M \), and therefore, it does not require the introduction of a metric or affine structure. 

However, one can establish a relationship between the Lie derivative and the covariant derivative. In Riemannian geometry, for example, it is well known that replacing the partial derivative \( \partial_{\mu} \) with the covariant derivative \( D_{\mu} \) in the expression for the Lie derivative does not alter the outcome. This equivalence arises because all terms involving the connection coefficients cancel identically during computation.

Despite these cancellations, expressing the Lie derivative in terms of the covariant derivative offers computational and conceptual advantages. For instance, the Lie derivative of the metric tensor can be expressed as
\begin{equation}
\mathcal{L}_{v} g_{\mu \nu} = v^{\lambda} \partial_{\lambda} g_{\mu \nu} + g_{\lambda \nu} \partial_{\mu} v^{\lambda} + g_{\mu \lambda} \partial_{\nu} v^{\lambda}.
\end{equation}
Substituting the partial derivatives by covariant derivatives yields
\begin{equation}
\mathcal{L}_{v} g_{\mu \nu} = v^{\lambda} D_{\lambda} g_{\mu \nu} + g_{\lambda \nu} D_{\mu} v^{\lambda} + g_{\mu \lambda} D_{\nu} v^{\lambda}.
\end{equation}
Since the covariant derivative \( D_{\lambda} \) is metric-compatible (\( D_{\lambda} g_{\mu \nu} = 0 \)), the expression simplifies to
\begin{equation}
\mathcal{L}_{v} g_{\mu \nu} = 2 D_{(\mu} v_{\nu)}.
\end{equation}
This compact form demonstrates the intrinsic geometric meaning of the Lie derivative when expressed via covariant differentiation.

When extending this framework to metric-affine geometries, it is often necessary to evaluate the Lie derivatives of both the metric and the affine connection in the presence of torsion and non-metricity. For a general affine connection \( \nabla \), these can be expressed as \citep{Hehl:1994ue, Obukhov2006, BeltranJimenez:2019esp}:
\begin{align}
\mathcal{L}_{v} g_{\mu \nu} &= 2 g_{\lambda (\mu} \nabla_{\nu)} v^{\lambda} + \left( Q_{\lambda \mu \nu} - 2 T_{(\mu \nu) \lambda} \right) v^{\lambda} = 2 D_{(\mu} v_{\nu)}, \nonumber \\
\mathcal{L}_{v} \Gamma^{\alpha}{}_{\mu \nu} &= \nabla_{\mu} \nabla_{\nu} v^{\alpha} - T^{\alpha}{}_{\nu \lambda} \nabla_{\mu} v^{\lambda} - \left( R^{\alpha}{}_{\nu \mu \lambda} + \nabla_{\mu} T^{\alpha}{}_{\nu \lambda} \right) v^{\lambda}.
\end{align}
Here, \( T^{\alpha}{}_{\mu \nu} \), \( Q_{\alpha \mu \nu} \), and \( R^{\alpha}{}_{\nu \mu \lambda} \) denote the torsion, non-metricity, and curvature tensors, respectively. 

It is crucial to apply the correct form when evaluating the Lie derivative of the connection, as the connection is not a tensor. The full expression reads
\begin{equation}
\mathcal{L}_{v} \Gamma^{\alpha}{}_{\mu \nu} = v^{\lambda} \partial_{\lambda} \Gamma^{\alpha}{}_{\mu \nu}
 - \Gamma^{\lambda}{}_{\mu \nu} \partial_{\lambda} v^{\alpha}
 + \Gamma^{\alpha}{}_{\lambda \nu} \partial_{\mu} v^{\lambda}
 + \Gamma^{\alpha}{}_{\mu \lambda} \partial_{\nu} v^{\lambda}
 + \partial_{\mu} \partial_{\nu} v^{\alpha}.
\end{equation}
This form follows from the inhomogeneous transformation property of the affine connection under coordinate changes. The first four terms correspond to the homogeneous part of the transformation, while the last term originates from the inhomogeneous contribution. Importantly, although the connection itself is not a tensor, its Lie derivative \( \mathcal{L}_{v} \Gamma^{\alpha}{}_{\mu \nu} \) transforms as a true tensor of type \( (1,2) \), as is evident from the tensorial structure of the right-hand side.

The expressions above play a central role in defining symmetries within metric-affine geometries. Following the approach of Refs. \citep{yano2020theory, Hehl:1976kj, Haghani:2012bt, BeltranJimenez:2017tkd,}, a one-parameter family of diffeomorphisms 
\[
\phi_{s}: \mathbb{R} \times M \rightarrow M, 
\]
is said to generate a symmetry of the metric-affine geometry \( (M, g, \Gamma) \) if it satisfies:
\begin{equation}
\begin{cases}
\phi_{s}^{*} g_{\mu \nu} = g_{\mu \nu}, \\
\phi_{s}^{*} \Gamma^{\alpha}{}_{\mu \nu} = \Gamma^{\alpha}{}_{\mu \nu}.
\end{cases}
\end{equation}
These relations, known as the \emph{symmetry conditions}, ensure that both the metric and the affine connection remain invariant under the diffeomorphism, implying that derived quantities such as curvature, torsion, and non-metricity are also preserved.

Expanding the symmetry condition to first order in the parameter \( s \) around \( s=0 \) yields the infinitesimal symmetry condition:
\begin{equation}
\begin{cases}
\mathcal{L}_{\xi} g_{\mu \nu} = 0, \\
\mathcal{L}_{\xi} \Gamma^{\alpha}{}_{\mu \nu} = 0,
\end{cases}
\qquad 
\xi := \left. \frac{d \phi_{s}}{ds} \right|_{s=0},
\end{equation}
where \( \xi \) is the vector field generating the flow \( \phi_{s} \). This vector field is referred to as the \emph{generating vector field} of the symmetry. These symmetry conditions and their associated Lie derivatives play a crucial role in the study of invariant geometrical structures, particularly in the contexts of cosmology and black hole solutions.

\vspace{1em}
\subsection{Collection of Geometric Identities}

In the context of a general affine connection $\Gamma^{\alpha}{}_{\mu\nu}$, computations involving the covariant derivative $\nabla_{\mu}$ frequently require commuting two such operators in order to simplify expressions or expose their underlying geometric structure. Since $\nabla_{\mu}$ can act on scalars, vectors, tensors, and tensor densities, it is instructive to begin with the simplest case: a scalar field $f$.

Using the definitions introduced earlier, the commutator of two covariant derivatives acting on a scalar field takes the form
\begin{equation}
[\nabla_{\mu}, \nabla_{\nu}] f = - T^{\lambda}{}_{\mu\nu} \, \partial_{\lambda} f ,
\label{eq:comm_scalar}
\end{equation}
where $T^{\lambda}{}_{\mu\nu}$ denotes the torsion tensor. Consequently, this commutator vanishes if and only if the torsion tensor itself is zero.

Next, consider the action of the commutator on a vector field $v^{\alpha}$. Employing the definitions of the covariant derivative and affine connection yields
\begin{equation}
[\nabla_{\mu}, \nabla_{\nu}] v^{\alpha} = R^{\alpha}{}_{\lambda\mu\nu} v^{\lambda} - T^{\lambda}{}_{\mu\nu} \nabla_{\lambda} v^{\alpha},
\label{eq:comm_vector}
\end{equation}
where $R^{\alpha}{}_{\lambda\mu\nu}$ is the curvature tensor associated with $\Gamma^{\alpha}{}_{\mu\nu}$. In the absence of torsion, equation (2.65) reduces to the familiar Riemannian expression
\begin{equation}
[\nabla_{\mu}, \nabla_{\nu}] v^{\alpha} = R^{\alpha}{}_{\lambda\mu\nu} v^{\lambda}.
\label{eq:comm_vector_riemann}
\end{equation}
It should be noted, however, that in a metric-affine setting, $R^{\alpha}{}_{\lambda\mu\nu}$ need not correspond to the curvature of the Levi-Civita connection, since non-metricity may be nonzero.

An analogous identity holds for a covector (1-form) $\omega_{\alpha}$:
\begin{equation}
[\nabla_{\mu}, \nabla_{\nu}] \omega_{\alpha} = - R^{\lambda}{}_{\alpha\mu\nu} \omega_{\lambda} - T^{\lambda}{}_{\mu\nu} \nabla_{\lambda} \omega_{\alpha}.
\label{eq:comm_oneform}
\end{equation}
The negative sign preceding the curvature term arises naturally due to the covariant nature of $\omega_{\alpha}$.

Combining the above, the commutator acting on a general tensor field $S^{\mu_{1} \dots \mu_{p}}{}_{\nu_{1} \dots \nu_{q}}$ of type $(p, q)$ is given by
\begin{equation}
\begin{split}
& [\nabla_{\mu}, \nabla_{\nu}] S^{\mu_{1} \dots \mu_{p}}{}_{\nu_{1} \dots \nu_{q}} =
\sum_{i=1}^{p} R^{\mu_{i}}{}_{\lambda\mu\nu} S^{\mu_{1} \dots \lambda \dots \mu_{p}}{}_{\nu_{1} \dots \nu_{q}} \\
 & - \sum_{j=1}^{q} R^{\lambda}{}_{\nu_{j}\mu\nu} S^{\mu_{1} \dots \mu_{p}}{}_{\nu_{1} \dots \lambda \dots \nu_{q}}
 - T^{\lambda}{}_{\mu\nu} \nabla_{\lambda} S^{\mu_{1} \dots \mu_{p}}{}_{\nu_{1} \dots \nu_{q}} .
\label{eq:comm_tensor}
\end{split}
\end{equation}
This result is fully general and encompasses the cases of most practical interest. The remaining case involves tensor densities, which require additional consideration of the metric determinant.

For the metric tensor $g_{\alpha\beta}$, by invoking the definition of the non-metricity tensor and equation (2.68), one obtains
\begin{equation}
[\nabla_{\mu}, \nabla_{\nu}] g_{\alpha\beta}
= 2 \nabla_{[\mu} Q_{\nu] \alpha \beta}
= -2 R_{(\alpha\beta)\mu\nu} - T^{\lambda}{}_{\mu\nu} Q_{\lambda\alpha\beta},
\label{eq:comm_metric}
\end{equation}
where $Q_{\nu\alpha\beta} := -\nabla_{\nu} g_{\alpha\beta}$ denotes the non-metricity tensor.

For a scalar density of weight $w$, defined as $|g|^{w/2}$, the commutator acts as
\begin{equation}
[\nabla_{\mu}, \nabla_{\nu}] |g|^{w/2}
= w |g|^{w/2} \nabla_{[\mu} Q_{\nu]} .
\label{eq:comm_density}
\end{equation}
Combining this with equation (2.68), the general formula for the commutator acting on a tensor density of type $(p,q)$ and weight $w$ reads
\begin{equation}
[\nabla_{\mu}, \nabla_{\nu}]
\left( |g|^{w/2} S^{\mu_{1} \dots \mu_{p}}{}_{\nu_{1} \dots \nu_{q}} \right)
= |g|^{w/2} \left( [\nabla_{\mu}, \nabla_{\nu}] S^{\mu_{1} \dots \mu_{p}}{}_{\nu_{1} \dots \nu_{q}}
+ w \, \nabla_{[\mu} Q_{\nu]} \right).
\label{eq:comm_tensordensity}
\end{equation}
Setting $w = 0$ recovers the earlier results for tensor fields, while $p = q = 0$ and $w = 0$ reproduce the scalar case.

Furthermore, the covariant derivatives defined by a general affine connection satisfy the Jacobi identity,
\begin{equation}
[\nabla_{\alpha}, [\nabla_{\beta}, \nabla_{\gamma}]]
+ [\nabla_{\beta}, [\nabla_{\gamma}, \nabla_{\alpha}]]
+ [\nabla_{\gamma}, [\nabla_{\alpha}, \nabla_{\beta}]] = 0.
\label{eq:jacobi}
\end{equation}

\vspace{1em}
\subsubsection{Identities Involving the Curvature Tensor}

From the definitions of the curvature tensor, torsion tensor, and the covariant derivative, several important identities follow:
\begin{align}
R^{\alpha}{}_{\mu(\nu\rho)} &= 0, \nonumber \\
R^{\mu}{}_{[\alpha\beta\gamma]} - \nabla_{[\alpha} T^{\mu}{}_{\beta\gamma]} + T^{\lambda}{}_{[\alpha\beta} T^{\mu}{}_{\gamma]\lambda} &= 0, \nonumber \\
\nabla_{[\alpha} R^{\mu}{}_{|\nu|\beta\gamma]} - T^{\lambda}{}_{[\alpha\beta} R^{\mu}{}_{|\nu|\gamma]\lambda} &= 0.
\label{eq:curvature_identities}
\end{align}

To relate the curvature tensor $R^{\alpha}{}_{\mu\nu\rho}(\Gamma)$ of a general affine connection to the curvature tensor $R^{\alpha}{}_{\mu\nu\rho}(g)$ of the Levi-Civita connection, we note that adding a $(1,2)$ tensor $\Omega^{\alpha}{}_{\mu\nu}$ to a connection defines a new one,
\begin{equation}
\widehat{\Gamma}^{\alpha}{}_{\mu\nu} = \Gamma^{\alpha}{}_{\mu\nu} + \Omega^{\alpha}{}_{\mu\nu}.
\label{eq:newconnection}
\end{equation}
The curvature tensors corresponding to $\Gamma^{\alpha}{}_{\mu\nu}$ and $\widehat{\Gamma}^{\alpha}{}_{\mu\nu}$ are related by
\begin{equation}
\widehat{R}^{\alpha}{}_{\beta\mu\nu}
= R^{\alpha}{}_{\beta\mu\nu}
+ T^{\lambda}{}_{\mu\nu} \Omega^{\alpha}{}_{\lambda\beta}
+ 2 D_{[\mu} \Omega^{\alpha}{}_{\nu]\beta}
+ 2 \Omega^{\alpha}{}_{[\mu|\lambda|} \Omega^{\lambda}{}_{\nu]\beta},
\label{eq:curvature_relation}
\end{equation}
where $D_{\mu}$ denotes the Levi-Civita covariant derivative.

Choosing $\Gamma^{\alpha}{}_{\mu\nu} = \{^{\alpha}_{\mu\nu}\}$ and $\Omega^{\alpha}{}_{\mu\nu} = K^{\alpha}{}_{\mu\nu} + L^{\alpha}{}_{\mu\nu}$, where $K^{\alpha}{}_{\mu\nu}$ and $L^{\alpha}{}_{\mu\nu}$ are the contortion and disformation tensors respectively, leads to
\begin{align}
\begin{split}
R^{\alpha}{}_{\mu\nu\rho}(\Gamma) & = R^{\alpha}{}_{\mu\nu\rho}(g)
+ T^{\lambda}{}_{\nu\rho} K^{\alpha}{}_{\lambda\mu}
+ 2 D_{[\nu} K^{\alpha}{}_{\rho]\mu}
+ T^{\lambda}{}_{\nu\rho} L^{\alpha}{}_{\lambda\mu}
\\ 
& + 2 D_{[\nu} L^{\alpha}{}_{\rho]\mu} \quad + 2 K^{\alpha}{}_{[\nu|\lambda|} K^{\lambda}{}_{\rho]\mu}
+ 2 L^{\alpha}{}_{[\nu|\lambda|} K^{\lambda}{}_{\rho]\mu} \\
& + 2 K^{\alpha}{}_{[\nu|\lambda|} L^{\lambda}{}_{\rho]\mu}
+ 2 L^{\alpha}{}_{\nu[\lambda} L^{\lambda}{}_{\rho]\mu}.
\label{eq:curvature_decomposition}
\end{split}
\end{align}

The corresponding Ricci scalars are related as
\begin{equation}
R(\Gamma) = R(g) + T + Q + T^{\rho\mu\nu} Q_{\mu\nu\rho}
- T^{\mu} Q_{\mu} + T^{\mu} \bar{Q}_{\mu}
+ D_{\alpha} ( Q^{\alpha} - \bar{Q}^{\alpha} + 2 T^{\alpha} ),
\label{eq:ricci_relation}
\end{equation}
where $T$ and $Q$ denote the torsion and non-metricity scalars, respectively, given by
\begin{align}
T &= \frac{1}{2} \left( \frac{1}{4} T_{\alpha\mu\nu} + \frac{1}{2} T_{\mu\alpha\nu} - g_{\alpha\mu} T_{\nu} \right) T^{\alpha\mu\nu}, \nonumber \\
Q &= \frac{1}{4} Q_{\alpha\mu\nu} Q^{\alpha\mu\nu}
- \frac{1}{2} Q_{\alpha\mu\nu} Q^{\mu\alpha\nu}
- \frac{1}{4} Q_{\alpha} Q^{\alpha}
+ \frac{1}{2} Q_{\alpha} \bar{Q}^{\alpha}.
\label{eq:torsion_nonmetricity_scalars}
\end{align}

Two important specializations of equation (2.77) are:
\begin{align}
R(\Gamma) &= R(g) + T + 2 D_{\alpha} T^{\alpha}, && \text{(Teleparallel case: } Q_{\alpha\mu\nu} = 0), \label{eq:TEGR}\\
R(\Gamma) &= R(g) + Q + D_{\alpha}(Q^{\alpha} - \bar{Q}^{\alpha}), && \text{(Symmetric Teleparallel case: } T^{\alpha}{}_{\mu\nu} = 0).
\label{eq:STEGR}
\end{align}

Lastly, for a general affine connection, the curvature tensor need not be antisymmetric in its first two indices. The corresponding homothetic curvature tensor is defined as $H_{\mu\nu} := R^{\lambda}{}_{\lambda\mu\nu}$. Taking traces of the Bianchi identity gives
\begin{equation}
H_{\mu\nu} = 2 R_{[\mu\nu]} + 2 \nabla_{[\mu} T_{\nu]} + \nabla_{\lambda} T^{\lambda}{}_{\mu\nu} + T^{\lambda} T_{\lambda\mu\nu},
\label{eq:homothetic}
\end{equation}
indicating that $H_{\mu\nu}$ is not an independent curvature component. Similarly, the co-Ricci tensor
\begin{equation}
P_{\mu\nu} := R_{\mu\lambda\nu}{}^{\lambda}
\label{eq:coricci}
\end{equation}
is also not independent, as one finds
\begin{equation}
P_{\mu\nu} = R_{\mu\nu} - 2 \nabla_{[\nu} Q_{\lambda]}{}^{\mu\lambda}.
\label{eq:coricci_relation}
\end{equation}
In the Levi-Civita limit, metric compatibility ensures $P_{\mu\nu} = R_{\mu\nu}$ and $H_{\mu\nu} = 0$, implying that the Ricci tensor is symmetric, as in standard Riemannian geometry.

\vspace{1em}
\section{The Geometrical Trinity of General Relativity}
In 1915, Einstein formulated his General Theory of Relativity, grounding it in Riemannian geometry. He recognized that this then-developing mathematical framework provided an appropriate language to 
\begin{enumerate}
    \item Construct a field-theoretic description of gravity that resolved the action-at-a-distance issue in Newtonian gravity,
    \item Fully express the implications of the equivalence principle, and 
    \item Impose the requirement that the laws of physics should be independent of arbitrary coordinate choices.
\end{enumerate}
The latter idea, now known as the principle of general covariance, was highly unconventional and revolutionary at the time.

Although Einstein did not aim to ``geometrize gravity''~\citep{Lehmkuhl:2014qhu}, his formulation naturally allows 
gravity to be interpreted as an effect of spacetime curvature. This viewpoint has dominated our understanding of gravity for more than a century. However, Riemannian geometry represents only a special case within the broader framework of metric-affine geometry. There is no known physical principle that uniquely mandates the Riemannian description. In fact, gravity admits three mathematically distinct yet physically equivalent formulations within metric-affine geometry, in which gravitational phenomena arise from non-zero curvature, torsion, 
or non-metricity. Together, these three viewpoints constitute the \emph{geometric trinity of gravity}~\cite{Heisenberg:2018vsk, BeltranJimenez:2019esp}.

The following three subsections correspond to the corners of the triangle illustrated in Figure~2.5- \gls{gr}, \gls{tegr} and \gls{stegr}. We also outline how CGR can be viewed as a gauge-fixed version of \gls{stegr}, and how GTEGR gives rise to both \gls{tegr} and \gls{stegr}, effectively forming the lower edge of the triangle 
in Figure~12. We begin by reviewing Einstein's original formulation to establish notation and highlight foundational concepts. This also facilitates comparison with the alternative formulations presented in 
Subsections~2.5.2 (\gls{tegr}) and 2.5.3 (\gls{stegr}). In each formulation, we start from a general metric-affine geometry 
and adopt a systematic approach to constructing the theory and analysing its key features:

\begin{enumerate}
    \item Geometric Postulates,
    \item Form of the Connection,
    \item Construction of the Action Functional,
    \item Metric and Connection Field Equations,
    \item Palatini Formulation of the Action Principle,
    \item Bianchi Identities, and
    \item Counting of Degrees of Freedom.
\end{enumerate}

We depart from this structure in Subsection~2.5.4, where CGR is introduced as the gauge-fixed limit of \gls{stegr}. A different strategy is also adopted in Subsection~2.5.5, which focuses on GTEGR, highlights its distinctive features, and explains how \gls{tegr} and \gls{stegr} emerge from this more general theory.

\begin{figure}[htp!]
\begin{center}
\includegraphics[width=1\textwidth]{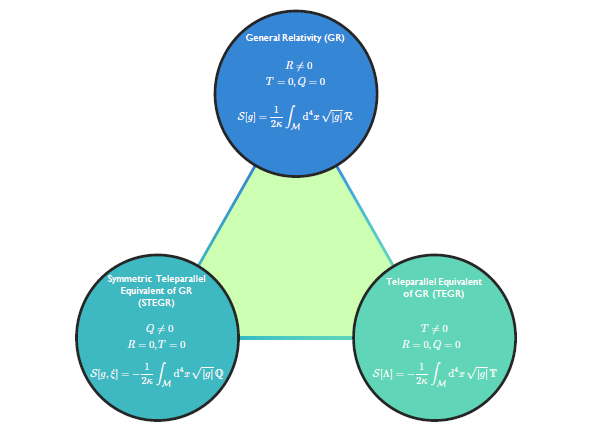}
\caption{Three equivalent formulations of gravity: GR described by curvature, STEGR by non-metricity, and TEGR by torsion.}
\end{center}
\end{figure}

\vspace{1em}
\subsection{Einstein’s Original Formulation of General Relativity}
\vspace{1em}
\subsubsection{Geometric Postulates}
We begin with a metric–affine geometric setup $(M,g,\Gamma)$, without fixing either the underlying manifold $M$ or the metric $g$. Both will later emerge as solutions to Einstein’s field equations. 
However, constructing the theory requires choosing a connection $\Gamma$ (or a suitable class of connections). The fundamental postulate is that the connection satisfies  
\begin{equation}
T^{\alpha}{}_{\mu\nu} = 0, \qquad Q_{\alpha\mu\nu} = 0 ,
\label{eq:torsion_nonmetricity_zero}
\end{equation}
so that curvature $R^{\alpha}{}_{\mu\nu\rho}$ becomes the only non-vanishing geometric quantity relevant 
for the description of spacetime.

\vspace{1em}
\subsubsection{Torsionless and Metric-Compatible Connection}
Using equation(2.84) and relation (2.53), one finds that the unique connection satisfying these conditions is the Levi-Civita connection,
\begin{equation}
\Gamma^{\alpha}{}_{\mu\nu} \equiv 
\left\{ \begin{matrix}
\alpha \\ \mu \nu
\end{matrix}\right\}
= \frac{1}{2} g^{\alpha\lambda}
\left( \partial_{\mu} g_{\nu\lambda} 
+ \partial_{\nu} g_{\mu\lambda}
- \partial_{\lambda} g_{\mu\nu} \right).
\label{eq:LC_connection}
\end{equation}
Since $\Gamma$ is fixed completely by $g_{\mu\nu}$, we will often write the geometry simply as $(M,g)$, with the understanding that the Levi-Civita connection \eqref{eq:LC_connection} is used.

\vspace{1em}
\subsubsection{Action Functional}
General Relativity is defined through the Einstein–Hilbert (EH) action together with the \gls{ghy} boundary term. Adding a cosmological constant and minimally coupled matter fields~$\Psi$, the action reads
\begin{align}
S_{\mathrm{GR}}[g,\Psi] 
&:= S_{\mathrm{EH}}[g] + S_{\mathrm{GHY}}[h] + S_{\mathrm{matter}}[g,\Psi] \nonumber \\
&= \frac{1}{2\kappa} 
\int_{M} d^{4}x \sqrt{|g|}\,(R - 2\Lambda)
+ \frac{1}{\kappa} 
\int_{\partial M} d^{3}y\sqrt{|h|}\,\varepsilon K
+ S_{\mathrm{matter}}[g,\Psi],
\label{eq:GR_action}
\end{align}
where $\kappa = 8\pi G$, $h$ is the induced metric on the boundary~$\partial M$, $K$ is the extrinsic curvature, and $\varepsilon=n_{\mu}n^{\mu} = \pm 1$ depending on whether the boundary is timelike or spacelike. The \gls{ghy} term is required to ensure a well-posed variational principle when $\partial M$ is nonempty.

\vspace{1em}
\subsubsection{Field Equations}
We seek second-order metric field equations of the form
\begin{equation}
E_{\mu\nu}(g) = 8\pi G\, T_{\mu\nu}, \qquad D_{\mu}E^{\mu}{}_{\nu} = 0 ,
\label{eq:EFE_general}
\end{equation}
where $T_{\mu\nu}$ is the energy–momentum tensor. Lovelock’s theorem shows that the only such tensor constructed solely from $g_{\mu\nu}$ is
\begin{equation}
E_{\mu\nu}(g) = G_{\mu\nu} + \Lambda g_{\mu\nu},
\label{eq:LHS_EFE}
\end{equation}
with
\begin{equation}
G_{\mu\nu} = R_{\mu\nu} - \tfrac12 R g_{\mu\nu}.
\end{equation}
Cosmological observations indicate $\Lambda > 0$ and extremely small in magnitude ($\Lambda \sim 10^{-52}\,\mathrm{m}^{-2}$ in SI units). Thus the Einstein field equations take the familiar form
\begin{equation}
R_{\mu\nu} - \frac12 R g_{\mu\nu} + \Lambda g_{\mu\nu}
= \kappa\, T_{\mu\nu}.
\label{eq:EFE_full}
\end{equation}
These equations determine the spacetime geometry $(M,g)$ up to diffeomorphisms, analogous to the gauge freedom of electromagnetism.

\vspace{1em}
\subsubsection{Palatini Formulation}

GR may also be formulated on a general metric–affine geometry $(M,g,\Gamma)$ in which curvature, torsion, and non-metricity may all be nonzero a priori. In the Palatini approach, the metric and connection are treated as independent variables. The action without a cosmological constant is
\begin{equation}
S_{\mathrm{GR}}[g,\Gamma]
= \frac{1}{2\kappa}
\int_{M} d^{4}x \sqrt{|g|}\, g^{\mu\nu} R_{\mu\nu}(\Gamma)
+ S_{\mathrm{matter}} ,
\label{eq:Palatini_action}
\end{equation}
and contains no second derivatives, so no boundary term is necessary.

Varying equation (2.91) with respect to the metric yields
\begin{equation}
R_{(\mu\nu)}(\Gamma)
- \tfrac12 R(\Gamma) g_{\mu\nu}
= \kappa\, T_{\mu\nu}.
\label{eq:Palatini_metric_EFE}
\end{equation}
Variation with respect to the connection, using the Palatini identity,
\begin{equation}
\delta R_{\mu\nu}(\Gamma)
= \nabla_{\alpha}\delta\Gamma^{\alpha}{}_{\nu\mu}
- \nabla_{\nu}\delta\Gamma^{\alpha}{}_{\alpha\mu}
- T^{\alpha}{}_{\beta\nu}\delta\Gamma^{\beta}{}_{\alpha\mu},
\end{equation}
produces algebraic equations fixing~$\Gamma$. Assuming vanishing torsion and hypermomentum, the connection equations reduce to
\begin{equation}
\nabla_{\alpha}\!\left( \sqrt{|g|}\, g^{\mu\nu} \right)
= 0,
\end{equation}
which is equivalent to
\begin{equation}
Q_{\alpha\mu\nu}=0.
\end{equation}
Thus $\Gamma$ is forced to be the Levi-Civita connection. In the general case (with torsion/hypermomentum), the connection is fixed up to a projective transformation, see~\citep{BeltranJimenez:2017doy}.

The important point is that an EH-type action always yields GR, even in the Palatini approach. Constructing teleparallel equivalents therefore requires altering both the geometric setting and the 
action so that it is dynamically equivalent but not identical in form.

\vspace{1em}
\subsubsection{Bianchi Identities}
The integrand $\sqrt{|g|}\,R$ is a scalar density of weight $+1$:
\begin{equation}
\sqrt{|g(x)|}\,R(x)
= \det(J)\,\sqrt{|g(x')|}\,R(x'),
\end{equation}
so the EH action is invariant under arbitrary diffeomorphisms. Considering a one-parameter family of diffeomorphisms $\phi_s$ generated by a vector field $v^\mu$, the Lie derivative gives
\begin{equation}
\phi_s^\ast g_{\mu\nu} = g_{\mu\nu} + s\,\mathcal{L}_v g_{\mu\nu}.
\end{equation}
Diffeomorphism invariance implies
\begin{equation}
\delta_v S_{\mathrm{EH}} = 0
\end{equation}
and using $\mathcal{L}_v g_{\mu\nu}=2\nabla_{(\mu} v_{\nu)}$, one obtains
\begin{equation}
\nabla_\mu G^{\mu}{}_{\nu} = 0,
\end{equation}
the contracted Bianchi identities.

\vspace{1em}
\subsubsection{Counting Degrees of Freedom}
The metric contains ten independent components.  General covariance provides four gauge freedoms (diffeomorphisms), and the Bianchi identities imply that four Einstein equations are constraints rather than dynamical equations. Thus the number of propagating degrees of freedom is
\begin{equation}
10 - 4 - 4 = 2,
\end{equation}
corresponding to the two polarizations of the gravitational field.

\vspace{1em}
\subsection{The Teleparallel Equivalent of General Relativity}

In the previous subsection, we showed that \gls{gr} follows from an action principle together with the assumptions of vanishing torsion and vanishing non-metricity. We also noted that these assumptions can be relaxed: within the Palatini approach, where the connection $\Gamma$ is treated as an independent variable, the field equations enforce that $\Gamma$ reduces to the Levi-Civita connection. The key point is that as long as the action retains the Einstein-Hilbert (EH) form, \gls{gr} naturally re-emerges.

To construct a formulation of gravity that is dynamically equivalent to \gls{gr} but geometrically distinct, one must modify both the geometric setting and the action principle. This leads to the so-called \emph{Teleparallel Equivalent of General Relativity}, abbreviated as \gls{tegr}. In what follows, we outline how \gls{tegr} is obtained.

\vspace{1em}
\subsubsection{Geometric Postulates}
In \gls{tegr}, gravitational effects are attributed to torsion rather than curvature. We begin with a metric-affine geometry $(\mathcal{M},g,\Gamma)$ and impose the postulates
\begin{equation}
    R^{\alpha}{}_{\mu\nu\rho}=0, \qquad Q_{\alpha\mu\nu}=0 .
    \label{eq:TEGR_postulates}
\end{equation}
At first sight, the requirement of vanishing curvature seems to contradict the equivalence with \gls{gr}. The resolution lies in recognizing that the curvature tensor in \eqref{eq:TEGR_postulates} refers to the connection $\Gamma$, not to the curvature of the Levi-Civita connection. The two curvature tensors are related by the identity (2.86). When non-metricity is set to zero, this relation reduces to the expression
\begin{equation}
    R(\Gamma) = R(g) + T + 2 D_{\alpha} T^{\alpha},
    \label{eq:curvature_identity}
\end{equation} where $T$ is the torsion scalar and $D_{\alpha}$ denotes the Levi-Civita covariant derivative. This identity will later allow us to replace curvature with torsion in the action.

The torsion scalar appearing in equation (2.102) is given by
\begin{equation}
    T :=
    \frac{1}{2}
    \left(
    \frac{1}{4} T_{\alpha\mu\nu}
    + \frac{1}{2} T_{\mu\alpha\nu}
    - g_{\alpha\mu} T_{\nu}
    \right)
    T^{\alpha\mu\nu}.
    \label{eq:TorsionScalar}
\end{equation}

\vspace{1em}
\subsubsection{Form of a Flat, Metric-Compatible Connection}
The conditions of flatness and metric-compatibility do not uniquely determine the connection. Instead, they admit an entire class of acceptable connections.

Consider the trivial connection $\hat{\Gamma}^{\alpha}{}_{\mu\nu}=0$, which is manifestly flat. Under a coordinate transformation $\hat{x}^{\mu}\mapsto x^{\mu}(\hat{x})$, the connection transforms
inhomogeneously, giving
\begin{equation}
    \Gamma^{\alpha}{}_{\mu\nu}
    =
    \frac{\partial x^{\alpha}}{\partial \hat{x}^{\lambda}}
    \frac{\partial^{2} \hat{x}^{\lambda}}{\partial x^{\mu}\partial x^{\nu}}.
\end{equation}
If we identify $\partial \hat{x}^{\mu}/\partial x^{\nu} = \Lambda^{\mu}{}_{\nu}$, this becomes
\begin{equation}
    \Gamma^{\alpha}{}_{\mu\nu}
    =
    -(\Lambda^{-1})^{\alpha}{}_{\lambda}\,\partial_{\mu}\Lambda^{\lambda}{}_{\nu}.
    \label{eq:flat_connection}
\end{equation}
Two crucial observations follow:
\begin{enumerate}
    \item If a connection is flat in one coordinate system, it is flat in all coordinate systems.
    \item Any matrix $\Lambda\in GL(4,\mathbb{R})$ inserted into equation (2.105) produces a flat connection.
\end{enumerate}
Next, imposing metric-compatibility $Q_{\alpha\mu\nu}=0$ yields
\begin{equation}
    g_{\lambda(\mu}\partial_{\alpha}\Lambda^{\nu)}{}_{\rho}(\Lambda^{-1})^{\rho}{}_{\lambda}
    = \frac{1}{2}\partial_{\alpha}g_{\mu\nu},
    \label{eq:metricity_compatibility}
\end{equation}
which allows the metric to be written entirely in terms of $\Lambda$.
The torsion tensor corresponding to the flat connection equation (2.105) becomes
\begin{equation}
    T^{\alpha}{}_{\mu\nu}
    =
    2\,(\Lambda^{-1})^{\alpha}{}_{\lambda}\,\partial_{[\mu}\Lambda^{\lambda}{}_{\nu]}.
    \label{eq:torsion_matrix_form}
\end{equation}

\subsubsection{Construction of the Action Functional}
Imposing flatness in equation (2.102) gives
\begin{equation}
    R(g) = -T(\Lambda) - 2 D_{\alpha} T^{\alpha}.
\end{equation}
Substituting this into the EH action would reproduce the original action exactly, since the boundary term $D_{\alpha}T^{\alpha}$ ensures cancellation of the torsion contribution. To obtain a genuinely different but dynamically equivalent action, we simply drop this total derivative.

Thus, the \gls{tegr} action is defined as
\begin{equation}
    S_{\mathrm{TEGR}}[\Lambda]
    :=
    -\frac{1}{2\kappa}
    \int_{\mathcal{M}}
    d^{4}x \,\sqrt{|g|}\,T(\Lambda)
    + S_{\mathrm{matter}}.
    \label{eq:TEGR_action}
\end{equation}
Since $\Lambda$ contains 16 components while $g_{\mu\nu}$ has only 10, the variation of the action yields additional equations. Ten of these reproduce Einstein’s equations, while the remaining six encode Bianchi identities associated with local Lorentz symmetry. An alternative construction relies on building scalars from the torsion tensor. The most general torsion scalar quadratic in $T^{\alpha}{}_{\mu\nu}$ is
\begin{equation}
    \hat{T}
    = c_{1}T_{\alpha\mu\nu}T^{\alpha\mu\nu}
    + c_{2}T_{\mu\alpha\nu}T^{\alpha\mu\nu}
    + c_{3}T_{\mu}T^{\mu},
\end{equation}
and demanding two propagating degrees of freedom fixes
\begin{equation}
    c_{2}=2c_{1}, \qquad c_{3}=-4c_{1},
\end{equation}
reproducing the scalar in equation (2.103).

\vspace{1em}
\subsubsection{Palatini Formulation}
A manifestly covariant form of the \gls{tegr} action is
\begin{equation}
    S_{\mathrm{TEGR}}[g,\Gamma;\tilde{\Pi},\tilde{\chi}]
    =
    -\int_{\mathcal{M}} d^{4}x
    \left[
        \frac{1}{2\kappa}\sqrt{|g|}\,T
        + \tilde{\Pi}^{\alpha}{}_{\mu\nu\rho} R^{\alpha}{}_{\mu\nu\rho}
        + \tilde{\chi}^{\alpha}{}_{\mu\nu} Q_{\alpha\mu\nu}
    \right]
    + S_{\mathrm{matter}},
\end{equation}
where $\tilde{\Pi}$ enforces flatness and $\tilde{\chi}$ enforces metric compatibility.

\vspace{1em}
\subsubsection{Metric and Connection Field Equations}
Variation of the action with respect to $g_{\mu\nu}$, $\Gamma^{\alpha}{}_{\mu\nu}$, $\tilde{\Pi}$, and $\tilde{\chi}$ gives:
\begin{align}
    \frac{\delta S}{\delta g_{\mu\nu}} &= 0, \\
    \frac{\delta S}{\delta \tilde{\Pi}^{\alpha}{}_{\mu\nu\rho}} &= 0, \\
    \frac{\delta S}{\delta \Gamma^{\alpha}{}_{\mu\nu}} &= 0, \\
    \frac{\delta S}{\delta \tilde{\chi}^{\alpha}{}_{\mu\nu}} &= 0.
\end{align} 
The resulting field equations \citep{BeltranJimenez:2018vdo} are
\[
(\nabla_{\alpha}+T_{\alpha}) S^{\alpha}{}_{(\mu\nu)} + t_{\mu\nu}
- \frac{1}{2}T g_{\mu\nu}
= \kappa T_{\mu\nu},
\]
and the connection equation
\[
(\nabla_{\alpha} + T_{\alpha})
\left(
\sqrt{|g|} S^{[\mu}{}_{\alpha}{}^{\nu]}
\right) = 0.
\]
The torsion conjugate is
\[
S^{\alpha}{}_{\mu\nu}
=
\frac{1}{4}T^{\alpha}{}_{\mu\nu}
+ \frac{1}{2}T_{[\mu}{}^{\alpha}{}_{\nu]}
- \delta^{\alpha}{}_{[\mu}T_{\nu]},
\]
and
\[
t_{\mu\nu}
=
\frac{1}{2}S_{(\mu}{}^{\lambda\kappa}T_{\nu)\lambda\kappa}
- T_{\lambda\kappa(\mu} S^{\lambda\kappa}{}_{\nu)}.
\]
The metric equations reproduce Einstein’s equations, while the connection equations reduce to Bianchi identities and carry no dynamical content.

\vspace{1em}
\subsubsection{Bianchi Identities}
For a diffeomorphism-invariant action $S[g,\Gamma]$, the metric-affine Bianchi identity reads \citep{BeltranJimenez:2020sih}
\begin{equation}
    2 D_{\mu}\tilde{M}^{\mu}{}_{\lambda}
    - \hat{\nabla}_{\nu}\hat{\nabla}_{\mu}\tilde{C}^{\lambda\mu\nu}
    + T^{\alpha}{}_{\lambda\nu}\hat{\nabla}_{\mu}\tilde{C}_{\alpha}{}^{\mu\nu}
    + (R^{\alpha}{}_{\nu\mu\lambda}-T_{\mu} T^{\alpha}{}_{\nu\lambda})
    \tilde{C}_{\alpha}{}^{\mu\nu}
    \equiv 0.
\end{equation}
For flat, metric-compatible connections this becomes
\begin{equation}
    \hat{\nabla}_{\nu}\hat{\nabla}_{\mu}\tilde{C}^{\lambda\mu\nu}
    - T^{\alpha}{}_{\lambda\nu}\hat{\nabla}_{\mu}\tilde{C}_{\alpha}{}^{\mu\nu}
    + T_{\mu}T^{\alpha}{}_{\nu\lambda}\tilde{C}_{\alpha}{}^{\mu\nu}
    \equiv 0,
\end{equation}
reflecting the fact that only the connection equations remain nontrivial, and these are identically identities.

\vspace{1em}
\subsubsection{Counting Degrees of Freedom}
A metric $g_{\mu\nu}$ together with a general connection $\Gamma^{\alpha}{}_{\mu\nu}$ carries $10 + 64$ components. Imposing flatness restricts the connection to the form equation (2.105), reducing it to at most $16$ components. Metric-compatibility further constrains $g_{\mu\nu}$ to the form
\begin{equation}
    g_{\mu\nu}
    =
    \Lambda^{\alpha}{}_{\mu}
    \Lambda^{\beta}{}_{\nu}
    c_{\alpha\beta},
\end{equation}
with $c_{\alpha\beta}$ constant. Choosing $c_{\alpha\beta} = \eta_{\alpha\beta}$ leaves the Lorentzian signature
intact.
Flat connections admit the gauge freedom
\[
    \Lambda \rightarrow \Lambda U, \qquad U\in SO^{+}(1,3),
\]
which removes six components. Thus, at most $16-6 = 10$ independent components remain. Finally, diffeomorphism invariance removes $2\times4$ components, leaving precisely two propagating degrees of
freedom-the same number as in \gls{gr}.

\vspace{1em}
\subsection{The Symmetric Teleparallel Equivalent of General Relativity }
\vspace{1em}
\subsubsection{Geometric Postulates}
We now turn to the third geometrical approach to describing gravity, in which gravitational effects are attributed to non-metricity \citep{Nester:1998mp}. This framework is known as the \emph{Symmetric Teleparallel} Equivalent of General Relativity (\gls{stegr}) \citep{BeltranJimenez:2017tkd,}. As before, we begin with a metric-affine space $(M,g,\Gamma)$, but we impose the additional geometric conditions
\begin{equation}
    R^{\alpha}{}_{\mu\nu\rho} = 0, 
    \qquad 
    T^{\alpha}{}_{\mu\nu} = 0.
    \label{eq:ST-postulates}
\end{equation}
The vanishing of the curvature tensor might seem at odds with the well-known role of curvature in \gls{gr}. However, just as in \gls{tegr}, what is eliminated here is the curvature of the independent affine connection $\Gamma$, not that of the Levi-Civita connection, which continues to encode the standard curvature of spacetime.

\vspace{1em}
\subsubsection{Form of a Flat, Torsionless Connection and the Coincident Gauge}
Before introducing an action for \gls{stegr}, let us derive the explicit form of a connection that is both flat and torsion-free. From the previous section, a flat connection can always be written as
\begin{equation}
    \Gamma^{\alpha}{}_{\mu\nu}
      = -\left(\Lambda^{-1}\right)^{\alpha}{}_{\lambda}\,\partial_{\mu}\Lambda^{\lambda}{}_{\nu},
    \label{eq:FlatConnection}
\end{equation}
where $\Lambda^{\mu}{}_{\nu}$ is an element of $\mathrm{GL} (4,\mathbb{R})$. Imposing the vanishing of torsion gives
\begin{equation}
    T^{\alpha}{}_{\mu\nu}=0 
    \quad\Longrightarrow\quad
    \partial_{[\mu}\Lambda^{\alpha}{}_{\nu]}=0.
\end{equation}
This condition implies that $\Lambda^{\alpha}{}_{\nu}$ can be written locally as the Jacobian of four arbitrary functions:
\begin{equation}
    \Lambda^{\alpha}{}_{\nu} = \partial_{\nu}\xi^{\alpha}.
\end{equation}
Thus a flat, torsionless connection takes the form
\begin{equation}
    \Gamma^{\alpha}{}_{\mu\nu}
      = \frac{\partial x^{\alpha}}{\partial \xi^{\lambda}}\,\partial_{\mu}\partial_{\nu}\xi^{\lambda},
    \label{eq:ConnectionFinal}
\end{equation}
where the Jacobian $\partial \xi^{\lambda}/\partial x^{\alpha}$ is assumed to be invertible. A remarkable feature follows immediately from equation (2.124), by choosing coordinates such that
\[
    x^{\mu} = \xi^{\mu},
\]
the connection coefficients vanish identically. This coordinate choice is known as the \emph{coincident gauge}. Importantly, this gauge may always be adopted as long as the conditions $R^\alpha{}_{\mu\nu\rho}=0$ and $T^\alpha{}_{\mu\nu}=0$ hold; no action principle is required to access it. 

\vspace{1em}
\subsubsection{Construction of the Action Functional}
To construct the \gls{stegr} action, we use the identity relating the curvature of the independent connection to that of the Levi-Civita connection:
\begin{equation}
    R(\Gamma)
      = R(g)
      + Q
      + D_{\alpha}\!\left(Q^{\alpha}-\bar{Q}^{\alpha}\right),
    \label{eq:CurvatureIdentity}
\end{equation}
where $D_{\alpha}$ is the Levi-Civita covariant derivative and where the two traces of the non-metricity tensor are
\begin{equation}
    Q_{\alpha} = Q_{\alpha\lambda}{}^{\lambda},
    \qquad
    \bar{Q}_{\alpha} = Q_{\lambda\alpha}{}^{\lambda}.
\end{equation}
The non-metricity scalar is defined as
\begin{equation}
    Q = \frac{1}{4} Q_{\alpha\mu\nu}Q^{\alpha\mu\nu}
      - \frac{1}{2} Q_{\alpha\mu\nu}Q^{\mu\alpha\nu}
      - \frac{1}{4} Q_{\alpha}Q^{\alpha}
      + \frac{1}{2} Q_{\alpha}\bar{Q}^{\alpha}.
    \label{eq:Qscalar}
\end{equation}
In terms of the disformation tensor $L^\alpha{}_{\mu\nu} = \frac{1}{2}\left(Q^\alpha{}_{\mu\nu} - Q_{(\mu}{}^{\alpha}{}_{\nu)}\right)$, it can also be written as
\begin{equation}
    Q 
    = g^{\mu\nu}\left(
      L^{\alpha}{}_{\alpha\beta} L^{\beta}{}_{\mu\nu}
      - L^{\alpha}{}_{\beta\mu} L^{\beta}{}_{\nu\alpha}
    \right).
\end{equation}

Since \gls{stegr} imposes $R(\Gamma)=0$, the above identity gives
\begin{equation}
    R(g)
      = -Q - D_{\alpha}\!\left(Q^{\alpha}-\bar{Q}^{\alpha}\right).
\end{equation}
Replacing $R(g)$ in the Einstein-Hilbert action and discarding the total divergence yields the \gls{stegr} action:
\begin{equation}
    S_{\mathrm{STEGR}}[g,\xi]
      = -\frac{1}{2\kappa}\int_{M}\! d^{4}x\,\sqrt{|g|}\,Q(g,\xi)
      + S_{\mathrm{matter}}.
    \label{eq:STEGRAction}
\end{equation}
The variables of this action are the metric $g_{\mu\nu}$ and the four functions $\xi^{\alpha}$ which encode the flat, torsionless connection.

\vspace{1em}
\subsubsection{Palatini Formulation}
As in \gls{tegr}, one may formulate the action in a Palatini form:
\begin{equation}
\begin{split}
\begin{aligned}
    S_{\mathrm{STEGR}}[g,\Gamma] 
    := -\int_{M} d^{4}x\,
    \Bigg[
        \frac{1}{2\kappa}\sqrt{|g|}\,Q(g,\Gamma)
        + \tilde{\Pi}^{\alpha}{}_{\mu\nu\rho}\,R^{\mu\nu\rho}{}_{\alpha}
        + \tilde{\chi}^{\alpha}{}_{\mu\nu}\,T^{\mu\nu}{}_{\alpha}
    \Bigg]\\
    + S_{\mathrm{matter}},
\end{aligned}
\end{split}
\end{equation}
where the Lagrange multipliers enforce the flatness and torsionlessness of the connection. Variation with respect to these multipliers imposes the geometric postulates, leaving four arbitrary functions $\xi^{\alpha}$ that parametrize the remaining gauge freedom of the connection.

\vspace{1em}
\subsubsection{Field Equations}
The variation of the action yields the metric field equations
\begin{equation}
    2\sqrt{|g|}\,\nabla_{\alpha}
    \left(\sqrt{|g|} P^{\alpha}{}_{\mu\nu}\right)
    + q_{\mu\nu}
    - \frac{1}{2}Q g_{\mu\nu}
    = \kappa T_{\mu\nu},
    \label{eq:MetricEOM}
\end{equation}
where the non-metricity conjugate is
\begin{equation}
    P^{\alpha}{}_{\mu\nu}
      = -\frac{1}{4}Q^{\alpha}{}_{\mu\nu}
        + \frac{1}{2}Q_{(\mu}{}^{\alpha}{}_{\nu)}
        + \frac{1}{4}g_{\mu\nu}Q^{\alpha}
        - \frac{1}{4}\left(
            g_{\mu\nu}\bar{Q}^{\alpha}
            - \delta^{\alpha}{}_{(\mu}Q_{\nu)}
        \right),
\end{equation}
and $q_{\mu\nu}$ is defined by
\begin{equation}
    q_{\mu\nu}
      = P_{(\mu|\alpha\beta}Q_{\nu)}{}^{\alpha\beta}
        - 2P^{\alpha\beta}{}_{(\nu} Q_{\alpha\beta|\mu)}.
\end{equation}
Variation with respect to the affine connection gives
\begin{equation}
    \nabla_{\mu}\nabla_{\nu}
    \left(\sqrt{|g|} P^{\mu\nu}{}_{\alpha}\right)=0.
\end{equation}

\vspace{1em}
\subsubsection{Bianchi Identities}
Diffeomorphism invariance of the action leads to the identities
\begin{equation}
    D_{\mu} M^{\mu}{}_{\nu} + C_{\nu} \equiv 0,
\end{equation}
where
\begin{align}
    M_{\mu\nu} &= 
    2\sqrt{|g|}\,\nabla_{\alpha}
    \left(\sqrt{|g|} P^{\alpha}{}_{\mu\nu}\right)
    + q_{\mu\nu} - \frac{1}{2}Qg_{\mu\nu},\\[4pt]
    C_{\alpha} &= 
    \nabla_{\mu}\nabla_{\nu}
    \left(\sqrt{|g|} P^{\mu\nu}{}_{\alpha}\right).
\end{align}
When the metric field equations are satisfied, the connection equation reduces to an identity:
\[
    C_{\nu}=0.
\]
A notable result is that $M_{\mu\nu}$ coincides with the Einstein tensor $G_{\mu\nu}$ (without cosmological constant), and therefore the connection field equations are identically satisfied, independent of the choice of $\xi^{\alpha}$.

\vspace{1em}
\subsubsection{Counting Degrees of Freedom}
\gls{stegr} propagates the same two physical degrees of freedom as \gls{gr}. Although the formulation uses both $g_{\mu\nu}$ (10 fields) and $\Gamma^{\alpha}{}_{\mu\nu}$ (64 fields), the geometric constraints reduce the independent components of the connection to four functions $\xi^{\alpha}$. These do not satisfy dynamical equations and can be freely specified.  
Hence, only the metric equations carry physical content, and as in GR, diffeomorphism invariance removes $2\times 4$ components, leaving two true degrees of freedom. In this sense, the fields $\xi^{\alpha}$ act as St\"uckelberg variables ensuring covariance.

\vspace{1em}
\subsection{Coincident General Relativity}
\gls{cgr} is a particular realisation of \gls{stegr} obtained by adopting the \emph{coincident gauge}. In this gauge, the flat and torsionless affine connection vanishes everywhere: $\Gamma^{\alpha}{}_{\mu\nu}=0$. Although this may appear to be a simple choice of coordinates, \gls{cgr} has become an influential framework in applications such as $f(Q)$ cosmology~\citep{DAmbrosio:2020nev, Bajardi:2020fxh, Ayuso2021, Frusciante:2021sio, BeltranJimenez:2019tme, Atayde:2021pgb, Anagnostopoulos2021, Capozziello:2022wgl}. Contrasting \gls{cgr} with the full \gls{stegr} theory, as well as with what is often called the Einstein action, helps illuminate the role played by the flat, torsion-free connection or, equivalently, the four functions $\xi^{\mu}$ that parametrise it.

To introduce \gls{cgr}, we start by noting that the coincident gauge implies
\begin{equation}
    \Gamma^{\alpha}{}_{\mu\nu}
    = 
    \left\{^{\alpha}_{\mu\nu}\right\}
    + L^{\alpha}{}_{\mu\nu}
    \Big|_{\mathrm{CG}}
    = 0
    \quad\Longrightarrow\quad
    L^{\alpha}{}_{\mu\nu}\Big|_{\mathrm{CG}}
    = -\left\{^{\alpha}_{\mu\nu}\right\}.
    \label{eq:CGRdisformation}
\end{equation}
The star or gauge label reminds us that this relation is valid only once the coincident gauge is imposed.

Since the \gls{stegr} action can be expressed entirely in terms of the disformation tensor $L^{\alpha}{}_{\mu\nu}$, the \gls{cgr} action becomes
\begin{equation}
\begin{aligned}
    S_{\mathrm{CGR}}[g]
    &\equiv S_{\mathrm{STEGR}}[g,\Gamma = 0] \\
    &= \frac{1}{2\kappa}\!\int_{M} d^{4}x \sqrt{|g|}\,
    g^{\mu\nu}
    \left(
        L^{\alpha}{}_{\alpha\beta} L^{\beta}{}_{\mu\nu}
        - L^{\alpha}{}_{\beta\mu} L^{\beta}{}_{\nu\alpha}
    \right) \\
    &\stackrel{\mathrm{CG}}{=}
    \frac{1}{2\kappa}\!\int_{M} d^{4}x \sqrt{|g|}\,
    g^{\mu\nu}
    \left(
        \left\{^{\alpha}_{\alpha\beta}\right\}
        \left\{^{\beta}_{\mu\nu}\right\}
        -
        \left\{^{\alpha}_{\beta\mu}\right\}
        \left\{^{\beta}_{\nu\alpha}\right\}
    \right).
\end{aligned}
\label{eq:CGRaction}
\end{equation}

The first expression holds for any choice of connection, whereas the final expression applies only in coincident gauge. An important observation is that the \gls{cgr} action depends solely on the metric. This simplification makes \gls{cgr} an attractive entry point for phenomenological studies. Nevertheless, fixing the
coincident gauge from the start may introduce subtleties in nonlinear generalisations such as $f(Q)$ theories, particularly when implementing symmetry assumptions (e.g.\ spherical symmetry or \gls{flrw} symmetry).

A further key feature of \gls{cgr} is that its action coincides with the \emph{Einstein action}, i.e.\ the Einstein-Hilbert action without the second-derivative contributions.  
The Ricci tensor of the Levi-Civita connection is
\begin{equation}
    R_{\mu\nu}
    =
    \partial_{\alpha}\left\{^{\alpha}_{\nu\mu}\right\}
    - \partial_{\nu}\left\{^{\alpha}_{\alpha\mu}\right\}
    + \left\{^{\alpha}_{\alpha\beta}\right\}
      \left\{^{\beta}_{\mu\nu}\right\}
    - \left\{^{\alpha}_{\beta\mu}\right\}
      \left\{^{\beta}_{\nu\alpha}\right\}.
    \label{eq:Ricci_decomposition}
\end{equation}
Comparing equation (2.141) with equation (2.140), the Einstein action corresponds to the Ricci scalar minus the total derivative
\begin{equation}
    g^{\mu\nu}
    \bigg(
        \partial_{\alpha}\left\{^{\alpha}_{\nu\mu}\right\}
        - \partial_{\nu}\left\{^{\alpha}_{\alpha\mu}\right\}
    \bigg).
\end{equation}
Consequently, neither \gls{cgr} nor the Einstein action requires the \gls{ghy} boundary term; both yield a well-defined variational principle.

However, despite their identical appearance, an important distinction exists: the Einstein action is \emph{not} fully diffeomorphism invariant because the connection does not transform as a tensor, and boundary terms arise under coordinate changes. By contrast, \gls{cgr} is derived from the fully covariant \gls{stegr} action after gauge fixing.  From this viewpoint, the affine connection acts as a St\"uckelberg field that restores the diffeomorphism invariance of the Einstein action.

\vspace{1em}
\subsection{General Teleparallel Equivalent of General Relativity}
We have now encountered three geometric formulations of gravity:  
\begin{enumerate}
    \item the curvature-based description of \gls{gr};
    \item \gls{tegr}, which uses torsion in a curvature-free, metric-compatible geometry; and  
    \item \gls{stegr}, which relies on non-metricity in a flat, torsionless geometry.
\end{enumerate}  
These can be visualised as the three vertices of a triangle (Fig.~12). Moving along the lower edge between \gls{tegr} and \gls{stegr} leads to a more general framework known as the \gls{gtegr} \citep{BeltranJimenez:2019odq, Heisenberg:2022mbo, Hohmann:2022mlc}. \gls{tegr} and \gls{stegr} arise as gauge-fixed limits of this broader theory.

To construct \gls{gtegr}, we again begin with a metric-affine manifold $(M,g,\Gamma)$ and impose only one
geometric condition:
\begin{equation}
    R^{\alpha}{}_{\mu\nu\rho} = 0.
    \label{eq:GTEGRflat}
\end{equation}
Using the identity, the Levi-Civita Ricci scalar can be expressed as
\begin{align} \nonumber
-R(g) & = T + Q
    + T^{\rho\mu\nu} Q_{\mu\nu\rho}
    - T_{\mu}Q^{\mu}
    + T_{\mu}\bar{Q}^{\mu}
    + D_{\alpha}\!\left(Q^{\alpha} - \bar{Q}^{\alpha} + 2 T^{\alpha}\right)\\ 
    &= G + D_{\alpha}\!\left(Q^{\alpha} - \bar{Q}^{\alpha} + 2 T^{\alpha}\right),
\end{align}
where we introduced the scalar
\begin{equation}
    G := T + Q + T^{\rho\mu\nu}Q_{\mu\nu\rho}
    - T_{\mu}Q^{\mu} + T_{\mu}\bar{Q}^{\mu}.
\end{equation}

The action of \gls{gtegr} is then defined as
\begin{equation}
    S_{\mathrm{GTEGR}}[g,\Lambda]
    :=
    -\frac{1}{2\kappa}
    \int_{M} d^{4}x \sqrt{|g|}\, G(g,\Lambda)
    + S_{\mathrm{matter}},
\end{equation}
where $\Lambda \in \mathrm{GL}(4,\mathbb{R})$ parametrises the flat connection. Because this differs from the Einstein-Hilbert action only by a total derivative, \gls{gtegr} and \gls{gr} are
dynamically equivalent. The presence of $\Lambda$ does not enlarge the number of propagating degrees of freedom: \begin{center}
 $\delta_{\Lambda} S_{\mathrm{GTEGR}} = 0$   
\end{center} 
holds off-shell, so the connection is non-dynamical and purely gauge. Thus the metric alone carries the two degrees of freedom of \gls{gr}.

If we additionally impose vanishing non-metricity, $Q^{\alpha}{}_{\mu\nu}=0$, then $G$ reduces to the torsion
scalar $T$ and \gls{tegr} is recovered. Likewise, setting torsion to zero yields \gls{stegr}. Hence \gls{tegr} and \gls{stegr} may be viewed as partially gauge-fixed versions of \gls{gtegr}.

\vspace{1em}
\subsection{Matter Coupling}
Up to this point, the equivalence between the various teleparallel theories and \gls{gr} has been discussed only
for the vacuum sector. A natural question is whether the equivalence persists when matter fields are included.

In \gls{gr}, matter couples to gravity via the minimal coupling prescription: replace $\eta_{\mu\nu}\to g_{\mu\nu}$ and $\partial_{\mu}\to D_{\mu}$ in a flat-space matter Lagrangian, provided the matter
fields couple only to $g_{\mu\nu}$, $g^{\mu\nu}$, and $\sqrt{|g|}$, and do not involve derivatives of the metric.

Applying this naively to \gls{tegr} leads to difficulties. For instance, the Maxwell field strength becomes
\begin{equation}
    F_{\mu\nu}
    =
    \nabla_{\mu}A_{\nu}
    -
    \nabla_{\nu}A_{\mu}
    =
    \partial_{\mu}A_{\nu}
    -
    \partial_{\nu}A_{\mu}
    -
    T^{\alpha}{}_{\mu\nu}A_{\alpha},
\end{equation}
and therefore picks up explicit torsion contributions.  
This spoils the equivalence between \gls{tegr} and \gls{gr}. Fermionic fields behave similarly: the Dirac Lagrangian acquires torsion-dependent modifications via the spin connection.

In \gls{stegr}, the situation is different. Because the non-metricity tensor is symmetric in its lower indices, its contribution to the Maxwell field
strength cancels:
\begin{equation}
    F_{\mu\nu}
    =
    \nabla_{\mu}A_{\nu}
    -
    \nabla_{\nu}A_{\mu}
    =
    \partial_{\mu}A_{\nu}
    -
    \partial_{\nu}A_{\mu},
\end{equation}
exactly as in \gls{gr}.  
The coupling of fermions is likewise unaffected by non-metricity. A detailed treatment can be found in~\citep{BeltranJimenez:2020sih}.

In summary, once matter is included, only \gls{stegr} remains fully equivalent to \gls{gr}. \gls{tegr} coupled to matter generally leads to departures from Einsteinian gravity.

\vspace{1em}
\section{The Geometrical Trinity of Modified Gravity Theories}
In this Section, we introduced three geometric formulations of \gls{gr}. This \emph{geometrical trinity} provides several conceptual benefits. For example, both \gls{tegr} and \gls{stegr} admit well-defined variational principles \citep{BeltranJimenez:2019esp, BeltranJimenez:2018vdo} without requiring a GHY boundary term. Moreover, \gls{stegr} and \gls{cgr} have motivated new proposals for defining the long-sought gravitational energy-momentum tensor \citep{BeltranJimenez:2019bnx, Koivisto:2022nar, BeltranJimenez:2017tkd,, BeltranJimenez:2021kpj}. They also allow the computation of black hole entropy without introducing boundary counterterms in the action \citep{BeltranJimenez:2019bnx, BeltranJimenez:2018vdo, Heisenberg:2022nvs}, and the coincident gauge may offer promising insights toward the 
quantization of gravity.

Although \gls{tegr} and \gls{stegr} reproduce Einstein’s field equations exactly, they cannot address phenomena beyond standard \gls{gr}, such as cosmic acceleration or galactic rotation curves. Such issues motivate modified gravity theories, and the geometric trinity may serve as a natural starting point for constructing them. In the literature, two broad strategies are commonly followed:
\begin{enumerate}
    \item Since the actions of \gls{tegr} and \gls{stegr} consist of quadratic contractions of the torsion and non-metricity tensors, respectively, one can generalize them by constructing the most general scalar quadratic in these tensors. For torsion, this leads to a three-parameter family of theories; for non-metricity, a five-parameter class of quadratic actions arises. These generalizations are discussed in subsections~2.6.1 and~2.6.2.

    \item A second widely studied direction is to introduce non-linear generalizations of the form $f(T)$ or $f(Q)$, where $f$ is an arbitrary differentiable function with non-vanishing first derivative. These extensions are reviewed in subsection~2.6.3. Section~2.7 provides a detailed discussion of $f(Q)$ gravity, including its cosmological applications, black hole solutions, and the question of propagating degrees of freedom.
\end{enumerate}
Because these theories are developed within the broader framework of metric-affine geometry, we refer to them collectively as the \emph{geometrical trinity of modified gravity}.

\vspace{1em}
\subsection{Quadratic Actions in Torsion-Based Theories}

Recall from Subsection~2.5.2 that the \gls{tegr} action is constructed from quadratic contractions of the torsion tensor. The torsion scalar is defined as
\begin{equation}
    T :=
    \frac{1}{2}
    \left(
        \frac{1}{4} T^{\alpha}{}_{\mu\nu}
        + \frac{1}{2} T_{\mu\alpha\nu}
        - g_{\alpha\mu} T_{\nu}
    \right)
    T^{\alpha}{}_{\mu\nu}.
    \label{eq:TEGRscalar}
\end{equation}
To construct the most general quadratic scalar from $T^{\alpha}{}_{\mu\nu}$, we analyze the index symmetries. Although a general rank-three tensor allows six distinct self-contractions, the antisymmetry in the lower indices reduces the independent contractions to
\begin{equation}
    T_{\alpha\mu\nu} T^{\alpha\mu\nu},
    \qquad
    T_{\mu\alpha\nu} T^{\alpha\mu\nu}.
\end{equation}
The torsion tensor also has a single non-vanishing trace:
\begin{equation}
    T_{\mu} := T^{\alpha}{}_{\mu\alpha},
\end{equation}
from which we may construct an additional quadratic contraction,
\begin{equation}
    T_{\mu} T^{\mu}.
\end{equation}
Thus, the most general quadratic torsion scalar is a linear combination of the three independent terms:
\begin{equation}
    \widehat{T}
    :=
    c_{1}\, T_{\alpha\mu\nu}T^{\alpha\mu\nu}
    + c_{2}\, T_{\mu\alpha\nu}T^{\alpha\mu\nu}
    + c_{3}\, T_{\mu}T^{\mu},
    \label{eq:generalT}
\end{equation}
with real constants $c_{1}, c_{2}, c_{3}$. \gls{tegr} corresponds to the parameter choice
\[
    c_{1} = \frac{1}{4}, \qquad
    c_{2} = \frac{1}{2}, \qquad
    c_{3} = -1.
\]
Using $\widehat{T}$, the action of \gls{tg} \footnote{\gls{tg} here denotes the three-parameter generalization of \gls{tegr}} takes the form

\begin{equation}
\begin{split}
 S_{\mathrm{TG}}[g,\Gamma,\Psi]
 :=
 -\int_{M} d^{4}x\,
 \left[
   \frac{1}{\kappa} \sqrt{|g|}\,\widehat{T}
   + \widetilde{\Pi}^{\alpha}{}_{\mu\nu\rho} R^{\mu\nu\rho}{}_{\alpha}
   + \widetilde{\chi}^{\alpha}{}_{\mu\nu} Q^{\mu\nu}{}_{\alpha}
 \right]
 \\
 + S_{\mathrm{matter}}[g,\Psi].
\end{split}
\label{eq:TGaction}
\end{equation}

where matter couples minimally and carries vanishing hypermomentum.
The resulting field equations closely resemble those of \gls{tegr}:

\emph{Metric field equations}
\begin{equation}
    (\nabla_{\alpha} + T_{\alpha}) \widehat{S}^{(\mu\nu)\alpha}
    + \widehat{t}^{\mu\nu}
    - \frac{1}{2}\, \widehat{T}\, g^{\mu\nu}
    = \kappa\, T^{\mu\nu}.
\end{equation}

\emph{Connection field equations}
\begin{equation}
    (\nabla_{\alpha} + T_{\alpha})
    \left[
        \sqrt{|g|}\, \widehat{S}^{[\mu|\alpha|\nu]}
    \right]
    = 0,
\end{equation}
where
\begin{align}
    \widehat{S}^{\alpha}{}_{\mu\nu}
    &:= \frac{\partial \widehat{T}}{\partial T_{\alpha}{}^{\mu\nu}}
     = c_{1} T^{\alpha}{}_{\mu\nu}
       + c_{2} T_{[\mu}{}^{\alpha}{}_{\nu]}
       + c_{3} \delta^{\alpha}{}_{[\mu} T_{\nu]}, \\
    \widehat{t}_{\mu\nu}
    &:=
     \frac{\partial \widehat{T}}{\partial g^{\mu\nu}}
     = 
     \frac{1}{2}\widehat{S}_{(\mu|\lambda\kappa|} T_{\nu)}{}^{\lambda\kappa}
     - T_{\lambda\kappa(\mu} \widehat{S}^{\lambda\kappa}{}_{\nu)}.
\end{align}
The Bianchi identities retain the same structure as in \gls{tegr}. Furthermore, the parametrization of the teleparallel connection discussed in Subsection~2.5.2 remains valid, since it is independent of the action.

Although \gls{tg} resembles \gls{tegr}, a crucial distinction lies in their dynamical content. Whereas \gls{tegr} propagates the expected two degrees of freedom, \gls{tg} can propagate additional modes depending on the parameter choices $(c_{1}, c_{2}, c_{3})$. In certain combinations, specific second-order time derivatives of the metric may cancel, turning dynamical equations into constraints or vice versa. Additional integrability conditions may also arise, further modifying the number of degrees of freedom.

A Hamiltonian analysis offers insight into these issues. Studies in \citep{Blixt:2018znp, Blixt:2019ene} (summarized in \citep{Blixt:2020ekl}) examined the primary constraints that appear when particular combinations of $(c_{1}, c_{2}, c_{3})$ vanish. The three-parameter family of theories divides into nine \emph{primary sectors}, each defined by distinct parameter degeneracies. The number of primary constraints in each sector is listed in Table \ref{tab:primary}.

\begin{table}[h]
\centering
\begin{tabular}{|c|c|c|c|c|c|}
\hline
Sector &
$2c_{1}+c_{2}+c_{3}$ &
$2c_{1}-c_{2}$ &
$2c_{1}+c_{2}$ &
$2c_{1}+c_{2}+3c_{3}$ &
PC \\
\hline
0   & $\neq 0$ & $\neq 0$ & $\neq 0$ & $\neq 0$ & 0 \\
\hline
I   & $= 0$     & $\neq 0$ & $\neq 0$ & $\neq 0$ & 3 \\
\hline
II  & $\neq 0$ & $= 0$     & $\neq 0$ & $\neq 0$ & 3 \\
\hline
III & $\neq 0$ & $\neq 0$ & $= 0$     & $\neq 0$ & 5 \\
\hline
IV  & $\neq 0$ & $\neq 0$ & $\neq 0$ & $= 0$     & 1 \\
\hline
V (TEGR) & $= 0$ & $= 0$   & $\neq 0$ & $\neq 0$ & 6 \\
\hline
VI  & $\neq 0$ & $= 0$     & $= 0$     & $\neq 0$ & 8 \\
\hline
VII & $\neq 0$ & $= 0$     & $\neq 0$ & $= 0$     & 4 \\
\hline
VIII& $= 0$     & $\neq 0$ & $= 0$     & $= 0$     & 9\\
\hline
\end{tabular}
\caption{Primary sectors defined by the vanishing of particular parameter combinations in equation (2.153). The number of associated Primary Constraints (PC) is shown in the last column. Section 2.5.2 corresponds to \gls{tegr}. Table adapted from \citep{Blixt:2020ekl} with modified notation following \citep{Dambrosio:2020wbi}.}
\label{tab:primary}
\end{table}
Primary constraints reduce the number of propagating degrees of freedom. However, a complete Hamiltonian classification remains unknown: the full Dirac-Bergmann algorithm cannot be generally applied to teleparallel theories \citep{DAmbrosio:2023asf}.

Finally, \gls{tegr} occupies a special position within the \gls{tg} family. Perturbative analysis around Minkowski spacetime \citep{BeltranJimenez:2019esp, Koivisto:2018loq} shows that consistency requires
\[
    2c_{1}+c_{2}+c_{3}=0,
\]
which reduces the theory to a one-parameter family propagating one extra mode in addition to the graviton. Imposing the further condition
\[
    2c_{1}-c_{2}=0
\]
introduces an additional symmetry, eliminating the extra mode and recovering \gls{tegr}. Relaxing either condition breaks the symmetry and generally introduces unwanted degrees of freedom.

\vspace{1em}
\subsection{Quadratic Actions for Non-Metricity Theories}

In subsection~2.5.3, we introduced the action for \gls{stegr} by constructing the non-metricity scalar
\begin{equation}
Q = \frac{1}{4} Q_{\alpha\mu\nu} Q^{\alpha\mu\nu}
     - \frac{1}{2} Q_{\alpha\mu\nu} Q^{\mu\alpha\nu}
     - \frac{1}{4} Q_\alpha Q^\alpha
     + \frac{1}{2} Q_\alpha \bar{Q}^\alpha .
\end{equation}
We now broaden the framework and define a wider family of theories, collectively referred to as \gls{stg}, by employing the most general scalar quantity that is quadratic in the non-metricity tensor. To construct this, we explore all independent contractions of $Q_{\alpha\mu\nu}$ with itself. Although there are six possible index permutations, the symmetry of the tensor in its last two indices reduces the number of independent contractions to three. Using this symmetry, one finds
that only the following two non-redundant contractions exist:
\begin{equation}
Q_{\alpha\mu\nu} Q^{\alpha\mu\nu},
\qquad
Q_{\mu\alpha\nu} Q^{\alpha\mu\nu}.
\end{equation}
Next, we examine the traces of the non-metricity tensor. Owing to its symmetry properties, there are two distinct traces:
\begin{equation}
Q_\alpha := Q_{\alpha\lambda}{}^{\lambda},
\qquad
\bar{Q}_\alpha := Q^{\lambda}{}_{\lambda\alpha}.
\end{equation}
With these, the remaining quadratic contractions take the form
\begin{equation}
Q_\mu Q^\mu, \qquad
\bar{Q}_\mu \bar{Q}^\mu, \qquad
Q_\mu \bar{Q}^\mu .
\end{equation}
Collecting all such contributions, the most general quadratic scalar that can be formed from the non-metricity tensor is
\begin{equation}
\hat{Q} :=
c_1 Q_{\alpha\mu\nu} Q^{\alpha\mu\nu}
+ c_2 Q_{\mu\alpha\nu} Q^{\alpha\mu\nu}
+ c_3 Q_\mu Q^\mu
+ c_4 \bar{Q}_\mu \bar{Q}^\mu
+ c_5 Q_\mu \bar{Q}^\mu ,
\end{equation}
where $c_1, c_2, c_3, c_4,$ and $c_5$ are arbitrary real constants.

Analogous to the teleparallel case, the action for \gls{stg} is obtained by replacing $Q$ with $\hat{Q}$ in the \gls{stegr} action.
This yields
\begin{equation}
\begin{split}
S_{\rm STG}[g,\Gamma,\Psi]
=
-
\int_{\mathcal{M}} d^4 x
\left[
\frac{1}{\kappa} \sqrt{|g|}\, \hat{Q}
+ \tilde{\Pi}^{\alpha}{}_{\mu\nu\rho} R_{\alpha}{}^{\mu\nu\rho}
+ \tilde{\chi}^{\alpha}{}_{\mu\nu} T_{\alpha}{}^{\mu\nu}
\right]\\
+ S_{\rm matter}[g,\Psi],
\end{split}
\end{equation}
where matter fields $\Psi$ couple minimally and have vanishing hypermomentum $H^{\alpha}{}_{\mu\nu}$.
The structure of the resulting field equations closely resembles that of \gls{stegr}. The metric and connection equations read:
\begin{align}
2 \sqrt{|g|}\, \nabla_\alpha \!\left( \sqrt{|g|}\, \hat{P}^{\alpha}{}_{\mu\nu} \right)
+ \hat{q}_{\mu\nu}
- \frac{1}{2} \hat{Q} g_{\mu\nu}
& = \kappa\, T_{\mu\nu}, \\
 \nabla_\mu \nabla_\nu \left( \sqrt{|g|}\, \hat{P}^{\mu\nu}{}_{\alpha} \right) & = 0.
\end{align}
The conjugate tensor $\hat{P}^{\alpha}{}_{\mu\nu}$ and the symmetric tensor $\hat{q}_{\mu\nu}$ are defined by

\begin{equation}
\begin{split}
\hat{P}^{\alpha}{}_{\mu\nu}=
\frac{1}{2} \frac{\partial \hat{Q}}{\partial Q_{\alpha}{}^{\mu\nu}} 
& =
c_1 Q^{\alpha}{}_{\mu\nu}
+ c_2 Q_{(\mu}{}^{\alpha}{}_{\nu)}
+ c_3 g_{\mu\nu} Q^\alpha \\
& + c_4 \delta^\alpha{}_{(\mu} \bar{Q}_{\nu)}
+ \frac{1}{2} c_5
   \left(
   g_{\mu\nu} \bar{Q}^\alpha
   + \delta^\alpha{}_{(\mu} Q_{\nu)}
   \right),
   \end{split}
   \end{equation}

   \begin{align}
\hat{q}_{\mu\nu}
&=
\frac{\partial \hat{Q}}{\partial g^{\mu\nu}}
=
P_{(\mu|\lambda\kappa} Q_{\nu)}{}^{\lambda\kappa}
- 2 P_{\lambda\kappa(\mu} Q^{\lambda\kappa}{}_{\nu)} .
\end{align}
The scalar $\hat{Q}$ can also be expressed compactly as
\begin{equation}
\hat{Q} = \hat{P}^{\alpha\mu\nu} Q_{\alpha\mu\nu}.
\end{equation}
Because the action in equation (2.164) is diffeomorphism invariant, its Bianchi identities take the form
\begin{equation}
D_\nu \hat{M}^{\nu}{}_{\mu} + \hat{C}_{\mu} \equiv 0,
\end{equation}
where
\begin{align}
\hat{M}_{\mu\nu}
&=
2 \sqrt{|g|}\, \nabla_\alpha
\!\left( \sqrt{|g|}\, \hat{P}^{\alpha}{}_{\mu\nu} \right)
+ \hat{q}_{\mu\nu}
- \frac{1}{2} \hat{Q} g_{\mu\nu}, \\
\hat{C}_\alpha
&=
\nabla_\mu \nabla_\nu
\left( \sqrt{|g|}\, \hat{P}^{\mu\nu}{}_{\alpha} \right).
\end{align}
Thus, whenever the metric equations are satisfied,
\begin{equation}
\hat{M}_{\mu\nu} = \kappa T_{\mu\nu}
\quad \Longrightarrow \quad
D_\nu \hat{M}^{\nu}{}_{\mu} = 0
\quad \Longrightarrow \quad
\hat{C}_\mu \equiv 0,
\end{equation}
meaning the connection equations follow automatically.

A key difference between \gls{stegr} and generic \gls{stg} models lies in their physical degrees of freedom. \gls{stegr} propagates the same two modes as \gls{gr}. In contrast, the number of degrees of freedom in \gls{stg} depends on the specific choice of parameters $c_i$. As in \gls{tg}, certain combinations of $c_i$ eliminate second-order time derivatives in the field equations, thereby converting some dynamical equations into constraints and reducing the number of propagating modes.
A Hamiltonian analysis was initiated in~\citep{Dambrosio:2020wbi}. After performing an ADM decomposition and adopting the coincident gauge, the kinetic part of the \gls{stg} Lagrangian becomes
\begin{align}
\begin{split}
L_{\rm kinetic}
=
- \sqrt{h}
\Bigg[
\frac{2\tilde{c}}{N^3} \dot{N}^2
+ \frac{c_{35}}{N^2} \dot{N}\, h^{ab}\dot{h}_{ab}
- \frac{\hat{c}}{2 N^3} h_{ab} \dot{N}^a \dot{N}^b\\
+ \frac{1}{2N}
\left(
c_1 h^{ac} h^{bd} \dot{h}_{ad} \dot{h}_{cb}
+ c_3 h^{ac} h^{bd} \dot{h}_{ac} \dot{h}_{bd}
\right)
\Bigg],
\end{split}
\end{align}
where $\tilde{c} = c_1 + c_2 + c_3 + c_4 + c_5$, $\hat{c} = 2c_1 + c_2 + c_4$, and $c_{35} = 2c_3 + c_5$.
The canonical momenta take the form
\begin{align}
\tilde{\pi}
&=
-\frac{\sqrt{h}}{N^2}
\left(
4\tilde{c}\, \dot{N}
+ c_{35} N\, h^{ab}\dot{h}_{ab}
\right)
+ \text{(no time derivatives)}, \\
\tilde{\pi}_a
&=
\frac{\sqrt{h}}{N^3}
\hat{c}\, h_{ab} \dot{N}^b
+ \text{(no time derivatives)}, \\
\tilde{\pi}^{ab}
&=
-\frac{\sqrt{h}}{N^2}
\left[
c_1 h^{ac} h^{bd} \dot{h}_{cd} N
+ h^{ab}
\big( c_2 h^{cd} \dot{h}_{cd} N + c_{35} \dot{N} \big)
\right]
+ \text{(no time derivatives)}.
\end{align}
The momentum conjugate to $N$ becomes a primary constraint when
\[
\tilde{c} = 0,
\qquad
c_{35} = 0.
\]
Likewise, the momentum conjugate to the shift becomes a constraint when
\[
\hat{c} = 0.
\]
These correspond to sectors in Table 2.1 and Table 2.2.
A full classification of the kinetic matrix shows that its determinant is
\begin{equation}
\det K
=
\frac{8 h^2}{N^{18}}
c_5^1
\hat{c}^3
\left[ 3 c_{35}^2 - 4 (c_1 + 3 c_3) \tilde{c} \right].
\end{equation}
Setting $\det K = 0$ gives four primary sectors:
\begin{align}
\text{Sector I:} & \qquad \qquad  \tilde{c} = 0, \ c_{35}=0 \\
\text{Sector II:} & \qquad \qquad  \hat{c}=0 \\
\text{Sector III:} & \qquad \qquad  c_1 = 0 \\
\text{Sector IV:} & \qquad \qquad  c_3 = - \frac{c_1}{3} + \frac{c_{35}^2}{4\tilde{c}} .
\end{align}

\begin{table}[h]
\centering
\begin{tabular}{|c|c|c|c|c|c|c|}
\hline
\textbf{Sector} & $\tilde{c}$ & $\hat{c}$ & $c_{35}$ & $c_{1}$ & $\dfrac{c_1 c_3 - c_2}{35}$ & \textbf{Constraints} \\
\hline
0   & $\neq 0$ & $\neq 0$ & $\neq 0$ & $\neq 0$ & $\neq 0$ & 0 \\
\hline
I   & $=0$    & $\neq 0$ & $=0$     & $\neq 0$ & N/A       & 1 \\
\hline
II  & $\neq 0$ & $=0$    & $\neq 0$ & $\neq 0$ & $\neq 0$ & 3 \\
\hline
III & $\neq 0$ & $\neq 0$ & $\neq 0$ & $=0$     & $\neq 0$ & 5 \\
\hline
IV  & $\neq 0$ & $\neq 0$ & $\neq 0$ & $\neq 0$ & $=0$     & 1 \\
\hline
V (STEGR) & $=0$ & $=0$ & $=0$ & $\neq 0$ & N/A & 4 \\
VI  & $=0$    & $\neq 0$ & $=0$     & $=0$     & N/A       & 6 \\
\hline
VII & $\neq 0$ & $=0$    & $\neq 0$ & $=0$     & $\neq 0$ & 8 \\
\hline
VIII & $\neq 0$ & $=0$    & $\neq 0$ & $\neq 0$ & $=0$     & 4 \\
\hline
IX   & $\neq 0$ & $\neq 0$ & $\neq 0$ & $=0$     & $=0$     & 6 \\
\hline
X    & $=0$    & $=0$    & $=0$      & $=0$     & N/A       & 10 \\
\hline
XI   & $\neq 0$ & $=0$    & $\neq 0$ & $=0$     & $=0$     & 9 \\
\hline
\end{tabular}
\caption{Primary sectors defined by the vanishing of specific parameter combinations $\tilde{c}$, $\hat{c}$, and $c_{35}$. The final column lists the number of independent primary constraints. \gls{stegr} corresponds to Section 2.5.}
\label{tab:primary_sectors}
\end{table}

Evaluating $10 - \text{rank}(K)$ yields the number of primary constraints in each case: $1, 3, 5,$ and $1$, respectively. Combinations of these conditions generate the full set of sectors summarized in Table 2.2, with \gls{stegr} sitting in section 2.5, which reproduces the correct two degrees of freedom of \gls{gr}.

However, just as in teleparallel gravity, a full constraint analysis remains incomplete. The Dirac-Bergmann algorithm faces technical obstacles, making the identification of secondary constraints challenging. As a result, the precise number of propagating modes in many sectors remains unknown. A particularly interesting feature of \gls{stegr} is its enlarged symmetry structure. Expanding around
Minkowski spacetime,
\[
g_{\mu\nu} = \eta_{\mu\nu} + h_{\mu\nu}, \qquad |h_{\mu\nu}| \ll 1,
\]
leads to a quadratic Lagrangian
\begin{equation}
L =
c_1 \partial_\alpha h_{\mu\nu} \partial^\alpha h^{\mu\nu}
+ (c_2 + c_4) \partial_\alpha h_{\mu\nu} \partial^\mu h^{\alpha\nu}
+ c_3 \partial_\alpha h \partial^\alpha h
+ c_5 \partial_\mu h^{\mu\nu} \partial_\nu h.
\end{equation}

Requiring invariance under transverse diffeomorphisms,
\[
h_{\mu\nu} \rightarrow h_{\mu\nu} + 2 \partial_{(\mu} \xi_{\nu)}, \qquad \partial_\mu \xi^\mu = 0,
\]
imposes the relation
\begin{equation}
2 c_1 + c_2 + c_4 = 0,
\end{equation}
which is satisfied by the \gls{stegr} parameters. Full linearized diffeomorphism symmetry,
\[
h_{\mu\nu} \rightarrow h_{\mu\nu} + 2 \partial_{(\mu} \xi_{\nu)},
\]
further leads to the conditions
\begin{equation}
2c_1 = -2c_3 = c_5.
\end{equation}
Together, these imply
\begin{equation}
c_3 = -c_1,
\qquad
c_4 = -2c_1 - c_2,
\qquad
c_5 = 2c_1,
\end{equation}
or equivalently
\begin{equation}
\tilde{c} = 0,
\qquad
\hat{c} = 0,
\qquad
c_{35} = 0,
\end{equation}
which precisely define section 2.5 in Table~2.2. Thus \gls{stegr} is singled out by its symmetry properties and its healthy spectrum of degrees of freedom.

Imposing instead a linearized Weyl symmetry,
\[
h_{\mu\nu} \rightarrow h_{\mu\nu} + \phi\, \eta_{\mu\nu},
\]
along with transverse diffeomorphisms, gives
\begin{equation}
c_3 = -\frac{3}{8} c_1,
\qquad
c_5 = 2c_1.
\end{equation}
This corresponds to linearized unimodular gravity, equivalent to GR with the additional constraint $\sqrt{|g|}=1$, in which the cosmological constant appears as an integration constant. The
classification in Table 2.2, however, was obtained without imposing the unimodular constraint $\sqrt{|g|}=1$.

\vspace{1em}
\section{Non-Linear Extensions: \texorpdfstring{$f(R)$, $f(T)$, $f(Q)$ and $f(G)$}{f(R), f(T), f(Q), f(G)} Theories}

As outlined in Section 2.5, gravitational dynamics can be formulated through a geometric trinity: Einstein's curvature-based description (\gls{gr}), the torsion-based (\gls{tegr}), and the non-metricity-based (\gls{stegr}). Although these arise from different geometric structures-curvature, torsion, and non-metricity, they all reproduce Einstein's field equations and propagate the same physical degrees of freedom. The \gls{gtegr} unifies the torsional and non-metricity formulations and remains dynamically equivalent to \gls{gr}.

Each element of the trinity follows from a variational principle. The corresponding gravitational actions are
\begin{align}
S_{\mathrm{EH}}[g] &= \frac{1}{2\kappa} \int_{M} \sqrt{|g|}\, R\, d^4x, \\
S_{\mathrm{TEGR}}[\Lambda] &= -\frac{1}{2\kappa} \int_{M} \sqrt{|g|}\, T\, d^4x, \\
S_{\mathrm{STEGR}}[g,\xi] &= -\frac{1}{2\kappa} \int_{M} \sqrt{|g|}\, Q\, d^4x, \\
S_{\mathrm{GTEGR}}[g,\Lambda] &= -\frac{1}{2\kappa} \int_{M} \sqrt{|g|}\, G\, d^4x .
\end{align}

Although these actions are dynamically equivalent, they are not identical: they differ by boundary terms and depend on different dynamical fields. This observation motivates a broader generalization in which the scalar quantities $R$, $T$, $Q$, and $G$ are replaced by arbitrary functions. One therefore considers the non-linear actions
\begin{align}
S_{f(R)}[g] &:= \frac{1}{2\kappa} \int_{M} \sqrt{|g|}\, f(R)\, d^4x, \\
S_{f(T)}[\Lambda] &:= -\frac{1}{2\kappa} \int_{M} \sqrt{|g|}\, f(T)\, d^4x, \\
S_{f(Q)}[g,\xi] &:= -\frac{1}{2\kappa} \int_{M} \sqrt{|g|}\, f(Q)\, d^4x, \\
S_{f(G)}[g,\Lambda] &:= -\frac{1}{2\kappa} \int_{M} \sqrt{|g|}\, f(G)\, d^4x .
\end{align}

The underlying idea is that the additional freedom provided by the functional form of $f$ may account for phenomena such as cosmic acceleration or structure formation without invoking dark matter or dark energy within the standard \gls{gr} framework. Since the original actions differ only by boundary terms, their non-linear extensions are no longer equivalent and each produces distinct field equations and degrees of freedom. 

\vspace{1em}
\subsection{\texorpdfstring{$f(R)$}{f(R)} Gravity}

Among the nonlinear generalizations, $f(R)$ gravity is the most extensively studied, dating back to Buchdahl's work \citep{Buchdahl:1970ynr}. Varying the action yields
\begin{equation}
f'(R) R_{\mu\nu} - \frac{1}{2} f(R) g_{\mu\nu}
+ \left( g_{\mu\nu} \Box - D_\mu D_\nu \right) f'(R)
= \kappa\, T_{\mu\nu},
\end{equation}
where $f'(R) = \frac{df}{dR}$ and $\Box = g^{\mu\nu} D_\mu D_\nu$. Setting $f(R)=R$ retrieves 
Einstein's field equations. For $f''(R)\neq 0$, these equations become fourth-order in the metric, but the theory remains free of pathological modes: it propagates two tensorial modes (the graviton) and an additional scalar degree of freedom \citep{Sotiriou:2008rp, Capozziello2011}.

\vspace{1em}
\subsection{\texorpdfstring{$f(T)$}{f(T)} Gravity}

The action for $f(T)$ gravity coupled to matter is
\begin{equation}
S_{f(T)}[g,\Gamma] := -\frac{1}{2\kappa} \int_{M} \sqrt{|g|}\, f(T)\, d^4x + S_{\mathrm{matter}}[g,\Psi].
\end{equation}
The resulting field equations split into metric and connection parts:
\begin{align}
(\nabla_\alpha + T_\alpha)\!\left[f'(T) S_{(\mu\nu)}{}^{\alpha}\right]
+ f'(T) t_{\mu\nu} - \frac{1}{2} f(T) g_{\mu\nu} &= \kappa T_{\mu\nu}, \\
(\nabla_\mu + T_\mu)\!\left[f'(T) S^{[\alpha}{}_{\mu}{}^{\beta]}\right] &= 0 .
\end{align}
Unlike $f(R)$ gravity, these equations are second order. In the limit $f(T)=T$, \gls{tegr} is recovered. A commonly used rearrangement is
\begin{equation}
f'(T)G_{\mu\nu}
-\frac{1}{2}\!\left[f(T)-f'(T)T\right] g_{\mu\nu}
+ f''(T) S_{(\mu\nu)}{}^{\alpha} \partial_{\alpha} T
= \kappa T_{\mu\nu}.
\end{equation}
The number of degrees of freedom is still under debate, with estimates ranging between three \citep{Ferraro:2018tpu} and five \citep{Li:2011rn, Blagojevic:2020dyq}.

\vspace{1em}
\subsection{\texorpdfstring{$f(Q)$}{f(Q)} Gravity}

The action for $f(Q)$ gravity is
\begin{equation}
S_{f(Q)}[g,\Gamma] := -\frac{1}{2\kappa} \int_{M} \sqrt{|g|}\, f(Q)\, d^4x + S_{\mathrm{matter}}[g,\Psi].
\end{equation}
The field equations take the form \citep{DAmbrosio:2021zpm, DAmbrosio:2021pnd}
\begin{align}
2\sqrt{|g|}\, \nabla_{\alpha}\!\left[ \sqrt{|g|}\, f'(Q) P^{\alpha}{}_{\mu\nu} \right]
+ f'(Q) q_{\mu\nu}
- \frac{1}{2} f(Q) g_{\mu\nu}
&= \kappa T_{\mu\nu}, \\
\nabla_{\mu}\nabla_{\nu}\!\left( \sqrt{|g|}\, f'(Q) P^{\mu\nu}{}_{\alpha} \right) &= 0.
\end{align}
These equations resemble those of \gls{stegr}, but can be rewritten as
\begin{equation}
f'(Q) G_{\mu\nu}
- \frac{1}{2}\!\left[f(Q)-f'(Q)Q\right] g_{\mu\nu}
+ 2 f''(Q) P^{\alpha}{}_{\mu\nu} \partial_{\alpha} Q
= \kappa T_{\mu\nu}.
\end{equation}
The last term shows explicitly how the dynamics deviate from \gls{stegr} and \gls{gr} when $f''(Q)\neq 0$. The number of propagating degrees of freedom is presently unknown. 

A useful simplification arises when the non-metricity scalar is constant,
\begin{equation}
Q = Q_0 = \text{const},
\end{equation}
which reduces the metric equations to
\begin{equation}
G_{\mu\nu}
= \Lambda_{\mathrm{eff}} g_{\mu\nu}
+ \kappa\, \widehat{T}_{\mu\nu},
\end{equation}
where
\begin{equation}
\Lambda_{\mathrm{eff}}
= -\frac{1}{2}\frac{f(Q_0) - f'(Q_0)Q_0}{f'(Q_0)}, 
\qquad
\widehat{T}_{\mu\nu}
= \frac{1}{f'(Q_0)} T_{\mu\nu}.
\end{equation}
Thus \gls{gr}-like solutions can emerge even in the non-linear regime.

A generalized Bianchi identity for $f(Q)$ gravity reads
\begin{equation}
D_{\mu} M^{\mu}{}_{\nu} + C_{\nu} \equiv 0,
\end{equation}
with
\begin{align}
M_{\mu\nu} &= 2\sqrt{|g|}\,\nabla_{\alpha}\!\left[ \sqrt{|g|}\, f'(Q) P^{\alpha}{}_{\mu\nu} \right]
+ f'(Q) q_{\mu\nu}
- \frac{1}{2} f(Q) g_{\mu\nu}, \\
C_{\alpha} &= \nabla_{\mu}\nabla_{\nu}\!\left( \sqrt{|g|}\, f'(Q) P^{\mu\nu}{}_{\alpha} \right).
\end{align}
When the metric field equations hold ($M_{\mu\nu}=\kappa T_{\mu\nu}$), the connection equations follow automatically.

\vspace{1em}
\subsection{\texorpdfstring{$f(G)$}{f(G)} Gravity}

\gls{gtegr} contains \gls{tegr} and \gls{stegr} as gauge-fixed sectors. Therefore, a non-linear extension $f(G)$ naturally generalizes both $f(T)$ and $f(Q)$ theories. Following the formulation i n~\cite{Heisenberg:2022mbo, Hohmann:2022mlc},
one introduces the auxiliary tensors
\begin{align}
M^{\alpha}{}_{\mu\nu} &= \Gamma^{\alpha}{}_{\mu\nu} - \{^{\alpha}{}_{\mu\nu}\}
= K^{\alpha}{}_{\mu\nu} + L^{\alpha}{}_{\mu\nu}, \\
Z^{\alpha}{}_{\mu\nu} &= -M^{\alpha}{}_{\mu\nu}
- M_{\nu\mu}{}^{\alpha}
+ M^{\rho}{}_{\alpha\rho} g_{\mu\nu}
+ M_{\mu}{}^{\rho}{}_{\rho} \delta^{\alpha}_{\nu}.
\end{align}

The field equations of $f(G)$ gravity then take the compact form
\begin{align}
\begin{split}
f'(G) G_{\mu\nu}
- \frac{1}{2}\!\left[f(G)-f'(G)G\right] g_{\mu\nu}
+ D_{(\mu}\!\left[ f'(G) M^{\sigma}{}_{\nu)\sigma} \right]\\
+ f''(G)\Big(
M^{[\rho\sigma]}{}_{\sigma} g_{\mu\nu} - M_{\rho(\mu\nu)}
\Big)\partial_{\rho}G
&= \kappa T_{\mu\nu},
\end{split}
\end{align}
\begin{align}
\nabla_{\rho}\!\left[f'(G) Z^{\mu}{}_{\nu\rho}\right]
- f'(G) M^{\lambda}{}_{\rho\lambda} Z^{\mu}{}_{\nu\rho}
&= 0.
\end{align}

Setting either $T^{\alpha}{}_{\mu\nu}=0$ or $Q^{\alpha}{}_{\mu\nu}=0$ retrieves the $f(T)$ or $f(Q)$ field equations, respectively. As before, if $G=G_0$ is constant, one recovers the \gls{gr}-like form
\begin{equation}
G_{\mu\nu}
= \Lambda_{\mathrm{eff}} g_{\mu\nu}
+ \kappa\, \widehat{T}_{\mu\nu}.
\end{equation}
Since $f(G)$ gravity does not rely on gauge-fixing conditions, it may allow for richer classes of beyond \gls{gr} solutions. Preliminary cosmological solutions have been explored in \citep{Heisenberg:2022mbo}.

\vspace{1em}
\section{$f(Q)$ gravity and its extensions}
This section is dedicated to exploring the geometric structure associated with non-metricity, which plays a central role in symmetric teleparallel gravity. In a manifold equipped with a metric $g_{\mu\nu}$ and a general affine connection, one can define the inner product of two vectors $u^\mu$ and $v^\mu$ as $u \cdot v = u^\mu v^\nu g_{\mu\nu}$. During parallel transport of these vectors along a curve $C$ parameterized by $x^\mu(\lambda)$, the presence of non-metricity affects the inner product, as revealed by taking the covariant derivative of the metric tensor. If the vectors are parallel transported such that 
\begin{equation}
 \frac{dx^\alpha}{d\lambda} \tilde{\nabla}_\alpha u^\mu = 0  \quad \text{and}  \quad  \frac{dx^\alpha}{d\lambda} \tilde{\nabla}_\alpha v^\mu = 0 ,
\end{equation}
then the derivative of the inner product simplifies to 
\begin{equation}
    \tilde{\nabla}_\lambda(u \cdot v) = Q_{\alpha\mu\nu} \frac{dx^\alpha}{d\lambda} u^\mu v^\nu,
\end{equation} 
where $Q_{\alpha\mu\nu} = \tilde{\nabla}_\alpha g_{\mu\nu}$ denotes the non-metricity tensor. This indicates that in spacetimes with non-metricity, the inner product and hence the magnitude of a vector is not conserved under parallel transport.

In metric-affine geometry, the affine connection $\Gamma^\lambda_{\mu\nu}$ and the metric $g_{\mu\nu}$ are treated independently, giving rise to a broader class of geometric and physical models than those allowed by the metric-compatible connections of \gls{gr}. The covariant derivative of a mixed tensor $T^\lambda_\nu$ takes the general form:
\begin{equation}
\nabla_\mu T^\lambda_\nu = \partial_\mu T^\lambda_\nu + \Gamma^\lambda_{\mu\alpha} T^\alpha_\nu - \Gamma^\alpha_{\mu\nu} T^\lambda_\alpha.
\end{equation}
The affine connection $\Gamma^\lambda_{\mu\nu}$ can be decomposed into three components:
\begin{equation}
\Gamma^\lambda_{\mu\nu} = \{^\lambda_{\mu\nu}\} + K^\lambda_{\mu\nu} + L^\lambda_{\mu\nu},
\end{equation}
where $\{^\lambda_{\mu\nu}\}$ is the Levi-Civita connection (derived from the metric), $K^\lambda_{\mu\nu}$ is the contortion tensor representing torsion, and $L^\lambda_{\mu\nu}$ is the disformation tensor associated with non-metricity. These components are defined respectively as:
\begin{itemize}
    \item \textit{Levi-Civita connection}
    \begin{equation}
      \{^\lambda_{\mu\nu}\} = \frac{1}{2} g^{\lambda\beta} (\partial_\mu g_{\beta\nu} + \partial_\nu g_{\beta\mu} - \partial_\beta g_{\mu\nu}),  
    \end{equation}
    \item \textit{Contortion}
    \begin{equation}
        K^\lambda_{\mu\nu} = \frac{1}{2} g^{\lambda\beta}(T_{\mu\beta\nu} + T_{\nu\beta\mu} + T_{\beta\mu\nu}),
    \end{equation}
    \item \textit{Disformation}
    \begin{equation}
      L^\lambda_{\mu\nu} = \frac{1}{2} g^{\lambda\beta}(-Q_{\mu\beta\nu} - Q_{\nu\beta\mu} + Q_{\beta\mu\nu}).  
    \end{equation}
\end{itemize}

In this framework, three independent geometric quantities arise:
\begin{itemize}
    \item \textit{Torsion:} $T^\lambda_{\mu\nu} = \Gamma^\lambda_{\mu\nu} - \Gamma^\lambda_{\nu\mu}$, which measures the antisymmetry of the connection,
    \item \textit{Curvature:} $R^\sigma_{\rho\mu\nu}$, which describes the failure of parallel transport to preserve vector orientation and is given by:
    \begin{equation}
        R^\sigma_{\rho\mu\nu} = \partial_\mu \Gamma^\sigma_{\nu\rho} - \partial_\nu \Gamma^\sigma_{\mu\rho} + \Gamma^\alpha_{\nu\rho} \Gamma^\sigma_{\mu\alpha} - \Gamma^\alpha_{\mu\rho} \Gamma^\sigma_{\nu\alpha},
    \end{equation}
    \item \textit{Non-metricity:} arising from the failure of the metric to be covariantly constant, $\nabla_\alpha g_{\mu\nu} \neq 0$.
\end{itemize}

When the geometry is defined primarily by non-metricity, as in symmetric teleparallel gravity, the torsion and curvature are assumed to vanish, and the gravitational dynamics are governed by the non-metricity scalar $Q$. The most general even-parity, second-order scalar constructed from non-metricity can be written as:
\begin{equation}
Q = \frac{1}{4} Q_{\alpha\beta\gamma} Q^{\alpha\beta\gamma} - \frac{1}{2} Q_{\alpha\beta\gamma} Q^{\beta\alpha\gamma} - \frac{1}{4} Q_\alpha Q^\alpha + \frac{1}{2} Q_\alpha \tilde{Q}^\alpha,
\end{equation}
where $Q_\alpha = Q_{\alpha\mu}^{\ \ \ \mu}$ and $\tilde{Q}_\alpha = Q_{\mu\alpha}^{\ \ \ \mu}$.
The general action for symmetric teleparallel theories of gravity, incorporating Lagrange multipliers to enforce vanishing curvature and torsion, takes the form:
\begin{equation}
S_Q = \int d^4x \left[ \frac{\sqrt{-g}}{16\pi G_N} Q + \lambda^\alpha_{\ \beta\mu\nu} R^\beta_{\ \alpha\mu\nu} + \lambda^\alpha_{\ \mu\nu} T^\mu_{\ \nu\alpha} \right].
\end{equation}

A key result in the formulation of \gls{stegr} is the relation:
\begin{equation}
R(\Gamma) = R(g) + Q + \nabla_\alpha(Q^\alpha - \tilde{Q}^\alpha),
\end{equation}
which shows that the curvature scalar of the affine connection $R(\Gamma)$ differs from the Riemannian Ricci scalar $R(g)$ by a combination of the non-metricity scalar and a total derivative. By imposing the condition $R(\Gamma) = 0$, one obtains:
\begin{equation}
R(g) = -Q - \nabla_\alpha(Q^\alpha - \tilde{Q}^\alpha),
\end{equation}
leading to the STEGR action:
\begin{equation}
S_{\text{STEGR}} = \frac{1}{16\pi G_N} \int d^4x \sqrt{-g} Q.
\end{equation}

Due to the divergence term being a total derivative, it does not affect the field equations, rendering the dynamics of \gls{stegr} equivalent to those of \gls{gr}, though realized through a torsion-free and curvature-free geometry. This formulation highlights the unique symmetry properties of the non-metricity scalar $Q$, distinguishing it from the Ricci scalar by a boundary term. As such, \gls{stegr} belongs to a broader class of symmetric teleparallel gravity theories that provide an alternative but dynamically equivalent description of gravitation compared to \gls{gr}.

\vspace{1em}
\subsection{$f(Q)$ gravity} 
The $f(Q)$ theory of gravity extends the \gls{stegr} by promoting the non-metricity scalar $Q$ to a general function $f(Q)$ in the action. The modified gravitational action is given by
\begin{equation}
S = \int d^4x \, \sqrt{-g} \left[ \frac{1}{2\kappa} f(Q) + \mathcal{L}_m \right],
\end{equation}
where $\kappa$ is the gravitational constant, $f(Q)$ is an arbitrary function of the non-metricity scalar $Q$, and $\mathcal{L}_m$ is the matter Lagrangian. For $f(Q) = Q$, the action reduces to the form of \gls{stegr}, recovering \gls{gr}.
Variation of this action with respect to the metric $g_{\mu\nu}$ leads to the field equations:
\begin{equation}
\begin{split}
2 \nabla_\alpha \left( \sqrt{-g} f_Q P^\alpha_{\ \mu\nu} \right) + \sqrt{-g} \left[ \frac{1}{2} g_{\mu\nu} f + f_Q \left( P_{\mu\alpha\beta} Q^{\alpha\beta}_{\ \ \nu} - 2 Q_{\alpha\beta\mu} P^{\alpha\beta}_{\ \ \nu} \right) \right]\\
= \kappa \sqrt{-g} T_{\mu\nu},
\end{split}
\end{equation}
where $f_Q = \partial f / \partial Q$, $P^\alpha_{\ \mu\nu}$ is the superpotential, and $T_{\mu\nu}$ is the energy-momentum tensor.

A covariant form of these equations, applicable in cosmological contexts, can be written as:
\begin{equation}
f_Q \, \mathring{G}_{\mu\nu} + \frac{1}{2} g_{\mu\nu}(Q f_Q - f) + 2 f_{QQ} P^\lambda_{\ \mu\nu} \mathring{\nabla}_\lambda Q = \kappa T_{\mu\nu},
\end{equation}
where $\mathring{G}_{\mu\nu}$ is the Einstein tensor associated with the Levi-Civita connection.

Variation of the action with respect to the connection yields:
\begin{equation}
\nabla_\mu \nabla_\nu \left( \sqrt{-g} f_Q P^\lambda_{\ \mu\nu} + \mathcal{H}^\lambda_{\ \mu\nu} \right) = 0,
\end{equation}
where $\mathcal{H}^\lambda_{\ \mu\nu} = -\frac{1}{2} \frac{\delta(\sqrt{-g} \mathcal{L}_m)}{\delta \Gamma^\lambda_{\ \mu\nu}}$ is the hypermomentum tensor density. If $\nabla_\mu \nabla_\nu \mathcal{H}^\lambda_{\ \mu\nu} = 0$, the equation reduces to:
\begin{equation}
\nabla_\mu \nabla_\nu \left( \sqrt{-g} f_Q P^\lambda_{\ \mu\nu} \right) = 0.
\end{equation}

Two main approaches exist for incorporating symmetric teleparallelism:
\begin{enumerate}
    \item \textit{Inertial Variation:} The connection is treated as a pure gauge and substituted before variation.
    \item \textit{Lagrange Multiplier Approach:} Constraints are introduced to enforce vanishing curvature and torsion explicitly.
\end{enumerate}

\vspace{1em}
\subsubsection{Coincident Gauge}
In the coincident gauge, the connection can be parameterized by functions $\xi^\lambda$:
\begin{equation}
\Gamma^\alpha_{\mu\nu} = \frac{\partial x^\alpha}{\partial \xi^\lambda} \partial_\mu \partial_\nu \xi^\lambda.
\end{equation}
Choosing coordinates such that $x^\mu = \xi^\mu$ results in $\Gamma^\alpha_{\mu\nu} = 0$. Consequently, the covariant derivative becomes a partial derivative, and the non-metricity tensor simplifies to $Q_{\alpha\mu\nu} = \partial_\alpha g_{\mu\nu}$. The Levi-Civita connection is then written as
\begin{equation}
\{^\lambda_{\mu\nu}\} = -L^\lambda_{\mu\nu}.
\end{equation}
This formulation involves only first-order derivatives of the metric, eliminating the need for Gibbons–Hawking–York boundary terms and supporting a well-posed variational principle. The coordinate freedom introduces a gauge symmetry akin to Stueckelberg fields in field theory, enhancing the gauge structure beyond that of \gls{tegr}.

For a spatially flat \gls{flrw} metric and perfect fluid, the modified Friedmann equations in $f(Q)$ gravity are:
\begin{align}
Q f_Q - \frac{f}{2} &= \rho, \\
f_Q \dot{H} + H \dot{f}_Q &= -\frac{1}{2} (\rho + p),
\end{align}
where $\rho$ and $p$ are energy density and pressure. Setting $f(Q) = Q$ retrieves the standard \gls{gr} equations.

\vspace{1em}
\subsubsection{Non-Coincident Gauge}
In non-coincident gauges, $\Gamma^\alpha_{\mu\nu} \neq 0$, and the spacetime deviates from the Riemannian structure. The non-metricity scalar $Q$ depends on connection parameters $K_1$, $K_2$, and $K_3$:
\begin{equation}
Q = -6H^2 + 9HK_3 + 3K_3(K_1 - K_3) + \frac{3K_2 H}{a^2} - \frac{3K_2(K_1 + K_3)}{a^2}.
\end{equation}

The curvature-free conditions yield:
\begin{align}
K_3(K_1 - K_3) - \dot{K}_3 &= 0, \\
K_2(K_1 - K_3) + \dot{K}_2 &= 0, \\
\kappa + K_2 K_3 &= 0.
\end{align}

Three notable choices for $\kappa = 0$ include:

\paragraph{Connection I:} $K_1 = \gamma(t),\quad K_2 = K_3 = 0$
\begin{eqnarray}
Q = -6H^2, \\
3H^2 f_Q + \frac{1}{2}(f - Qf_Q) = \rho, \\
-2 \frac{d}{dt}(f_Q H) - 3H^2 f_Q - \frac{1}{2}(f - Qf_Q) = p.
\end{eqnarray}

\paragraph{Connection II:} $K_1 = \gamma(t) + \frac{\dot{\gamma}(t)}{\gamma(t)},\quad K_2 = 0,\quad K_3 = \gamma(t)$
\begin{align}
Q = -6H^2 + 9\gamma H + 3\dot{\gamma}, \\
3H^2 f_Q + \frac{1}{2}(f - Qf_Q) + \frac{3\gamma}{2} \dot{Q} f_{QQ} = \rho, \\
-2 \frac{d}{dt}(f_Q H) - 3H^2 f_Q - \frac{1}{2}(f - Qf_Q) + \frac{3\gamma}{2} \dot{Q} f_{QQ} = p.
\end{align}

\paragraph{Connection III:} $K_1 = -\frac{\dot{\gamma}(t)}{\gamma(t)},\quad K_2 = \gamma(t),\quad K_3 = 0$
\begin{align}
Q = -6H^2 + \frac{3\gamma H}{a^2} + \frac{3\dot{\gamma}}{a^2}, \\
3H^2 f_Q + \frac{1}{2}(f - Qf_Q) - \frac{3\gamma}{2a^2} \dot{Q} f_{QQ} = \rho, \\
-2 \frac{d}{dt}(f_Q H) - 3H^2 f_Q - \frac{1}{2}(f - Qf_Q) + \frac{\gamma}{2a^2} \dot{Q} f_{QQ} = p.
\end{align}

These different connections highlight the role of affine structure in modifying cosmological dynamics. Notably, Connection I reproduces the coincident gauge case.

\vspace{1em}
\subsection{$f(Q, \mathcal{T})$ gravity}
The $f(Q, \mathcal{T})$ gravity theory generalizes the symmetric teleparallel gravity by considering a gravitational action that depends not only on the non-metricity scalar $Q$ but also on the trace $\mathcal{T}$ of the energy-momentum tensor $T_{\mu\nu}$. This coupling allows for non-minimal interaction between matter and geometry, potentially giving rise to modified dynamics and effective \gls{de} behavior.

The action for $f(Q, \mathcal{T})$ gravity is given by
\begin{equation}
S = \int d^4x \sqrt{-g} \left[ \frac{1}{2\kappa} f(Q, \mathcal{T}) + \mathcal{L}_m \right],
\end{equation}
where:
\begin{itemize}
    \item $f(Q, \mathcal{T})$ is a general function of the non-metricity scalar $Q$ and the trace of the energy-momentum tensor $\mathcal{T} = g^{\mu\nu}T_{\mu\nu}$,
    \item $\mathcal{L}_m$ is the matter Lagrangian density,
    \item $\kappa = 8\pi G$ is the gravitational constant.
\end{itemize}

Varying the action with respect to the metric tensor $g^{\mu\nu}$ gives the field equations:
\begin{equation}
\begin{split}
\frac{2}{\sqrt{-g}} \nabla_\alpha \left( \sqrt{-g} f_Q P^\alpha_{\ \mu\nu} \right)
& + \frac{1}{2} g_{\mu\nu} f(Q, \mathcal{T}) 
+ f_Q \left( P_{\mu\alpha\beta} Q^{\alpha\beta}_{\ \ \nu} - 2 Q_{\alpha\beta\mu} P^{\alpha\beta}_{\ \ \nu} \right) \\
& - f_{\mathcal{T}} (T_{\mu\nu} + \Theta_{\mu\nu}) = \kappa T_{\mu\nu},
\end{split}
\end{equation}
where:
\begin{itemize}
    \item $f_Q = \partial f / \partial Q$, \quad $f_{\mathcal{T}} = \partial f / \partial \mathcal{T}$,
    \item $P^\alpha_{\ \mu\nu}$ is the non-metricity conjugate (superpotential), defined in symmetric teleparallel gravity,
    \item $\Theta_{\mu\nu} \equiv g^{\alpha\beta} \delta T_{\alpha\beta} / \delta g^{\mu\nu}$ accounts for the variation of $T_{\mu\nu}$ with respect to the metric.
\end{itemize}

The energy-momentum tensor is not generally conserved in $f(Q, \mathcal{T})$ gravity due to the coupling between geometry and matter. This yields a non-vanishing covariant divergence:
\begin{equation}
\nabla^\mu T_{\mu\nu} = \frac{f_{\mathcal{T}}}{\kappa - f_{\mathcal{T}}} \left[ (T_{\mu\nu} + \Theta_{\mu\nu}) \nabla^\mu \ln f_{\mathcal{T}} + \nabla^\mu \Theta_{\mu\nu} - \frac{1}{2} g_{\mu\nu} \nabla^\mu \mathcal{T} \right].
\end{equation}

For a spatially flat \gls{flrw} universe with metric:
\begin{equation}
ds^2 = -dt^2 + a^2(t) (dx^2 + dy^2 + dz^2),
\end{equation}
the non-metricity scalar becomes:
\begin{equation}
Q = 6H^2,
\end{equation}
where $H = \dot{a}/a$ is the Hubble parameter.
Assuming a perfect fluid with energy-momentum tensor:
\begin{equation}
T_{\mu\nu} = (\rho + p)u_\mu u_\nu + p g_{\mu\nu},
\end{equation}
and choosing $\mathcal{L}_m = -p$, the trace is $\mathcal{T} = -\rho + 3p$.

The modified Friedmann equations become:
\begin{align}
6H^2 f_Q - \frac{1}{2} f + f_{\mathcal{T}} (\rho + p) &= \kappa \rho, \\
\left( 12H^2 f_{QQ} + f_Q \right)\dot{H} + \frac{1}{2} f - f_{\mathcal{T}}(\rho + p) &= -\frac{1}{2} (\kappa + f_{\mathcal{T}})(\rho + p).
\end{align}

These equations demonstrate that the dynamics of the universe in $f(Q, \mathcal{T})$ gravity are influenced by the choice of the function $f(Q, \mathcal{T})$ and the form of matter coupling, potentially leading to late-time acceleration and deviations from \gls{lcdm} cosmology.

\vspace{1em}
\subsection{$f(Q, B)$ gravity} 
The $f(Q, B)$ gravity theory extends the symmetric teleparallel framework by including both the non-metricity scalar $Q$ and a boundary term $B$, analogous to how $f(R)$ theories generalize \gls{gr}. The boundary term $B$ relates $Q$ to the Ricci scalar $R$ through:
\begin{equation}
R = -Q + B,
\end{equation}
where $R$ is the Ricci scalar built from the Levi-Civita connection. This relationship allows $f(Q, B)$ gravity to interpolate between $f(Q)$ gravity and $f(R)$ gravity, thus providing a richer structure to explore modified gravitational dynamics.

The action for $f(Q, B)$ gravity is defined as:
\begin{equation}
S = \int d^4x \sqrt{-g} \left[ \frac{1}{2\kappa} f(Q, B) + \mathcal{L}_m \right],
\end{equation}
where:
\begin{itemize}
    \item $f(Q, B)$ is a smooth function of the non-metricity scalar $Q$ and the boundary term $B$,
    \item $\mathcal{L}_m$ is the matter Lagrangian density,
    \item $\kappa = 8\pi G$ is the gravitational constant.
\end{itemize}

Varying the action with respect to the metric $g^{\mu\nu}$ leads to the field equations:
\begin{align}
&\frac{2}{\sqrt{-g}} \nabla_\alpha\left( \sqrt{-g} f_Q P^\alpha_{\ \mu\nu} \right) 
+ \frac{1}{2} g_{\mu\nu} f 
+ f_Q \left( P_{\mu\alpha\beta} Q^{\alpha\beta}_{\ \ \nu} - 2 Q_{\alpha\beta\mu} P^{\alpha\beta}_{\ \ \nu} \right) \nonumber \\
& + f_B \left( R_{\mu\nu} - \frac{1}{2} g_{\mu\nu} R \right)
+ \left( g_{\mu\nu} \Box - \nabla_\mu \nabla_\nu \right) f_B = \kappa T_{\mu\nu},
\end{align}
where:
\begin{itemize}
    \item $f_Q \equiv \partial f/\partial Q$, \quad $f_B \equiv \partial f/\partial B$,
    \item $P^\alpha_{\ \mu\nu}$ is the superpotential defined in the symmetric teleparallel gravity,
    \item $T_{\mu\nu}$ is the energy-momentum tensor.
\end{itemize}

For a spatially flat \gls{flrw} universe:
\begin{equation}
ds^2 = -dt^2 + a^2(t)(dx^2 + dy^2 + dz^2),
\end{equation}
the non-metricity scalar is given by:
\begin{equation}
Q = 6H^2,
\end{equation}
and the boundary term $B$ is:
\begin{equation}
B = 6(\dot{H} + 3H^2),
\end{equation}
where $H = \dot{a}/a$ is the Hubble parameter.
Assuming a perfect fluid energy-momentum tensor:
\begin{equation}
T_{\mu\nu} = (\rho + p)u_\mu u_\nu + p g_{\mu\nu},
\end{equation}
the modified Friedmann equations for $f(Q, B)$ gravity are obtained as:
\begin{align}
3H^2 f_Q - \frac{1}{2} f + 3H \dot{f}_B &= \kappa \rho, \\
\left( -2\dot{H} - 3H^2 \right) f_Q + \frac{1}{2} f - \ddot{f}_B - 2H \dot{f}_B &= \kappa p,
\end{align}
where overdots denote derivatives with respect to cosmic time $t$.

These equations show that the inclusion of the boundary term $B$ introduces higher-order derivatives into the dynamics, allowing for richer cosmological behavior, including possible inflationary and late-time accelerated expansion phases.

\vspace{1em}
\subsection{$f(Q, \phi)$ gravity} 
The $f(Q, \phi)$ gravity theory generalizes symmetric teleparallel gravity by including a coupling between the non-metricity scalar $Q$ and a scalar field $\phi$. This framework allows for dynamical modifications of gravity through scalar degrees of freedom and can model various cosmological phenomena such as inflation and late-time acceleration.

The general action for $f(Q, \phi)$ gravity is given by
\begin{equation}
S = \int d^4x \sqrt{-g} \left[ \frac{1}{2\kappa} f(Q, \phi) - \frac{1}{2} g^{\mu\nu} \partial_\mu \phi \partial_\nu \phi - V(\phi) + \mathcal{L}_m \right],
\end{equation}
where:
\begin{itemize}
    \item $f(Q, \phi)$ is a general function of the non-metricity scalar $Q$ and scalar field $\phi$,
    \item $V(\phi)$ is the scalar field potential,
    \item $\mathcal{L}_m$ is the matter Lagrangian,
    \item $\kappa = 8\pi G$ is the gravitational constant.
\end{itemize}

Variation of the action with respect to the metric $g^{\mu\nu}$ yields the modified field equations:
\begin{equation}
\begin{split}
\frac{2}{\sqrt{-g}} \nabla_\alpha \left( \sqrt{-g} f_Q P^\alpha_{\ \mu\nu} \right)+ \frac{1}{2} g_{\mu\nu} f 
& + f_Q \left( P_{\mu\alpha\beta} Q^{\alpha\beta}_{\ \ \nu} - 2 Q_{\alpha\beta\mu} P^{\alpha\beta}_{\ \ \nu} \right) \\
& - \frac{1}{2} T^{(\phi)}_{\mu\nu} = \kappa T_{\mu\nu},
\end{split}
\end{equation}
where:
\begin{itemize}
    \item $f_Q = \partial f / \partial Q$,
    \item $P^\alpha_{\ \mu\nu}$ is the superpotential associated with non-metricity,
    \item $T_{\mu\nu}$ is the energy-momentum tensor of matter,
    \item $T^{(\phi)}_{\mu\nu}$ is the energy-momentum tensor of the scalar field:
    \begin{equation}
    T^{(\phi)}_{\mu\nu} = \partial_\mu \phi \partial_\nu \phi - g_{\mu\nu} \left( \frac{1}{2} \partial^\alpha \phi \partial_\alpha \phi + V(\phi) \right).
    \end{equation}
\end{itemize}

Varying the action with respect to the scalar field $\phi$ gives the scalar field equation:
\begin{equation}
\square \phi - V'(\phi) + \frac{1}{2\kappa} f_\phi = 0,
\end{equation}
where:
\begin{itemize}
    \item $\square \phi = \nabla^\mu \nabla_\mu \phi$ is the d'Alembert operator,
    \item $f_\phi = \partial f / \partial \phi$,
    \item $V'(\phi) = dV / d\phi$.
\end{itemize}

In a flat \gls{flrw} background,
\begin{equation}
ds^2 = -dt^2 + a^2(t)(dx^2 + dy^2 + dz^2),
\end{equation}
the non-metricity scalar is given by:
\begin{equation}
Q = 6H^2,
\end{equation}
where $H = \dot{a}/a$ is the Hubble parameter.

Assuming a perfect fluid for matter and a homogeneous scalar field $\phi(t)$, the modified Friedmann equations become:
\begin{align}
3H^2 f_Q - \frac{1}{2} f + \frac{1}{2} \dot{\phi}^2 + V(\phi) = \kappa \rho, \\
\left(2\dot{H} + 3H^2\right) f_Q + 2H \dot{f}_Q + \ddot{f}_Q + \frac{1}{2} f + \frac{1}{2} \dot{\phi}^2 - V(\phi) = -\kappa p.
\end{align}

The scalar field equation reduces to:
\begin{equation}
\ddot{\phi} + 3H \dot{\phi} + V'(\phi) = \frac{1}{2\kappa} f_\phi.
\end{equation}

These equations demonstrate the rich cosmological dynamics offered by $f(Q, \phi)$ gravity, including possibilities for inflationary and \gls{de}-driven expansion, depending on the forms of $f(Q, \phi)$ and $V(\phi)$.

\vspace{1em}
\section{Summary of Literature and Research Gaps}

From the reviewed literature, several key conclusions can be drawn:
\begin{itemize}
    \item $f(Q)$ gravity provides a geometrically motivated, second-order theory of gravity that preserves Lorentz invariance.
    \item Cosmological models in $f(Q)$ gravity can successfully describe late-time acceleration without a cosmological constant.
    \item Extensions such as $f(Q, T)$ and $f(Q, \phi)$ introduce matter–geometry and scalar–geometry couplings, offering richer cosmological dynamics.
    \item The Noether symmetry approach serves as a powerful method to constrain the functional forms of $f(Q)$ and obtain exact solutions.
    \item Dynamical system methods offer insights into the stability and evolution of cosmological solutions, connecting theory with observations.
\end{itemize}

However, despite these advances, several open issues remain:
\begin{enumerate}
    \item A systematic classification of cosmological solutions within $f(Q)$ and its extensions remains incomplete.
    \item The nature of perturbations and the role of non-metricity in structure formation require further investigation.
    \item The coupling between geometry and matter fields in extended $f(Q)$ models introduces complexities that demand deeper dynamical and observational analysis.
\end{enumerate}

\vspace{1em}
\section{Concluding Remarks}

In conclusion, the symmetric teleparallel formulation and its extensions represent a rapidly evolving domain of gravitational research. The $f(Q)$ gravity theory unifies geometric elegance with dynamical richness, providing an appealing alternative to curvature- and torsion-based modifications. Its second-order nature, combined with successful cosmological applications, makes it a promising candidate for exploring the Universe’s past, present, and future dynamics.




\chapter{Dynamical System Analysis in Symmetric Teleparallel Gravity} \label{c3}
\section{Introduction}

This chapter presents a comprehensive dynamical systems analysis within the framework of $f(Q)$ gravity, focusing on both background and perturbative levels \citep{Vishwakarma:2024bnd, Duchaniya:2022fmc}. The primary objective is to explore the mechanisms driving the late-time acceleration of the Universe and to characterize the evolution through different cosmological epochs, including phases dominated by \gls{dm} and \gls{de}. In the context of dynamical systems, an attractor corresponds to a region in phase space toward which the system's trajectories asymptotically approach over time. In cosmology, such attractor behavior signifies that the Universe's dynamical evolution converges to a stable configuration, largely independent of initial conditions.Additional research on the late-time cosmic expansion concerns related to $f(Q)$ gravity can be found in \citep{Khyllep:2021wjd, BeltranJimenez:2017tkd, BeltranJimenez:2019tme, Anagnostopoulos:2022gej, Vishwakarma:2023brw}.

\vspace{1em}
\section{$f(Q)$ gravity field equations}
In this study, we investigate the theoretical structure of $f(Q)$ gravity, with particular emphasis on the action formalism described by equation (1.46). The analysis is conducted within the context of a spatially flat Friedmann–Lemaître–Robertson–Walker (\gls{flrw}) metric, as defined in equation (1.14), and the associated tetrad configuration provided by $e^A{}_\mu = \text{diag}(1, a(t), a(t), a(t)),$. By performing a variation of the action (equation (1.46)) with respect to the tetrad field, we simplify the equation by assuming that $ 8\pi G_N=1 $. We also impose the splitting condition $ f(Q)=Q+F(Q) $ and utilise the \gls{flrw} metric.we obtain the modified field equations that characterize the dynamics of $f(Q)$ gravity,
\begin{equation}
3H^{2}= \rho + \frac{F}{2}-QF_{Q},
\end{equation}
\begin{equation}
(2QF_{QQ}+F_{Q}+1)\dot{H}+\frac{1}{4}(Q+2QF_{Q}-F)=-2p,
\end{equation}
where $ \rho $ and $ p $ denote the energy density and pressure, respectively. In this context, $ F_{Q} = \frac{dF}{dQ} $ and $ F_{QQ} = \frac{d^2 F}{dQ^2} $. These equations characterize the evolution of the Hubble parameter $H$ and the effective energy density $ \rho $, incorporating contributions from the non-metricity scalar $ Q $ and the function $ F(Q) $. This formulation offers a comprehensive framework for understanding the dynamics of the universe within the context of $ f(Q) $ gravity.

To define the field equations of $F(Q)$ gravity in the dark energy sector ($ p_{\mathrm{de}} $ and $ \rho_{\mathrm{de}} $) as follows:
\begin{equation}
\rho_{\mathrm{de}}=\frac{1}{16\pi G_N}\left[ F-2QF_{Q} \right] ,
\end{equation}
\begin{equation}
p_{\mathrm{de}}=\frac{1}{16\pi G_N}\left[ 4(F_{Q}+2QF_{QQ})\dot{H} -F+ 2QF_{Q}\right].
\end{equation}
 Substituting the non-metricity Scalar $ Q=6H^{2} $, the EoS parameter of the dark energy sector ($ \displaystyle w_{\mathrm{de}}$) can be obtained as, 
 \begin{equation}
  \displaystyle w_{\mathrm{de}}= -1+ \frac{4(F_{Q}+2QF_{QQ}) }{F-2QF_{Q}}\frac{\dot{H}}{H^{2}}.  
 \end{equation}
Furthermore, the total EoS ($ \displaystyle w_{\mathrm{tot}}$) and the deceleration parameter ($ q $) can be defined as,
\begin{equation}
\displaystyle w_{\mathrm{tot}}= -1- \frac{2\dot{H}}{3 H^{2}}\equiv \frac{p_{\mathrm{m}}+p_{\mathrm{de}}}{\rho_{\mathrm{m}}+\rho_{\mathrm{de}}}, 
\end{equation}
\begin{equation}
q= -1 -\frac{\dot{H}}{H^{2}}.
\end{equation}

We are intending to study the interacting cosmology by performing the dynamical system analysis. From the expressions of equation (3.3)--(3.7), we can see that the functional form of $f(Q)$ is required for further study. Therefore, we shall consider three distinct forms of  $f(Q)$ in the following section.

\vspace{1em}
\section{Phase Space Analysis}

In this section, we formulate the dynamical system corresponding to the background and perturbed cosmological equations. To incorporate the evolution of matter perturbations on sub-horizon scales, we consider the following differential equation:
\begin{equation}
    \ddot{\delta}_m + 2H\dot{\delta}_m = 4\pi G_{\text{eff}} \rho_m \delta_m, \label{eq:perturb_growth}
\end{equation}
where \( \delta_m = \frac{\delta \rho_m}{\rho_m} \) represents the matter overdensity, and \( G_{\text{eff}}(a) = G Y(a) \) is the effective Newtonian gravitational constant, with \( G \) being the standard Newton's constant. Typically, \( G_{\text{eff}}(a) \) varies with the scale factor, and its form is determined by the underlying gravitational theory.

For the perturbative analysis in \( f(Q) \) gravity, we adopt a general expression for \( Y(a) \) as given in \citep{Gannouji:2008wt}:
\begin{equation}
    Y(a) = \frac{G_{\text{eff}}(a)}{G} = \frac{1}{1 + F_Q}, \label{eq:Y_function}
\end{equation}
Substituting equation (3.9) into equation (3.8), we obtain:
\begin{equation}
    \ddot{\delta}_m + 2H\dot{\delta}_m = \frac{4\pi G \rho_m \delta_m}{1 + F_Q}. \label{eq:perturb_modified}
\end{equation}

Using equations (3.1), (3.2), and \eqref{eq:perturb_modified}, we define the following dimensionless dynamical variables for a general \( f(T) \) gravity model:
\begin{equation}
    x = -\frac{F}{6H^2}, \quad y = -2F_Q, \quad \xi = \frac{d(\ln \delta)}{d(\ln a)}, \label{eq:dynamical_vars}
\end{equation}
where \( x \) and \( y \) describe the background evolution of the Universe, while \( \xi \) characterizes the growth rate of matter perturbations. A positive value of \( \xi \) implies the amplification of matter overdensities, whereas a negative value indicates suppression.

Rewriting equation (3.1) in terms of these variables yields the constraint:
\begin{equation}
    x + y + \Omega_m = 1. \label{eq:friedmann_constraint}
\end{equation}
Here, the dark energy density parameter is given by \( \Omega_{\text{DE}} = x + y \).
Similarly, equation (3.2) becomes:
\begin{equation}
\frac{\dot{H}}{H^{2}} = \frac{3(x+y-1)}{4QF_{QQ} - y + 2}.\label{eq:H_dot}
\end{equation}
With the set of variables \( (x, y, \xi) \), we can perform a three-dimensional phase space analysis. In this formulation, the full cosmological dynamics, both at the background and perturbative levels-can be analyzed systematically through the autonomous system formed by these variables given by
\begin{align}
x' &= -\frac{\dot{H}}{H^{2}}(y+2x), \\
y' &= -\frac{\dot{H}}{H^{2}}4QF_{QQ}, \\
\xi' &= -\xi(\xi+2)+\frac{3(1-x-y)}{2-y}-\frac{\dot{H}}{H^{2}}\xi,
\end{align}

The critical points of the autonomous system described by equations (3.14)--(3.16) are determined to analyze the dynamical evolution of the system. The stability of these critical points is then assessed. From a physical perspective, it is well understood that a stable critical point with \( \xi > 0 \) indicates a continuous growth of matter perturbations. However, such a scenario suggests that the system remains dynamically unstable with respect to perturbations. On the other hand, a stable critical point with \( \xi < 0 \) corresponds to a decay in matter perturbations. If \( \xi = 0 \) at a stable point, it implies that matter perturbations remain constant over time, indicating a state of equilibrium in their evolution.

To solve the dynamical system defined by equations (3.14)--(3.16), a specific form of the function \( f(Q) \) must be chosen. In the subsequent sections, we will examine three distinct functional forms of \( f(Q) \), each offering potentially significant insights into the late-time evolution and overall dynamics of the Universe.

\vspace{1em}
\subsection{Model I}
The class of models we will investigate is based on the form of f(Q) given by Refs. \citep{Anagnostopoulos:2022gej, Ayuso2021, Goncalves:2024sem}
\begin{equation}
f(Q) = \frac{\alpha}{2}\sqrt{Q_{0} Q} \ln \left( \gamma \frac{Q}{Q_{0}} \right) + \beta Q,
\end{equation}
where the parameters $\alpha$, $\beta$, and $\gamma$, are arbitrary constants.

Therefore the autonomous systems given by equations (3.14)--(3.16) becomes,
\begin{align}
x' &= \frac{3(x+y-1)}{x+y-2}(y+2x), \\
y' &=  \frac{3(x+y-1)}{x+y-2}(-x), \\
\xi' &= -\xi(\xi+2) + \frac{3(1-x-y)}{2-y} +  \frac{3(x+y-1)}{x+y-2}\xi.
\end{align}
The singularities of the given system occur at $ y = 2 $ and $ x = 0 $, and the EoS parameters and deceleration parameter becomes,
\begin{align}
\displaystyle w_{\mathrm{de}} &= -1 + \frac{(x-2y)(x+y-1)}{2(x+y)(2-x-y)}, \\
\displaystyle w_{\mathrm{tot}} &= -1 + \frac{2(x+y-1)}{(x+y-2)}, \\
q &= -1 + \frac{3(x+y-1)}{(x+y-2)}. 
\end{align}

Four critical points have been identified and are detailed in Table 3.1, along with their associated cosmological characteristics. Table 3.2 presents the eigenvalues of the Jacobian matrix for both background and perturbation levels.

\begin{table}[htp!]
\centering
\begin{tabular}{ ||p{1.5cm}||p{2.6cm}||p{0.8cm}||p{0.8cm}||p{0.8cm}||p{0.8cm}||p{0.8cm}||p{0.8cm}|| }
 \hline
 \textit{CP} & $\mathbf{(x_{c}, y_{c}, \xi_{c})}$ & $\mathbf{\Omega_{\mathrm{m}}}$ & $\mathbf{\Omega_{\mathrm{de}}}$ & $ \mathbf {\xi} $ & $ \mathbf{\displaystyle w_{\mathrm{de}}} $ & $\mathbf{ \displaystyle w_{\mathrm{tot}}} $ & $ \mathbf{q}$ \\
 \hline
 $ A_{1} $ & $ (1-y,y,0) $ & $ 0 $ & $ 1 $ & $ 0 $ & $ -1 $ & $ -1 $ & $ -1 $  \\
 \hline
 $ A_{2} $ & $ (1-y,y,-2) $ & $ 0 $ & $ 1 $ & $ -2 $ & $ -1 $ & $ -1 $ & $ -1 $  \\
 \hline
  $ A_{3} $ & $ (0,0,-\frac{3}{2}) $ & $ 1 $ & $ 0 $ & $ -\frac{3}{2} $ & $ --- $ & $ 0 $ & $ \frac{1}{2} $ \\
 \hline
  $ A_{4} $ & $ (0,0,1) $ & $ 1 $ & $ 0 $ & $ 1 $ & $ --- $ & $ 0 $ & $ \frac{1}{2} $ \\
 \hline
\end{tabular}
\caption{Critical points (CP), matter, dark energy, perturbation and EoS parameters}
\end{table} 

\begin{table}[htp!]
\centering
\begin{tabular}{ ||p{1.5cm}||p{2.4cm}||p{3.2cm}|| }
 \hline
 \textit{CP} & \textit{Eigen-Values} & \textit{Stability Condition}  \\
 \hline
 $ A_{1} $ & $ (0,-3,-2) $ & Stable  \\
 \hline
 $ A_{2} $ & $ (0,-3,2) $ & Saddle  \\
 \hline
  $ A_{3} $ & $ (\frac{5}{2},\frac{3}{2},\frac{3}{2}) $ & Unstable \\
 \hline
  $ A_{4} $ & $ (-\frac{5}{2},\frac{3}{2},\frac{3}{2}) $ & Saddle \\
 \hline
\end{tabular}
\caption{Eigen values and Stability conditions}
\end{table} 

\vspace{1em}
\subsection*{Summary of the critical points (Model-I):}
\begin{itemize}
   \item  \textit{Critical Point \( A_1 = (1 - y, y, 0) \):} 
This critical point corresponds to a de Sitter universe entirely dominated by dark energy, where \( \Omega_{\mathrm{m}} = 0 \) and \( \Omega_{\mathrm{de}} = 1 \). The equation of state parameters \( w_{\mathrm{de}} = w_{\mathrm{tot}} = -1 \) and the deceleration parameter \( q = -1 \) indicate an accelerated expansion consistent with a cosmological constant. The eigenvalues \( (0, -3, -2) \) reveal that the point is non-hyperbolic but asymptotically stable, as confirmed via center manifold theory. The perturbation variable \( \xi = 0 \) implies that matter perturbations remain constant, and the reduced dynamics further support the stability of this point. Overall, \( A_1 \) is a late-time attractor in the phase space, signifying a universe evolving toward dark energy domination.

   \item  \textit{Critical Point \( A_2 = (1 - y, y, -2) \):} 
This point also represents a dark energy-dominated de Sitter universe, with \( \Omega_{\mathrm{de}} = 1 \) and \( \Omega_{\mathrm{m}} = 0 \), yielding \( w_{\mathrm{de}} = w_{\mathrm{tot}} = -1 \) and \( q = -1 \). However, the associated eigenvalues \( (0, -3, 2) \) indicate a saddle point, signifying that the system is unstable along at least one direction. At the perturbation level, \( \xi = -2 \) shows decay in matter perturbations. Although it shares a similar background behavior with \( A_1 \), the saddle nature of \( A_2 \) implies that it cannot serve as a late-time attractor but may correspond to an early inflationary phase in cosmic evolution.

   \item  \textit{Critical Point \( A_3 = (0, 0, -\frac{3}{2}) \):} 
This critical point describes a matter-dominated universe with \( \Omega_{\mathrm{m}} = 1 \) and \( \Omega_{\mathrm{de}} = 0 \), leading to \( w_{\mathrm{tot}} = 0 \) and a deceleration parameter \( q = \frac{1}{2} \). The perturbation parameter \( \xi = -\frac{3}{2} \) indicates a decay of matter overdensities. The eigenvalues \( \left(\frac{5}{2}, \frac{3}{2}, \frac{3}{2}\right) \) are all positive, making this point unstable in all directions. Thus, although \( A_3 \) captures a transient matter-dominated phase, it cannot represent the Universe's late-time behavior. It plays a crucial role in the intermediate cosmological era before the onset of dark energy domination.

   \item  \textit{Critical Point \( A_4 = (0, 0, 1) \):} 
This point also represents a matter-dominated universe (\( \Omega_{\mathrm{m}} = 1 \), \( \Omega_{\mathrm{de}} = 0 \)), with \( w_{\mathrm{tot}} = 0 \) and \( q = \frac{1}{2} \). The positive perturbation variable \( \xi = 1 \) implies growing matter perturbations, making this point relevant for structure formation. The eigenvalues \( \left(-\frac{5}{2}, \frac{3}{2}, \frac{3}{2}\right) \) indicate a saddle point with both stable and unstable directions. Therefore, \( A_4 \) is unstable but essential in describing the transition from matter dominance to dark energy dominance. It highlights the era of large-scale structure growth in the Universe and acts as a precursor to the late-time attractor behavior.
\end{itemize}

The analysis reveals the existence of two dark energy-dominated critical points (($A_{1}$ and $A_{2}$) and two matter-dominated critical points ($A_{3}$ and $A_{4}$) within the framework of the logarithmic form of $f(Q)$ gravity. The matter-dominated points, ($A_{3}$ and $A_{4}$, are found to be inherently unstable. Specifically, critical point $A_{3}$, which functions as a saddle point, is indicative of a defined growth rate in matter perturbations. In contrast, critical point $A_{4}$, identified as an unstable node, signifies the decay of matter perturbations. Notably, $A_{4}$ exhibits accelerated expansion behavior at the background level only. On the other hand, critical point $A_{1}$ demonstrates this accelerated expansion behavior consistently at both the background and perturbation levels, distinguishing it as a stable configuration.

\begin{figure}[htp!]
    \centering
    \includegraphics[width=0.48\textwidth]{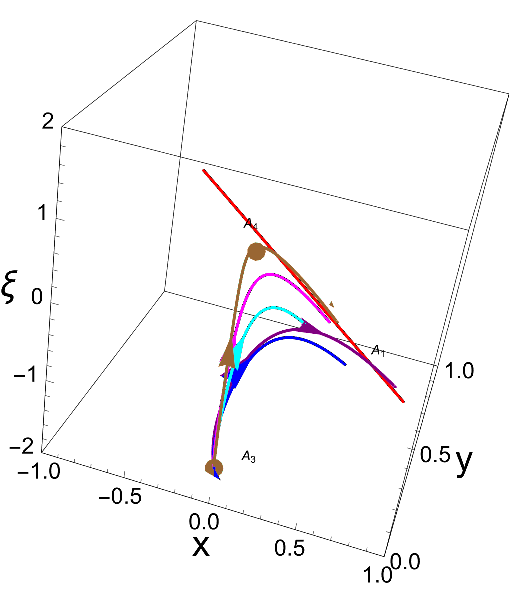}
    \caption{3D phase  portrait for Model-I.}
\end{figure}

\begin{figure}[htp!]
   \centering 
   \mbox{\includegraphics[width=0.48\textwidth]{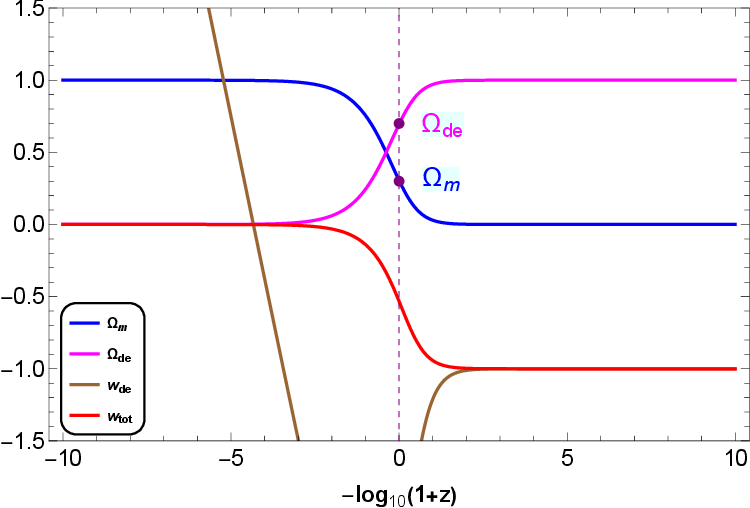}} 
    \hspace{10pt}
    \mbox{\includegraphics[width=0.48\textwidth]{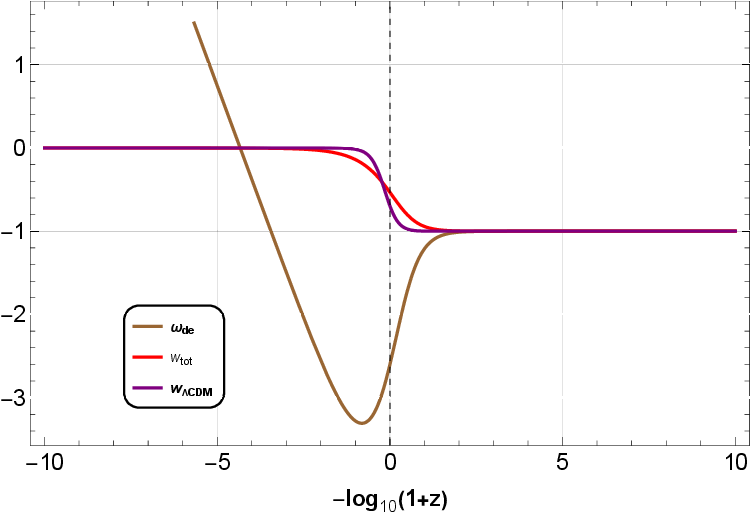}}
    \hspace{10pt}
    \mbox{\includegraphics[width=0.48\textwidth]{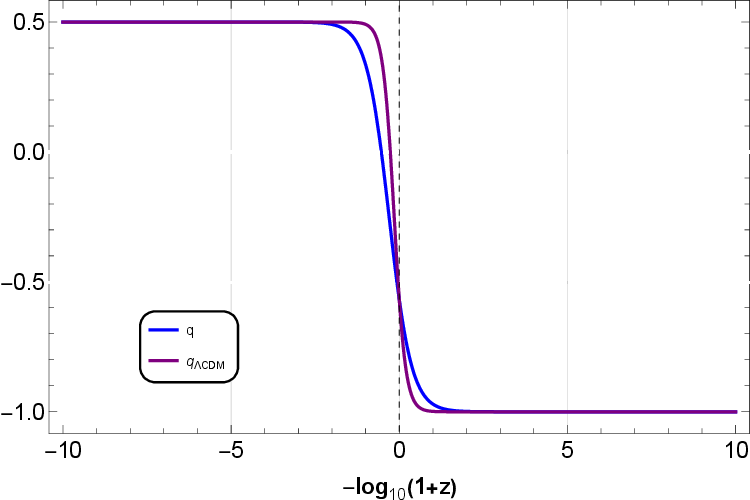}}
    \caption{ The evolution of the density parameters (shown in the \textit{Upper left panel}), the EoS parameters (illustrated in the \textit{Upper right panel}), and the deceleration parameter (depicted in the \textit{Lower panel}) are presented for Model-I. The initial conditions are set as $x = 10^{-2}$, $y = 10^{-6}$. The vertical dashed line represents the present time.}
   \label{Phase Portrait for the dynamical system of Model I}
\end{figure}

Figure 3.1 presents the phase portrait within a three-dimensional space, illustrating the evolution of the selected trajectory as it transitions from matter-dominated to dark-energy-dominated critical points. The diagram clearly depicts the sequential progression of the trajectory, transitioning from the unstable node at critical point $A_{3}$ to the saddle instability at $A_{4}$, and finally settling at the stable node represented by critical point $A_{1}$.

Figure 3.2 (Upper panels) illustrates the evolutionary history of the density parameter and equation of state (EoS) parameter. In upper left panel, the universe undergoes a transition from a matter-dominated phase to an accelerated expansion era at the late times. The current density parameters for the matter and dark energy sectors are approximately $ \Omega_{\mathrm{m}} \approx 0.3 $ and $\Omega_{\mathrm{de}} \approx 0.7$, respectively. The upper right panel shows the the total EoS parameter, which begins in a matter dominated era ($ \displaystyle w_{\mathrm{tot}}=0 $) and gradually transitions to the dark energy sector ($ \displaystyle w_{\mathrm{tot}} \approx -1 $) as the universe evolves. Concurrently, the dark energy EoS parameter approached $ -1 $ in the late stages of evolution, with the present value of $ \displaystyle w_{\mathrm{de}}=-1 $ aligniing with the current observational range of $ \displaystyle w_{\mathrm{de}}= -1.028 \pm 0.032 $ \citep{Planck:2018fzr}. In the lower panel, the deceleration parameter exhiboits a transition from a decelerating to an accelerating phase, with the transition point accurring at $ z= 0.59 $ and the current deceleration parameter value being $ q_{0}= -0.57 $ \citep{Camarena:2019moy}.

\vspace{1em}
\subsection{Model II}
The polynomial model, as introduced in \citep{BeltranJimenez:2019tme, Anagnostopoulos:2022gej} given by
\begin{equation}
f(Q)=Q-6 \lambda M^{2} \left(  \frac{Q}{6 M^{2}} \right) ^{\alpha}
\end{equation} serves as a generalization of the square-root model. In this context, $\lambda$ and $\alpha$ are dimensionless parameters, while $M$ represents a mass scale, typically of the order of $\sqrt{\Lambda}$, where $\Lambda$ denotes the standard cosmological constant.

The dynamical system can be presented as
\begin{align}
x'&=\frac{3(1-x-y)(y+2x)}{y(1-2\alpha)+2}, \\
y'&=\frac{6y(1-x-y)(1-\alpha)}{y(1-2\alpha)+2}, \\
\xi' &= -\xi(\xi+2) + \frac{3(1-x-y)}{2-y} + \frac{3(1-x-y)\xi}{y(1-2\alpha)+2}.
\end{align}
The singularities of the given system are located at $ y = 2 $ or $ y = -\frac{2}{1 - 2\alpha} $.
The corresponding Eos and deceleration parameters reduces to, 
\begin{align}
\displaystyle w_{\mathrm{de}} &= -1 - \frac{y(2-\alpha)(x+y-1)}{(x+y)(y(1-2\alpha)+2)}, \\
\displaystyle w_{\mathrm{tot}} &= -1 - \frac{2(x+y-1)}{y(1-2\alpha)+2}, \\
q &= -1 - \frac{3(x+y-1)}{y(1-2\alpha)+2}.
\end{align}

Four critical points have been identified and are detailed in Table 3.3, which outlines their corresponding cosmological characteristics. Table 3.4 provides the eigenvalues of the Jacobian matrix at both the background and perturbation levels.

\begin{table}[htp!]
\centering
\begin{tabular}{ ||p{1.5cm}||p{2.6cm}||p{0.8cm}||p{0.8cm}||p{0.8cm}||p{0.8cm}||p{0.8cm}||p{0.8cm}|| }
 \hline
 \textit{CP} & $\mathbf{(x_{c}, y_{c}, \xi_{c})}$ & $\mathbf{\Omega_{\mathrm{m}}}$ & $\mathbf{\Omega_{\mathrm{de}}}$ & $ \mathbf {\xi} $ & $ \mathbf{\displaystyle w_{\mathrm{de}}} $ & $\mathbf{ \displaystyle w_{\mathrm{tot}}} $ & $ \mathbf{q }$ \\
 \hline
 $ B_{1} $ & $ (1-y,y,0) $ & $ 0 $ & $ 1 $ & $ 0 $ & $ -1 $ & $ -1 $ & $ -1 $  \\
 \hline
 $ B_{2} $ & $ (1-y,y,-2) $ & $ 0 $ & $ 1 $ & $ -2 $ & $ -1 $ & $ -1 $ & $ -1 $  \\
 \hline
  $ B_{3} $ & $ (0,0,-\frac{3}{2}) $ & $ 1 $ & $ 0 $ & $ -\frac{3}{2} $ & $ --- $ & $ 0 $ & $ \frac{1}{2} $ \\
 \hline
  $ B_{4} $ & $ (0,0,1) $ & $ 1 $ & $ 0 $ & $ 1 $ & $ --- $ & $ 0 $ & $ \frac{1}{2} $ \\
 \hline
\end{tabular}
\caption{Critical points (CP), matter, dark energy, perturbation and EoS parameters}
\end{table} 

\begin{table}[htp!]
\centering
\begin{tabular}{ ||p{1.5cm}||p{2.4cm}||p{3.2cm}|| }
 \hline
 \textit{CP} & \textit{Eigen-Values} & \textit{Stability Condition}  \\
 \hline
 $ B_{1} $ & $ (0,-3,-2) $ & Stable  \\
 \hline
 $ B_{2} $ & $ (0,-3,2) $ & Saddle  \\
 \hline
  $ B_{3} $ & $ (3,\frac{29997}{10000},\frac{5}{2}) $ & Unstable \\
 \hline
  $ B_{4} $ & $ (3,\frac{29997}{10000},-\frac{5}{2}) $ & Saddle \\
 \hline
\end{tabular}
\caption{Eigen values and Stability conditions}
\end{table} 

\vspace{1em}
\subsection*{Summary of the critical points (Model-II):}
\begin{itemize}
    \item  \textit{Critical Point \( B_1 = (1 - y, y, 0) \):} 
The critical point \( B_1 \) corresponds to a Universe entirely dominated by dark energy, with \( \Omega_{\mathrm{m}} = 0 \) and \( \Omega_{\mathrm{de}} = 1 \). The associated parameters \( w_{\mathrm{de}} = w_{\mathrm{tot}} = -1 \) and \( q = -1 \) characterize a de Sitter-like accelerating phase. The eigenvalues \( (0, -3, -2) \) indicate non-hyperbolic stability, with two negative eigenvalues supporting stability and a zero eigenvalue suggesting a marginally stable direction. The application of center manifold theory confirms asymptotic stability near this point. The reduced equation \( x' = -\frac{3}{2} + \mathcal{O}(x^2) \) shows exponential decay in the center direction. Thus, \( B_1 \) represents a late-time attractor where the Universe asymptotically approaches accelerated expansion driven by dark energy.

    \item  \textit{Critical Point \( B_2 = (1 - y, y, -2) \):} 
The point \( B_2 \) depicts a Universe completely dominated by dark energy, with \( \Omega_{\mathrm{m}} = 0 \) and \( \Omega_{\mathrm{de}} = 1 \), and equation of state parameters \( w_{\mathrm{de}} = w_{\mathrm{tot}} = -1 \), consistent with a cosmological constant. The deceleration parameter \( q = -1 \) also supports an accelerated expansion. The eigenvalues \( (0, -3, 2) \) indicate that this point is a saddle, where stability exists in some directions while instability is present in others. The perturbation variable \( \xi = -2 \) indicates a decaying mode, and the dynamical configuration resembles an inflationary epoch. Hence, \( B_2 \) represents a transient dark energy–dominated state not suitable as a late-time attractor.

    \item  \textit{Critical Point \( B_3 = \left(0, 0, -\frac{3}{2} \right) \):} 
The critical point \( B_3 \) corresponds to a matter-dominated universe with \( \Omega_{\mathrm{m}} = 1 \) and \( \Omega_{\mathrm{de}} = 0 \). Here, \( w_{\mathrm{tot}} = 0 \) and \( q = \frac{1}{2} \), indicating a decelerating expansion phase. Dark energy is absent, so \( w_{\mathrm{de}} \) is undefined. The eigenvalues \( \left(3, \frac{29997}{10000}, \frac{5}{2} \right) \) are all positive, suggesting that the critical point is an unstable node. The perturbation parameter \( \xi = -\frac{3}{2} \) shows a decaying mode. Therefore, \( B_3 \) represents a transient phase of matter domination, unstable under small perturbations and not a final attractor in the cosmic evolution.

    \item  \textit{Critical Point \( B_4 = (0, 0, 1) \):} 
The point \( B_4 \) also signifies a matter-dominated universe, with \( \Omega_{\mathrm{m}} = 1 \) and \( \Omega_{\mathrm{de}} = 0 \). The parameters \( w_{\mathrm{tot}} = 0 \) and \( q = \frac{1}{2} \) reflect a decelerating expansion. Dark energy is absent, making \( w_{\mathrm{de}} \) undefined. The eigenvalues \( \left(3, \frac{29997}{10000}, -\frac{5}{2} \right) \) include both positive and negative values, identifying \( B_4 \) as a saddle point. The perturbation variable \( \xi = 1 \) indicates growing matter perturbations. This critical point plays a crucial role in the structure formation era, but its instability implies the Universe will eventually evolve away from this phase toward a dark energy–dominated regime.
\end{itemize}

The analysis identifies two dark energy-dominated critical points ($B_{1}$ and $B_{2}$) and two matter-dominated critical points ($B_{3}$ and $B_{4}$) within the context of the polynomial form of $f(Q) $ gravity. Both matter-dominated points, $B_{3}$ and $B_{4}$, are characterized by inherent instability. Specifically, critical point $B_{3}$, acting as a saddle point, reflects a defined growth rate in matter perturbations. Conversely, critical point $B_{4}$, identified as an unstable node, indicates the decay of matter perturbations. It is important to note that $B_{4}$ exhibits accelerated expansion behavior solely at the background level. In contrast, critical point $B_{1}$ consistently demonstrates this accelerated expansion behavior at both the background and perturbation levels, thereby establishing it as a stable configuration.

\begin{figure}[htp!]
    \centering
    \includegraphics[width=0.48\textwidth]{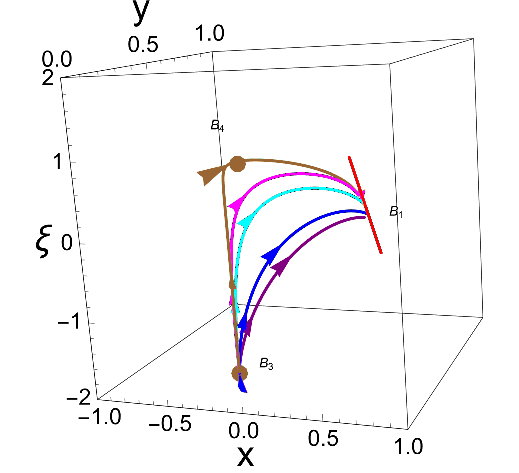}
    \caption{3D phase  portrait for Model-II.}
\end{figure}

\begin{figure}[htp!] 
   \centering 
   \mbox{\includegraphics[width=0.48\textwidth]{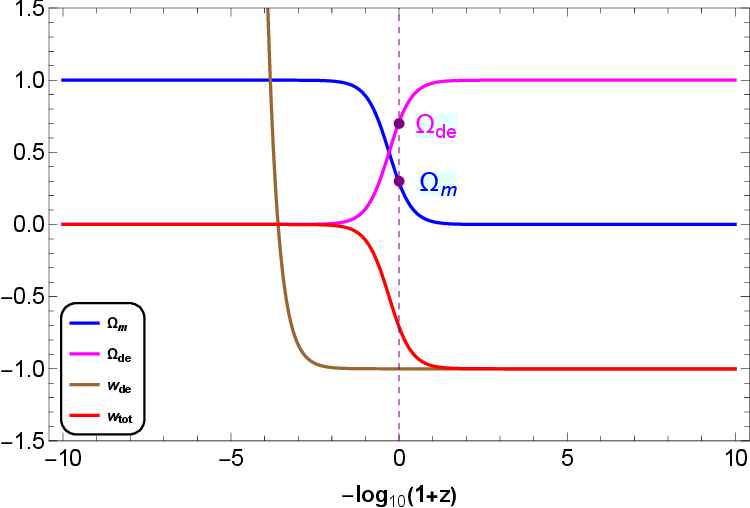}} 
    \hspace{10pt}
    \mbox{\includegraphics[width=0.48\textwidth]{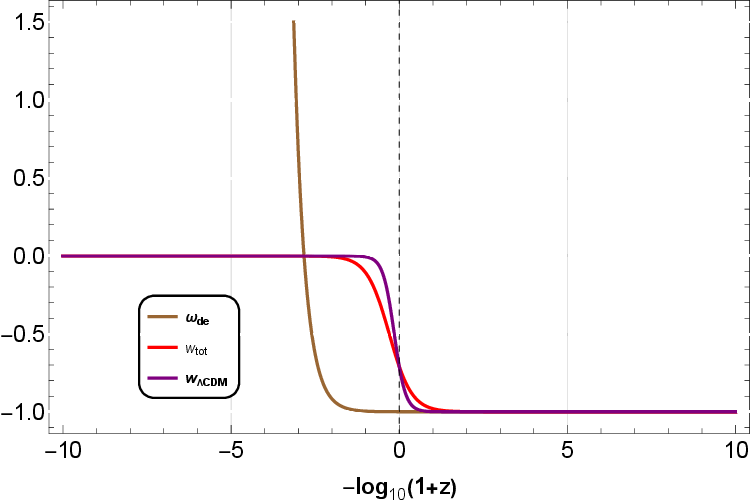}}
    \hspace{10pt}
    \mbox{\includegraphics[width=0.48\textwidth]{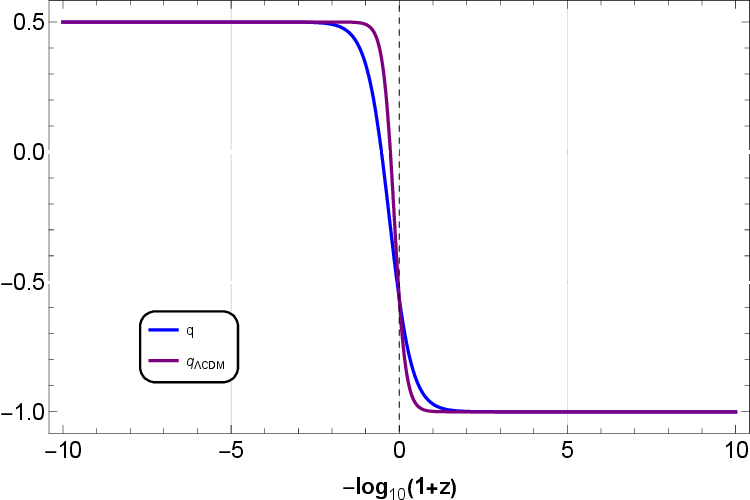}}
    \caption{The evolution of the density parameters (shown in the \textit{Upper left panel}), the EoS parameters (illustrated in the \textit{Upper right panel}), and the deceleration parameter (depicted in the \textit{Lower panel}) are presented for Model-II. The initial conditions are set as $x = 10^{-3}$, $y = 10^{-6}$, and $\alpha=0.001$. The vertical dashed line represents the present time.}
   \label{Phase Portrait for the dynamical system of Model II}
\end{figure}

Figure 3.3 illustrates the phase portrait in three-dimensional space, depicting the trajectory's evolution as it transitions from matter-dominated to dark-energy-dominated critical points. The diagram clearly shows the sequential progression of the trajectory, beginning at the unstable node corresponding to critical point $B_{3}$, passing through the saddle instability at $B_{4}$, and ultimately stabilizing at the node represented by critical point $B_{1}$.

Figure 3.4 (Upper panels) provides an illustration of the evolutionary history of the density and equation of state (EoS) parameters as functions of redshift $ z $. The initial conditions are calibrated to correspond to present-day values (at $ z = 0 $). In the upper left panel, the transition of the universe from a matter-dominated phase to an accelerated expansion era is depicted. The current density parameters are approximately $ \Omega_{\mathrm{m}} \approx 0.3 $ for matter and $ \Omega_{\mathrm{de}} \approx 0.7 $ for dark energy. The upper right panel shows the evolution of the total EoS parameter, which starts in a matter-dominated era with $ \displaystyle w_{\mathrm{tot}} = 0 $ and evolves towards the dark energy sector, where $ \displaystyle w_{\mathrm{tot}} \approx -1 $. The dark energy EoS parameter approaches $ -1 $ in the later stages of evolution, with the present value of $ \displaystyle w_{\mathrm{de}} = -1 $ aligning with the current observational constraint of $ \displaystyle w_{\mathrm{de}} = -1.028 \pm 0.032 $ \citep{Planck:2018fzr}. In the lower panel, the deceleration parameter transitions from a decelerating phase to an accelerating phase, with the transition occurring at $ z = 0.64 $ and the current value of the deceleration parameter being $ q_{0} = -0.56 $ \citep{Camarena:2019moy}.  

\vspace{1em}
\subsection{Model III}
In this subsection we consider 
\begin{equation}
F(Q)=\kappa Q \ln \left( \frac{Q}{Q_{0}} \right)- Q ,
\end{equation}
 The autonomous systems are  
 \begin{equation}
    x'=\frac{3(1-x-y)(y+2x)}{4 \kappa-y+2},  
 \end{equation}
 \begin{equation}
    y'=\frac{12 \kappa(1-x-y)}{4 \kappa-y+2},  
 \end{equation}
\begin{equation}
\xi' = -\xi(\xi+2) + \frac{3(1-x-y)}{2-y} + \frac{3(1-x-y)\xi}{4 \kappa-y+2}.
\end{equation}
The given system exhibits singularities only at $ y=2 $.
The corresponding Eos and deceleration parameters reduces to,  
\begin{align}
\displaystyle w_{\mathrm{de}} &= -1 + \frac{(4 \kappa-y)(x+y-1)}{(x+y)(4 \kappa-y+2)}, \\
\displaystyle w_{\mathrm{tot}} &= -1 - \frac{2(x+y-1)}{4 \kappa-y+2}, \\
q &= -1 - \frac{3(x+y-1)}{4 \kappa-y+2}.
\end{align}

Four critical points have been identified and are presented in Table 3.5, which details their associated cosmological characteristics. Table 3.6 provides the eigenvalues of the Jacobian matrix for both the background and perturbation levels.

\begin{table}[htp!]
\centering
\begin{tabular}{ ||p{1.5cm}||p{2.6cm}||p{0.8cm}||p{0.8cm}||p{0.8cm}||p{2cm}||p{0.8cm}||p{0.8cm}|| }
 \hline
 \textit{CP} & $\mathbf{(x_{c}, y_{c}, \xi_{c})}$ & $\mathbf{\Omega_{\mathrm{m}}}$ & $\mathbf{\Omega_{\mathrm{de}}}$ & $ \mathbf {\xi} $ & $ \mathbf{\displaystyle w_{\mathrm{de}}} $ & $\mathbf{ \displaystyle w_{\mathrm{tot}}} $ & $ \mathbf{q }$ \\
 \hline
 $ C_{1} $ & $ (x,1-x,0) $ & $ 0 $ & $ 1 $ & $ 0 $ & $ -1+\frac{8\kappa}{1+x+8\kappa} $ & $ -1 $ & $ -1 $  \\
 \hline
 $ C_{2} $ & $ (x,1-x,-2) $ & $ 0 $ & $ 1 $ & $ -2 $ & $ -1+\frac{8\kappa}{1+x+8\kappa} $ & $ -1 $ & $ -1 $  \\
 \hline
  $ C_{3} $ & $ (x,-2x,-\frac{3}{2}) $ & $ 1+x $ & $ -x $ & $ -\frac{3}{2} $ & $ 0 $ & $ 0 $ & $ \frac{1}{2} $ \\
 \hline
  $ C_{4} $ & $ (x,-2x,1) $ & $ 1+x $ & $ -x $ & $ 1 $ & $ 0 $ & $ 0 $ & $ \frac{1}{2} $ \\
 \hline
\end{tabular}
\caption{Critical points (CP), matter, dark energy, perturbation and EoS parameters}
\end{table} 

\begin{table}[htp!]
\centering
\begin{tabular}{ ||p{1.5cm}||p{2.4cm}||p{3.2cm}|| }
 \hline
 \textit{CP} & \textit{Eigen-Values} & \textit{Stability Condition}  \\
 \hline
 $ C_{1} $ & $ (0,-3,-2) $ & Stable  \\
 \hline
 $ C_{2} $ & $ (0,-3,2) $ & Saddle  \\
 \hline
  $ C_{3} $ & $ (0,\frac{5}{2},3) $ & Unstable \\
 \hline
  $ C_{4} $ & $ (0,-\frac{5}{2},3) $ & Saddle \\
 \hline
\end{tabular}
\caption{Eigen values and Stability conditions}
\end{table} 

\vspace{1em}
\subsection*{Summary of the critical points (Model-II):}
\begin{itemize}

    \item  \textit{Critical Point \( C_1 = (x, 1 - x, 0) \):} 
This critical point represents a universe completely dominated by dark energy, with \( \Omega_{\mathrm{de}} = 1 \) and \( \Omega_{\mathrm{m}} = 0 \). The effective dark energy equation of state is given by \( w_{\mathrm{de}} = -1 + \frac{8\kappa}{1 + x + 4\kappa} \), showing dependence on \( x \) and \( \kappa \), while the total equation of state and deceleration parameters are \( w_{\mathrm{tot}} = -1 \) and \( q = -1 \), respectively, indicating exponential expansion. The eigenvalues \( (0, -3, -2) \) imply non-hyperbolic stability, and center manifold theory is applied. The reduced equation suggests asymptotic stability in the central direction. Thus, \( C_1 \) corresponds to a late-time accelerating attractor driven by dynamic dark energy.

    \item  \textit{Critical Point \( C_2 = (1 - y, y, -2) \):} 
This critical point also represents a dark energy–dominated universe, with \( \Omega_{\mathrm{m}} = 0 \), \( \Omega_{\mathrm{de}} = 1 \), and perturbation parameter \( \xi = -2 \). The dynamic equation of state for dark energy is \( w_{\mathrm{de}} = -1 + \frac{8\kappa}{1 + x + 4\kappa} \), leading to an overall \( w_{\mathrm{tot}} = -1 \) and \( q = -1 \), consistent with de Sitter expansion. However, the eigenvalues \( (0, -3, 2) \) indicate a saddle-type instability, with perturbations growing in certain directions. Therefore, \( C_2 \) corresponds to an inflationary-like phase, but is not a stable attractor.

\item  \textit{Critical Point \( C_3 = (x, -2x, -\frac{3}{2}) \):} 
The point \( C_3 \) describes a matter-dominated universe where \( \Omega_{\mathrm{m}} = 1 + x \), \( \Omega_{\mathrm{de}} = -x \), with \( x \in [-1, 0] \). The dark energy behaves unusually like pressureless dust (\( w_{\mathrm{de}} = 0 \)), and the total equation of state is \( w_{\mathrm{tot}} = 0 \), with \( q = \frac{1}{2} \), indicating decelerating expansion. The eigenvalues \( (0, \frac{5}{2}, 3) \) confirm that the point is unstable. Thus, \( C_3 \) is a transient matter-dominated state, with instability leading to future evolution away from this point.

    \item \textit{Critical Point \( C_4 = (0, 0, 1) \):} 
This critical point also lies in a matter-dominated regime with \( \Omega_{\mathrm{m}} = 1 + x \), \( \Omega_{\mathrm{de}} = -x \), for \( x \in [-1, 0] \). Both \( w_{\mathrm{de}} = 0 \) and \( w_{\mathrm{tot}} = 0 \), with deceleration parameter \( q = \frac{1}{2} \), indicating a standard decelerating phase. The eigenvalues \( (0, -\frac{5}{2}, 3) \) suggest that the point is a saddle, with one stable and one unstable direction. Hence, \( C_4 \) is important for understanding cosmic structure formation but is not a final attractor.
\end{itemize}

The analysis reveals the presence of two matter-dominated critical points ($C_{3}$ and $C_{4}$) and two dark energy-dominated critical points ($C_{1}$ and $C_{2}$) within the framework of the logarithmic form of $ f(Q) $ gravity. The matter-dominated critical points, $C_{3}$ and $C_{4}$, exhibit inherent instability. In particular, critical point $C_{3}$, functioning as a saddle point, signifies a well-defined growth rate in matter perturbations. In contrast, critical point $C_{4}$, identified as an unstable node, represents the decay of matter perturbations, with accelerated expansion observed only at the background level. On the other hand, critical point $C_{1}$ displays consistent accelerated expansion at both the background and perturbation levels, establishing it as a stable configuration.

\begin{figure}[htp!]
    \centering
    \includegraphics[width=0.48\textwidth]{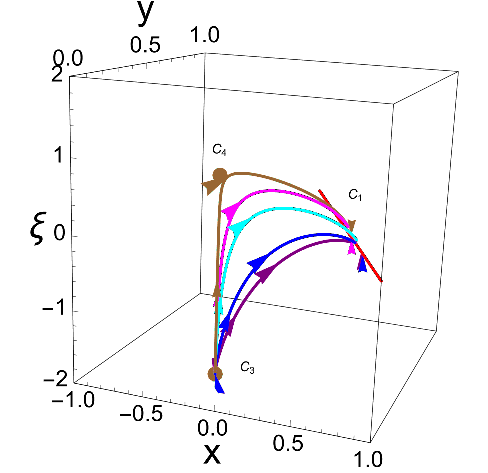}
    \caption{3D phase  portrait for Model-III.}
\end{figure}

\begin{figure}[htp!]
   \centering 
   \mbox{\includegraphics[width=0.48\textwidth]{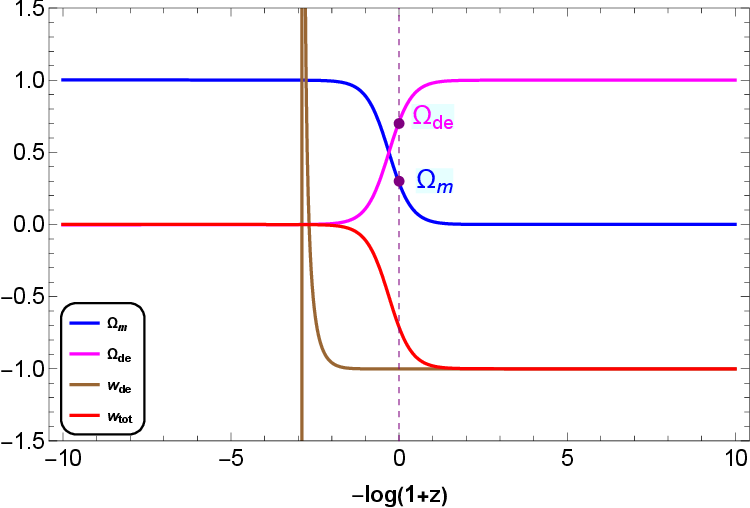}}
    \hspace{10pt}
    \mbox{\includegraphics[width=0.48\textwidth]{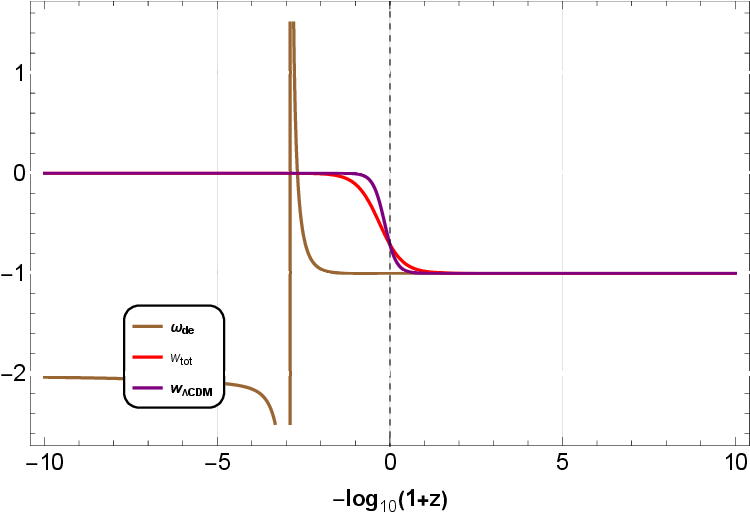}}
    \hspace{10pt}
    \mbox{\includegraphics[width=0.48\textwidth]{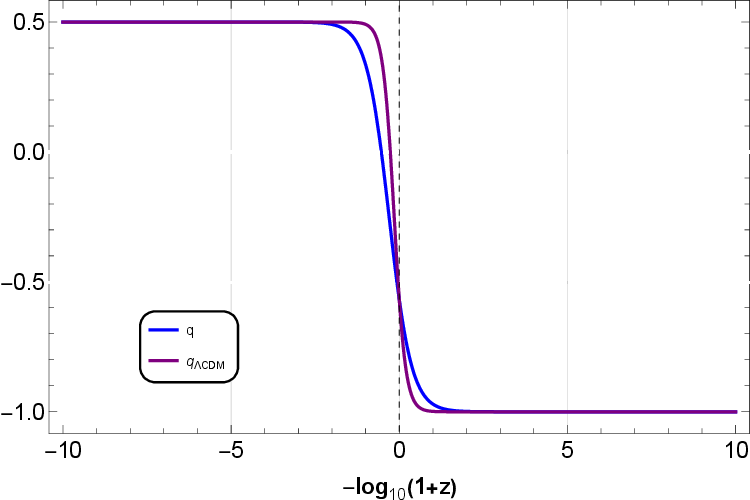}}
    \caption{ The evolution of the density parameters (shown in the \textit{Upper left panel}), the EoS parameters (illustrated in the \textit{Upper right panel}), and the deceleration parameter (depicted in the \textit{Lower panel}) are presented for Model-III. The initial conditions are set as $x = 10^{-3}$, $y = 10^{-6}$, and $\kappa = 0.0001$. The vertical dashed line represents the present time.}
   \label{Phase Portrait for the dynamical system of Model III}
\end{figure}

Figure 3.5 presents the phase portrait in three-dimensional space, illustrating the trajectory's evolution as it progresses from matter-dominated to dark-energy-dominated critical points. The diagram clearly delineates the sequential transition of the trajectory, commencing at the unstable node associated with critical point $C_{3}$, moving through the saddle point instability at $C_{4}$, and ultimately converging to the stable node represented by critical point $C_{1}$.

Figure 3.6 (Upper panels) presents the evolutionary trajectories of the density and equation of state (EoS) parameters as functions of redshift $ z $. The initial conditions are adjusted to match the current values at $ z = 0 $. The \textit{upper left panel} illustrates the universe's progression from a matter-dominated phase to a phase of accelerated expansion. At present, the density parameters are approximately $\Omega_{\mathrm{m}} \approx 0.3$ for matter and $\Omega_{\mathrm{de}} \approx 0.7$ for dark energy. The \textit{upper right panel} depicts the evolution of the total equation of state (EoS) parameter, which begins in the matter-dominated regime with $\displaystyle w_{\mathrm{tot}} = 0$ and gradually transitions towards the dark energy regime, where $\displaystyle w_{\mathrm{tot}}$ approaches $-1$. Simultaneously, the EoS parameter for dark energy converges towards $-1$ in the later stages of evolution, aligning with the current observational constraint of $\displaystyle w_{\mathrm{de}} = -1.028 \pm 0.032$ \citep{Planck:2018fzr}. The \textit{lower panel} shows the deceleration parameter's shift from a decelerating phase to an accelerating phase, with the transition occurring at $z = 0.60$. The current value of the deceleration parameter is $q_{0} = -0.57$ \citep{Camarena:2019moy}.

\vspace{1em}
\section{Discussion and Conclusion}

The $f(Q)$ gravity framework provides a promising alternative to General Relativity, allowing for the explanation of various cosmological phenomena including late-time acceleration. Through dynamical system analysis, we explored the qualitative behavior of the Universe by converting the cosmological field equations into an autonomous system using variables $(x, y, \xi)$ that encode the background and perturbative dynamics. Three distinct models based on logarithmic and polynomial forms of $f(Q)$ were studied, and each yielded four critical points representing dark energy and matter-dominated phases.

In all models, the deceleration parameter showed a clear transition from early-time deceleration to late-time acceleration, with present values of $q_0 \approx -0.56$ to $-0.57$, consistent with observations. The dark energy equation of state at present is $w_{\mathrm{de}} = -1$, and the density parameters satisfy $\Omega_{\mathrm{m}} \approx 0.3$ and $\Omega_{\mathrm{de}} \approx 0.7$. Phase-space trajectories demonstrate the system’s evolution from unstable matter-dominated states to stable dark energy attractors, highlighting the viability of $f(Q)$ gravity to reproduce late-time cosmic acceleration without invoking exotic matter.

This study reaffirms that dynamical system techniques, when extended to perturbation levels, provide significant insight into the stability and evolution of cosmological models in modified gravity. The consistency across different functional forms of $f(Q)$ strengthens its position as a viable framework for explaining the accelerated expansion of the Universe.

\chapter{Cosmological Constraints from Big Bang Nucleosynthesis and Dynamical Analysis in Symmetric Teleparallel Gravity} \label{c4}
\section{Introduction}
In this chapter, we analyze two prominent $f(Q)$ gravity models with boundary terms namely, the Log-square-root model and the Hyperbolic tangent-power model within the framework of higher-curvature gravity based on non-metricity. The study employs Big Bang Nucleosynthesis (\gls{bbn}) constraints to bound deviations in the freeze-out temperature $T_f$ from the standard $\Lambda$CDM scenario, establishing limits on model parameters through observational bounds on $\left| \frac{\delta T_f}{T_f} \right|$.
Using dynamical system analysis, we investigate both the background evolution and perturbative behavior of the models. Critical points are identified and their stability properties examined, including non-hyperbolic cases through Center Manifold Theory. The results reveal a matter-dominated saddle phase with appropriate perturbation growth, followed by a stable dark-energy-dominated phase featuring late-time cosmic acceleration. These findings support the observational viability of $f(Q)$ gravity with boundary terms as an effective alternative to the $\Lambda$CDM model \citep{Vishwakarma:2024qvw}.

\vspace{1em}
\section{Mathematical Formalism}
Jimenez et al.\citep{BeltranJimenez:2019tme, BeltranJimenez:2017tkd} introduced the concept of symmetric teleparallel gravity, also identified as $f(Q)$ gravity. This modified gravitational framework is characterized by an action defined as follows:
\begin{equation}
S=\int [-\frac{1}{16\pi G}f(Q)+{\mathcal{L}}_{m}]\sqrt{-g} \,d^{4}x\ ,
\end{equation}
wherein $g$ represents the determinant of the metric tensor $ g_{\mu \nu} $, while $ {\mathcal{L}}_{m} $ denotes the Lagrangian density associated with matter. The function $ f(Q) $ in this formulation stands as an arbitrary expression of $ Q $, the non-metricity scalar that governs gravitational interactions within this context.
The non-metricity scalar $Q$ is written as  
\begin{equation}
Q=-\frac{1}{4}Q_{\alpha \beta \gamma}Q^{\alpha \beta \gamma}+\frac{1}{2}Q_{\alpha \beta \gamma}Q^{\gamma \beta \alpha} + \frac{1}{4}Q_{\alpha}Q^{\alpha}-\frac{1}{2}Q_{\alpha}\bar{Q}^{\alpha} ,
\end{equation}
where the quantities $ Q_{\alpha}\equiv Q_{\alpha^{\mu}\mu} $ and $ \bar{Q}^{\alpha}\equiv Q_{\mu}^{\mu \alpha} $ represent two independent traces derived by contracting the non-metricity tensor, expressed as \begin{center} $ Q_{\alpha \mu \nu}\equiv \bigtriangledown_{\alpha}g_{\mu \nu} $ \end{center}. This tensorial manipulation allows for the definition of the non-metricity scalar, given by $ Q=-Q_{\alpha \mu \nu } Q $.

To facilitate the analysis, assuming the decomposition $ f(Q)=Q+F(Q) $ and setting $ 8\pi G_N=1 $ for simplicity, the corresponding field equations can be expressed as:
\begin{equation}
3H^{2}= \rho + \frac{F}{2}-QF_{Q} ,
\end{equation}
\begin{equation}
(2QF_{QQ}+F_{Q}+1)\dot{H}+\frac{1}{4}(Q+2QF_{Q}-F)=-2p
\end{equation}
Here, $ F_{Q}=\frac{dF}{dQ} $ and $ F_{QQ}=\frac{d^{2}F}{dQ^{2}} $. These equations describe the evolution of the Hubble parameter $ H $ and the effective energy density $ \rho $, encompassing contributions from the non-metricity scalar $ Q $ and the function $ F(Q) $, providing a comprehensive understanding of the universe's dynamics within this theoretical framework.

Hence the Friedmann's equation (4.3) can be simply written as
\begin{equation}
\Omega_{m}+\Omega_{Q}=1 ,
\end{equation}
where \begin{center}
$\Omega_{m}=\dfrac{\rho}{3H^{2}} , \qquad \Omega_{Q}=\dfrac{\dfrac{F}{2}-QF_{Q}}{3H^{2}} $, 
\end{center}
Introducing the effective total energy density $ \rho_{eff} $ and total energy pressure $ p_{eff} $, as \citep{Anagnostopoulos:2022gej}
\begin{equation}
\rho_{eff}\equiv \rho+\frac{F}{2}-QF_{Q}  ,
\end{equation}
\begin{equation}
p_{eff}\equiv \frac{\rho (1+\omega)}{2QF_{QQ}+F_{Q}+1}-\frac{Q}{2}  ,
\end{equation}
the corresponding total equation of state $  \displaystyle w_{eff} $ can be written as
\begin{equation}
 \displaystyle w_{eff}=\frac{p_{eff}}{\rho_{eff}}=-1+\frac{\Omega_{m}(1+\omega)}{2QF_{QQ}+F_{Q}+1} .
\end{equation}

In the investigation of linear perturbations, our focus lies on the matter density contrast $ \delta=\frac{\delta _{\rho}}{\rho} $, wherein $ \delta_{ \rho} $ denotes the perturbation in the matter energy density. The governing equation for the evolution of the matter over density in the quasi-static regime, as presented in \citep{Khyllep:2022spx}, is expressed as:
\begin{equation}
\ddot{\delta}+2H\dot{\delta}=\frac{\rho \delta}{2(1+F_{Q})},
\end{equation}
where the denominator on the right-hand side elucidates the emergence of an effective Newtonian constant. Notably, within the context of minute scales significantly confined within the cosmic horizon, the terms associated with temporal derivatives in the perturbation equations are neglected, thereby reducing the formulation to predominantly encompass spatial derivative terms \citep{Anagnostopoulos:2022gej, Lazkoz:2019sjl}.

\vspace{1em}
\section{Big-Bang Nucleosynthesis (\gls{bbn}) Constraints}

The \gls{bbn} constraints formalism is discussed in this section. The radiation period sees the realisation of \gls{bbn} \citep{Bernstein:1988ad, Kolb:1990vq, Olive:1999ij, Cyburt:2015mya}. 
In the context of \gls{bbn} and standard cosmology within the framework of the Standard Model of particle physics and general relativity, an approximation is made to the first Friedmann equation as detailed in reference \citep{Asimakis:2021yct}:
\begin{equation}
H^{2}\approx \frac{M_{P}^{-2}}{3}\rho_{r}\equiv H^{2}_{GR}
\end{equation}
During the radiation-dominated epoch, crucial in \gls{bbn}, the consideration of the energy density of relativistic particles becomes imperative. This density, denoted by $\rho_{r}$, is expressed as:
\begin{equation}
\rho_{r}=\frac{\pi^{2}}{30}g_{\ast}T^{4}
\end{equation}
Here, $T$ signifies the temperature, while the parameter $g_{\ast}$, approximately valued at $10$, represents the effective number of degrees of freedom accounting for relativistic particles \citep{Cyburt:2015mya, Barrow:2020kug}.
Hence from equation (4.10) and (4.11), we obtain
\begin{equation}
H(T) \approx (\frac{4\pi^{3}g_{\ast}}{45})^{\frac{1}{2}}\frac{T^{2}}{M_{Pc}},
\end{equation}
where \begin{center}
$ M_{Pc}=(8\pi)^{\frac{1}{2}}M_{P}=1.22\times 10^{19} GeV $,
\end{center} is the Planck mass.
We can extract the expression between temperature and time, since the radiation conservation equation finally results in a scale factor evolution of the form $ a\sim t^{\frac{1}{2}} $, namely 
\begin{center}
$ \frac{1}{t}\simeq (\frac{32 \pi^{3}g_{\ast}}{90})^{\frac{1}{2}}\frac{T^{2}}{M_{Pc}} \;\;\; (or \;\; T(t)\simeq (\frac{t}{sec})^{\frac{-1}{2}} MeV ) $.
\end{center} 

In the context of \gls{bbn}, the calculation of neutron abundance involves considering the rate of conversion between protons and neutrons, as referenced in \citep{Olive:1999ij, Cyburt:2015mya}. This conversion rate $\lambda_{pn}(T)$ is defined as a combination of various processes:
\begin{equation}
\lambda_{pn}(T)=\lambda_{(n+\nu_{e}\rightarrow p+e^{1})} + \lambda_{(n+e^{+}\rightarrow p+\bar{\nu}_{e})} +\lambda_{(n\rightarrow p+e^{-}+\bar{\nu}_{e})}
\end{equation}
The total rate, denoted as $\lambda_{tot}(T)$, is the summation of $\lambda_{np}(T)$ and $\lambda_{pn}(T)$, where $\lambda_{np}(T)$ represents the inverse of $\lambda_{pn}(T)$. Upon straightforward calculations, an expression for $\lambda_{tot}(T)$ is obtained, as referenced in \citep{Torres:1997sn, Lambiase:2005kb, Lambiase:2011zz, Capozziello:2017bxm, Barrow:2020kug}:
\begin{equation}
\lambda_{tot}(T)= 4AT^{3}(4! T^{2}+2\times 3!MT+2!M^{2});
\end{equation}
Here, $M=m_{n}-m_{p}=1.29\times 10^{-3} \text{GeV}$ signifies the mass difference between the proton and neutron, while $A=1.02\times 10^{-11} \text{GeV}^{-4}$. The assumptions underlying this calculation entail equality in temperatures among different particles (neutrinos, electrons, and photons) and the application of Boltzmann distribution, assuming sufficiently low temperatures to use this distribution instead of the Fermi-Dirac distribution. The comprehensive framework of BBN details can be found in the Appendix of \citep{Barrow:2020kug}.

The determination of the freeze-out temperature ($T_{f}$) involves a comparison between the expansion rate of the universe ($\frac{1}{H}$) and the total interaction rate $\lambda_{tot}(T)$. A significant distinction arises based on the relative magnitudes: if $\frac{1}{H}$ is notably smaller than $\lambda_{tot}(T)$, it signifies that all processes occur within a state of thermal equilibrium, as cited in \citep{Bernstein:1988ad, Kolb:1990vq}. Conversely, when $\frac{1}{H}$ significantly exceeds $\lambda_{tot}(T)$, particles lack adequate time intervals for interactions to occur. The freeze-out temperature $T_{f}$ delineates the temperature at which particles undergo a disconnection phase, and it corresponds to the moment when $H = \lambda_{tot}(T) \approx qT^{5}$, with $q=4A4!\simeq 9.8\times10^{-10} \text{GeV}^{-4}$, as referenced in \citep{Torres:1997sn, Lambiase:2005kb, Lambiase:2011zz, Capozziello:2017bxm, Barrow:2020kug}. Combining equations (29) and (31), the expression for $T_{f}$ is derived:
\begin{equation}
 T_{f}= \left( \frac{4\pi^{3}g_{\ast}}{45 M^{2}_{Pc}q^{2}}\right)^{\frac{1}{6}}\sim 0.0006 GeV
 \end{equation}
Here, $M_{P}$ denotes the Planck mass, and $g_{\ast}$ represents the effective number of degrees of freedom. This determination holds significance in understanding the temperature regime where particle interactions cease, leading to the transition from equilibrium to non-equilibrium conditions in the early universe.

In the epoch of \gls{bbn}, Modified gravity introduces an effective dark energy density, denoted as $\rho_{DE}$, which coexists alongside other components, particularly radiation density ($\rho_{r}$). Considering $\rho_{DE}$ to be smaller than $\rho_{r}$, it is deemed a first-order deviation. Consequently, the Hubble function is modified as follows:
\begin{center}
$ H= H_{GR}\sqrt{1+\frac{\rho_{DE}}{\rho_{r}}}= H_{GR}+\delta H $,
\end{center}
where $H_{GR}$ represents the Hubble parameter in standard cosmology, and the deviation from this standard scenario is encapsulated by:
\begin{center}
$ \delta H=\left(\sqrt{1+\frac{\rho_{DE}}{\rho_{r}}}-1\right) H_{GR} $
\end{center}
This deviation, $\delta H$, induces a consequential shift in the freeze-out temperature, denoted as $\delta T_{f}$. This alteration signifies the impact of modified gravity, particularly the introduced effective dark energy, on the critical temperature where particle interactions cease during \gls{bbn}.
However, the relationship $H_{GR}=\lambda_{tot}\approx qT_{f}^{5}$ leads to an expression for the deviation $\delta H$ as \begin{center}
$ \delta H=\left(\sqrt{1+\frac{\rho_{DE}}{\rho_{r}}}-1\right) H_{GR}=5qT_{f}^{4}\delta T_{f} $.
\end{center}. Considering $\rho_{DE}$ to be significantly smaller than $\rho_{r}$ during this period, the equation simplifies to:
\begin{equation}
\frac{\delta T_{f}}{T_{f}}\simeq \frac{\rho_{DE}}{\rho_{r}}\frac{H_{GR}}{10qT_{f}^{5}}.
\end{equation}
This theoretically derived ratio $\frac{\delta T_{f}}{T_{f}}$ needs comparison with an observational constraint:
\begin{equation}
\vert \frac{\delta T_{f}}{T_{f}} \vert < 4.7\times 10^{-4} 
\end{equation}
This constraint is established through observational assessments of the baryon mass fraction converted into $^{4}H_{e}$, as documented in \citep{Coc:2003ce, Olive:1996zu, Izotov:1998mj, Fields:1998gv, Izotov:1999wa, Kirkman:2003uv, Izotov:2003xn}. In forthcoming subsections, this formalism, specifically expressed by equation (4.16), is applied to impose limitations on $\rho_{DE}$ and consequently on specific models of modified gravity. This process aids in understanding the bounds of the effective dark energy and its implications for various scenarios within modified gravity frameworks.

\vspace{1em}
\section{\gls{bbn} Constraints on $f(Q)$~ gravity}
In this section, the models, i.e the Log-square-root model, $ f(Q)=Q+nQ_{0}\sqrt{\frac{Q}{\lambda Q_{0}}}ln(\lambda\frac{Q_{0}}{Q}) $ and the Hyperbolic tangent-power model, $ f(Q)=Q+\lambda Q_{0}(\frac{Q}{Q_{0}})^{n}tanh(\frac{Q_{0}}{Q}) $ will be tested against the constraint (4.17). This will be realized by calculating (4.16) for each of the models.

\vspace{1em}
\subsection{Log-square-root model}
In accordance with the methodology outlined in \citep{Anagnostopoulos:2022gej, Bamba:2010wb}, we introduce a logarithmic-square-root model represented by $f(Q)$:
\begin{center}
$ f(Q)=Q+nQ_{0}\sqrt{\frac{Q}{\lambda Q_{0}}}ln(\lambda\frac{Q_{0}}{Q}) $
\end{center}
Here, $n$ and $\lambda > 0$ are the parameters characterizing the model. Correspondingly, the associated effective energy density is expressed as:
\begin{equation}
\rho_{DE}=\frac{\Omega_{F0}}{16\pi G}\sqrt{Q_{0}Q} 
\end{equation}
The first Friedmann equation, evaluated at the present time, leads to a relationship determining the parameter $n$:
\begin{equation}
n=\frac{\Omega_{F0}}{2}\sqrt{\lambda}
\end{equation}
Here, $\Omega_{F0}=1-\Omega_{m0}-\Omega_{r0}$, denoting the present value of a quantity, with subscripts indicating the present time. Substituting equations (4.18) and (4.19) into (4.16) yields:
\begin{equation}
\frac{\delta T_{f}}{T_{f}}\simeq \frac{8\pi^{3}}{9}\sqrt{\frac{2g_{\ast}}{5}}\frac{\Omega_{F0}T_{0}^{2}}{5qT_{f}^{5}M_{P}}.
\end{equation}
Upon incorporating equation (4.20) into equation (4.15), it's observed that within the range $0 \leq \Omega_{F0}\leq1$, the constraint (4.15) remains consistently satisfied. Consequently, this model aligns with the requirements of \gls{bbn} and offers a potential explanation for the acceleration observed in the late-time universe.

\vspace{1em}
\subsection{Hyperbolic tangent-power model}

Following a similar approach to \citep{Anagnostopoulos:2022gej, Wu:2010av}, we explore the hyperbolic tangent power model represented by:
\begin{center}
$ f(Q)=Q+Q_{0}(\frac{Q}{Q_{0}})^{n}tanh(\frac{Q_{0}}{Q}) $
\end{center} 
Here, the parameters $n$ and $\lambda > 0$ are introduced. The corresponding effective energy density is formulated as:
\begin{equation}
\rho_{DE} =  \frac{1}{16\pi G} \left( \frac{Q}{Q_{0}}\right)^{(n-1)}  \left[ Qtanh(\frac{Q_{0}}{Q})(1-2n)+2Q_{0}sech^{2}(\frac{Q_{0}}{Q})\right] 
\end{equation}
The expression for the parameter $n$ is derived from the first Friedmann equation at the present epoch:
\begin{equation}
n=\frac{1}{2}\left[-coth(1)\Omega_{F0}+1+4csch(2)\right] 
\end{equation}
Here, $\Omega_{F0}=1-\Omega_{m0}-\Omega_{r0}$ denotes the present value of a quantity. Substituting equations (4.21) and (4.22) into equation (4.15) yields:
\begin{equation}
\begin{split}
& \frac{\delta T_{f}}{T_{f}}\simeq \sqrt{\frac{3g_{\ast}}{5}}\frac{8\pi^{2}}{5qT_{f}^{7}M_{P}}\left( \frac{T_{f}}{T_{0}}\right)^{(4n-4)} \\ & 
\left\lbrace (1-2n)T_{f}^{4}tanh\left[ \left( \frac{T_{0}}{T_{f}}\right) ^{4}\right]+2T_{0}^{4}sech^{2}\left[ \left( \frac{T_{0}}{T_{f}}\right) ^{4}\right]\right\rbrace  .
\end{split}
\end{equation}
Upon imposing the constraint (4.16) on the parameter space of the hyperbolic tangent-power model, we obtained Figure 1. Through the plot of $\frac{\delta T_{f}}{T_{f}}$ against the model parameter $n$, it's observed that the \gls{bbn} constraints (4.16) are satisfied within the range $n\leq 1.9$. This validation indicates that this model aligns with the requirements of \gls{bbn} and presents a potential explanation for the acceleration observed in the late-time universe.
\begin{figure}
    \centering
    \includegraphics[width=0.48\textwidth]{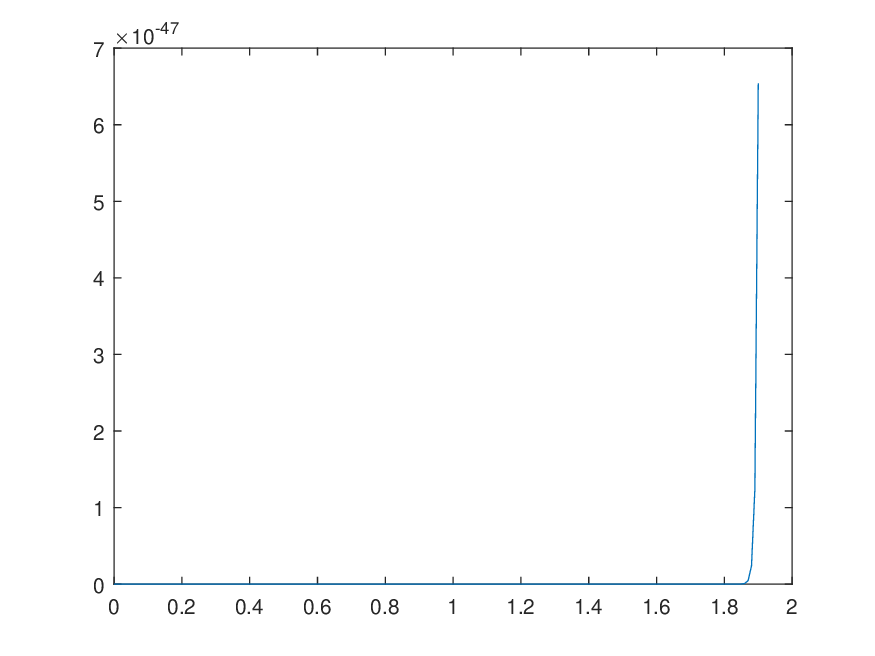}
    \caption{ $\frac{\delta T_{f}}{T_{f}}$ vs the model parameter $ n $ }
\end{figure}

\vspace{1em}
\section{Dynamical System for Background and Perturbation Analysis}

In this section, we construct a dynamical system incorporating both the background and perturbation equations for a general function $F(Q)$. Using the dynamical variables
\begin{equation}
x = \frac{F}{6H^2}, \qquad y = -2F_Q, \qquad u = \frac{d\ln\delta}{d\ln a},
\end{equation}
we transform the field equations into a first-order autonomous system. Here, $x$ and $y$ characterize the background evolution, while $u$ quantifies the growth or decay of matter perturbations.

The cosmic parameters are expressed as:
\begin{align*}
\Omega_m &= 1 - x - y, \qquad \Omega_Q = x + y, \\
w_{\mathrm{eff}} &= -1 + \frac{(1 - x - y)(1 + \omega)}{2QF_{QQ} - \frac{y}{2} + 1},
\end{align*}
with the evolution equations:
\begin{align}
x' &= -\frac{\dot{H}}{H^2}(y + 2x), \\
y' &= -\frac{\dot{H}}{H^2} 4QF_{QQ}, \\
u' &= -u(u+2) + \frac{3(1 - x - y)}{2 - y} - \frac{\dot{H}}{H^2}u,
\end{align}
and
\begin{equation}
\frac{\dot{H}}{H^2} = \frac{3(x + y - 1)}{4QF_{QQ} - y + 2}.
\end{equation}

The sign of $u$ determines perturbation behavior: $u > 0$ implies growth (instability), $u < 0$ implies decay (stability), and $u = 0$ indicates constant perturbations. Unstable or saddle points with $u > 0$ typically precede a stable late-time attractor with $u = 0$, marking the onset of accelerated expansion. We now proceed to analyze two specific models by specifying $F(Q)$ and evaluating $QF_{QQ}$.

\vspace{1em}
\subsection{Log-square-root model} 
The Log-square-root model, represented by the function 
\begin{equation}
F(Q)=nQ_{0}\sqrt{\frac{Q}{\lambda Q_{0}}}ln(\lambda\frac{Q_{0}}{Q}) 
\end{equation}, yields insights into the system's behavior. Specifically, when evaluating the second derivative of $F(Q)$ with respect to $ Q $ ($QF_{QQ}$), it follows that 
\begin{center}
$ QF_{QQ}=\frac{-n}{4}x $,
\end{center}.

The state variables evolve according to the differential equations:
\begin{align}
x' &= \frac{3(1-x-y)}{-nx-y+2}(y+2x), \\
y' &= \frac{3(1-x-y)}{-nx-y+2}(-nx), \\
u' &= -u(u+2) + \frac{3(1-x-y)}{2-y} + \frac{3(1-x-y)}{-nx-y+2}u.
\end{align}

The formulated dynamical system yields four critical points, which are presented in Table 4.1.
\begin{table}[htp!]
\centering
\begin{tabular}{ |p{2.3cm}|p{4.2cm}|p{0.6cm}|p{0.6cm}|p{0.6cm}|p{0.7cm}|p{2.6cm}| }
 \hline
 Critical Points & Eigen-Values & $\mathbf{\Omega_{m}}$ & $\mathbf{\Omega_{Q}}$ & $ u $ & $ \displaystyle w_{eff} $ & Stability Condition\\
 \hline
 $ (1-y,y,0) $ & $ (0,-3,-2) $ & $ 0 $ & $ 1 $ & $ 0 $ & $ -1 $ & Stable \\
 \hline
 $ (1-y,y,-2) $ & $ (0,-3,2) $ & $ 0 $ & $ 1 $ & $ -2 $ & $ -1 $ & Saddle\\
 \hline
  $ (0,0,-\frac{3}{2}) $ & $ (\frac{5}{2},\frac{3}{2}(1-\sqrt{1-n}),\frac{3}{2}(1+\sqrt{1-n})) $ & $ 1 $ & $ 0 $ & $ -\frac{3}{2} $ & $ 0 $ & Unstable spiral for $ n>1 $ and Unstable $ n\leq 1 $\\
 \hline
  $ (0,0,1) $ & $ (-\frac{5}{2},\frac{3}{2}(1-\sqrt{1-n}),\frac{3}{2}(1+\sqrt{1-n})) $ & $ 1 $ & $ 0 $ & $ 1 $ & $ 0 $ & Saddle for all values of $ n $ \\
 \hline
\end{tabular}
\caption{Critical Points, Stability Conditions, Matter perturbation and EoS parameter}
\end{table} 

\vspace{1em}
\subsection*{Summary of Critical Points of Log-square-root model:}
\begin{itemize}
    
\item  \textit{Critical Point $(1 - y, y, 0)$:}  
This point corresponds to a dark energy-dominated universe with $\Omega_{m} = 0$ and $\Omega_{Q} = 1$, representing accelerated expansion driven solely by dark energy. The effective equation of state is $w_{\text{eff}} = -1$, mimicking a cosmological constant. At the perturbation level, $u = 0$ indicates constant matter perturbations. The system is non-hyperbolic due to a zero eigenvalue, and center manifold analysis reveals asymptotic stability. This critical point effectively models the late-time universe.

\item \textit{Critical Point $(1 - y, y, -2)$:}  
This point also represents a dark energy-dominated universe with $\Omega_{Q} = 1$ and $w_{\text{eff}} = -1$. However, unlike the previous case, $u = -2$ denotes decaying matter perturbations. It is a saddle point with eigenvalues $(0, -3, 2)$, making it unsuitable for describing the final cosmic phase, but it can correspond to an intermediate inflationary stage.

\item \textit{Critical Point $(0, 0, -\frac{3}{2})$:}  
Here, $\Omega_{m} = 1$ and $w_{\text{eff}} = 0$, signifying a matter-dominated universe. The perturbative component $u = -\frac{3}{2}$ implies decaying overdensities. The point is unstable, with eigenvalues depending on the parameter $n$. For $n > 1$, it exhibits an unstable spiral behavior, and for $n \leq 1$, it is purely unstable. This point highlights early matter-era dynamics but is not a late-time attractor.

\item \textit{Critical Point $(0, 0, 1)$:}  
This matter-dominated solution also has $\Omega_{m} = 1$ and $w_{\text{eff}} = 0$. The perturbation variable $u = 1$ implies growing matter overdensity ($\rho \propto a$), making it relevant for structure formation. It is a saddle point for all $n$, acting as a transient phase in cosmic evolution before transitioning to dark energy domination.

\end{itemize}

\begin{figure}[htp!]  
   \mbox{\includegraphics[scale=0.47]{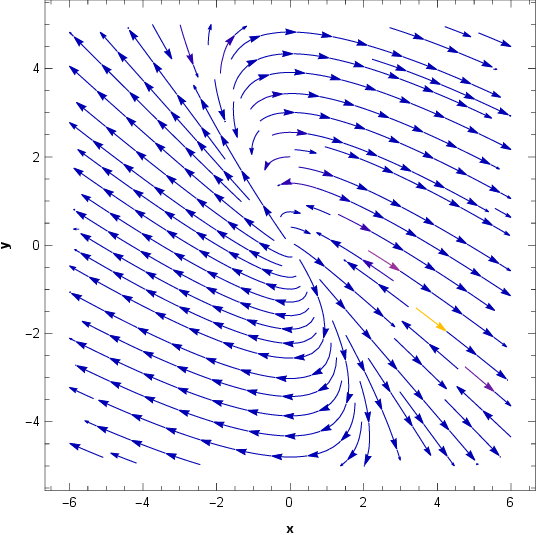}}   
    \hspace{10pt}
    \mbox{\includegraphics[scale=0.47]{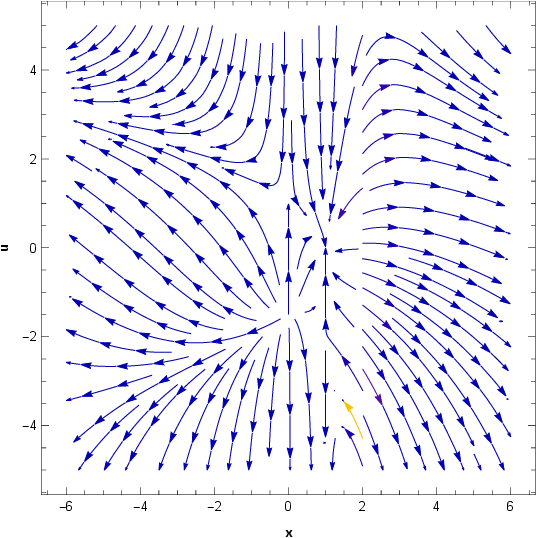}}
    \hspace{10pt}
    \mbox{\includegraphics[scale=0.47]{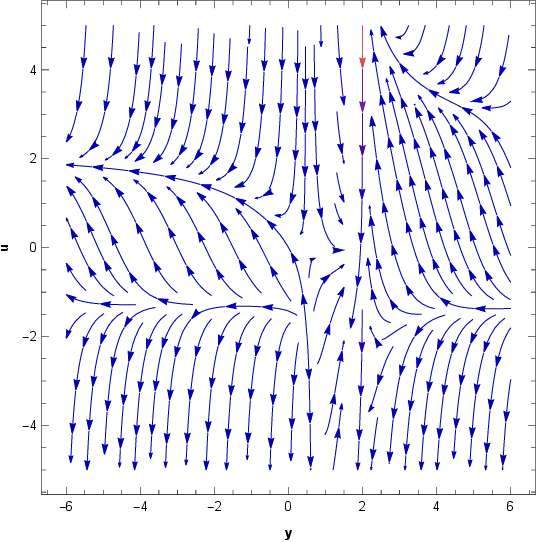}}
    \caption{ Phase Portrait for the dynamical system of Model I, (i)left panel($u=0$ and $n=1$); (ii)middle panel($y=0$ and $n=1$); (iii)right panel($x=0.5$ and $n=1$)}
   \label{Phase Portrait for the dynamical system of Model I}
\end{figure}

\begin{figure}[htp!]  
   \mbox{\includegraphics[scale=0.47]{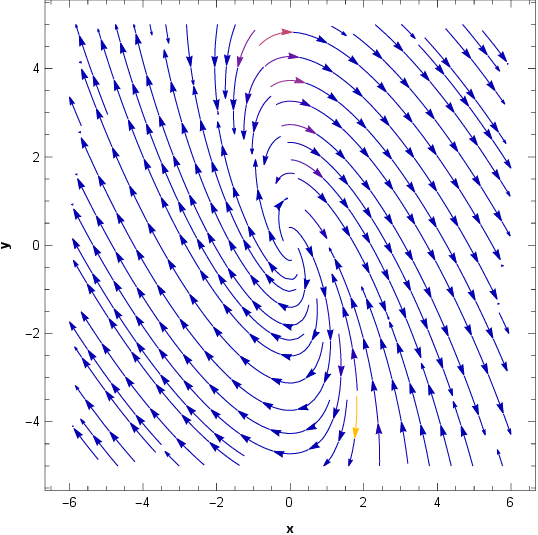}}   
    \hspace{10pt}
    \mbox{\includegraphics[scale=0.47]{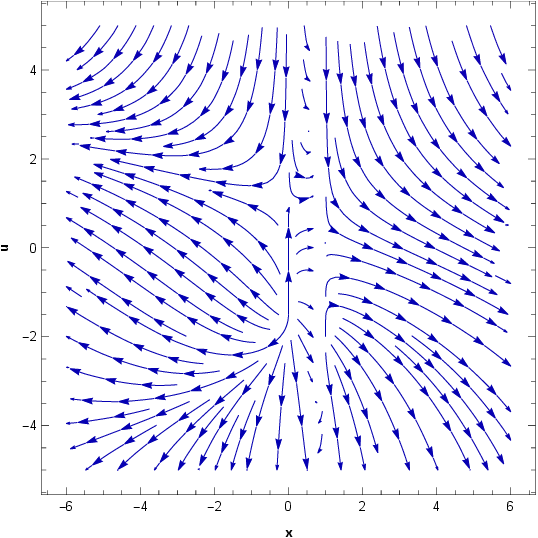}}
    \hspace{10pt}
    \mbox{\includegraphics[scale=0.47]{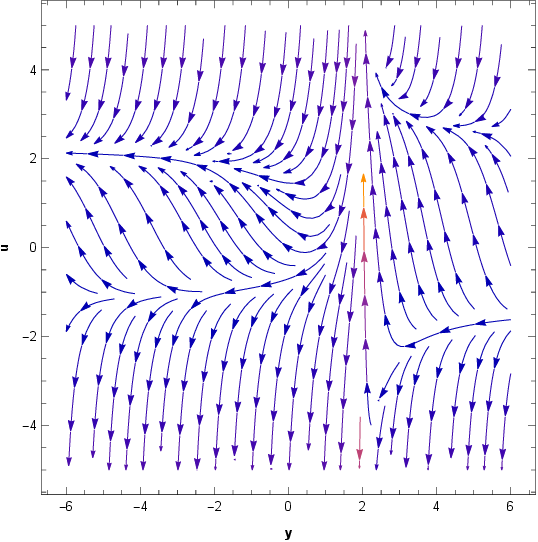}}
    \caption{ Phase Portrait for the dynamical system of Model I, (i)left panel($u=0$ and $n=3$); (ii)middle panel($y=0$ and $n=3$); (iii)right panel($x=0.5$ and $n=3$)}
   \label{Phase Portrait for the dynamical system of Model I}
\end{figure}

\vspace{1em}
\subsection{Hyperbolic tangent-power model} 
In this subsection we consider the hyperbolic tangent-power model
\begin{equation}
F(Q)=Q_{0}(\frac{Q}{Q_{0}})^{n}tanh(\frac{Q_{0}}{Q}) ,
\end{equation}
As
\begin{center}
$ QF_{QQ}=(1-n)(y+nx) $,
\end{center}
The autonomous system becomes
\begin{align}
 x' & =\frac{3(1-x-y)}{4(1-n)(y+nx)-y+2}(y+2x) \\ 
 y'& =\frac{3(1-x-y)}{4(1-n)(y+nx)-y+2}4(1-n)(y+nx) \\
 u' & = -u(u+2)+  \frac{3(1-x-y)}{2-y} +\frac{3(1-x-y)}{4(1-n)(y+nx)-y+2}u 
\end{align}
In this case, the system has four curves of critical points which are presented in Table 4.2.

\begin{table}[htp!]
\centering
\begin{tabular}{|p{2.3cm}|p{6.1cm}|p{0.7cm}|p{0.7cm}|p{0.7cm}|p{0.8cm}| }
 \hline
 Critical Points & Eigen-Values & $\mathbf{\Omega_{m}}$ & $\mathbf{\Omega_{Q}}$ & $ u $ & $\displaystyle w_{eff} $ \\
 \hline
 $ (1-y,y,0) $ & $ \left( 0,-2,-\frac{3}{2}\left( \frac{2+4n-4n^{2}+3y-8ny+4n^{2}y}{1+2n-2n^{2}-4ny+2n^{2}y} \right)\right) $ & $ 0 $ & $ 1 $ & $ 0 $ & $ -1 $  \\
 \hline
 $ (1-y,y,-2) $ & $ \left(0,2,-\frac{3}{2}\left( \frac{2+4n-4n^{2}+3y-8ny+4n^{2}y}{1+2n-2n^{2}-4ny+2n^{2}y} \right) \right) $ & $ 0 $ & $ 1 $ & $ -2 $ & $ -1 $ \\
 \hline
 $ (0,0,-\frac{3}{2}) $ & $ (\frac{5}{2},-3(n-2),-3(n-1)) $ & $ 1 $ & $ 0 $ & $ -\frac{3}{2} $ & $ 0 $  \\
 \hline
 $ (0,0,1) $ & $ (-\frac{5}{2},-3(n-2),-3(n-1)) $ & $ 1 $ & $ 0 $ & $ -\frac{3}{2} $ & $ 0 $ \\
 \hline
\end{tabular}
\caption{Critical Points, Stability Conditions, Matter perturbation and EoS parameter}
\end{table} 

\vspace{1em}
\subsection*{Summary of Critical Points of Hyperbolic tangent-power model}
\begin{itemize}
 \item  \textit{Critical Point $(1 - y, y, 0)$:}  
This critical point describes a dark energy-dominated solution with $\Omega_{Q} = 1$ and $w_{\text{eff}} = -1$, replicating the behavior of a cosmological constant. At the perturbation level, $u = 0$ indicates constant matter fluctuations. Three special cases are examined: (i) for $y = 0$, the point becomes $(1, 0, 0)$, asymptotically stable via center manifold theory; (ii) for $y = \frac{1}{2}$, the point $(\frac{1}{2}, \frac{1}{2}, 0)$ is stable for $n \neq 1$; and (iii) for $y = 1$, the point $(0, 1, 0)$ also maintains asymptotic stability for $n \neq 1$. In all cases, this point reflects the late-time universe dominated by dark energy with negligible matter contribution and non-growing perturbations.

 \item  \textit{Critical Point $(1 - y, y, -2)$:}  
This point also describes a dark energy-dominated universe, with $\Omega_{Q} = 1$ and $w_{\text{eff}} = -1$. However, $u = -2$ signifies decaying matter perturbations. Despite its accelerated expansion, the eigenvalues indicate a saddle point. Therefore, this configuration does not describe the late universe but is better interpreted as a transient inflationary phase in early cosmic evolution.

 \item  \textit{Critical Point $(0, 0, -\frac{3}{2})$:}  
This point corresponds to a matter-dominated era with $\Omega_{m} = 1$ and $w_{\text{eff}} = 0$. The perturbation parameter $u = -\frac{3}{2}$ suggests a decaying matter overdensity, $\delta \propto a^{-3/2}$, which is incompatible with structure formation. The eigenvalues of the Jacobian are parameter-dependent: for $n < 2$, the point is unstable; for $n \geq 2$, it becomes a saddle point. This point characterizes early matter-domination but lacks structure-forming features.

 \item  \textit{Critical Point $(0, 0, 1)$:}  
This point also represents a matter-dominated solution with $\Omega_{m} = 1$ and $w_{\text{eff}} = 0$. Unlike the previous case, $u = 1$ indicates growing matter perturbations, $\delta \propto a$, making it relevant for describing structure formation. The point remains a saddle for all $n < 2$, allowing trajectories to pass through before transitioning to a stable dark energy-dominated state. It therefore captures the essence of cosmic evolution during the structure formation epoch.
\end{itemize}

\begin{figure}[htp!]
   \mbox{\includegraphics[scale=0.47]{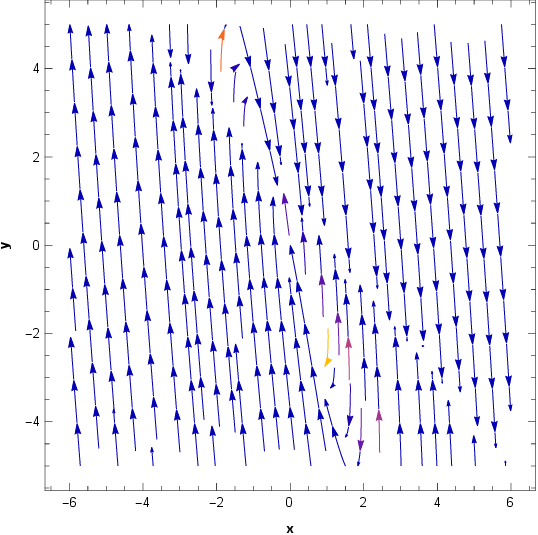}}  \hspace{10pt}
    \mbox{\includegraphics[scale=0.47]{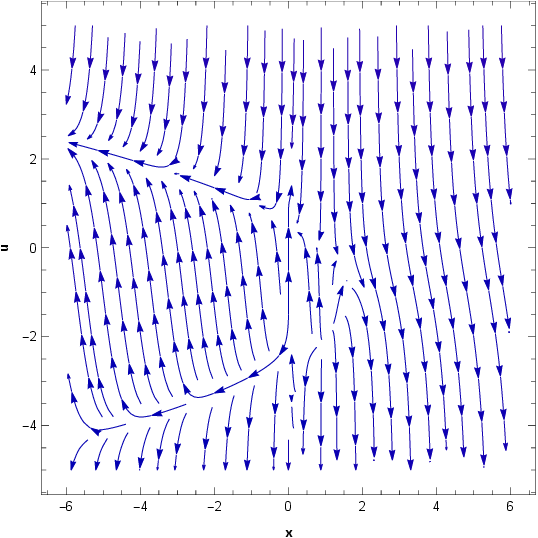}}
    \hspace{10pt}
    \mbox{\includegraphics[scale=0.47]{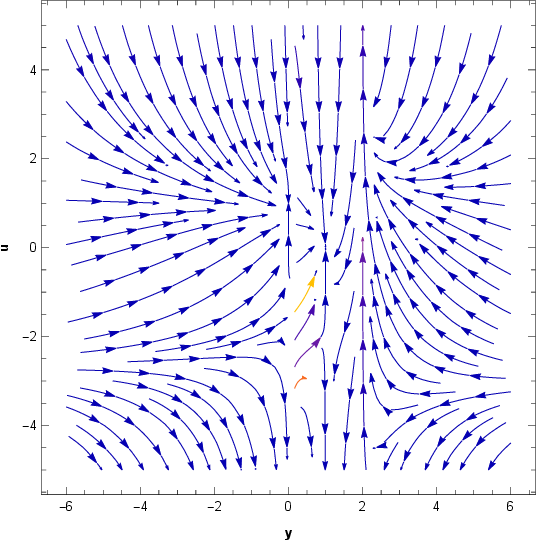}}
    \caption{ Phase Portrait for the dynamical system of Model II, (i)left panel($u=0$ and $n=3$); (ii)middle panel($y=0$ and $n=3$); (iii)right panel($x=0.5$ and $n=3$)}
   \label{Phase Portrait for the dynamical system of Model II}
\end{figure}

\vspace{1em}
\section{Discussion and Conclusion}

In this work, we performed a comprehensive dynamical systems analysis of both background and perturbation equations in the framework of $f(Q)$ gravity, focusing on two specific models: the Log-square-root model and the Hyperbolic tangent power model. Employing the flat \gls{flrw} metric, we reformulated the equations into autonomous systems and analyzed the critical points using center manifold theory. In both models, the critical point $(1-y, y, 0)$ emerged as asymptotically stable, corresponding to a dark energy-dominated universe with $\Omega_{Q}=1$, $w_{\text{eff}}=-1$, and constant matter perturbation ($u=0$), resembling a cosmological constant. Contrarily, the curve $(1-y, y, -2)$, although similar at the background level, revealed a saddle nature due to decaying matter perturbations, and hence does not describe a viable late-time cosmological phase.

Matter-dominated critical points such as $(0, 0, -\frac{3}{2})$ and $(0, 0, 1)$ exhibited different dynamical behaviors depending on the model parameter $n$. Notably, $(0, 0, 1)$ consistently assumed a saddle-like nature and allowed growing matter perturbations ($u=1$), making it suitable for explaining structure formation. The perturbative analysis highlighted how background-level similarities can diverge at the perturbation level, as seen in the Log-square-root model. Phase portraits further supported the critical point trajectories, and constraints from \gls{bbn} were satisfied within the allowed range of $n \leq 1.9$ for the Hyperbolic tangent power model. In this context, the \gls{bbn} constraint provides an important consistency test of the modified-gravity framework, since the expansion rate during the early radiation-dominated epoch directly affects the synthesis of the light elements. Therefore, compatibility with \gls{bbn} places restrictions on the deviations of the modified gravitational dynamics from the standard cosmological evolution at early times. The allowed range $n\leq1.9$ obtained for the Hyperbolic tangent power model indicates that the model can remain consistent with the early-universe expansion history within the considered parameter space. This agreement with \gls{bbn}, together with the dynamical-system results, strengthens the cosmological viability of the model by demonstrating that its successful late-time behavior does not come at the expense of the well-established early-universe constraints. Thus, \gls{bbn} serves as a complementary constraint to the background and perturbation analyses, linking the early-time cosmological dynamics with the late-time accelerated evolution. Additionally, we found no critical points at infinity, suggesting a closed and finite phase space structure.

In conclusion, the dynamical behavior of these $f(Q)$ models, validated through both background and perturbation analysis and further constrained by \gls{bbn}, demonstrates their potential as viable alternatives to the $\Lambda$CDM paradigm, achieving stability and acceleration without invoking an exotic dark energy component. The compatibility with \gls{bbn} is particularly significant because a viable modified-gravity model must reproduce the observed early-universe expansion sufficiently well while simultaneously providing an appropriate description of the late-time accelerated phase. The combined consideration of the phase-space dynamics, matter perturbations, and \gls{bbn} constraints therefore provides a more comprehensive assessment of the cosmological viability of the models considered in this chapter. Future work may consider more complex $f(Q)$ or higher-curvature models with boundary terms, or extend the framework to $f(Q,T)$ gravity and Weyl-type geometries. These explorations could further illuminate the cosmological viability and predictive power of modified gravity theories.

\chapter{Scalar-Nonmetricity Interactions and Exact Solutions through Noether Symmetry} \label{c5}
\section{Introduction}

This chapter explores the application of the Noether symmetry approach within the framework of extended symmetric teleparallel $ f(Q, \phi) $ gravity, which integrates the non-metricity scalar $ Q $ and a scalar field $ \phi $, in the context of a Friedmann-Lemaître-Robertson-Walker (FLRW) spacetime. By employing this approach, we investigate a specific form of $ f(Q, \phi) $ and examine the existence of a nontrivial Noether vector. The Noether symmetry method proves instrumental in constructing models and deriving exact solutions for the Lagrangian. We demonstrate how this method determines the functional form of $ f(Q, \phi) $ and yields precise cosmological solutions. These solutions align with the symmetries identified and suggest an accelerated expansion of the Universe. Furthermore, we compute key cosmological parameters, including the Hubble parameter, deceleration parameter, and effective equation of state parameter, which are consistent with current observational data \citep{Vishwakarma:2025yph, Vishwakarma:2026zmm, Duchaniya:2022hiy}.

\vspace{1em}
\section{Field Equations of the Scalar-Non-Metricity $ f(Q, \phi) $ Gravity}
The \gls{stegr} action is composed of the linear non-metricity scalar, which can be generalized to $ f(Q) $ gravity. By extending the non-metricity scalar $ Q $ to an arbitrary function and introducing a scalar field $ \phi $, the action can be written as \citep{Gonzalez-Espinoza:2021mwr, DAmbrosio:2021pnd, Bahamonde:2017ize}:
\begin{equation}
S = \int d^4x \, e \left[ f(Q, \phi) + P(\phi)X \right] + S_{m} + S_{r} 
\end{equation}
where $ e = \sqrt{-g} $ is the determinant. The matter action is denoted by $ S_{m} $, while the radiation action is characterized by $ S_{r} $.

The function $ f(Q, \phi) $ denotes an arbitrary function of the scalar field $ \phi $ and the non-metricity scalar $ Q $. The kinetic term of the scalar field is $ X = -\frac{1}{2} \partial_{\mu} \phi \partial^{\mu} \phi $. This general action includes non-minimally coupled scalar-non-metricity gravity models, $ f(Q) $ gravity, and a minimally coupled scalar field.

The non-metricity scalar is given by:
\begin{equation}
Q = -\frac{1}{4} Q_{\alpha \beta \mu} Q^{\alpha \beta \mu} + \frac{1}{2} Q_{\alpha \beta \mu} Q^{\beta \alpha \mu} + \frac{1}{4} Q_{\alpha \beta}^{\ \ \alpha} Q^{\gamma \beta}_{\ \ \gamma},
\end{equation}
where $ Q_{\alpha \beta \mu} $ denotes the non-metricity tensor. This can be decomposed into traces and irreducible components.

For a homogeneous and isotropic universe, the \gls{flrw} metric is given by:
\begin{equation}
ds^2 = -dt^2 + a^2(t) \left( dx^2 + dy^2 + dz^2 \right),
\end{equation}
where $ a(t) $ is the scale factor representing cosmic expansion in the spatial directions.

By varying the action, the gravitational field equations of $ f(Q, \phi) $ gravity for cosmological applications in the homogeneous and isotropic flat \gls{flrw} metric are obtained as:
\begin{equation}
f(Q, \phi) - P(\phi)\frac{\dot{\phi}^{2}}{2} - 2Q f_{Q} = \rho_{m}  
\end{equation}
\begin{equation}
f(Q, \phi) + P(\phi)\frac{\dot{\phi}^{2}}{2} - 2Q f_{Q} - 4\dot{H} f_{Q} - 4H \dot{f}_{Q} = -p_{m} 
\end{equation}
\begin{equation}
P_{\phi}\frac{\dot{\phi}^{2}}{2} - 3P(\phi)H \dot{\phi} - P(\phi) \ddot{\phi} + f_{\phi} = 0
\end{equation}
where $ f_{Q} = \frac{\partial f(Q, \phi)}{\partial Q} $, and $ \rho_{m} $ denote the energy densities of matter, respectively. $ p_{m} $ represents the pressure in the matter era, and $ H = \frac{\dot{a}}{a} $ is the Hubble parameter. The overdot represents differentiation with respect to cosmic time $ t $. $ P_{\phi} = \frac{\partial P(\phi)}{\partial \phi} $ and $ f_{\phi} = \frac{\partial f(Q, \phi)}{\partial \phi} $.

The Friedmann equations and the Klein-Gordon equation in the context of $f(T, \phi)$ gravity are thoroughly discussed in References \citep{Gonzalez-Espinoza:2021mwr, Gonzalez-Espinoza:2020jss, Gonzalez-Espinoza:2021qnv, Gonzalez-Espinoza:2020azh}.

\vspace{1em}
\section{Lagrangian Formalism Of $ f(Q, \phi) $ Theory}

In this section, we present the Lagrangian formulation of $ f(Q, \phi) $ theory. The point-like Lagrangian plays a crucial role in the analysis of Noether symmetries, which are tied to the Friedmann equations. This Lagrangian can be derived from equation (5.1) \citep{Capozziello:1999xr}. By defining the canonical Lagrangian $ \mathcal{L} = \mathcal{L}(a, \dot{a}, Q, \dot{Q}, \phi, \dot{\phi}) $, one can obtain the cosmological equations within the \gls{flrw} framework. The configuration space for this system is given by $ \mathcal{Q} = (a, Q, \phi) $, and from it, the tangent space $ \mathcal{T}\mathcal{Q} = (a, \dot{a}, Q, \dot{Q}, \phi, \dot{\phi}) $ is constructed, where the Lagrangian $ \mathcal{L} $ is defined as a function on this space. In this formalism, the scale factor $ a(t) $, the non-metricity scalar $ Q $, and the scalar field $ \phi(t) $ are treated as independent dynamical variables. To impose the constraint $ Q - 6 \frac{\dot{a}^2}{a^2} = 0 $, one can apply the method of Lagrange multipliers, and after integrating by parts, the resulting Lagrangian $ \mathcal{L} $ becomes similar to the expression found in Reference \citep{Capozziello:1999xr}.

\begin{equation}
S = \int dt \, a^{3} \left[f(Q, \phi) + P(\phi)\frac{\dot{\phi}^2}{2} - \lambda \left(Q - 6\frac{\dot{a}^2}{a^2} \right) -\frac{\rho_{m0}}{a^{3}} \right],
\end{equation}
Here, $ \lambda $ represents the Lagrange multiplier introduced to enforce the constraint, and $ \rho_{m0} $ denotes the matter energy density at the current time. Varying the action given in equation (5.7) with respect to $Q$, the non-metricity scalar, leads to the following condition for $ \lambda $:  
\begin{center}
$\lambda = f_{Q}(Q, \phi),$
\end{center}
where $ f_{Q} = \frac{\partial f}{\partial Q}$. Therefore the above action can be written as
\begin{equation}
S = \int dt \, a^{3} \left[f(Q, \phi) + P(\phi)\frac{\dot{\phi}^2}{2} - f_{Q}(Q, \phi) \left(Q - 6\frac{\dot{a}^2}{a^2} \right) -\frac{\rho_{m0}}{a^{3}} \right],
\end{equation}
The point-like Lagrangian is:
\begin{equation}
\mathcal{L}(a, \dot{a}, Q, \phi, \dot{\phi}) = a^3 \left[f(Q, \phi) + P(\phi)\frac{\dot{\phi}^2}{2} - Q f_{Q}(Q, \phi)\right] + 6a\dot{a}^2 f_{Q}(Q, \phi).
\end{equation}

The equations of motion are derived using the Euler-Lagrange equations:
\begin{equation}
\frac{d}{dt} \left(\frac{\partial \mathcal{L}}{\partial \dot{q}_i}\right) - \frac{\partial \mathcal{L}}{\partial q_i} = 0,
\end{equation}
where $ q_{i} = \{a, Q, \phi\} $.

In this case, the equations of motion are derived using the Euler-Lagrange formalism and can be expressed as follows:
\begin{enumerate}
\item For the scale factor $ a $:
\begin{equation}
\frac{d}{dt} \left( \frac{\partial \mathcal{L}}{\partial \dot{a}} \right) - \frac{\partial \mathcal{L}}{\partial a} = 0,
\end{equation}
leading to the equation:
\begin{equation}
f(Q, \phi) + P(\phi)\frac{\dot{\phi}^{2}}{2} - 2Q f_{Q} - 4\dot{H} f_{Q} - 4H \dot{f}_{Q} = 0.
\end{equation}

\item For the non-metricity scalar $Q$:
\begin{equation}
\frac{d}{dt} \left( \frac{\partial \mathcal{L}}{\partial \dot{Q}} \right) - \frac{\partial \mathcal{L}}{\partial Q} = 0,
\end{equation}
which simplifies to:
\begin{equation}
a^3 f_{QQ} \left( Q- 6\frac{\dot{a}^{2}}{a^{2}} \right) = 0.
\end{equation}

\item For the scalar field $ \phi $:
\begin{equation}
\frac{d}{dt} \left( \frac{\partial \mathcal{L}}{\partial \dot{\phi}} \right) - \frac{\partial \mathcal{L}}{\partial \phi} = 0,
\end{equation}
resulting in:
\begin{equation}
P_{\phi} \frac{\dot{\phi}^{2}}{2} - 3P(\phi)H \dot{\phi} - P(\phi) \ddot{\phi} + f_{\phi} = 0
\end{equation}
where $ H = \frac{\dot{a}}{a} $ is the Hubble parameter.
\end{enumerate}

From equation (5.14), if $ f_{QQ} \neq 0 $, it follows that $ Q = 6\frac{\dot{a}^2}{a^2} = 6H^2 $, where $ Q $ represents the non-metricity scalar in the FLRW metric. Similarly, from equations (5.12) and equation (5.16), it can be seen that these equations correspond to the same relations as equations (5.5)--(5.6), indicating that the modified Friedmann equations are recovered through the Lagrangian $ \mathcal{L} $.

The energy conditions for the Lagrangian $ \mathcal{L} $ are defined as:
\begin{equation}
E_{\mathcal{L}} = \frac{\partial \mathcal{L}}{\partial \dot{a}} \dot{a} + \frac{\partial \mathcal{L}}{\partial \dot{Q}} \dot{Q} + \frac{\partial \mathcal{L}}{\partial \dot{\phi}} \dot{\phi} - \mathcal{L}.
\end{equation}

Substituting the point-like Lagrangian from equation (5.17) into this expression yields:
\begin{equation}
E_{\mathcal{L}}(a, \dot{a}, Q, \dot{Q}, \phi, \dot{\phi}) = 12H^2 f_{Q} + P(\phi) \frac{\dot{\phi}^2}{2} - f(Q, \phi) + \frac{\rho_{m0}}{a^3}.
\end{equation}

Imposing the total energy condition $ E_{\mathcal{L}} = 0 $, we arrive at:
\begin{equation}
f(Q, \phi) - 12H^2 f_{Q} - P(\phi) \frac{\dot{\phi}^2}{2} = \frac{\rho_{m0}}{a^3}.
\end{equation}

This equation is equivalent to equation (5.4), demonstrating that the point-like Lagrangian in equation (5.9) is sufficient to derive all the cosmological equations governing $ f(Q, \phi) $ gravity.

\vspace{1em}
\section{Noether Symmetries}

Noether symmetries play a crucial role in physics, particularly in analyzing the integrability of differential equations and simplifying complex dynamical systems. Their existence is often associated with conserved quantities that have physical significance. In the context of cosmology, the Noether symmetry approach proves to be a powerful tool for deriving exact solutions to governing equations. 

To illustrate this, consider a system characterized by $ n $ generalized coordinates $ x_{j} $ and an independent variable $ t $, with its dynamics described by a Lagrangian $ \mathcal{L} $. An infinitesimal one-parameter point transformation can be expressed as:
\begin{equation}
\bar{t} = \psi(t, x_{k}, \epsilon), \quad \bar{x}^{A} = \eta(t, x_{k}, \epsilon),
\end{equation}
where $ \psi $ and $ \eta $ define the transformation. The generator for such a transformation is written as:
\begin{equation}
Y = \xi(t, x_{k}, \epsilon) \frac{\partial}{\partial t} + \alpha_{j}(t, x_{k}, \epsilon) \frac{\partial}{\partial x_{k}},
\end{equation}
with the coefficients $ \xi(t, x_{k}) $ and $ \alpha_{j}(t, x_{k}) $ determined by:
\begin{equation}
\xi(t, x_{k}) = \frac{\partial \psi(t, x_{k}, \epsilon)}{\partial \epsilon} \bigg|_{\epsilon \to 0}, \quad \alpha_{j}(t, x_{k}) = \frac{\partial \eta(t, x_{k}, \epsilon)}{\partial \epsilon} \bigg|_{\epsilon \to 0}.
\end{equation}

The $ n $-th prolongation of the generator vector is defined as:
\begin{equation}
Y^{[n]} = Y + \alpha_{j}^{[1]} \frac{\partial}{\partial \dot{x}_{j}} + \dots + \alpha_{j}^{[n]} \frac{\partial}{\partial x_{j}^{(n)}},
\end{equation}
where the coefficients $ \alpha_{j}^{[1]} $ and $ \alpha_{j}^{[n]} $ are obtained recursively:
\begin{equation}
\alpha_{j}^{[1]} = \frac{d\alpha_{j}}{dt} - \dot{x}_{j} \frac{d\xi}{dt}, \quad \alpha_{j}^{[n]} = \frac{d\alpha_{j}^{[n-1]}}{dt} - x_{j}^{(n)} \frac{d\xi}{dt}.
\end{equation}

For the Euler-Lagrange equations of a dynamical system with Lagrangian $ \mathcal{L} = \mathcal{L}(t, x_{j}, \dot{x}_{j}) $ to remain invariant under the transformation, there must exist a function $ g = g(t, x_{j}) $ such that the Rund-Trautman identity is satisfied:
\begin{equation}
Y^{[1]}\mathcal{L} + \mathcal{L} \frac{d\xi(t, x_{j})}{dt} = \frac{dg(t, x_{j})}{dt},
\end{equation}
where $ Y^{[1]} $ is the first prolongation of $ Y $. If the generator $ Y $ satisfies this condition, it qualifies as a Noether symmetry for the system described by $ \mathcal{L} $.

According to Noether’s theorem, the presence of a symmetry ensures the existence of a conserved quantity, or Noether charge, given by:
\begin{equation}
\mathcal{Q}_{0} = \sum_{j} \alpha_{j} \frac{\partial \mathcal{L}}{\partial \dot{q}_{j}} = \text{constant},
\end{equation}
where $ q_{j} $ represents the coordinates in the configuration space, and $ \alpha_{j} $ are the components of the Noether factors. This conserved quantity provides critical insights into the dynamics of the system.

\vspace{1em}
\section{Noether Symmetries in $ f(Q, \phi) $ Gravity}

This section explores Noether symmetry within the context of scalar-nonmetricity $ f(Q, \phi) $ theory. The Noether symmetry approach is employed to identify potential symmetries in the dynamical system described by the Lagrangian in equation (5.9). This method is particularly effective for obtaining exact solutions to the equations of motion and provides a fundamental justification for the models derived. The Noether symmetry is represented by a generator vector $Y$, which is defined within the tangent space of the Lagrangian $L $. 

To examine the symmetries, we consider a one-parameter point transformation in the configuration space $(t, a, Q, \phi)$. The generator of this transformation is expressed as:
\begin{equation}
Y = \xi(t, a, Q, \phi)\frac{\partial}{\partial t} + \alpha_1(t, a, Q, \phi)\frac{\partial}{\partial a} + \alpha_2(t, a, Q, \phi)\frac{\partial}{\partial Q} + \alpha_3(t, a, Q, \phi)\frac{\partial}{\partial \phi},
\end{equation}
where $\xi$, $\alpha_1 $, $ \alpha_2 $, and $ \alpha_3 $ are functions of the variables $ t, a, Q, \phi $. The first prolongation of the generator is given by:
\begin{equation}
Y^{[1]} = Y + \alpha_1^{[1]} \frac{\partial}{\partial \dot{a}} + \alpha_3^{[1]} \frac{\partial}{\partial \dot{\phi}},
\end{equation}
where the prolongation terms are:
\begin{equation}
\alpha_1^{[1]} = \frac{\partial \alpha_1}{\partial t}  + \dot{a} \frac{\partial \alpha_1}{\partial a} + \dot{\phi} \frac{\partial \alpha_1}{\partial \phi} + \dot{Q} \frac{\partial \alpha_1}{\partial Q} - \dot{a} \frac{\partial \xi}{\partial t} - \dot{a}^2 \frac{\partial \xi}{\partial a} - \dot{a} \dot{\phi} \frac{\partial \xi}{\partial \phi} - \dot{a} \dot{Q} \frac{\partial \xi}{\partial Q},
\end{equation}
\begin{equation}
\alpha_3^{[1]} = \frac{\partial \alpha_3}{\partial t}  + \dot{a} \frac{\partial \alpha_3}{\partial a} + \dot{\phi} \frac{\partial \alpha_3}{\partial \phi} + \dot{Q} \frac{\partial \alpha_3}{\partial Q} - \dot{\phi} \frac{\partial \xi}{\partial t} - \dot{a} \dot{\phi} \frac{\partial \xi}{\partial a}  - \dot{\phi}^2 \frac{\partial \xi}{\partial \phi} - \dot{\phi} \dot{Q} \frac{\partial \xi}{\partial Q}.
\end{equation}

The terms of the symmetry condition from equation (5.25) are calculated as follows. The first term, $ Y^{[1]}\mathcal{L} $, is given by:
\begin{equation}
\begin{split}
 Y^{[1]}\mathcal{L} & =
3a^2\alpha_1 f  - 3\alpha_1a^2Q f_Q - a^3\alpha_2Q f_{QQ} + \alpha_3a^3 f_\phi  - a^3\alpha_3Q f_{Q \phi} \\
& + 12a\dot{a}\dot{Q} f_Q \frac{\partial\alpha_1}{\partial Q} - 12a\dot{a}^3 f_Q \frac{\partial\xi}{\partial a} - 12a\dot{a}^2\dot{\phi} f_Q \frac{\partial\xi}{\partial \phi}  - 12a\dot{a}^2\dot{Q} f_Q \frac{\partial\xi}{\partial Q} \\ &  +\left( \frac{3}{2} \alpha_1a^2P(\phi) + \frac{1}{2} a^3\alpha_3P_\phi + a^3P(\phi) \frac{\partial\alpha_3}{\partial \phi} - a^3P(\phi) \frac{\partial\xi}{\partial t} \right) \dot{\phi}^2 \\  
& + \left( 6\alpha_1 f_Q + 6\alpha_2a f_{QQ} + 6\alpha_3a f_{Q\phi} + 12a f_Q \frac{\partial\alpha_1}{\partial a} - 12a f_Q \frac{\partial\xi}{\partial t} \right) \dot{a}^2 \\ 
&  + 12a\dot{a} f_Q \frac{\partial\alpha_1}{\partial t} + \left( 12a f_Q \frac{\partial\alpha_1}{\partial \phi} + a^3P(\phi) \frac{\partial\alpha_3}{\partial a} \right) \dot{a}\dot{\phi}  + a^3P(\phi) \frac{\partial\alpha_3}{\partial t} \dot{\phi} \\ 
& + a^3P(\phi)\dot{\phi}\dot{Q} \frac{\partial\alpha_3}{\partial Q}  - a^3P(\phi)\dot{\phi}^3 \frac{\partial\xi}{\partial \phi} - a^3P(\phi)\dot{\phi}^2\dot{Q} \frac{\partial\xi}{\partial Q} - a^3P(\phi)\dot{a}\dot{\phi}^2 \frac{\partial\xi}{\partial a}  
\end{split} 
\end{equation}

The second term, $ \mathcal{L} \dot{\xi} $, is:
\begin{equation} 
\begin{split}
 \mathcal{L} \dot{\xi} & =
\left[ a^3 f - a^3Q f_Q + a^3P(\phi) \frac{\dot{\phi}^2}{2}  
+ 6a \dot{a}^2 f_Q \right] \frac{\partial\xi}{\partial t} \\ 
&  + \left[ a^3 \dot{a} f - a^3 \dot{a}Q f_Q + a^3 \dot{a}P(\phi) \frac{\dot{\phi}^2}{2} + 6a \dot{a}^3 f_Q \right] \frac{\partial\xi}{\partial a} \\ 
& + \left[ a^3 \dot{Q} f - a^3 \dot{Q} Q f_Q + a^3 \dot{Q}P(\phi) \frac{\dot{\phi}^2}{2} + 6a \dot{Q} \dot{a}^2 f_Q  
\right] \frac{\partial\xi}{\partial Q} \\ 
&  + \left[ a^3 \dot{\phi} f - a^3 \dot{\phi} Q f_Q + a^3 \dot{\phi} P(\phi) \frac{\dot{\phi}^2}{2}+ 6 \dot{\phi} a \dot{a}^2 f_Q \right] \frac{\partial\xi}{\partial \phi}.  
\end{split}
\end{equation}

The right-hand side of equation (5.25) becomes:
\begin{equation}
\dot{g} = \frac{\partial g}{\partial t} + \dot{a} \frac{\partial g}{\partial a} + \dot{Q} \frac{\partial g}{\partial Q} + \dot{\phi} \frac{\partial g}{\partial \phi}.
\end{equation}

Substituting these results into equation (5.25) and equating terms with powers of $ \dot{a}^2 $, $\dot{Q}^2 $, $ \dot{\phi}^2 $, $ \dot{a} \dot{Q} $, $ \dot{a} \dot{\phi} $, and $ \dot{\phi} \dot{Q} $ to zero, the following Noether symmetry conditions are obtained:
\begin{equation}
\begin{split}
3a^2 \alpha_1 f - 3\alpha_1 a^2 Q f_Q & - a^3 \alpha_2 Q f_{QQ} + \alpha_3 a^3 f_\phi  - a^3 \alpha_3 Q f_{Q\phi} \\ 
& + a^3 f \frac{\partial \xi}{\partial t} - a^3 Q f_Q \frac{\partial \xi}{\partial t} = \frac{\partial g}{\partial t},
\end{split}
\end{equation}
\begin{equation}
3\alpha_1 a^2 P(\phi) + a^3 \alpha_3 P_\phi + 2a^3 P(\phi) \frac{\partial \alpha_3}{\partial \phi} - a^3 P(\phi) \frac{\partial \xi}{\partial t} = 0,
\end{equation}
\begin{equation}
6\alpha_1 f_Q + 6\alpha_2 a f_{QQ} + 6\alpha_3 a f_{Q\phi} + 12a f_Q \frac{\partial \alpha_1}{\partial a} - 6a f_Q \frac{\partial \xi}{\partial t} = 0,
\end{equation}
\begin{equation}
12a f_Q \frac{\partial \alpha_1}{\partial t} + a^3 f \frac{\partial \xi}{\partial a} - a^3 Q f_Q \frac{\partial \xi}{\partial a} = \frac{\partial g}{\partial a},
\end{equation}
\begin{equation}
12a f_Q \frac{\partial \alpha_1}{\partial \phi} + a^3 P(\phi) \frac{\partial \alpha_3}{\partial a} = 0,
\end{equation}
\begin{equation}
a^3 P(\phi) \frac{\partial \alpha_3}{\partial t} + a^3 f \frac{\partial \xi}{\partial \phi} - a^3 Q f_Q \frac{\partial \xi}{\partial \phi} = \frac{\partial g}{\partial \phi},
\end{equation}
\begin{equation}
a^3 f \frac{\partial \xi}{\partial Q} - a^3 Q f_Q \frac{\partial \xi}{\partial Q} = \frac{\partial g}{\partial Q}
\end{equation}
\begin{equation}
12a f_Q \frac{\partial \alpha_1}{\partial Q} = 0, \quad a^3 P(\phi) \frac{\partial \alpha_3}{\partial Q} = 0,
\end{equation}
\begin{equation}
6a f_Q \frac{\partial \xi}{\partial a} = 0, \quad 6a f_Q \frac{\partial \xi}{\partial \phi} = 0, \quad 6a f_Q \frac{\partial \xi}{\partial Q} = 0,
\end{equation}
\begin{equation}
a^3 P(\phi) \frac{\partial \xi}{\partial a} = 0, \quad a^3 P(\phi) \frac{\partial \xi}{\partial \phi} = 0, \quad a^3 P(\phi) \frac{\partial \xi}{\partial Q} = 0.
\end{equation}

The unknowns $ \xi(t, a, Q, \phi) $, $ \alpha_1(t, a, Q, \phi) $, $ \alpha_2(t, a, Q, \phi) $, and $ \alpha_3(t, a, Q, \phi) $ and the function $ f(Q, \phi) $ can be determined by solving this system of partial differential equations. Alternatively, specific forms of $ f(Q, \phi) $ can be assumed to simplify the problem. For instance, from the symmetry conditions, it can be inferred that $ \alpha_1 $ and $ \alpha_3 $ are independent of $ Q $, and $ \xi $ does not depend on $ a $, $ Q $, or $ \phi $. Therefore, we set $ \alpha_1 = \alpha_1(t, a, \phi) $, $ \alpha_3 = \alpha_3(t, a, \phi) $, and $ \xi = \xi(t) $. 

This work adopts the second strategy, wherein specific forms of $ f(Q, \phi) $ are assumed, and the resulting system of equations is solved. For simplicity, $ P(\phi) = 1 $ is also considered in the solution process.

\vspace{1em}
\section{Cosmological Model $ f(Q, \phi)= -QF(\phi)+V(\phi) $}

In the above formulation of $ f(Q, \phi) $, $ F(\phi) $ represents the non-minimal coupling function of the scalar field $ \phi $, while $ V(\phi) $ denotes the scalar potential function. For this study, we consider $ F(\phi) = F_0 \phi^2 $ and $ V(\phi) = V_0 \phi^k $, allowing $ f(Q, \phi) $ to be expressed as $ f(Q, \phi) = -Q F_0 \phi^2 + V_0 \phi^k $, where $ F_0 $, $ V_0 $, and $ k $ are arbitrary constants. This particular choice of $ f(Q, \phi)$ is motivated by references \citep{Gonzalez-Espinoza:2021mwr, Gonzalez-Espinoza:2020jss, Gonzalez-Espinoza:2021qnv, Gonzalez-Espinoza:2020azh, Hohmann:2018rwf, Faraoni:2004pi}. Substituting this form of $ f(Q, \phi) $ into the system of partial differential equations (5.39)--(5.47) and employing the method of separation of variables, we obtain the Noether coefficients for the Noether vector (5.27). In this context, the function $ g $ remains constant and is denoted by $ G_0 $, while $ \alpha_0 $ serves as an integration constant. The coefficients are given as follows:  

\begin{eqnarray}
\alpha_1(t, a, Q, \phi) = \alpha_0 \frac{a}{a^3} \\
\alpha_2(t, a, Q, \phi) = \alpha_2(t, a, Q, \phi) \\
\alpha_3(t, a, Q, \phi) = -\alpha_0 \frac{\phi}{k} \\
\xi(t, a, Q, \phi) = \alpha_0 t \\
g(t, a, Q, \phi) = G_0.
\end{eqnarray}

From equations (5.44)--(5.48), the condition for the model parameter $ k \neq 0 $ is obtained. Substituting the derived Noether coefficients into equation (5.27), the Noether vector is expressed as:  

\begin{equation}
Y = \alpha_0 t \frac{\partial}{\partial t} + \frac{\alpha_0}{a^3} \frac{\partial}{\partial a} + \alpha_2 \frac{\partial}{\partial Q} - \frac{\alpha_0 \phi^k}{k} \frac{\partial}{\partial \phi}.
\end{equation}

Next, we seek a cosmological solution for this class of functions. The point-like Lagrangian from equation (5.9) takes the form:  
\begin{equation}
\mathcal{L} = a^3 V_0 \phi^k + \frac{a^3 \dot{\phi}^2}{2} - 6 a F_0 \phi^2 \dot{a}^2 - \rho_{m0}.
\end{equation}

The Euler-Lagrange equations for the scale factor, derived from equation (5.12), and for the energy density, derived from equation (5.19), are given as:  
\begin{equation}
2 F_0 \phi^2 \frac{\dot{a}^2}{a^2} + 4 F_0 \phi^2 \frac{\ddot{a}}{a} + V_0 \phi^k + \frac{\dot{\phi}^2}{2} = 0,
\end{equation}
\begin{equation}
6 F_0 \phi^2 \frac{\dot{a}^2}{a^2} + V_0 \phi^k - \frac{\dot{\phi}^2}{2} - \frac{\rho_{m0}}{a^3} = 0. 
\end{equation}

Finding solutions to these dynamical equations, equations (5.51)--(5.52), proves challenging due to their non-linear nature. To address this, additional variables are introduced into the Lagrangian (5.9). When Noether symmetry is present, it is possible to employ cyclic variables for coordinate transformation. Following the approach in references \citep{Capozziello:1999xr, deRitis:1990ba, Demianski:1992tu, Capozziello:1993vr, Kucukakca:2013mya}, the transformation $(a, \phi) \to (r, s)$ is performed, where $r$ is a cyclic variable. The partial differential equations associated with this transformation are:  
\begin{eqnarray}
\alpha_1 \frac{\partial r}{\partial a} + \alpha_3 \frac{\partial r}{\partial \phi} = 1,\\
\alpha_1 \frac{\partial s}{\partial a} + \alpha_3 \frac{\partial s}{\partial \phi} = 0,
\end{eqnarray}
where the new variables $r$ and $s$ are functions of the original variables $a$ and $\phi$. The solutions to equations (5.53)--(5.54) are found to be:  
\begin{equation}
r(a, \phi) = \frac{3 \ln(a)}{\alpha_0}, \quad s(a, \phi) = \phi a^{\frac{3}{k}}.
\end{equation}
Upon transforming $a$ and $\phi$ into the new variables $r$ and $s$, the inverse transformations are given by:  
\begin{equation}
a(r, s) = e^{\frac{\alpha_0 r}{3}}, \quad \phi(r, s) = s e^{-\frac{\alpha_0 r}{k}}
\end{equation}

It is essential to note that, under the transformation described above, the variable $ r $ does not appear explicitly in the Lagrangian (5.9), as $ r $ is treated as a cyclic variable in this study. From this point onward, we adopt the parameter $ F_0 = \frac{3}{16} $ and focus on the special case $ k = 2 $. This transformation allows the Lagrangian (5.50) to be expressed in the following form:  
\begin{equation}
\mathcal{L} = V_0 s^2 + \frac{1}{2} \dot{s}^2 - \frac{1}{2} s \dot{s} \dot{r} \alpha_0 - \rho_{m0}.
\end{equation}

It is evident from this Lagrangian that it is independent of the cyclic variable $ r $. Using this Lagrangian, the corresponding field equations are derived as:  
\begin{eqnarray}
s \dot{s} \alpha_0 = -2 Q_0,\\
\ddot{s} - \frac{1}{2} s \ddot{r} \alpha_0 - 2 V_0 s = 0,\\
\dot{s} s \dot{r} \alpha_0 - \dot{s}^2 + 2 V_0 s^2 - 2 \rho_{m0} = 0,
\end{eqnarray}
where $ Q_0 $ is a constant associated with a conserved quantity. From equation (5.58), the solution for $ s(t) $ is obtained as:  
\begin{equation}
s(t) = \left( -\frac{4 Q_0 t}{\alpha_0} + 2 s_1 \right)^{\frac{1}{2}},  
\end{equation}
where $ s_1 $ is an integration constant.

To determine $ r(t) $, the solution for $ s(t) $ obtained in equation (5.61) is substituted into equation (5.60). This yields:
\begin{equation}
r(t) = -\frac{2}{\alpha_0} \left[-\frac{1}{4} \ln\left(4 Q_0 t - 2 \alpha_0 s_1\right) - \frac{\alpha_0 t s_1 V_0}{Q_0} + t^2 V_0 \right] + r_0 t + r_1,
\end{equation}
where $ r_0 $ and $ r_1 $ are integration constants. From equation (5.60), the following constraint is derived:
\begin{equation}
-2 \left(\rho_{m0} + r_0 Q_0 \right) = 0.
\end{equation}
This condition clearly indicates that $ r_0 $ must be zero in the absence of standard matter ($ \rho_{m0} = 0 $). By substituting the solutions for $ s(t) $ and $ r(t) $ from equations (5.61) and (5.62) into equation (5.56), the cosmological solutions are determined as:
\begin{equation}
a(t) = e^{ \left[\frac{1}{3} \left( \alpha_0 r_1 + \frac{\ln\left(4 Q_0 t - 2 \alpha_0 s_1 - 4 t^2 V_0 \right)}{2 \alpha_0}  + \frac{2 t s_1 V_0}{Q_0} + r_0 t \right) \right]}, 
\end{equation}
\begin{equation}
\phi(t) = \sqrt{ \left(2 s_1 - \frac{4 Q_0 t}{\alpha_0}\right)} e^{-\frac{1}{4} \ln\left(4 Q_0 t - 2 \alpha_0 s_1\right) + t V_0 \left(t - \frac{\alpha_0 s_1}{Q_0}\right)-\frac{1}{2} \alpha_0 \left(r_0 t + r_1 \right)}.
\end{equation}
Here, $ a(t) $ represents the scale factor, and $ \phi(t) $ is the scalar field. These expressions provide cosmological solutions within the framework of the studied model.

The solutions presented above include three integration constants: $ s_1 $, $ r_0 $, and $ r_1 $. To determine these constants in the general solution of the field equations (equations (5.64)–(5.65)), the approach outlined in reference \citep{Rubano:2003et} is followed. First, we consider the scenario where $ a(0) = 0 $, which establishes the origin of time. This condition is chosen as an initial reference point in time and, when applied to the scale factor in equation (5.64), leads to $ s_1 = 0 $. 
Next, the present time is normalized to $ t_0 = 1 $. As such, we assume $ a(t_0 = 1) = 1 $ for convenience. This normalization yields the following relation for $ r_1 $:
\begin{equation}
r_1 = \frac{4 V_0 - 2 r_0 \alpha_0 - \ln(4 Q_0)}{2 \alpha_0^2}
\end{equation}
The final condition involves setting $ H(t_0 = 1) = H_0 $, where $ H(t) $ is the Hubble parameter. Note that the parameter $ H_0 $ here is distinct from the standard observational Hubble constant. Applying this condition to equation (5.64) gives:
\begin{equation}
r_0 = \frac{6 \alpha_0 H_0 - 1 - 8 V_0}{2 \alpha_0}.
\end{equation}
With these constraints applied, the scale factor $ a(t) $ and the scalar field $ \phi(t) $ are expressed as:
\begin{equation}
a(t) = (4t)^{\frac{1}{6 \alpha_0 }} Q_0^{-\frac{1}{2 \alpha_0}} e^{(t-1)\left(1 + 4 V_0 (3 + t) - 6 H_{0} \alpha_0 \right)},
\end{equation}
\begin{equation}
\phi(t) = 2 \sqrt{-\frac{Q_0 t}{\alpha_0}} e^{\frac{\alpha_0\left(H_0 (6 - 6 \alpha_0 t) - \ln(4 Q_0 t) + t\right) + 4 V_0 (\alpha_0 t (t + 2) - 3) - 1}{4 \alpha_0}}.
\end{equation}

The exact solutions presented in equation (5.68) and (5.69) can be utilized to derive essential physical quantities, such as the Hubble parameter, deceleration parameter, and the effective equation of state parameter. These quantities are expressed as follows:

The Hubble parameter is given by:
\begin{equation}
H(t) = -6 \alpha_0 H_0 + \frac{1}{6 \alpha_0 t} + 8 (t + 1) V_0 + 1, 
\end{equation}
The deceleration parameter takes the form:
\begin{equation}
q(t) = -1 + \frac{6 \alpha_0 \left[1 - 48 \alpha_0 t^2 V_0 \right] }
{\left[\frac{1}{6 \alpha_0 t} - \left(-6 \alpha_0 H_0 + 8 (t + 1) V_0 + 1 \right) + 1 \right]^2},
\end{equation}
The effective equation of state parameter is expressed as:
\begin{equation}
\displaystyle w_{eff}(t) = -1 +\frac{ 4 \alpha_0 \left[1 - 48 \alpha_0 t^2 V_0 \right]} 
{\left[6 \alpha_0 t \left(-6 \alpha_0 H_0 + 8 (t + 1) V_0 + 1 \right) + 1 \right]^2}. 
\end{equation}
These equations allow the computation of the cosmological dynamics and provide insights into the evolution of the universe under the proposed model.

\begin{figure} 
   \centering 
   \mbox{\includegraphics[width=0.48\textwidth]{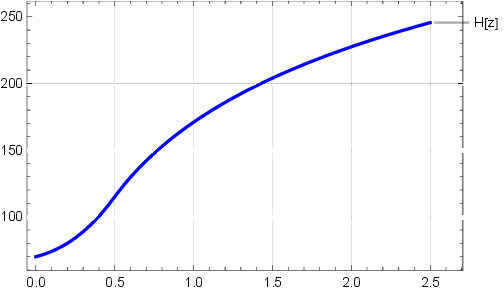}}   
    \hspace{10pt}
    \mbox{\includegraphics[width=0.48\textwidth]{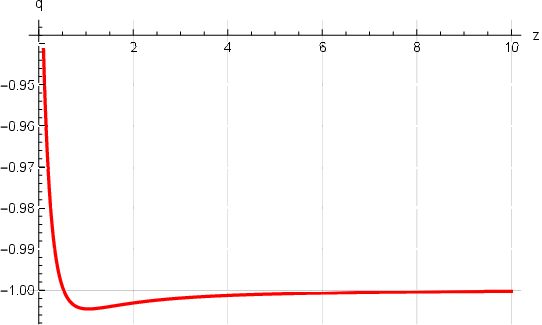}}
    \hspace{10pt}
    \mbox{\includegraphics[width=0.48\textwidth]{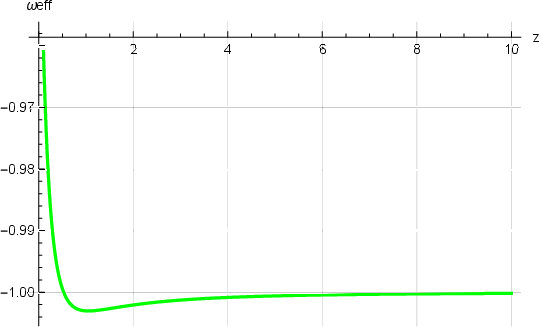}}
    \caption{ The evolution of the Hubble (shown in the \textit{Upper Left panel}), the deceleration parameters (illustrated in the \textit{Upper Right panel}), and the Effective EoS parameter (depicted in the \textit{Lower panel}) are presented for Model. The parameter scheme is defined as follows: $\beta_{0} = 0.012, \quad
     V_{0} = 3.75, \quad H_{0} = 2$.}
   \label{Hubble Parameter, Deceleration Parameter and Effective Equation of State Parameter}
\end{figure}

The graphical representation of the physical parameters is depicted in Figure 5.1. Figure 5.1 (Upper Left Panel) illustrates the evolution of the Hubble parameter over time, revealing that it increases as cosmic time progresses.
At $ t = 0 $, the behavior of the Hubble parameter reflects the dynamics during the early Universe. In this epoch, as $ t \to 0 $, the scale factor $ a(t) $ becomes negligibly small. Consequently, when $ a(t) \to 0 $, the Hubble parameter diverges to infinity, signifying an infinite expansion rate at $ t = 0 $. This indicates a singularity at the beginning of the Universe, consistent with the Big Bang model. The upper left panel of Figure 5.1 thus demonstrates that the Hubble parameter approaches infinity at $ t = 0 $, reflecting the rapid expansion characteristic of the Universe's initial stages. As time advances, the Hubble parameter continues to increase, suggesting that the rate of expansion, $ \dot{a}(t) $, surpasses the scale factor $ a(t) $. Cosmological observations imply that dark energy, presumed to dominate the Universe's energy density, will likely maintain its current behavior, akin to a cosmological constant or a similar phenomenon. This persistent behavior of dark energy drives the accelerated expansion of the Universe, potentially culminating in an exponential expansion. Such a scenario may eventually disrupt gravitationally bound structures, leading to a cosmological outcome often referred to as ``rip cosmology''.

The upper right panel of Figure 5.1 illustrates the deceleration parameter, which exhibits a negative phase, indicative of the Universe's current accelerated expansion. The present value of the deceleration parameter is calculated as $-1.0045$, confirming this accelerated phase. Lower panel of Figure 5.1 highlights the transition of the effective equation of state parameter, $ \displaystyle w_{eff} $, from the quintessence regime ($ \displaystyle w_{eff} > -1 $) to the phantom regime ($ \displaystyle w_{eff} < -1 $). The current value of the effective equation of state parameter is determined to be $-1.0085$. These results indicate a behavior consistent with observational cosmology, as the calculated values for the geometric and dynamical parameters align well with existing cosmological data and observations \citep{Camarena:2019moy, Denzel:2020zuq, Planck:2018vyg}.

\vspace{1em}
\section{Discussion and Conclusion}

This study investigates the Noether symmetry approach within the context of scalar-nonmetricity $ f(Q, \phi) $ gravity. The Noether symmetry method is a powerful mathematical framework for identifying conserved quantities and simplifying dynamical systems. It is an effective tool for classifying models and deriving exact cosmological solutions to field equations. In this framework, the Lagrangian plays a crucial role in describing symmetries and defining the Noether vector. Furthermore, the use of Lagrangian multipliers facilitates transforming the Lagrangian into canonical form.

In this work, we analyze the $ f(Q, \phi) $ theory, which incorporates a non-minimal coupling between the nonmetricity scalar $ Q $ and the scalar field $ \phi $. The nonmetricity scalar forms the foundation of the \gls{stegr} action, thereby allowing the scalar field to exhibit dynamic behavior during certain phases of the Universe's evolution. The Lagrangian for the \gls{flrw} space-time metric is formulated in equation (5.9). Subsequently, the Rund-Trautman identity, represented in equation (5.25), is utilized to derive the system of partial differential equations (5.34--5.43), which govern the dynamics of the system. The unknown variables in this system include $ \xi, \alpha_1, \alpha_2, \alpha_3 $, and the function $ f(Q, \phi) $. These variables serve as Noether coefficients, while $ f(Q, \phi) $ is treated as an arbitrary function of the nonmetricity scalar and the scalar field.

Also, the Noether symmetry in $ f(Q, \phi) $ gravity is explored for a specific choice of $ f(Q, \phi) = -Q F_0 \phi^2 + V_0 \phi^k $. For this particular form of $ f(Q, \phi) $, a non-trivial Noether vector, as shown in equation (5.49), is obtained. To simplify the dynamical equations of the model, a coordinate transformation is introduced, as described in equation (5.55), based on the Noether symmetry. This transformation introduces a cyclic coordinate among the new variables. Utilizing these transformations, a new set of field equations (5.58--5.60) is derived for the \gls{flrw} metric, enabling analytical solutions for the scale factor and the scalar field when the parameters $ k = 2 $ and $ F_0 = \frac{3}{16} $ are adopted, as outlined in equations (5.64--5.65). To gain insights into the evolution of the Universe, key cosmological parameters are examined. In figure 5.1, the evolution of significant cosmological parameters over cosmic time is depicted. From Lower Panel of figure 5.1, it is observed that the effective equation of state parameter transitions from the quintessence phase to the phantom phase, yielding a value of $ \displaystyle w_{eff}(t_0) = -1.0085 $ at the present time $ t_0 = 1 $. Similarly, the deceleration parameter, shown in the Upper Panel of figure 5.1, indicates an accelerating phase of the Universe, with present values of $ q(t_0) = -1.0045 $ and $ H(t_0) = 71.7 $. These values for the geometric parameters align well with current cosmological observations. 

The Noether symmetry approach is also extended to evaluate alternative functional forms of $ f(Q, \phi) $, which could simplify the system's dynamics and provide further understanding of cosmological solutions within this scalar-nonmetricity framework.

\chapter{Dynamical System Analysis for Scalar Field Potential in Symmetric Teleparallel Gravity} \label{c6}
\section{Introduction}
In this study, we explore the cosmological dynamics of $f(Q, \phi)$ gravity, a class of modified gravity theories involving the non-metricity scalar $Q$ and a scalar field $\phi$. Employing dynamical systems techniques, the field equations are reformulated into an autonomous system to examine the Universe’s late-time behavior. Critical points are analyzed for stability using eigenvalue analysis and \gls{cmt}, revealing that accelerated expansion can arise naturally, with at least one stable attractor in each model. Additionally, the evolution of cosmological parameters such as the \gls{eos} and density fractions across radiation, matter, and dark energy eras is presented. The results suggest that $f(Q, \phi)$ gravity provides a viable framework for explaining late-time cosmic acceleration and the sequence of cosmological epochs.

\vspace{1em}
\section{Field Equations of the Scalar-Non-Metricity $ f(Q, \phi) $ Gravity}
The \gls{stegr} action is composed of the linear non-metricity scalar, which can be generalized to $ f(Q) $ gravity. By extending the non-metricity scalar $ Q $ to an arbitrary function and introducing a scalar field $ \phi $, the action can be written as \citep{Gonzalez-Espinoza:2021mwr, DAmbrosio:2021pnd, Bahamonde:2017ize}:
\begin{equation}
S = \int d^4x \, e \left[ f(Q, \phi) + P(\phi)X \right] + S_{m} + S_{r} 
\end{equation}
where $ e = \sqrt{-g} $ is the determinant. The matter action is denoted by $ S_{m} $, while the radiation action is characterized by $ S_{r} $. 

The function $ f(Q, \phi) $ denotes an arbitrary function of the scalar field $ \phi $ and the non-metricity scalar $ Q $. The kinetic term of the scalar field is $ X = -\frac{1}{2} \partial_{\mu} \phi \partial^{\mu} \phi $. This general action includes non-minimally coupled scalar-non-metricity gravity models, $ f(Q) $ gravity, and a minimally coupled scalar field.

The non-metricity scalar is given by:
\begin{equation}
Q = -\frac{1}{4} Q_{\alpha \beta \mu} Q^{\alpha \beta \mu} + \frac{1}{2} Q_{\alpha \beta \mu} Q^{\beta \alpha \mu} + \frac{1}{4} Q_{\alpha \beta}^{\ \ \alpha} Q^{\gamma \beta}_{\ \ \gamma},
\end{equation}
where $ Q_{\alpha \beta \mu} $ denotes the non-metricity tensor. This can be decomposed into traces and irreducible components.

For a homogeneous and isotropic universe, the \gls{flrw} metric is given by:
\begin{equation}
ds^2 = -dt^2 + a^2(t) \left( dx^2 + dy^2 + dz^2 \right),
\end{equation}
where $ a(t) $ is the scale factor representing cosmic expansion in the spatial directions.

By varying the action, the gravitational field equations of $ f(Q, \phi) $ gravity for cosmological applications in the homogeneous and isotropic flat \gls{flrw} metric are obtained as:
\begin{equation}
f(Q, \phi) - P(\phi)\frac{\dot{\phi}^{2}}{2} - 2Q f_{Q} = \rho_{m} + \rho_{r} 
\end{equation}
\begin{equation}
f(Q, \phi) + P(\phi)\frac{\dot{\phi}^{2}}{2} - 2Q f_{Q} - 4\dot{H} f_{Q} - 4H \dot{f}_{Q} = -p_{r} 
\end{equation}
\begin{equation}
P_{\phi} \frac{\dot{\phi}^{2}}{2} - 3P(\phi)H \dot{\phi} - P(\phi) \ddot{\phi} + f_{\phi} = 0
\end{equation}
where $ f_{Q} = \frac{\partial f(Q, \phi)}{\partial Q} $, and $ \rho_{m} $ and $ \rho_{r} $ denote the energy densities of matter and radiation, respectively. $ p_{r} $ represents the pressure in the radiation era, and $ H = \frac{\dot{a}}{a} $ is the Hubble parameter. The overdot represents differentiation with respect to cosmic time $ t $. $ P_{\phi} = \frac{\partial P(\phi)}{\partial \phi} $ and $ f_{\phi} = \frac{\partial f(Q, \phi)}{\partial \phi} $.

Assume a specific form for the coupling function $ f(Q, \phi) $ as \citep{Hohmann:2018rwf, Duchaniya:2022fmc}:
\begin{equation}
f(Q, \phi) = -\frac{Q}{2\kappa^{2}} - G(Q) - V(\phi),
\end{equation}
where $ V(\phi) $ represents the scalar potential, and $ G(Q) $ is an arbitrary function of the non-metricity scalar. In this form, $ \kappa^{2} = 8\pi G $ is the gravitational constant.
For a matter-dominated era $\displaystyle w_{m} = \frac{p_{m}}{\rho_{m}} = 0 $, and for a radiation-dominated era $ \displaystyle w_{r} = \frac{p_{r}}{\rho_{r}} = \frac{1}{3} $, the above equations reduce to:
\begin{equation}
\frac{3}{\kappa^{2}}H^{2} = P(\phi)\frac{\dot{\phi}^{2}}{2} + V(\phi) - 2Q G_{Q} + G(Q) + \rho_{m} + \rho_{r} 
\end{equation}
\begin{equation}
-\frac{2}{\kappa^{2}}\dot{H} = 2P(\phi)\frac{\dot{\phi}^{2}}{2} + 4\dot{H}(G_{Q} + 2Q G_{QQ}) + \rho_{m} + \frac{4}{3}\rho_{r}
\end{equation}
\begin{equation}
P(\phi)\ddot{\phi} - P_{\phi} \frac{\dot{\phi}^{2}}{2} + 3P(\phi)H\dot{\phi} + V_{\phi} = 0
\end{equation}

The total Friedmann equations can be rewritten as:
\begin{equation}
\frac{3}{\kappa^2}H^2 = \rho_m + \rho_r + \rho_{de} 
\end{equation}
\begin{equation}
-\frac{2}{\kappa^2}\dot{H} = \rho_m + \frac{4}{3}\rho_r + \rho_{de} + p_{de}.
\end{equation}
By comparing, we identify the energy density $ \rho_{de} $ and pressure $ p_{de} $ for the dark energy sector as:
\begin{equation}
\rho_{DE} =  P(\phi)\frac{\dot{\phi}^{2}}{2} + V(\phi) - 2Q G_{Q} + G(Q) ,
\end{equation}
\begin{equation}
p_{DE} = P(\phi)\frac{\dot{\phi}^{2}}{2} - V(\phi) + 2Q G_{Q} - G(Q) + 4\dot{H}(G_{Q} + 2Q G_{QQ}).
\end{equation}
Hence the \gls{eos} for dark energy is represented as
\begin{equation}
\displaystyle w_{DE}=\frac{p_{de}}{\rho_{de}}= -1 + \frac{2 P(\phi)\frac{\dot{\phi}^{2}}{2}+ 4\dot{H}(G_{Q} + 2Q G_{QQ})}{P(\phi)\frac{\dot{\phi}^{2}}{2} + V(\phi) - 2Q G_{Q} + G(Q)}
\end{equation}

For simplicity, let's take $ P(\phi) = 1 $, and the potential energy as $ V(\phi) = V_{0} e^{-\lambda \phi} $, where $ \lambda $ is a constant.
This framework allows us to study cosmological models in the dark energy sector within $ f(Q, \phi) $ gravity, along with a dynamical system analysis to develop various models of the universe. We will consider specific forms of $ G(Q) $ to explore different scenarios in subsequent sections.

\vspace{1em}
\section{Dynamical System Analysis}

The aim of this study is to investigate the cosmological dynamics of specific models within the broader class of scalar–non-metricity theories, incorporating nontrivial contributions from the non-metricity scalar. The theoretical framework of dynamical systems provides a framework for describing the evolution and potential future behavior of these cosmological models. A typical dynamical system can be expressed in the form $Y' = f(Y)$, where $Y$ is a column vector of auxiliary variables, and $f(Y)$ is the corresponding vector of autonomous equations. Here, the prime denotes differentiation with respect to the number of e-foldings, $N = \ln a$. Critical points satisfying $f(Y)=0$ characterize the possible cosmological phases. 
In this work, we introduce two models with well-known forms of $G(Q)$.

\vspace{1em}
\subsection{Model I: $ G(Q) = \frac{\alpha}{2}\sqrt{Q_{0} Q}\,\ln\!\left(\gamma \frac{Q}{Q_{0}}\right) + \beta Q $}

The $\sqrt{Q}\ln Q$ correction is motivated by effective field theory, where quantum–gravitational one-loop contributions naturally introduce logarithmic terms in the action. Following the class of models studied in \citep{Anagnostopoulos:2022gej, Goncalves:2024sem}, we adopt the general form  
\begin{equation}
G(Q) = \frac{\alpha'}{2}\sqrt{Q}\,\ln(\gamma' Q) + \beta Q,
\end{equation}
where $\alpha'$, $\beta$, and $\gamma'$ are constants. The standard GR limit is recovered for $\alpha'=0$ and $\beta=1$. The parameter $\beta$ is dimensionless, whereas $[\alpha']=[Q]^{1/2}$ and $[\gamma']=[Q]^{-1}$.
To work with dimensionless quantities, we define
\begin{equation*}
\alpha \equiv \frac{\alpha'}{\sqrt{Q_{0}}}, \qquad 
\gamma \equiv \gamma' Q_{0},
\end{equation*}
with $Q_0=6H_0^2$ denoting the present value of the non-metricity scalar. In terms of these, $G(Q)$ becomes
\begin{equation}
G(Q) = \frac{\alpha}{2}\sqrt{Q_{0}Q}\,\ln\!\left( \gamma \frac{Q}{Q_{0}} \right) + \beta Q,
\end{equation}
which is the form employed throughout this work.

This expression provides a minimal yet physically meaningful extension of \gls{stegr}: the logarithmic term encapsulates quantum-inspired corrections, while the linear term ensures a smooth recovery of the \gls{gr}/\gls{stegr} limit. Hence, the model incorporates classical behaviour, effective-field-theory corrections, and sufficient phenomenological freedom for cosmological applications.

The modified Friedmann equations take the form
\begin{equation}
\frac{3 H^{2}}{\kappa^2} = \frac{\dot{\phi}^2}{2} + V(\phi) - \alpha \sqrt{Q_{0} Q} - \beta Q + \rho_m + \rho_r,
\end{equation}
\begin{equation}
\frac{-2 \dot{H}}{\kappa^2}
= \dot{\phi}^2 
+ 4 \dot{H}\!\left[ \frac{\alpha}{2} \sqrt{\frac{Q_{0}}{Q}} + \beta \right]
+ \rho_m + \frac{4}{3}\rho_r .
\end{equation}
The scalar field (Klein–Gordon) equation becomes
\begin{equation}
\frac{d}{dt}\!\left( \frac{\dot{\phi}^2}{2} + V(\phi) \right)
= -3H\dot{\phi}^2.
\end{equation}
The effective dark energy density and pressure read
\begin{equation}
\rho_{\text{DE}}
=  \frac{\dot{\phi}^2}{2} + V(\phi) - \alpha \sqrt{Q_{0} Q} - \beta Q ,
\end{equation}
\begin{equation}
p_{\text{DE}}
=  \frac{\dot{\phi}^2}{2} - V(\phi)
 + \alpha \sqrt{Q_{0} Q} + \beta Q
 + 4\dot{H}\!\left[ \frac{\alpha}{2}\sqrt{\frac{Q_{0}}{Q}} + \beta \right].
\end{equation}
Thus, the dark energy equation of state is
\begin{equation}
w_{\text{DE}}
= -1 + 
\frac{\dot{\phi}^2 + 4\dot{H}\left[\frac{\alpha}{2}\sqrt{\frac{Q_0}{Q}}+\beta\right]}
{\frac{\dot{\phi}^2}{2} + V(\phi) -
\alpha \sqrt{Q_{0} Q} - \beta Q }.
\end{equation}
To perform a dynamical analysis, we define the dimensionless variables
\begin{equation}
\begin{split}
x = \frac{\kappa \dot{\phi}}{\sqrt{6}H}, \qquad  
y = \frac{\kappa \sqrt{V}}{\sqrt{3}H}, \qquad
\chi = -2\alpha \kappa^2\sqrt{\frac{Q_0}{Q}}, \\
\rho = \frac{\kappa \sqrt{\rho_r}}{\sqrt{3}H}, \qquad
\lambda = -\frac{V_\phi}{\kappa V}, \qquad
\Gamma = \frac{VV_{\phi\phi}}{V_\phi^2}.
\end{split}
\end{equation}
The autonomous system becomes
\begin{eqnarray}
\frac{dx}{dN} &=& -3x + \frac{\sqrt{6}}{2}\lambda y^2 - 3x \frac{\dot{H}}{H^{2}},\\
\frac{dy}{dN} &=& -\sqrt{\frac{3}{2}} \lambda xy - y\frac{\dot{H}}{H^{2}},\\
\frac{d \chi}{dN} &=& - \chi \frac{\dot{H}}{H^{2}},\\
\frac{d \rho}{dN} &=& -2\rho - \rho \frac{\dot{H}}{H^{2}},\\
\frac{d\lambda}{dN} &=& -\sqrt{6}(\Gamma - 1)x\lambda^{2}.
\end{eqnarray}
Here,
\begin{eqnarray}
\frac{\dot{H}}{H^{2}}
= \frac{3x^{2} + 5\rho^{2} + 3\left(1 - y^{2} - \chi - u\right)}
{2(\chi + u - 1)}, \\
u = -2\beta\kappa^2.
\end{eqnarray}

The cosmological parameters are
\begin{equation}
w_{\text{DE}}
= -1 + 
\frac{
2x - \frac{1}{3}\left(\frac{3x^{2}+5\rho^{2}+3(1-y^{2}-\chi -u)}{2(\chi+u-1)}\right)(\chi+u)
}{
x^{2}+y^{2}+\chi+u},
\end{equation}
\begin{equation}
q = \frac{(\chi+u-1)+ 3y^{2}-3x^{2}-5\rho^{2}}{2(\chi + u -1)},
\end{equation}
\begin{equation}
w_{\text{tot}}
= \frac{3y^{2}-3x^{2}-5\rho^{2}}{3(\chi + u -1)} .
\end{equation}
The density parameters are
\begin{eqnarray}
\Omega_{\text{DE}} = x^{2}+y^{2}+\chi+u, \\
\Omega_{r} = \rho^{2}, \\
\Omega_{m} = 1 - x^{2} - y^{2} - \chi - u - \rho^{2}.
\end{eqnarray}

Setting $V(\phi)=V_0 e^{-\lambda\phi}$ gives $\Gamma = 1$, allowing the critical points to be obtained from $x'=y'=\chi'=\rho'=0$. The critical points, their cosmological implications, and stability properties are summarised in Tables 6.1, 6.2, and 6.3.

\begin{table}[htp!]
\centering
\begin{tabular}{ ||p{0.7cm}||p{3cm}||p{0.8cm}||p{0.7cm}||p{0.6cm}||p{0.8cm}||p{0.5cm}||p{0.4cm}||p{1.6cm}|| }
 \hline
 \hline
 CP & $\mathbf{(x_{c}, y_{c}, \chi_{c}, \rho_{c})}$ & $\mathbf{\Omega_{m}}$ & $\mathbf{\Omega_{DE}}$ & $ \mathbf {\Omega_{r}} $ & $ \mathbf{\displaystyle w_{DE}} $ & $\mathbf{ \displaystyle w_{tot}} $ & $ \mathbf{q }$ & $ \text{Exist for }$ \\
 \hline
 $ A_{1} $ & $ (0, 0, 0, 0) $ & $ 1-u $ & $ u $ & $ 0 $ & $ -\frac{1}{2} $ & $ 0 $ & $ \frac{1}{2} $ & $ \text{for all} \quad \lambda $ \\
 \hline
 $ A_{2} $ & $ (0, y, 1-y^2-u, 0) $ & $ 0 $ & $ 1 $ & $ 0 $ & $ -1 $ & $ -1 $ & $ -1 $ & $ \text{for} \quad \lambda=0 $ \\
 \hline
  $ A_{3} $ & $ (0, 0, 0, \pm \sqrt{\frac{1-u}{5}}) $ & $ \frac{4-6u}{5} $ & $ u $ & $ \frac{1-u}{5} $ & $ -\frac{1}{3} $ & $ \frac{1}{3} $ & $ 1 $ & $ \text{for all} \quad \lambda $ \\
 \hline
 $ A_{4} $ & $ (0, \pm \sqrt{1-u}, 0, 0) $ & $ 0 $ & $ 1 $ & $ 0 $ & $ -1 $ & $ -1 $ & $ -1 $ & $ \text{for} \quad \lambda=0 $ \\
 \hline
 $ A_{5} $ & $ (\pm \sqrt{\frac{u-1}{3}}, 0, 0, 0) $ & $ \frac{4-4u}{3} $ & $ \frac{4u-1}{3} $ & $ 0 $ & $ \frac{u-3}{4u-1} $ & $ \frac{1}{3} $ & $ 0 $ & $ \text{for all} \quad \lambda $  \\
 \hline
 \hline
\end{tabular}
\caption{Critical points (CP), matter, dark energy, perturbation and EoS parameters}
\end{table} 

\begin{table}[htp!]
\centering
\begin{tabular}{ ||p{0.8cm}||p{5cm}||p{3.2cm}|| }
\hline
 \hline
 CP & Eigen-Values & Stability Condition  \\
 \hline
 $ A_{1} $ & $ (\frac{3}{2}, \frac{3}{2}, \frac{3}{2}, -\frac{1}{2}) $ & Saddle  \\
 \hline
 $ A_{2} $ & $ ( 0, -3, -2, \frac{-3}{2y^{2}}(1-u+y^{2})) $ & Stable  \\
 \hline
  $ A_{3} $ & $ (1, 2, 2, 3) $ & Unstable \\
 \hline
  $ A_{4} $ & $ (0, -3, -3, -2) $ & Stable \\
 \hline
 $ A_{5} $ & $ (-3, -1, 1, 1) $ & Saddle  \\
 \hline
 \hline
\end{tabular}
\caption{Eigen values and Stability conditions}
\end{table} 

\begin{table}[htp!]
\centering
\label{table:critical_points}
\begin{tabular}{||c|c|c|c||}
\hline
\hline
CP & Acceleration equation & Scale factor (Power law solution) & Universe phase \\ 
\hline
$ A_{1} $   & $\dot{H} = -\frac{3}{2} H^2$   &  $a(t) = t_0 \left(\frac{3}{2} t + c_2\right)^{\frac{2}{3}}$ & matter-dominated  \\ 
 \hline
 $ A_{2} $   & $\dot{H} = 0$  & $a(t) = t_0 e^{c_1 t}$  & de-Sitter phase   \\ 
 \hline
 $ A_{3} $   & $\dot{H} =-2H^{2}$   & $a(t) = t_0 \left(2 t + c_2\right)^{\frac{1}{2}}$ & radiation dominated \\ 
 \hline
$ A_{4} $   & $\dot{H} = 0$  & $a(t) = t_0 e^{c_1 t}$  & de-Sitter phase \\ 
 \hline
$ A_{5} $   & $\dot{H} = -3 H^2$  & $a(t) = t_0 \left(3 t + c_2\right)^{\frac{1}{3}}$  & stiff-matter \\           
\hline
\hline
\end{tabular}
\caption{Cosmological solutions of critical points}
\end{table}

\vspace{1em}
\subsection*{Summary of Critical Points}

\begin{itemize}
    \item \textit{Critical Point $A_1 = (0, 0, 0, 0)$:} This point corresponds to a universe dominated entirely by matter, with dark energy playing a negligible role and radiation absent. The total equation of state is $w_{\text{tot}} = 0$ and the deceleration parameter is $q = \frac{1}{2}$, consistent with a decelerating matter-dominated phase. The point exists for all values of $\lambda$ and has eigenvalues $(\frac{3}{2}, \frac{3}{2}, \frac{3}{2}, -\frac{1}{2})$, identifying it as a saddle point. This suggests that while the universe can pass through this phase, it will eventually evolve away from it. It serves as a suitable model for the intermediate epoch of structure formation.

    \item \textit{Critical Point $A_2 = (0, y, 1 - y^2 - u, 0)$:} Representing a dark energy-dominated de Sitter phase with $\Omega_{\text{DE}} = 1$, this point exhibits exponential expansion with $w_{\text{tot}} = -1$ and $q = -1$. The point exists under the condition $\lambda = 0$, and its eigenvalues are $(0, -3, -2, -\frac{3}{2y^2}(1 - u + y^2))$, implying non-hyperbolicity. A center manifold analysis confirms asymptotic stability. This makes $A_2$ an ideal candidate for describing the Universe's late-time accelerated expansion.

    \item \textit{Critical Point $A_3 = \left(0, 0, 0, \pm \sqrt{\frac{1 - u}{5}} \right)$:} This point captures a radiation-dominated phase where matter, radiation, and dark energy coexist, but radiation dominates. The total equation of state is $w_{\text{tot}} = \frac{1}{3}$, and the deceleration parameter is $q = 1$. The eigenvalues $(1, 2, 2, 3)$ indicate that it is an unstable node, making $A_3$ a past attractor. It provides a realistic description of the early universe, especially during the era of nucleosynthesis.

    \item \textit{Critical Point $A_4 = (0, \pm \sqrt{1 - u}, 0, 0)$:} This point describes a pure dark energy-dominated de Sitter universe with $\Omega_{\text{DE}} = 1$, $w_{\text{tot}} = -1$, and $q = -1$. It exists only for $\lambda = 0$ and has eigenvalues $(0, -3, -3, -2)$, indicating it is a non-hyperbolic but stable attractor. $A_4$ represents a viable late-time phase where the universe experiences exponential acceleration due to a cosmological constant-like dark energy.

    \item \textit{Critical Point $A_5 = \left(\pm \sqrt{\frac{u - 1}{3}}, 0, 0, 0\right)$:} This point signifies a mixed matter and dark energy phase with no radiation. The total equation of state is $w_{\text{tot}} = \frac{1}{3}$ and $q = 0$, placing the universe at the transition between deceleration and acceleration. The eigenvalues $(-3, -1, 1, 1)$ mark it as a saddle point. $A_5$ is relevant for describing an intermediate cosmological era where the scalar field starts contributing significantly alongside matter before the universe settles into a dark energy-dominated future.
\end{itemize}

\begin{figure}[htp!]
   \centering 
   \mbox{\includegraphics[width=0.35\textwidth]{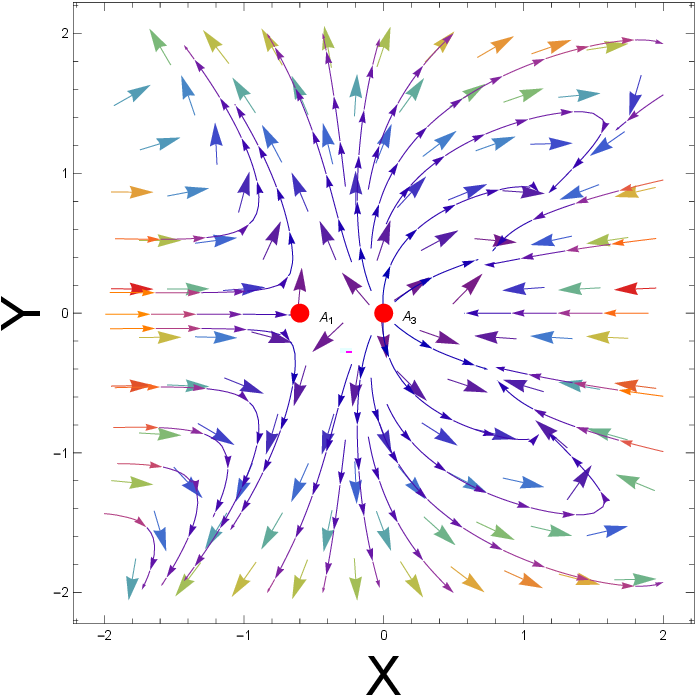}}   
    \hspace{10pt}
   \mbox{\includegraphics[width=0.35\textwidth]{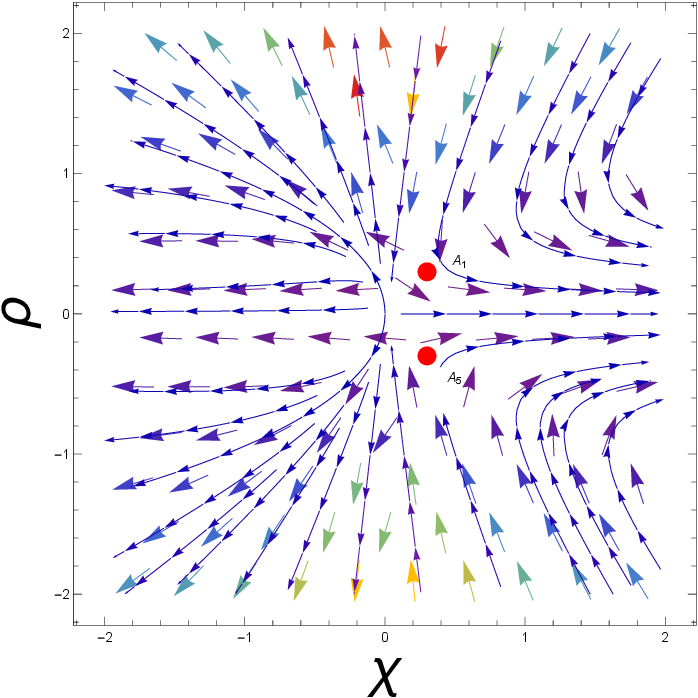}}
   \hspace{10pt}
   \mbox{\includegraphics[width=0.35\textwidth]{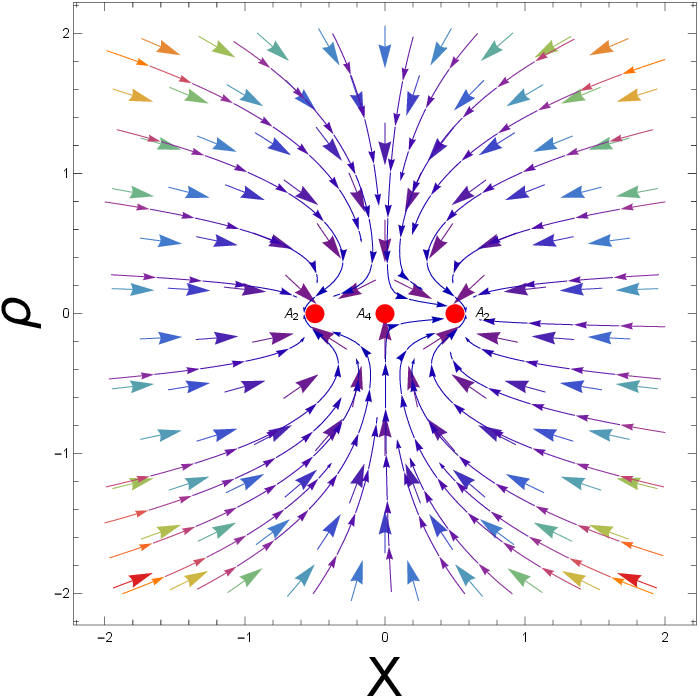}}
    \hspace{10pt}
   \mbox{\includegraphics[width=0.35\textwidth]{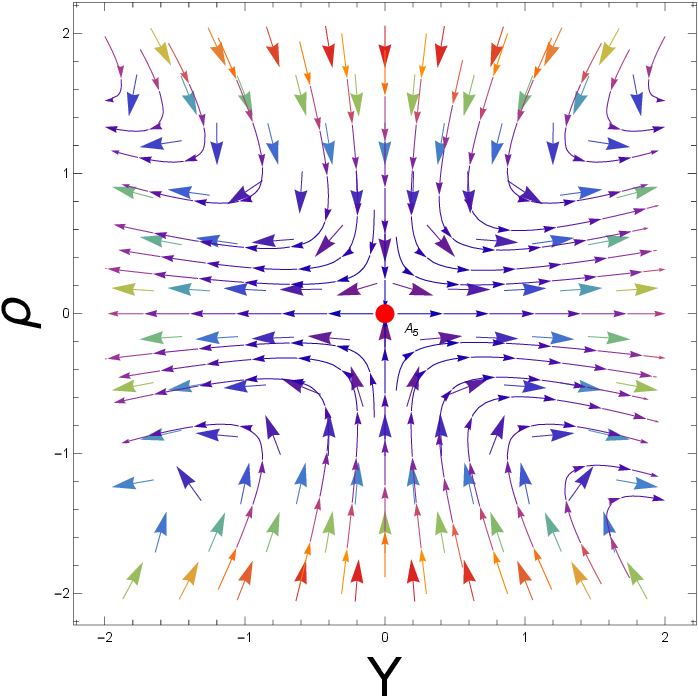}}
    \caption{The phase portraits for the dynamical system of Model are shown as follows: (i)Upper left panel: $ \chi = 0 $, $ \rho = 0 $; (ii) Upper right panel: $ x = 0 $, $ y = 0 $; (iii) Lower right panel: $ \chi = 0 $, $ y=0 $; (iv) Lower left panel: $ x = 0 $, $ \chi=0 $}
   \label{Phase Portrait for the dynamical system of Model}
\end{figure}

Figure~\ref{Phase Portrait for the dynamical system of Model} illustrates the phase portrait of the dynamical system in two dimensions. The critical points $A_2$ and $A_4$ represent stable, dark energy-dominated de Sitter solutions with $\displaystyle w_{DE} = -1$, acting as late-time attractors. In contrast, $A_1$ (matter-dominated) and $A_3$ (radiation-dominated) are saddle points, consistent with transient early-universe phases. The point $A_5$ exhibits unstable behavior, indicating a possible intermediate scaling or transition phase. Overall, the system captures a realistic cosmic evolution: from radiation domination ($A_3$), through matter domination ($A_1$), to accelerated expansion ($A_2$ or $A_4$), with $A_5$ offering additional transitional dynamics.

\begin{figure} [htp!]
   \centering 
   \mbox{\includegraphics[width=0.48\textwidth]{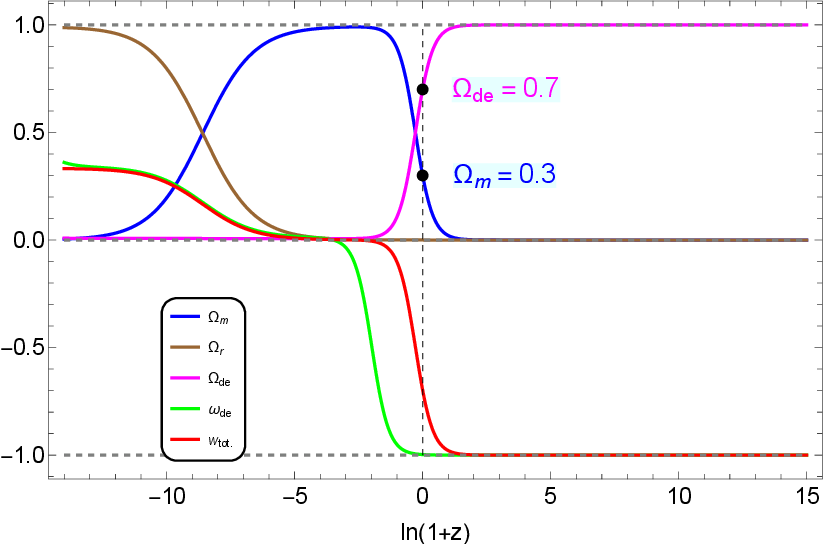}}   
    \hspace{10pt}
    \mbox{\includegraphics[width=0.48\textwidth]{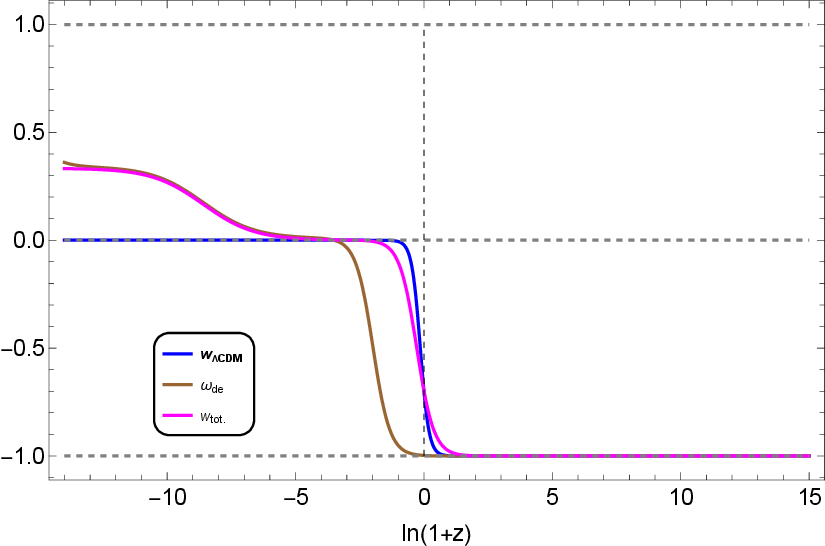}}
    \hspace{10pt}
   \mbox{\includegraphics[width=0.48\textwidth]{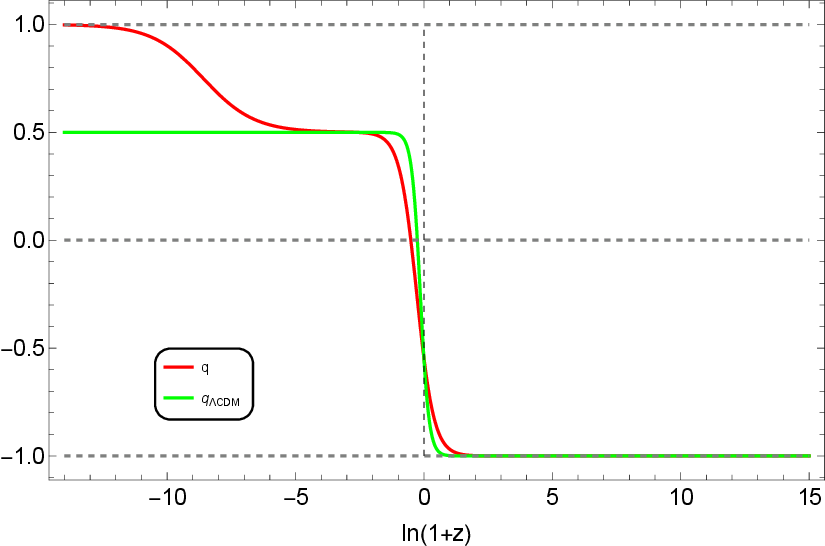}}
    \caption{The evolution of the density parameters (shown in the \textit{Upper Left Panel}), the equation-of-state parameter (shown in the \textit{Upper Right Panel}) and the deceleration parameter (depicted in the \textit{Lower panel}) as functions of redshift are presented for the case $ \lambda = 0.001 $, with the initial conditions of the dynamical system variables set as $ x = 10^{-4} $, $ y = 10^{-6} $, $ \chi = 0.7 \times 10^{-2} $, and $ \rho = 0.933254 $. The vertical dashed line indicates the present cosmological time ($ z = 0 $).}
   \label{Evolution Plot }
\end{figure}

Figure 6.2 (Upper Panels) shows the evolution of density and \gls{eos} parameters. The universe transitions from a matter-dominated era to late-time acceleration, with present-day values $\Omega_m \approx 0.3$ and $\Omega_{DE} \approx 0.7$. The total \gls{eos} evolves from $w_{tot} = 0$ to $w_{tot} \approx -1$, while $w_{DE} \rightarrow -1$, consistent with Planck 2018 \citep{Planck:2018fzr} and DESI observations \citep{DESI:2025euz}. The lower panel shows the deceleration parameter $q$, transitioning from deceleration to acceleration at $z = 0.59$, with a current value $q_0 = -0.57$, in agreement with observational data \citep{Camarena:2019moy}.

\vspace{1em}
\subsection{Model II: $G(Q)=6\eta M^{2}\!\left(\frac{Q}{6M^{2}}\right)^{\beta_{1}} + Q$}

The second class of models considered in this work is defined by the polynomial extension
\begin{equation}
G(Q)=6\eta M^{2}\left(\frac{Q}{6M^{2}}\right)^{\beta_{1}} + Q ,
\end{equation}
which represents the teleparallel analogue of the power-law modification $f(R)=R+\alpha R^{n}$. Such terms can modify both early- and late-time dynamics and generate scaling regimes or accelerated expansion. This model was introduced in \citep{BeltranJimenez:2019tme} as a generalization of the square-root case, and the linear term ensures recovery of the \gls{gr}/\gls{stegr} limit. Special limits include: $\beta_{1}=0$, yielding a cosmological constant term, and $\beta_{1}=1$, corresponding to STEGR with a rescaled gravitational coupling. Values $\beta_{1}>1$ mainly affect the early Universe, while $\beta_{1}<1$ are relevant to late-time acceleration and dark-energy phenomenology \citep{Ayuso2021}.

The effective dark energy density in this framework is
\begin{equation}
\rho_{\rm DE}=\frac{1}{16\pi G}\left[6 - \beta_{1}\mathcal{F}_{0}H_{0}^{-2\beta_{1}}Q^{\beta_{1}+1}\right],
\end{equation}
where $\mathcal{F}_{0}=1-\Omega_{m0}-\Omega_{r0}$. Using the Friedmann equation at $t=t_{0}$, the parameter $\eta$ may be written as
\begin{equation}
\eta=\frac{1}{12\beta_{1}+6}\left(\mathcal{F}_{0}H_{0}^{-2\beta_{1}}M^{2\beta_{1}-2}\right).
\end{equation}
For this model, equations (6.8)--(6.9) take the form
\begin{equation}
\frac{3H^{2}}{\kappa^{2}}=\frac{\dot{\phi}^{2}}{2}+V(\phi)+6\eta M^{2}\!\left(\frac{Q}{6M^{2}}\right)^{\beta_{1}}\!(1-2\beta_{1}) - Q + \rho_{m} + \rho_{r},
\end{equation}
\begin{equation}
\frac{-2\dot{H}}{\kappa^{2}}=\dot{\phi}^{2} + 4\dot{H}\!\left[\beta_{1}\eta\!\left(\frac{Q}{6M^{2}}\right)^{\beta_{1}-1}(2\beta_{1}-1)+1\right] + \rho_{m} + \frac{4}{3}\rho_{r},
\end{equation}
while the scalar-field equation becomes
\begin{equation}
\frac{d}{dt}\!\left(\frac{\dot{\phi}^{2}}{2}+V(\phi)\right)=-3H\dot{\phi}^{2}.
\end{equation}
The effective dark energy density and pressure read
\begin{equation}
\rho_{\rm DE}=\frac{\dot{\phi}^{2}}{2}+V(\phi)+6\eta M^{2}\!\left(\frac{Q}{6M^{2}}\right)^{\beta_{1}}\!(1-2\beta_{1}) - Q,
\end{equation}
\begin{equation}
\begin{split}
p_{\rm DE} &= \frac{\dot{\phi}^{2}}{2} - V(\phi)
- 6\eta M^{2}\!\left(\frac{Q}{6M^{2}}\right)^{\beta_{1}}\!(1-2\beta_{1}) + Q \\
&\quad + 4\dot{H}\!\left[\beta_{1}\eta\!\left(\frac{Q}{6M^{2}}\right)^{\beta_{1}-1}(2\beta_{1}-1)+1\right],
\end{split}
\end{equation}
with $w_{\rm DE}=p_{\rm DE}/\rho_{\rm DE}$.

To study the phase-space structure, we introduce the dimensionless variables
\begin{equation}
\begin{split}
& x=\frac{\kappa\dot{\phi}}{\sqrt{6}H}, \quad
y=\frac{\kappa\sqrt{V}}{\sqrt{3}H}, \quad
\rho=\frac{\kappa\sqrt{\rho_{r}}}{\sqrt{3}H}, \\
& \lambda=-\frac{V_{\phi}}{\kappa V}, \quad
\Gamma=\frac{VV_{\phi\phi}}{V_{\phi}^{2}}, \quad
\chi=-2(1-2\beta_{1})\eta\kappa^{2}\left(\frac{Q}{6M^{2}}\right)^{\beta_{1}-1}.
\end{split}
\end{equation}
Here we have 
\begin{equation}
\frac{\dot{H}}{H^{2}}=\frac{3x^{2}+5\rho^{2}+3(1-y^{2}-\chi-u)}{2(\beta_{1}\chi+u-1)}, \qquad
u=-2\kappa^{2}.
\end{equation}
The autonomous system is then given by
\begin{align}
\frac{dx}{dN} &= -3x + \frac{\lambda 3y^{2}}{\sqrt{6}} - 3x\frac{\dot{H}}{H^{2}},\\
\frac{dy}{dN} &= -\sqrt{\frac{3}{2}}\lambda xy - y\frac{\dot{H}}{H^{2}},\\
\frac{d\chi}{dN} &= 2(\beta_{1}-1)\chi\frac{\dot{H}}{H^{2}},\\
\frac{d\rho}{dN} &= -2\rho - \rho\frac{\dot{H}}{H^{2}},\\
\frac{d\lambda}{dN} &= -\sqrt{6}(\Gamma-1)x\lambda^{2},
\end{align}

The dark-energy equation of state, deceleration parameter, and total equation of state are
\begin{align}
w_{\rm DE} &= -1+\frac{2x-\frac{1}{3}\!\left(\frac{3x^{2}+5\rho^{2}+3(1-y^{2}-\chi-u)}{2(\beta_{1}\chi+u-1)}\right)(\chi+u)}{x^{2}+y^{2}+\chi+u},\\
q &= -1-\frac{3x^{2}+5\rho^{2}+3(1-y^{2}-\chi-u)}{2(\beta_{1}\chi+u-1)},\\
w_{\rm tot} &= -1-\frac{2}{3}\frac{3x^{2}+5\rho^{2}+3(1-y^{2}-\chi-u)}{2(\beta_{1}\chi+u-1)}.
\end{align}
The density parameters become
\begin{align}
\Omega_{\rm DE} &= x^{2}+y^{2}+\chi+u, \\
\Omega_{r} &= \rho^{2}, \\
\Omega_{m} &= 1 - x^{2} - y^{2} - \chi - u - \rho^{2}.
\end{align}

To determine the critical points, we solve the system
\[
x'=0,\quad y'=0,\quad \chi'=0,\quad \rho'=0,
\]
and assume an exponential potential $V(\phi)=V_{0}e^{-\lambda\phi}$, for which $\Gamma=1$.  
The resulting critical points, stability conditions, and cosmological interpretation are summarized in Tables (6.4)--(6.6).

\begin{table}[htp!]
\centering
\begin{tabular}{ ||p{0.6cm}|p{3.3cm}|p{0.9cm}|p{0.7cm}|p{0.7cm}|p{0.7cm}|p{0.5cm}|p{0.5cm}|p{1.6cm}|| }
 \hline
 \hline
 CP & $\mathbf{(x_{c}, y_{c}, \chi_{c}, \rho_{c})}$ & $\mathbf{\Omega_{m}}$ & $\mathbf{\Omega_{DE}}$ & $ \mathbf {\Omega_{r}} $ & $ \mathbf{\displaystyle w_{DE}} $ & $\mathbf{ \displaystyle w_{tot}} $ & $ \mathbf{q }$ & $ \text{Exist for }$ \\
 \hline
 $ B_{1} $ & $ \left( 0, 0, 1-u, 0 \right) $ & $ 0 $ & $ 1 $ & $ 0 $ & $ -1 $ & $ -1 $ & $ -1 $ & $ \text{for all} \quad \lambda $ \\
 \hline
 $ B_{2} $ & $ \left( 0, y, 1-u-y^{2}, 0 \right) $ & $ 0 $ & $ 1 $ & $ 0 $ & $ -1 $ & $ -1 $ & $ -1 $ & $ \text{for} \quad \lambda=0  $ \\
 \hline
  $ B_{3} $ & $ \left( 0, 0, 0, \pm \sqrt{\frac{1-u}{5}} \right)  $ & $ \frac{4-6u}{5} $ & $ u $ & $ \frac{1-u}{5} $ & $ -\frac{1}{3} $ & $ \frac{1}{3} $ & $ 1 $ & $ \text{for all} \quad \lambda $ \\
 \hline
  $ B_{4} $ & $ \left( 0, 0, 0, 0 \right) $ & $ 1-u $ & $ u $ & $ 0 $ & $ -\frac{1}{2} $ & $ 0 $ & $ \frac{1}{2} $ & $ \text{for all} \quad \lambda $ \\
 \hline
 $ B_{5} $ & $ \left( \pm \sqrt{\frac{u-1}{3}}, 0, 0, 0 \right) $ & $ \frac{4-4u}{3} $ & $ \frac{4u-1}{3} $ & $ 0  $ & $ \frac{1+u}{1-4u} $ & $ -\frac{1}{3} $ & $ 0 $ & $ \text{for} \quad \lambda=0 $ \\
 \hline
 $ B_{6} $ & $  \left( 0, \pm \sqrt{1-u}, 0, 0 \right) $ & $ 0 $ & $ 1 $ & $ 0 $ & $ -1 $ & $ -1 $ & $ -1 $ & $ \text{for} \quad \lambda=0 $  \\
 \hline
 \hline
\end{tabular}
\caption{Critical points (CP), matter, dark energy, perturbation and EoS parameters}
\end{table} 

\begin{table}[htp!]
\centering
\begin{tabular}{ ||p{1cm}||p{3cm}||p{3.2cm}|| }
\hline
 \hline
 CP & Eigen-Values & Stability Condition  \\
 \hline
 $ B_{1} $ & $ (-3, -3, -2, 0) $ & Stable  \\
 \hline
 $ B_{2} $ & $ ( 0, 0, -3, -3) $ & Stable  \\
 \hline
  $ B_{3} $ & $ (0, 1, 2, 3) $ & Unstable \\
 \hline
  $ B_{4} $ & $ (\frac{3}{2}, \frac{3}{2}, -\frac{1}{2}, 0) $ & Saddle \\
 \hline
 $ B_{5} $ & $ (0, -3, -1, 1) $ & Saddle  \\
 \hline
 $ B_{6} $ & $ ( 0, 0, -3, -3) $ & Stable  \\
 \hline
 \hline
\end{tabular}
\caption{Eigen values and Stability conditions}
\end{table} 

\vspace{1em}
\subsection*{Summary of Critical Points}

\begin{itemize}
    \item \textit{Critical Point $B_{1}=(0,0,1-u,0)$:}  
This point corresponds to a dark-energy–dominated de Sitter phase with 
\(\Omega_m=0\), \(\Omega_r=0\), \(\Omega_{de}=1\),  
\(w_{de}=w_{tot}=-1\), and \(q=-1\). The eigenvalues \((-3,-3,-2,0)\) render the point non-hyperbolic. Applying \gls{cmt}, after shifting the point to the origin, shows that the dynamics along the center variable are attracted toward the equilibrium. Hence, \(B_1\) is a \textit{stable late-time attractor} describing an asymptotic de Sitter state.

\item \textit{Critical Point $B_{2}=(0,y,1-u-y^{2},0)$:}  
This represents another de Sitter solution with \(\Omega_{de}=1\) and  
\(w_{de}=w_{tot}=-1\), \(q=-1\). It exists for \(\lambda=0\) and includes a free parameter \(y\). The eigenvalues \((0,0,-3,-3)\) indicate a two-dimensional center subspace. Using \gls{cmt}, the reduced dynamics confirm that \(B_2\) is also a \textit{stable attractor}, representing a family of late-time dark-energy–dominated solutions.

\item \textit{Critical Point $B_{3}=\left(0,0,0,\pm\sqrt{\frac{1-u}{5}}\right)$:}  
This point corresponds to a radiation-influenced epoch with  
\[
\Omega_m=\frac{4-6u}{5},\quad 
\Omega_r=\frac{1-u}{5},\quad 
\Omega_{de}=u,
\]
and \(w_{tot}=\frac{1}{3}\), \(q=1\), describing early-time radiation domination. All eigenvalues \((1,2,2,3)\) are positive, classifying \(B_3\) as an \textit{unstable node}. It acts as a past attractor representing the early radiation era.

\item \textit{Critical Point $B_{4}=(0,0,0,0)$:}  
This point represents a matter-dominated phase with  
\(\Omega_m=1-u\), \(\Omega_{de}=u\), \(\Omega_r=0\),  
\(w_{tot}=0\), and \(q=\frac{1}{2}\).  
The eigenvalues \(\left(\frac{3}{2},\frac{3}{2},-\frac{1}{2},0\right)\) indicate a saddle. Thus, \(B_4\) corresponds to a transient matter-dominated epoch essential for structure formation.

\item \textit{Critical Point $B_{5}=\left(\pm\sqrt{\frac{u-1}{3}},0,0,0\right)$:}  
This point describes a mixed matter-dark energy configuration with  
\[
\Omega_m=\frac{4-4u}{3},\qquad 
\Omega_{DE}=\frac{4u-1}{3}.
\]
Reality and positivity constraints restrict it to the degenerate case \(u=1\), reducing it to a pure dark-energy solution. The eigenvalues \((0,-3,-1,1)\) classify it as a \textit{saddle point}. Thus, \(B_5\) represents a transient state rather than a viable late-time attractor.

\item \textit{Critical Point $B_{6}=(0,\pm\sqrt{1-u},0,0)$:}  
This point has the same cosmological character as \(B_2\):  
\(\Omega_{de}=1\), \(w_{tot}=w_{de}=-1\), \(q=-1\).  
The eigenvalues \((0,0,-3,-3)\) indicate non-hyperbolicity, and \gls{cmt} confirms that it is a \textit{stable de Sitter attractor}.  
It represents another late-time dark-energy–dominated phase.

\end{itemize}

\begin{table}[htp!]
\centering
\label{table:critical_points}
\begin{tabular}{||c|c|c|c||}
\hline
\hline
CP & Acceleration equation & Scale factor (Power law solution) & Universe phase \\ 
\hline
$ B_{1} $   & $\dot{H} = 0$  & $a(t) = t_0 e^{c_1 t}$  & de-Sitter phase   \\
 \hline
 $ B_{2} $   & $\dot{H} = 0$  & $a(t) = t_0 e^{c_1 t}$  & de-Sitter phase   \\ 
 \hline
 $ B_{3} $   & $\dot{H} =-2H^{2}$   & $a(t) = t_0 \left(2 t + c_2\right)^{\frac{1}{2}}$ & radiation dominated \\ 
 \hline
 $ B_{4} $   & $\dot{H} = -\frac{3}{2} H^2$   &  $a(t) = t_0 \left(\frac{3}{2} t + c_2\right)^{\frac{2}{3}}$ & matter-dominated  \\  
 \hline
$ B_{5} $   & $\dot{H} = -3 H^2$  & $a(t) = t_0 \left(3 t + c_2\right)^{\frac{1}{3}}$  & stiff-matter \\ 
\hline
$ B_{6} $   & $\dot{H} = 0$  & $a(t) = t_0 e^{c_1 t}$  & de-Sitter phase \\         
\hline
\hline
\end{tabular}
\caption{Cosmological solutions of critical points}
\end{table}

In summary, the system admits multiple stable de Sitter attractors (\(B_1\), \(B_2\), \(B_6\)), a saddle matter point (\(B_4\)), and a past radiation attractor (\(B_3\)). These results demonstrate that the $f(Q,\phi)$ framework supports a realistic cosmological sequence: radiation $\rightarrow$ matter $\rightarrow$ dark energy domination.

\begin{figure}[htp!]
   \centering 
   \mbox{\includegraphics[width=0.35\textwidth]{E.eps}}   
    \hspace{10pt}
   \mbox{\includegraphics[width=0.35\textwidth]{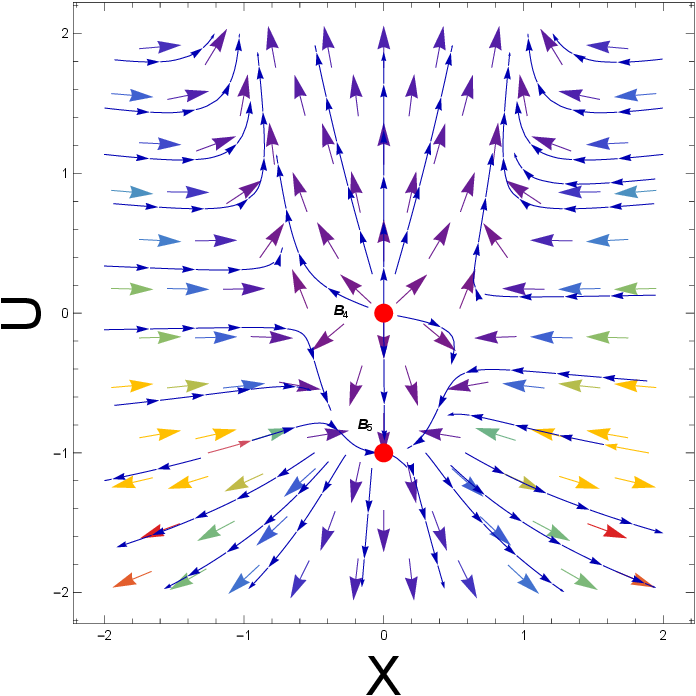}}
   \hspace{10pt}
   \mbox{\includegraphics[width=0.35\textwidth]{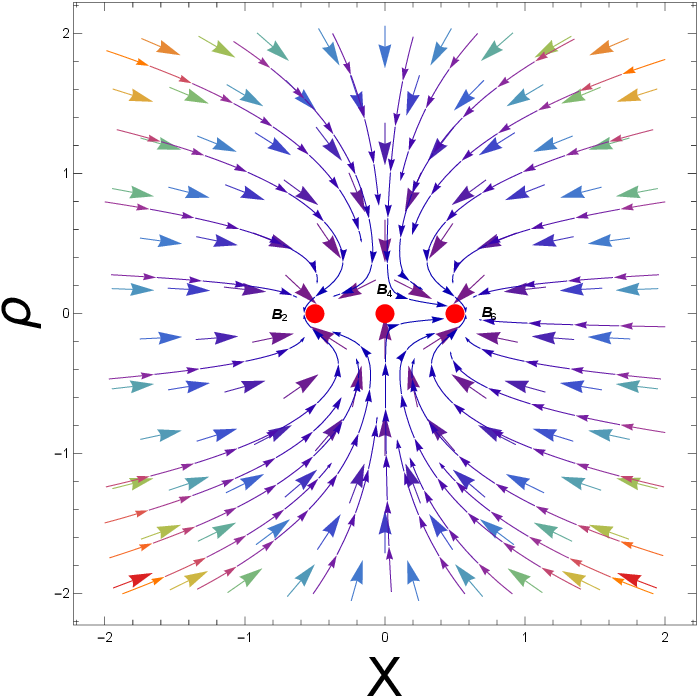}}
    \hspace{10pt}
   \mbox{\includegraphics[width=0.35\textwidth]{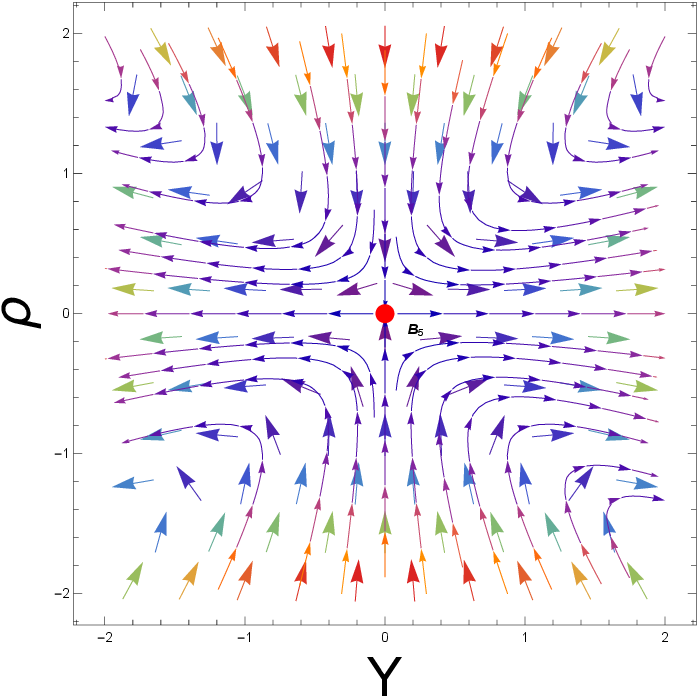}}
    \caption{The phase portraits for the dynamical system of Model are shown as follows: (i)Upper left panel: $ \chi = 0 $, $ \rho = 0 $; (ii) Upper right panel: $ x = 0 $, $ y = 0 $; (iii) Lower right panel: $ \chi = 0 $, $ y=0 $; (iv) Lower left panel: $ x = 0 $, $ \chi=0 $}
   \label{Phase Portrait for the dynamical system of Model-II}
\end{figure}

Figure~\ref{Phase Portrait for the dynamical system of Model-II} presents the phase portrait of the dynamical system in a two-dimensional space. The trajectories indicate that the critical points associated with matter and radiation eras are unstable, whereas the dark-energy–dominated points exhibit stable behavior. In particular, the critical points $B_{1}$, $B_{2}$, and $B_{6}$ correspond to de Sitter solutions with a constant Hubble parameter and an equation-of-state parameter $w_{DE}=-1$. Their stability confirms that they act as late-time attractors, providing viable candidates for describing the asymptotic evolution of the Universe within scalar-nonmetricity gravity.

The critical points $B_{4}$ and $B_{5}$ appear as saddle points associated with the matter-dominated epoch, while $B_{3}$ corresponds to the radiation-dominated era. Their saddle nature implies that these states function as transient phases, allowing the cosmological evolution to progress from radiation domination ($B_{3}$) to matter domination ($B_{4}$), and eventually to late-time accelerated expansion. This sequence aligns with the standard cosmological model and demonstrates that the present framework successfully reproduces the key evolutionary epochs. The point $B_{5}$ also behaves as a saddle, potentially representing an intermediate scaling regime or a transitional state.

\begin{figure} [htp!]
   \centering 
   \mbox{\includegraphics[width=0.48\textwidth]{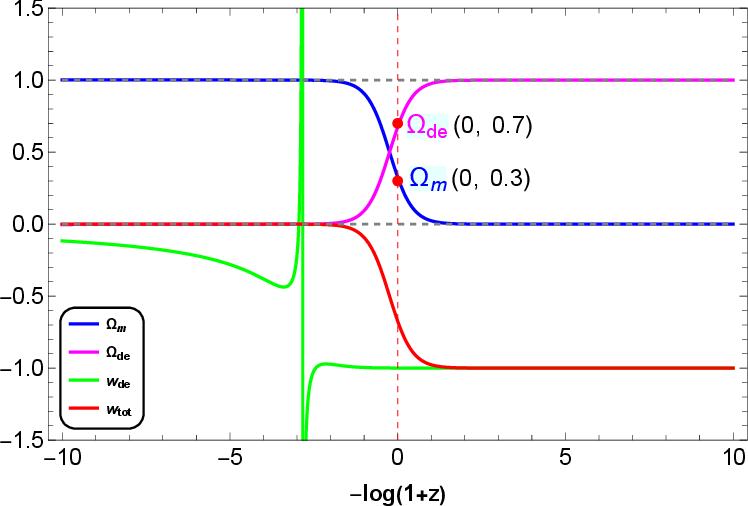}}   
    \hspace{10pt}
    \mbox{\includegraphics[width=0.48\textwidth]{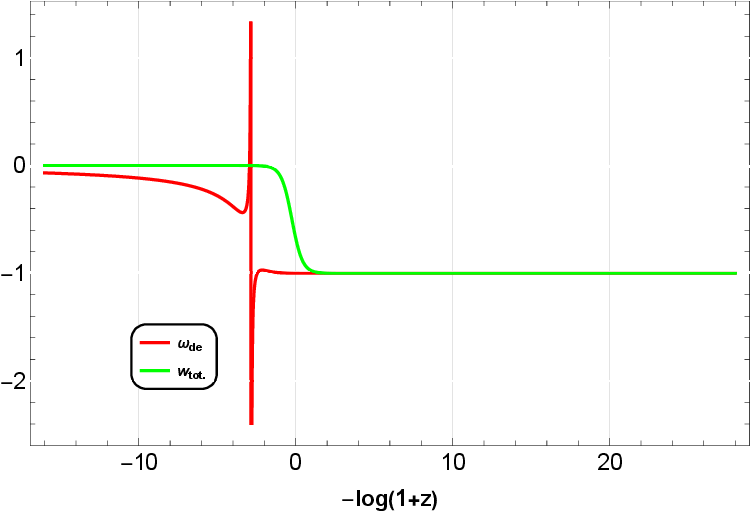}}
    \hspace{10pt}
   \mbox{\includegraphics[width=0.48\textwidth]{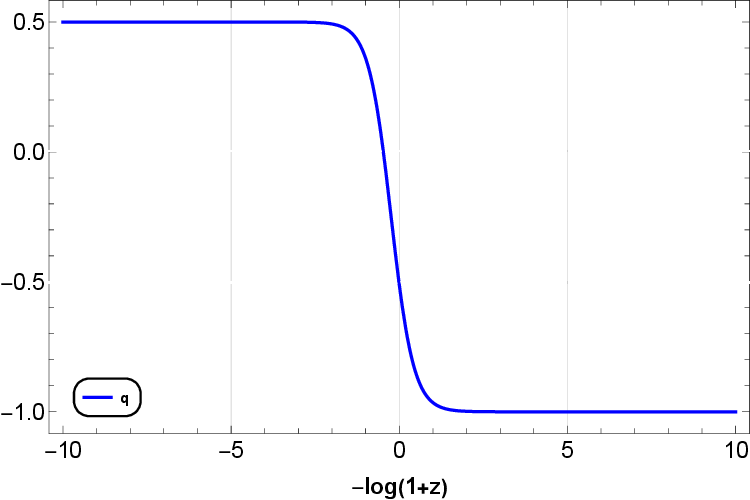}}
    \caption{The evolution of the density parameters (shown in the \textit{Upper Left Panel}), the equation-of-state parameter (shown in the \textit{Upper Right Panel}) and the deceleration parameter (depicted in the \textit{Lower panel}) as functions of redshift are presented for the case $ \lambda = 0.001 $, with the initial conditions of the dynamical system variables set as $ x = 10^{-4} $, $ y = 10^{-6} $, $ \chi = 0.7 \times 10^{-2} $, and $ \rho = 0.933254 $. The vertical dashed line indicates the present cosmological time ($ z = 0 $).}
   \label{Evolution Plot }
\end{figure}

Figure 6.4 (Upper Panels) illustrates the evolution of the density parameters and the \gls{eos} parameters for Model II. The upper left panel shows that the Universe evolves from a matter-dominated phase to a late-time accelerated phase, with present-day density parameters $\Omega_{m} \approx 0.3$ and $\Omega_{de} \approx 0.7$, consistent with current observations. The upper right panel displays the evolution of the total \gls{eos} parameter, which starts at $w_{tot}=0$ during the matter era and approaches $w_{tot} \approx -1$ as dark energy becomes dominant. The dark energy \gls{eos} parameter also tends toward $w_{DE} \approx -1$ at late times, with its present value $w_{DE} = -1$ aligning well with the Planck 2018 constraints ($w_{DE} = -1.028 \pm 0.032$~\citep{Planck:2018fzr}) and recent DESI results \citep{DESI:2025euz}. The lower panel presents the deceleration parameter $q$, which transitions from positive during the decelerating epoch to negative during the accelerating epoch. The transition occurs at redshift $z = 0.59$, and the current value $q_{0} = -0.57$ agrees with observational estimates~\citep{Camarena:2019moy}.

\vspace{1em}
\section{Discussion and Conclusion}

The dynamical systems approach provides a powerful and systematic framework for analyzing the background evolution of cosmological models, enabling the identification of critical points and the classification of their stability properties. By linking fixed points to distinct cosmological epochs, this method serves as an effective diagnostic tool for assessing the viability of modified gravity theories. When complemented with phase portraits, it offers a comprehensive visualization of the parameter space and the associated dynamical trajectories.

In this work, we employed this framework to investigate scalar-nonmetricity gravity described by the action in equation~(6.1). The effective Friedmann and Klein-Gordon equations, equations~(6.8)--(6.10), govern the background dynamics but are analytically challenging, which motivates the dynamical systems reformulation. The model incorporates a canonical scalar field with an exponential potential and a logarithmic $f(Q)$ correction following~\citep{Zhang:2011qp}, enriching the cosmological phenomenology and allowing the scalar field to influence the evolution across multiple epochs.

For Model~I, the critical points $A_{1}$, $A_{3}$, and $A_{5}$ correspond to matter, radiation, and mixed intermediate phases, respectively. The radiation point $A_{3}$ acts as an early-time repeller, while $A_{1}$ serves as a saddle point during the matter-dominated era. The dark-energy-dominated points $A_{2}$ and $A_{4}$ represent de Sitter attractors with $w_{\text{tot}}=-1$ and $q=-1$, ensuring stable late-time acceleration (with $A_{2}$ non-hyperbolic and $A_{4}$ asymptotically stable). Collectively, these features demonstrate that Model~I successfully reproduces the full cosmological sequence from radiation domination to late-time acceleration.

Model~II exhibits a comparable structure through its six critical points $B_{1}$--$B_{6}$. The radiation point $B_{6}$ is unstable, guiding the system toward the matter-era saddle $B_{5}$. The late-time evolution is controlled by the dark-energy attractors $B_{1}$, $B_{2}$, and $B_{3}$, with the stability of $B_{1}$ confirmed via Center Manifold Theory due to its non-hyperbolicity. The scaling-type point $B_{4}$ provides a transitional state between matter and dark-energy dominance, completing a realistic cosmological sequence.

Both models yield present-day observational values,
\[
\Omega_{m} \approx 0.3, \qquad 
\Omega_{de} \approx 0.7, \qquad 
w_{DE} = -1, \qquad 
q_{0} = -0.57,
\]
with a transition redshift of $z \simeq 0.59$, in agreement with current cosmological data. Furthermore, each model satisfies essential viability criteria for modified gravity theories, including: the existence of radiation and matter fixed points, stable de Sitter attractors, a smooth general relativity limit, and the absence of ghosts and gradient instabilities, as reported in the literature.

Overall, the combined analysis demonstrates that scalar-nonmetricity gravity with logarithmic corrections provides a consistent, observationally viable, and theoretically rich cosmological framework capable of reproducing the major evolutionary phases of the Universe. The dynamical system structure reveals sensitivity to the choice of the nonmetricity function $G(Q)$, with each model exhibiting distinct phase-space features, multiple stable late-time attractors, and scaling solutions that facilitate a smooth transition between epochs.

Future directions include constraining the scalar-nonmetricity parameters using large-scale structure and \gls{cmb} observations, as well as extending the analysis to cosmological perturbations to assess implications for structure formation. Further applications to astrophysical phenomena may also offer new insights into the role of nonmetricity in strong- and weak-gravity regimes. Altogether, scalar-nonmetricity gravity with logarithmic corrections emerges as a promising alternative to standard cosmology, combining mathematical consistency with rich dynamical behavior.

\chapter{Concluding Remarks and Future Perspectives} \label{c7}
This chapter summarized the main findings and provided concluding remarks. The collective results from this thesis demonstrate that symmetric teleparallel-based gravity theories-particularly those involving $f(Q)$ and $f(Q, \phi)$ extensions-offer a compelling and consistent framework for describing late-time cosmic acceleration while remaining compatible with early-universe constraints. The dynamical systems approach has proven to be a powerful tool for systematically classifying cosmological behavior across different models and epochs.

\vspace{1em}
\section{Conclusion}
This thesis has presented a comprehensive investigation into late-time cosmological dynamics within the framework of symmetric teleparallel gravity and its various extensions, with particular emphasis on dynamical system techniques and observational consistency.
The foundational groundwork laid in \emph{Chapter 1} established the mathematical and theoretical context necessary for analyzing cosmological models in the setting of symmetric teleparallel gravity. It offered a detailed exposition of \gls{gr}, its \gls{stegr}, and several modified gravity frameworks derived from \gls{stegr}, all of which serve as viable candidates for explaining the observed late-time acceleration of the Universe. 
\emph{Chapter 2} presents a comprehensive review of the existing literature relevant to the topic under investigation.
Building upon this framework, \emph{Chapter 3} examined the autonomous dynamical behavior of three distinct $f(Q)$ gravity models. The analysis demonstrated that each model naturally accommodates a transition from an early-time matter-dominated phase-characterized by unstable dynamics—to a stable, late-time dark energy-dominated regime. These results provide strong evidence for the capability of $f(Q)$ gravity theories to reproduce key features of cosmic evolution.
In \emph{Chapter 4}, we further refined the model selection by incorporating constraints from \gls{bbn}, a pivotal observational benchmark for the early Universe. Two $f(Q)$ models consistent with \gls{bbn} constraints were analyzed in detail, confirming that these models not only admit the required early- and late-time behavior but also exhibit robustness under perturbation and background-level analysis. The imposition of the \gls{bbn} constraint significantly restricted the parameter space, thus enhancing the physical viability of the models.
\emph{Chapter 5} explores the application of the Noether symmetry approach within the framework of extended symmetric teleparallel $f(Q,\phi)$ gravity, which incorporates both the non-metricity scalar $Q$ and a scalar field $\phi$ in a \gls{flrw} spacetime. Using this method, we analyze a specific functional form of $f(Q,\phi)$ and establish the existence of a nontrivial Noether vector. The Noether symmetry approach serves as a powerful tool for constructing viable models and deriving exact solutions for the corresponding Lagrangian. In particular, we show that the symmetry conditions uniquely determine the functional structure of $f(Q,\phi)$ and lead to precise cosmological solutions consistent with the underlying symmetries. These solutions indicate an accelerated expansion of the Universe. Moreover, we compute key cosmological quantities-such as the Hubble parameter, the deceleration parameter, and the effective \gls{eos} parameter and find them to be in agreement with current observational constraints. 
In \emph{Chapter 6}, the analysis was extended to $f(Q, \phi)$ gravity, a scalar-nonmetricity coupled theory, by exploring two distinct functional forms. The resulting dynamical system exhibited a rich set of critical points, corresponding to radiation, matter, stiff-matter, and dark energy-dominated phases. Both models captured the essential sequence of cosmic evolution, including a transition from early-time deceleration to late-time acceleration, highlighting the potential of scalar extensions of $f(Q)$ gravity in modeling a broad spectrum of cosmological epochs.

\vspace{1em}
\section{Future Scope}
Building on the findings of this thesis, several promising avenues for future research may be pursued to further advance our understanding of cosmological dynamics within modified gravity frameworks, particularly those based on symmetric teleparallelism.

Future investigations consider extending the current models by introducing interactions between dark energy and dark matter, potentially offering alternative explanations for late-time cosmic acceleration. Incorporating quantum corrections and exploring inflationary dynamics within teleparallel gravity could also yield important results regarding the primordial power spectrum and the reheating phase of the early Universe.
In addition, the Noether symmetry approach may be generalized to higher-order and more complex modified gravity theories, offering a powerful framework for discovering exact solutions and conserved quantities. Such extensions could deepen our understanding of the geometric and symmetry properties of gravitational theories beyond General Relativity.

A key direction involves conducting a comprehensive phase space analysis at both the background and perturbation levels, with a particular focus on the evolution of matter perturbations. Such an approach would enhance our understanding of structure formation and the stability properties of cosmological models across different epochs. Extending the perturbative analysis to include scalar, vector, and tensor modes is essential to probe the full implications of modified gravity on cosmological perturbations. Special attention should be given to the behavior of gravitational waves in symmetric teleparallel gravity, particularly in models incorporating boundary terms. These models may exhibit distinctive polarization modes, which could yield testable predictions for upcoming gravitational wave observatories such as LIGO-India, LISA, and the Einstein Telescope. Investigating the effects of parity violation and its connection to cosmic birefringence in this context may also provide valuable insights into the fundamental symmetries of gravity.

Collectively, these research directions hold the potential to establish a robust and unified framework for testing modifications to gravity across a wide range of cosmic epochs and observational modalities.

\addcontentsline{toc}{section}{\bf ~~~~~~~~~~~REFERENCES}
 \bibliographystyle{agsmemph}
\bibliography{references}

@article{einstein2005riemann,
  author={Einstein, Albert},
  title={Riemann geometry with maintenance of the concept of far parallelism},
  journal={Albert Einstein: Academy Lectures: Proceedings of the Prussian Academy of Sciences 1914--1932},
  pages={316--321},
  year={2005},
  publisher={Wiley Online Library}
}

@book{aldrovandi2012teleparallel,
    author = "Aldrovandi, Ruben and Pereira, Jos{\'e} Geraldo",
    title = "{Teleparallel Gravity}: {An Introduction}",
    doi = "10.1007/978-94-007-5143-9",
    isbn = "978-94-007-5142-2, 978-94-007-5143-9",
    publisher = "Springer",
    year = "2013"
}

@article{bahamonde2023teleparallel,
    author = "Bahamonde, Sebastian and others",
    title = "{Teleparallel gravity: from theory to cosmology}",
    eprint = "2106.13793",
    archivePrefix = "arXiv",
    primaryClass = "gr-qc",
    doi = "10.1088/1361-6633/ac9cef",
    journal = "Rept. Prog. Phys.",
    volume = "86",
    number = "2",
    pages = "026901",
    year = "2023"
}

@article{friedman1922krummung,
    author = "Friedman, A.",
    title = "{On the Curvature of space}",
    doi = "10.1007/BF01332580",
    journal = "Z. Phys.",
    volume = "10",
    pages = "377--386",
    year = "1922"
}

@article{friedmann1924moglichkeit,
  title={{\"U}ber die M{\"o}glichkeit einer Welt mit konstanten negativen Cur{\"cur}mung des Raum},
  author={Friedmann, Alexander},
  journal={Journal for Physics},
  volume={21},
  number={1},
  pages={326--332},
  year={1924},
  publisher={Springer}
}

@article{lemaitre4univers,
  title={A homogeneous universe of constant mass and increasing radius accounting for the radial velocity of extragalactic bulges},
  author={Lema{\^\i}tre, Georges},
  journal={Annales de la Soci{\'e}t{\'e} Scientifique de Bruxelles, A47, p. 49-59},
  volume={47},
  pages={49--59},
  year={1927}
}

@article{robertson1935kinematics,
  title={Kinematics and world-structure},
  author={Robertson, Howard Percy},
  journal={Astrophysical Journal, vol. 82, p. 284},
  volume={82},
  pages={284},
  year={1935}
}

@article{robertson1936kinematics,
  title={Kinematics and World-Structure II.},
  author={Robertson, Howard P},
  journal={Astrophysical Journal, vol. 83, p. 187},
  volume={83},
  pages={187},
  year={1936}
}

@article{walker1937milne,
  title={On Milne's theory of world-structure},
  author={Walker, Arthur Geoffrey},
  journal={Proceedings of the London Mathematical Society},
  volume={2},
  number={1},
  pages={90--127},
  year={1937},
  publisher={Wiley Online Library}
}

@article{Hubble:1931zz,
    author = "Hubble, Edwin and Humason, Milton L.",
    title = "{The Velocity-Distance Relation among Extra-Galactic Nebulae}",
    doi = "10.1086/143323",
    journal = "Astrophys. J.",
    volume = "74",
    pages = "43--80",
    year = "1931"
}

@book{Peebles:1980yev,
    author = "Peebles, P. James",
    title = "{The Large-Scale Structure of the Universe}",
    isbn = "978-0-691-08240-0, 978-0-691-20983-8, 978-0-691-20671-4",
    publisher = "Princeton University Press",
    month = "11",
    year = "1980"
}

@book{Peebles:1994xt,
    author = "Peebles, P. J. E.",
    title = "{Principles of Physical Cosmology}",
    isbn = "978-0-691-20981-4",
    publisher = "Princeton University Press",
    month = "9",
    year = "2020"
}

@article{Smoot1992,
    author = "Smoot, George F. and others",
    collaboration = "COBE",
    title = "{Structure in the COBE differential microwave radiometer first year maps}",
    reportNumber = "PRINT-92-0148 (UC,BERKELEY)",
    doi = "10.1086/186504",
    journal = "Astrophys. J. Lett.",
    volume = "396",
    pages = "L1--L5",
    year = "1992"
}

@article{Planck:2018vyg,
    author = "Aghanim, N. and others",
    collaboration = "Planck",
    title = "{Planck 2018 results. VI. Cosmological parameters}",
    eprint = "1807.06209",
    archivePrefix = "arXiv",
    primaryClass = "astro-ph.CO",
    doi = "10.1051/0004-6361/201833910",
    journal = "Astron. Astrophys.",
    volume = "641",
    pages = "A6",
    year = "2020",
    note = "[Erratum: Astron.Astrophys. 652, C4 (2021)]"
}

@article{Starobinsky1980,
    author = "Starobinsky, Alexei A.",
    editor = "Khalatnikov, I. M. and Mineev, V. P.",
    title = "{A New Type of Isotropic Cosmological Models Without Singularity}",
    doi = "10.1016/0370-2693(80)90670-X",
    journal = "Phys. Lett. B",
    volume = "91",
    pages = "99--102",
    year = "1980"
}

@article{Guth1981,
    author = "Guth, Alan H.",
    editor = "Fang, Li-Zhi and Ruffini, R.",
    title = "{The Inflationary Universe: A Possible Solution to the Horizon and Flatness Problems}",
    reportNumber = "SLAC-PUB-2576",
    doi = "10.1103/PhysRevD.23.347",
    journal = "Phys. Rev. D",
    volume = "23",
    pages = "347--356",
    year = "1981"
}

@article{Linde1982,
    author = "Linde, Andrei D.",
    editor = "Fang, Li-Zhi and Ruffini, R.",
    title = "{A New Inflationary Universe Scenario: A Possible Solution of the Horizon, Flatness, Homogeneity, Isotropy and Primordial Monopole Problems}",
    reportNumber = "LEBEDEV-81-229",
    doi = "10.1016/0370-2693(82)91219-9",
    journal = "Phys. Lett. B",
    volume = "108",
    pages = "389--393",
    year = "1982"
}

@article{Mukhanov1981,
    author = "Mukhanov, Viatcheslav F. and Chibisov, G. V.",
    title = "{Quantum Fluctuations and a Nonsingular Universe}",
    journal = "JETP Lett.",
    volume = "33",
    pages = "532--535",
    year = "1981"
}

@article{Mukhanov1992,
    author = "Mukhanov, Viatcheslav F. and Feldman, H. A. and Brandenberger, Robert H.",
    title = "{Theory of cosmological perturbations. Part 1. Classical perturbations. Part 2. Quantum theory of perturbations. Part 3. Extensions}",
    reportNumber = "BROWN-HET-796, BROWN-HET-800, BROWN-HET-780",
    doi = "10.1016/0370-1573(92)90044-Z",
    journal = "Phys. Rept.",
    volume = "215",
    pages = "203--333",
    year = "1992"
}

@article{Riess1998,
    author = "Riess, Adam G. and others",
    collaboration = "Supernova Search Team",
    title = "{Observational evidence from supernovae for an accelerating universe and a cosmological constant}",
    eprint = "astro-ph/9805201",
    archivePrefix = "arXiv",
    doi = "10.1086/300499",
    journal = "Astron. J.",
    volume = "116",
    pages = "1009--1038",
    year = "1998"
}

@article{Perlmutter1999,
    author = "Perlmutter, S. and others",
    collaboration = "Supernova Cosmology Project",
    title = "{Measurements of $\Omega$ and $\Lambda$ from 42 High Redshift Supernovae}",
    eprint = "astro-ph/9812133",
    archivePrefix = "arXiv",
    reportNumber = "LBNL-41801, LBL-41801",
    doi = "10.1086/307221",
    journal = "Astrophys. J.",
    volume = "517",
    pages = "565--586",
    year = "1999"
}

@article{Peebles2003,
    author = "Peebles, P. J. E. and Ratra, Bharat",
    editor = "Hsu, Jong-Ping and Fine, D.",
    title = "{The Cosmological Constant and Dark Energy}",
    eprint = "astro-ph/0207347",
    archivePrefix = "arXiv",
    reportNumber = "KSUPT-02-3",
    doi = "10.1103/RevModPhys.75.559",
    journal = "Rev. Mod. Phys.",
    volume = "75",
    pages = "559--606",
    year = "2003"
}

@article{Weinberg2013,
    author = "Weinberg, David H. and Mortonson, Michael J. and Eisenstein, Daniel J. and Hirata, Christopher and Riess, Adam G. and Rozo, Eduardo",
    title = "{Observational Probes of Cosmic Acceleration}",
    eprint = "1201.2434",
    archivePrefix = "arXiv",
    primaryClass = "astro-ph.CO",
    doi = "10.1016/j.physrep.2013.05.001",
    journal = "Phys. Rept.",
    volume = "530",
    pages = "87--255",
    year = "2013"
}

@article{Spergel2003,
    author = "Spergel, D. N. and others",
    collaboration = "WMAP",
    title = "{First year Wilkinson Microwave Anisotropy Probe (WMAP) observations: Determination of cosmological parameters}",
    eprint = "astro-ph/0302209",
    archivePrefix = "arXiv",
    doi = "10.1086/377226",
    journal = "Astrophys. J. Suppl.",
    volume = "148",
    pages = "175--194",
    year = "2003"
}

@article{Weinberg1989,
    author = "Weinberg, Steven",
    editor = "Hsu, Jong-Ping and Fine, D.",
    title = "{The Cosmological Constant Problem}",
    reportNumber = "UTTG-12-88",
    doi = "10.1103/RevModPhys.61.1",
    journal = "Rev. Mod. Phys.",
    volume = "61",
    pages = "1--23",
    year = "1989"
}

@article{Carroll2001,
    author = "Carroll, Sean M.",
    title = "{The Cosmological constant}",
    eprint = "astro-ph/0004075",
    archivePrefix = "arXiv",
    reportNumber = "EFI-2000-13",
    doi = "10.12942/lrr-2001-1",
    journal = "Living Rev. Rel.",
    volume = "4",
    pages = "1",
    year = "2001"
}

@article{Padmanabhan2003,
    author = "Padmanabhan, T.",
    title = "{Cosmological constant: The Weight of the vacuum}",
    eprint = "hep-th/0212290",
    archivePrefix = "arXiv",
    doi = "10.1016/S0370-1573(03)00120-0",
    journal = "Phys. Rept.",
    volume = "380",
    pages = "235--320",
    year = "2003"
}

@article{Clifton2012,
    author = "Clifton, Timothy and Ferreira, Pedro G. and Padilla, Antonio and Skordis, Constantinos",
    title = "{Modified Gravity and Cosmology}",
    eprint = "1106.2476",
    archivePrefix = "arXiv",
    primaryClass = "astro-ph.CO",
    doi = "10.1016/j.physrep.2012.01.001",
    journal = "Phys. Rept.",
    volume = "513",
    pages = "1--189",
    year = "2012"
}

@article{Nojiri2011,
    author = "Nojiri, Shin'ichi and Odintsov, Sergei D.",
    title = "{Unified cosmic history in modified gravity: from F(R) theory to Lorentz non-invariant models}",
    eprint = "1011.0544",
    archivePrefix = "arXiv",
    primaryClass = "gr-qc",
    doi = "10.1016/j.physrep.2011.04.001",
    journal = "Phys. Rept.",
    volume = "505",
    pages = "59--144",
    year = "2011"
}

@article{Capozziello2011,
    author = "Capozziello, Salvatore and De Laurentis, Mariafelicia",
    title = "{Extended Theories of Gravity}",
    eprint = "1108.6266",
    archivePrefix = "arXiv",
    primaryClass = "gr-qc",
    doi = "10.1016/j.physrep.2011.09.003",
    journal = "Phys. Rept.",
    volume = "509",
    pages = "167--321",
    year = "2011"
}

@article{Einstein1915,
    author = "Einstein, Albert",
    title = "{The Field Equations of Gravitation}",
    journal = "Sitzungsber. Preuss. Akad. Wiss. Berlin (Math. Phys. )",
    volume = "1915",
    pages = "844--847",
    year = "1915"
}

@article{Einstein1916,
    author = "Einstein, Albert",
    editor = "Hsu, Jong-Ping and Fine, D.",
    title = "{The foundation of the general theory of relativity.}",
    doi = "10.1002/andp.19163540702",
    journal = "Annalen Phys.",
    volume = "49",
    number = "7",
    pages = "769--822",
    year = "1916"
}

@article{Einstein1925,
  author       = {A. Einstein},
  title        = {Einheitliche Feldtheorie von Gravitation und Elektrizität},
  journal      = {Sitzungsberichte der Preussischen Akademie der Wissenschaften (Berlin)},
  pages        = {414--419},
  year         = {1925},
  note         = {Unified Field Theory of Gravitation and Electricity}
}

@article{Einstein1928,
  author       = {A. Einstein},
  title        = {Riemann-Geometrie mit Aufrechterhaltung des Begriffes des Fernparallelismus},
  journal      = {Sitzungsberichte der Preussischen Akademie der Wissenschaften (Berlin), Phys.-Math. Klasse},
  pages        = {217--221},
  year         = {1928},
  note         = {Riemannian Geometry with Preservation of the Notion of Distant Parallelism}
}

@article{Hehl1995,
    author = "Hehl, Friedrich W. and McCrea, J. Dermott and Mielke, Eckehard W. and Ne'eman, Yuval",
    title = "{Metric affine gauge theory of gravity: Field equations, Noether identities, world spinors, and breaking of dilation invariance}",
    eprint = "gr-qc/9402012",
    archivePrefix = "arXiv",
    reportNumber = "TAUP-N192-94, TAUP-192-94",
    doi = "10.1016/0370-1573(94)00111-F",
    journal = "Phys. Rept.",
    volume = "258",
    pages = "1--171",
    year = "1995"
}

@article{Obukhov2006,
    author = "Obukhov, Yuri N.",
    title = "{Poincare gauge gravity: Selected topics}",
    eprint = "gr-qc/0601090",
    archivePrefix = "arXiv",
    doi = "10.1142/S021988780600103X",
    journal = "Int. J. Geom. Meth. Mod. Phys.",
    volume = "3",
    pages = "95--138",
    year = "2006"
}

@article{BeltranJimenez:2019esp,
    author = "Beltr{\'a}n Jim{\'e}nez, Jose and Heisenberg, Lavinia and Koivisto, Tomi S.",
    title = "{The Geometrical Trinity of Gravity}",
    eprint = "1903.06830",
    archivePrefix = "arXiv",
    primaryClass = "hep-th",
    doi = "10.3390/universe5070173",
    journal = "Universe",
    volume = "5",
    number = "7",
    pages = "173",
    year = "2019"
}

@article{BeltranJimenez:2017tkd,
    author = "Beltr{\'a}n Jim{\'e}nez, Jose and Heisenberg, Lavinia and Koivisto, Tomi",
    title = "{Coincident General Relativity}",
    eprint = "1710.03116",
    archivePrefix = "arXiv",
    primaryClass = "gr-qc",
    reportNumber = "NORDITA-2017-100, IFT-UAM/CSIC-17-093, ITS-ETH-2017-10",
    doi = "10.1103/PhysRevD.98.044048",
    journal = "Phys. Rev. D",
    volume = "98",
    number = "4",
    pages = "044048",
    year = "2018"
}

@book{Aldrovandi2013,
    author = "Aldrovandi, Ruben and Pereira, Jos{\'e} Geraldo",
    title = "{Teleparallel Gravity}: {An Introduction}",
    doi = "10.1007/978-94-007-5143-9",
    isbn = "978-94-007-5142-2, 978-94-007-5143-9",
    publisher = "Springer",
    year = "2013"
}

@article{BeltranJimenez:2019tme,
    author = "Beltr{\'a}n Jim{\'e}nez, Jose and others",
    title = "{Cosmology in $f(Q)$ geometry}",
    eprint = "1906.10027",
    archivePrefix = "arXiv",
    primaryClass = "gr-qc",
    doi = "10.1103/PhysRevD.101.103507",
    journal = "Phys. Rev. D",
    volume = "101",
    number = "10",
    pages = "103507",
    year = "2020"
}

@article{Capozziello2022,
    author = "Capozziello, Salvatore and Cesare, Sara and Ferrara, Carmen",
    title = "{Extended geometric trinity of gravity}",
    eprint = "2503.08167",
    archivePrefix = "arXiv",
    primaryClass = "gr-qc",
    doi = "10.1140/epjc/s10052-025-14440-2",
    journal = "Eur. Phys. J. C",
    volume = "85",
    number = "9",
    pages = "932",
    year = "2025"
}

@article{Nojiri2017,
    author = "Nojiri, S. and Odintsov, S. D. and Oikonomou, V. K.",
    title = "{Modified Gravity Theories on a Nutshell: Inflation, Bounce and Late-time Evolution}",
    eprint = "1705.11098",
    archivePrefix = "arXiv",
    primaryClass = "gr-qc",
    reportNumber = "PHYS.REPT.-692-(2017)-1-104, Phys.Rept. 692 (2017) 1-104",
    doi = "10.1016/j.physrep.2017.06.001",
    journal = "Phys. Rept.",
    volume = "692",
    pages = "1--104",
    year = "2017"
}

@article{Xu2019,
    author = "Xu, Yixin and others",
    title = "{$f(Q,T)$ gravity}",
    eprint = "1908.04760",
    archivePrefix = "arXiv",
    primaryClass = "gr-qc",
    doi = "10.1140/epjc/s10052-019-7207-4",
    journal = "Eur. Phys. J. C",
    volume = "79",
    number = "8",
    pages = "708",
    year = "2019"
}

@article{Barros2020,
    author = "Barros, Bruno J. and others",
    title = "{Testing $F(Q)$ gravity with redshift space distortions}",
    eprint = "2004.07867",
    archivePrefix = "arXiv",
    primaryClass = "gr-qc",
    doi = "10.1016/j.dark.2020.100616",
    journal = "Phys. Dark Univ.",
    volume = "30",
    pages = "100616",
    year = "2020"
}

@article{Khyllep2021,
    author = "Khyllep, Wompherdeiki and Paliathanasis, Andronikos and Dutta, Jibitesh",
    title = "{Cosmological solutions and growth index of matter perturbations in $f(Q)$ gravity}",
    eprint = "2103.08372",
    archivePrefix = "arXiv",
    primaryClass = "gr-qc",
    doi = "10.1103/PhysRevD.103.103521",
    journal = "Phys. Rev. D",
    volume = "103",
    number = "10",
    pages = "103521",
    year = "2021"
}

@article{Mandal2020,
    author = "Mandal, Sanjay and Sahoo, P. K. and Santos, J. R. L.",
    title = "{Energy conditions in $f(Q)$ gravity}",
    eprint = "2008.01563",
    archivePrefix = "arXiv",
    primaryClass = "gr-qc",
    doi = "10.1103/PhysRevD.102.024057",
    journal = "Phys. Rev. D",
    volume = "102",
    number = "2",
    pages = "024057",
    year = "2020"
}

@article{Bajardi2020,
    author = "Koussour, M. and De, Avik",
    title = "{Observational constraints on two cosmological models of f(Q) theory}",
    eprint = "2304.11765",
    archivePrefix = "arXiv",
    primaryClass = "gr-qc",
    doi = "10.1140/epjc/s10052-023-11547-2",
    journal = "Eur. Phys. J. C",
    volume = "83",
    number = "5",
    pages = "400",
    year = "2023"
}

@article{Lazkoz2019,
    author = "Lazkoz, Ruth and others",
    title = "{Observational constraints of $f(Q)$ gravity}",
    eprint = "1907.13219",
    archivePrefix = "arXiv",
    primaryClass = "gr-qc",
    doi = "10.1103/PhysRevD.100.104027",
    journal = "Phys. Rev. D",
    volume = "100",
    number = "10",
    pages = "104027",
    year = "2019"
}

@article{Anagnostopoulos2021,
    author = "Anagnostopoulos, Fotios K. and others",
    title = "{First evidence that non-metricity f(Q) gravity could challenge {\ensuremath{\Lambda}}CDM}",
    eprint = "2104.15123",
    archivePrefix = "arXiv",
    primaryClass = "gr-qc",
    doi = "10.1016/j.physletb.2021.136634",
    journal = "Phys. Lett. B",
    volume = "822",
    pages = "136634",
    year = "2021"
}

@article{Hohmann2021,
    author = "Ferraro, Rafael and Guzm{\'a}n, Mar{\'\i}a Jos{\'e}",
    title = "{Pseudoinvariance and the extra degree of freedom in f(T) gravity}",
    eprint = "2001.08137",
    archivePrefix = "arXiv",
    primaryClass = "gr-qc",
    doi = "10.1103/PhysRevD.101.084017",
    journal = "Phys. Rev. D",
    volume = "101",
    number = "8",
    pages = "084017",
    year = "2020"
}

@article{Ayuso2021,
    author = "Ayuso, Ismael and Lazkoz, Ruth and Salzano, Vincenzo",
    title = "{Observational constraints on cosmological solutions of $f(Q)$ theories}",
    eprint = "2012.00046",
    archivePrefix = "arXiv",
    primaryClass = "astro-ph.CO",
    doi = "10.1103/PhysRevD.103.063505",
    journal = "Phys. Rev. D",
    volume = "103",
    number = "6",
    pages = "063505",
    year = "2021"
}

@article{DiValentino:2021izs,
    author = "Di Valentino, Eleonora and others",
    title = "{In the realm of the Hubble tension{\textemdash}a review of solutions}",
    eprint = "2103.01183",
    archivePrefix = "arXiv",
    primaryClass = "astro-ph.CO",
    reportNumber = "IPPP/20/108",
    doi = "10.1088/1361-6382/ac086d",
    journal = "Class. Quant. Grav.",
    volume = "38",
    number = "15",
    pages = "153001",
    year = "2021"
}

@article{Brout:2022vxf,
    author = "Brout, Dillon and others",
    title = "{The Pantheon+ Analysis: Cosmological Constraints}",
    eprint = "2202.04077",
    archivePrefix = "arXiv",
    primaryClass = "astro-ph.CO",
    doi = "10.3847/1538-4357/ac8e04",
    journal = "Astrophys. J.",
    volume = "938",
    number = "2",
    pages = "110",
    year = "2022"
}

@article{Bernal:2016gxb,
    author = "Bernal, Jose Luis and Verde, Licia and Riess, Adam G.",
    title = "{The trouble with $H_0$}",
    eprint = "1607.05617",
    archivePrefix = "arXiv",
    primaryClass = "astro-ph.CO",
    doi = "10.1088/1475-7516/2016/10/019",
    journal = "Journal of Cosmology and Astroparticle Physics ",
    volume = "10",
    pages = "019",
    year = "2016"
}

@article{Benisty:2021cmq,
    author = "Benisty, David and Davis, Anne-Christine",
    title = "{Dark energy interactions near the Galactic Center}",
    eprint = "2108.06286",
    archivePrefix = "arXiv",
    primaryClass = "astro-ph.CO",
    doi = "10.1103/PhysRevD.105.024052",
    journal = "Phys. Rev. D",
    volume = "105",
    number = "2",
    pages = "024052",
    year = "2022"
}

@article{verde2003first,
    author = "Verde, Licia and others",
    collaboration = "WMAP",
    title = "{First year Wilkinson Microwave Anisotropy Probe (WMAP) observations: Parameter estimation methodology}",
    eprint = "astro-ph/0302218",
    archivePrefix = "arXiv",
    doi = "10.1086/377335",
    journal = "Astrophys. J. Suppl.",
    volume = "148",
    pages = "195",
    year = "2003"
}

@article{vagnozzi2020new,
    author = "Vagnozzi, Sunny",
    title = "{New physics in light of the $H_0$ tension: An alternative view}",
    eprint = "1907.07569",
    archivePrefix = "arXiv",
    primaryClass = "astro-ph.CO",
    doi = "10.1103/PhysRevD.102.023518",
    journal = "Phys. Rev. D",
    volume = "102",
    number = "2",
    pages = "023518",
    year = "2020"
}

@article{Abdalla:2022yfr,
    author = "Abdalla, Elcio and others",
    title = "{Cosmology intertwined: A review of the particle physics, astrophysics, and cosmology associated with the cosmological tensions and anomalies}",
    eprint = "2203.06142",
    archivePrefix = "arXiv",
    primaryClass = "astro-ph.CO",
    reportNumber = "FERMILAB-CONF-22-192-SCD",
    doi = "10.1016/j.jheap.2022.04.002",
    journal = "Journal of High Energy Astrophysics",
    volume = "34",
    pages = "49--211",
    year = "2022"
}

@article{moresco2022unveiling,
    author = "Borghi, Nicola",
    title = "{Unveiling the expansion history of the universe with cosmic chronometers and gravitational waves}",
    doi = "10.48676/unibo/amsdottorato/11090",
    school = "Bologna U.",
    year = "2024"
}

@article{ade2014planck,
    author = "Ade, P. A. R. and others",
    collaboration = "Planck",
    title = "{Planck 2013 results. XVI. Cosmological parameters}",
    eprint = "1303.5076",
    archivePrefix = "arXiv",
    primaryClass = "astro-ph.CO",
    reportNumber = "CERN-PH-TH-2013-129",
    doi = "10.1051/0004-6361/201321591",
    journal = "Astron. Astrophys.",
    volume = "571",
    pages = "A16",
    year = "2014"
}

@article{Riess:2021jrx,
    author = "Riess, Adam G. and others",
    title = "{{A Comprehensive Measurement of the Local Value of the Hubble Constant with 1 km s$^{-1}$ Mpc$^{-1}$ Uncertainty from the Hubble Space Telescope and the SH0ES Team}}",
    eprint = "2112.04510",
    archivePrefix = "arXiv",
    primaryClass = "astro-ph.CO",
    doi = "10.3847/2041-8213/ac5c5b",
    journal = "Astrophys. J. Lett.",
    volume = "934",
    number = "1",
    pages = "L7",
    year = "2022"
}

@article{wong2017h0licow,
    author = "Wong, Kenneth C. and others",
    collaboration = "H0LiCOW",
    title = "{H0LiCOW {\textendash} IV. Lens mass model of HE 0435{\ensuremath{-}}1223 and blind measurement of its time-delay distance for cosmology}",
    eprint = "1607.01403",
    archivePrefix = "arXiv",
    primaryClass = "astro-ph.CO",
    doi = "10.1093/mnras/stw3077",
    journal = "Mon. Not. Roy. Astron. Soc.",
    volume = "465",
    number = "4",
    pages = "4895--4913",
    year = "2017"
}

@article{freedman2019carnegie,
    author = "Freedman, Wendy L. and others",
    title = "{The Carnegie-Chicago Hubble Program. VIII. An Independent Determination of the Hubble Constant Based on the Tip of the Red Giant Branch}",
    eprint = "1907.05922",
    archivePrefix = "arXiv",
    primaryClass = "astro-ph.CO",
    doi = "10.3847/1538-4357/ab2f73",
    journal = "Astrophys. J.",
    volume = "882",
    pages = "34",
    year = "2019"
}

@article{hakim2005inflationary,
    author = "Bamba, Kazuharu and Odintsov, Sergei D.",
    title = "{Inflationary cosmology in modified gravity theories}",
    eprint = "1503.00442",
    archivePrefix = "arXiv",
    primaryClass = "hep-th",
    reportNumber = "OCHA-PP-331",
    doi = "10.3390/sym7010220",
    journal = "Symmetry",
    volume = "7",
    number = "1",
    pages = "220--240",
    year = "2015"
}

@article{linde1983chaotic,
    author = "Linde, Andrei D.",
    title = "{Chaotic Inflation}",
    doi = "10.1016/0370-2693(83)90837-7",
    journal = "Phys. Lett. B",
    volume = "129",
    pages = "177--181",
    year = "1983"
}

@article{barrow1981inflation,
    author = "Linde, Andrei D.",
    title = "{The Inflationary Universe}",
    doi = "10.1088/0034-4885/47/8/002",
    journal = "Rept. Prog. Phys.",
    volume = "47",
    pages = "925--986",
    year = "1984"
}

@book{kolb1990early,
    author = "Kolb, Edward W. and Turner, Michael S.",
    title = "{The Early Universe}",
    reportNumber = "FERMILAB-BOOK-1990-01",
    doi = "10.1201/9780429492860",
    isbn = "978-0-429-49286-0, 978-0-201-62674-2",
    publisher = "Taylor and Francis",
    volume = "69",
    month = "5",
    year = "2019"
}

@book{mukhanov2005physical,
    author = "Mukhanov, V.",
    title = "{Physical Foundations of Cosmology}",
    doi = "10.1017/CBO9780511790553",
    isbn = "978-0-521-56398-7",
    publisher = "Cambridge University Press",
    address = "Oxford",
    year = "2005"
}

@article{rindler1956visual,
    author = "Rindler, W.",
    title = "{Visual Horizons in World Models}",
    doi = "10.1093/mnras/116.6.662",
    journal = "Mon. Not. Roy. Astron. Soc.",
    volume = "116",
    number = "6",
    pages = "662--677",
    year = "1956"
}

@article{Georgi:1974sy,
    author = "Georgi, H. and Glashow, S. L.",
    title = "{Unity of All Elementary Particle Forces}",
    doi = "10.1103/PhysRevLett.32.438",
    journal = "Phys. Rev. Lett.",
    volume = "32",
    pages = "438--441",
    year = "1974"
}

@article{goddard1978magnetic,
    author = "Goddard, P. and Olive, David I.",
    title = "{New Developments in the Theory of Magnetic Monopoles}",
    reportNumber = "CERN-TH-2445",
    doi = "10.1088/0034-4885/41/9/001",
    journal = "Rept. Prog. Phys.",
    volume = "41",
    pages = "1357",
    year = "1978"
}

@article{WMAP:2012nax,
    author = "Hinshaw, G. and others",
    collaboration = "WMAP",
    title = "{Nine-Year Wilkinson Microwave Anisotropy Probe (WMAP) Observations: Cosmological Parameter Results}",
    eprint = "1212.5226",
    archivePrefix = "arXiv",
    primaryClass = "astro-ph.CO",
    doi = "10.1088/0067-0049/208/2/19",
    journal = "Astrophys. J. Suppl.",
    volume = "208",
    pages = "19",
    year = "2013"
}

@article{DiValentino:2016hlg,
    author = "Di Valentino, Eleonora and others",
    title = "{Reconciling Planck with the local value of $H_0$ in extended parameter space}",
    eprint = "1606.00634",
    archivePrefix = "arXiv",
    primaryClass = "astro-ph.CO",
    doi = "10.1016/j.physletb.2016.08.043",
    journal = "Phys. Lett. B",
    volume = "761",
    pages = "242--246",
    year = "2016"
}

@book{misner1973gravitation,
  title        = {Gravitation},
  author       = {Misner, Charles W. and Thorne, Kip S. and Wheeler, John Archibald},
  year         = {1973},
  publisher    = {W. H. Freeman and Company},
  address      = {San Francisco, USA},
  isbn         = {978-0-7167-0344-0}
}

@article{faraoni2008f,
    author = "Faraoni, Valerio",
    title = "{f(R) gravity: Successes and challenges}",
    booktitle = "{18th SIGRAV Conference}",
    eprint = "0810.2602",
    archivePrefix = "arXiv",
    primaryClass = "gr-qc",
    month = "10",
    year = "2008"
}

@book{Weitzenbock1923,
    author = {Weitzenb{\"o}ck, Roland},
    title = "{Invariantentheorie}",
    publisher = "P. Noordhoff",
    address = "Groningen",
    year = "1923"
}

@article{Maluf:1994ji,
    author = "Maluf, J. W.",
    title = "{Hamiltonian formulation of the teleparallel description of general relativity}",
    doi = "10.1063/1.530774",
    journal = "J. Math. Phys.",
    volume = "35",
    pages = "335--343",
    year = "1994"
}

@article{Andrade2000MG9,
    author = "de Andrade, V. C. and Pereira, J. G.",
    title = "{Gravitational Lorentz force and the description of the gravitational interaction}",
    eprint = "gr-qc/9703059",
    archivePrefix = "arXiv",
    doi = "10.1103/PhysRevD.56.4689",
    journal = "Phys. Rev. D",
    volume = "56",
    pages = "4689--4695",
    year = "1997"
}

@article{Pereira:2019woq,
    author = "Pereira, Jos{\'e} G. and Obukhov, Yuri N.",
    title = "{Gauge Structure of Teleparallel Gravity}",
    eprint = "1906.06287",
    archivePrefix = "arXiv",
    primaryClass = "gr-qc",
    doi = "10.3390/universe5060139",
    journal = "Universe",
    volume = "5",
    number = "6",
    pages = "139",
    year = "2019"
}

@article{Chervon:2023gio,
    author = "Chervon, Sergei V. and Fomin, Igor V.",
    title = "{Reconstruction of Scalar-Torsion Gravity Theories from the Physical Potential of a Scalar Field}",
    doi = "10.3390/sym15020291",
    journal = "Symmetry",
    volume = "15",
    number = "2",
    pages = "291",
    year = "2023"
}

@article{Chevallier:2000qy,
    author = "Chevallier, Michel and Polarski, David",
    title = "{Accelerating universes with scaling dark matter}",
    eprint = "gr-qc/0009008",
    archivePrefix = "arXiv",
    doi = "10.1142/S0218271801000822",
    journal = "Int. J. Mod. Phys. D",
    volume = "10",
    pages = "213--224",
    year = "2001"
}

@article{Linder:2002et,
    author = "Linder, Eric V.",
    title = "{Exploring the expansion history of the universe}",
    eprint = "astro-ph/0208512",
    archivePrefix = "arXiv",
    doi = "10.1103/PhysRevLett.90.091301",
    journal = "Phys. Rev. Lett.",
    volume = "90",
    pages = "091301",
    year = "2003"
}

@article{jassal2005wmap,
    author = "Jassal, H. K. and Bagla, J. S. and Padmanabhan, T.",
    title = "{WMAP constraints on low redshift evolution of dark energy}",
    eprint = "astro-ph/0404378",
    archivePrefix = "arXiv",
    doi = "10.1111/j.1745-3933.2005.08577.x",
    journal = "Mon. Not. Roy. Astron. Soc.",
    volume = "356",
    pages = "L11--L16",
    year = "2005"
}

@article{Barboza:2008rh,
    author = "Barboza, Jr., E. M. and Alcaniz, J. S.",
    title = "{A parametric model for dark energy}",
    eprint = "0805.1713",
    archivePrefix = "arXiv",
    primaryClass = "astro-ph",
    doi = "10.1016/j.physletb.2008.08.012",
    journal = "Phys. Lett. B",
    volume = "666",
    pages = "415--419",
    year = "2008"
}

@article{Wetterich:1987fm,
    author = "Wetterich, C.",
    title = "{Cosmology and the Fate of Dilatation Symmetry}",
    eprint = "1711.03844",
    archivePrefix = "arXiv",
    primaryClass = "hep-th",
    reportNumber = "PRINT-87-0756, DESY-87-123",
    doi = "10.1016/0550-3213(88)90193-9",
    journal = "Nucl. Phys. B",
    volume = "302",
    pages = "668--696",
    year = "1988"
}

@article{Tsujikawa:2013fta,
    author = "Tsujikawa, Shinji",
    title = "{Quintessence: A Review}",
    eprint = "1304.1961",
    archivePrefix = "arXiv",
    primaryClass = "gr-qc",
    doi = "10.1088/0264-9381/30/21/214003",
    journal = "Class. Quant. Grav.",
    volume = "30",
    pages = "214003",
    year = "2013"
}

@article{Mao:2008ug,
    author = "Mao, Yi and others",
    title = "{How accurately can 21 cm tomography constrain cosmology?}",
    eprint = "0802.1710",
    archivePrefix = "arXiv",
    primaryClass = "astro-ph",
    doi = "10.1103/PhysRevD.78.023529",
    journal = "Phys. Rev. D",
    volume = "78",
    pages = "023529",
    year = "2008"
}

@article{muromachi2015search,
    author = "Muromachi, Yusuke and Okabayashi, Akira and Okada, Daiki and Hara, Tetsuya and Itoh, Yutaka",
    title = "{Search for dark energy potentials in quintessence}",
    eprint = "1503.03678",
    archivePrefix = "arXiv",
    primaryClass = "astro-ph.CO",
    doi = "10.1093/ptep/ptv115",
    journal = "Progress of Theoretical and Experimental Physics",
    volume = "2015",
    number = "9",
    pages = "093E01",
    year = "2015"
}

@article{Caldwell:1999ew,
    author = "Caldwell, R. R.",
    title = "{A Phantom menace?}",
    eprint = "astro-ph/9908168",
    archivePrefix = "arXiv",
    doi = "10.1016/S0370-2693(02)02589-3",
    journal = "Phys. Lett. B",
    volume = "545",
    pages = "23--29",
    year = "2002"
}

@article{Feng:2004ad,
    author = "Feng, Bo and Wang, Xiu-Lian and Zhang, Xin-Min",
    title = "{Dark energy constraints from the cosmic age and supernova}",
    eprint = "astro-ph/0404224",
    archivePrefix = "arXiv",
    doi = "10.1016/j.physletb.2004.12.071",
    journal = "Phys. Lett. B",
    volume = "607",
    pages = "35--41",
    year = "2005"
}

@article{Guo:2004fq,
    author = "Guo, Zong-Kuan and others",
    title = "{Cosmological evolution of a quintom model of dark energy}",
    eprint = "astro-ph/0410654",
    archivePrefix = "arXiv",
    doi = "10.1016/j.physletb.2005.01.017",
    journal = "Phys. Lett. B",
    volume = "608",
    pages = "177--182",
    year = "2005"
}

@article{Armendariz-Picon:1999hyi,
    title={k-Inflation},
  author={Armendariz-Picon, Christian and Damour, Thibault and Mukhanov, VF1999PhLB},
  journal={Phys. Lett. B},
  volume={458 },
  number={2-3},
  pages={209--218},
  year={1999},
  publisher={Elsevier}
}

@article{Armendariz-Picon:2000ulo,
    author = "Armendariz-Picon, C. and Mukhanov, Viatcheslav F. and Steinhardt, Paul J.",
    title = "{Essentials of k essence}",
    eprint = "astro-ph/0006373",
    archivePrefix = "arXiv",
    doi = "10.1103/PhysRevD.63.103510",
    journal = "Phys. Rev. D",
    volume = "63",
    pages = "103510",
    year = "2001"
}

@article{Chiba:1999ka,
    author = "Chiba, Takeshi and Okabe, Takahiro and Yamaguchi, Masahide",
    title = "{Kinetically driven quintessence}",
    eprint = "astro-ph/9912463",
    archivePrefix = "arXiv",
    reportNumber = "UTAP-352",
    doi = "10.1103/PhysRevD.62.023511",
    journal = "Phys. Rev. D",
    volume = "62",
    pages = "023511",
    year = "2000"
}

@article{Lazkoz:2019sjl,
    author = "Lazkoz, Ruth and others",
    title = "{Observational constraints of $f(Q)$ gravity}",
    eprint = "1907.13219",
    archivePrefix = "arXiv",
    primaryClass = "gr-qc",
    doi = "10.1103/PhysRevD.100.104027",
    journal = "Phys. Rev. D",
    volume = "100",
    number = "10",
    pages = "104027",
    year = "2019"
}

@article{Frusciante:2021sio,
    author = "Frusciante, Noemi",
    title = "{Signatures of $f(Q)$-gravity in cosmology}",
    eprint = "2101.09242",
    archivePrefix = "arXiv",
    primaryClass = "astro-ph.CO",
    doi = "10.1103/PhysRevD.103.044021",
    journal = "Phys. Rev. D",
    volume = "103",
    number = "4",
    pages = "044021",
    year = "2021"
}

@article{Xu:2020yeg,
    author = "Xu, Yixin and others",
    title = "{Weyl type $f(Q,T)$ gravity, and its cosmological implications}",
    eprint = "2005.04025",
    archivePrefix = "arXiv",
    primaryClass = "gr-qc",
    doi = "10.1140/epjc/s10052-020-8023-6",
    journal = "Eur. Phys. J. C",
    volume = "80",
    number = "5",
    pages = "449",
    year = "2020"
}

@article{Gonzalez-Espinoza:2021mwr,
    author = "Gonzalez-Espinoza, Manuel and Otalora, Giovanni and Saavedra, Joel",
    title = "{Stability of scalar perturbations in scalar-torsion f(T,{\ensuremath{\phi}}) gravity theories in the presence of a matter fluid}",
    eprint = "2101.09123",
    archivePrefix = "arXiv",
    primaryClass = "gr-qc",
    doi = "10.1088/1475-7516/2021/10/007",
    journal = "Journal of Cosmology and Astroparticle Physics ",
    volume = "10",
    pages = "007",
    year = "2021"
}

@article{Amendola:1999er,
    author = "Amendola, Luca",
    title = "{Coupled quintessence}",
    eprint = "astro-ph/9908023",
    archivePrefix = "arXiv",
    doi = "10.1103/PhysRevD.62.043511",
    journal = "Phys. Rev. D",
    volume = "62",
    pages = "043511",
    year = "2000"
}

@article{Copeland:2006wr,
    author = "Copeland, Edmund J. and Sami, M. and Tsujikawa, Shinji",
    title = "{Dynamics of dark energy}",
    eprint = "hep-th/0603057",
    archivePrefix = "arXiv",
    doi = "10.1142/S021827180600942X",
    journal = "Int. J. Mod. Phys. D",
    volume = "15",
    pages = "1753--1936",
    year = "2006"
}

@book{arrowsmith1990introduction,
  title={An introduction to dynamical systems},
  author={Arrowsmith, David K and Place, Colin M},
  year={1990},
  publisher={Cambridge university press}
}

@book{wiggins2003introduction,
  title={Introduction to applied nonlinear dynamical systems and chaos},
  author={Wiggins, Stephen},
  year={2003},
  publisher={Springer}
}

@book{perko2013differential,
  title={Differential equations and dynamical systems},
  author={Perko, Lawrence},
  volume={7},
  year={2013},
  publisher={Springer Science \& Business Media}
}

@article{bohmer2017dynamical,
    author = "Boehmer, Christian G. and Chan, Nyein",
    title = "{Dynamical systems in cosmology.}",
    eprint = "1409.5585",
    archivePrefix = "arXiv",
    primaryClass = "gr-qc",
    doi = "10.1142/9781786341044_0004",
    year = "2017"
}

@book{brauer1989qualitative,
  title={The qualitative theory of ordinary differential equations: an introduction},
  author={Brauer, Fred and Nohel, John A},
  year={1989},
  publisher={Courier Corporation}
}

@article{Charters:2001hi,
    author = "Charters, Tiago C. and Nunes, Ana and Mimoso, Jose P.",
    title = "{Stability analysis of cosmological models through Liapunov's method}",
    eprint = "gr-qc/0103060",
    archivePrefix = "arXiv",
    doi = "10.1088/0264-9381/18/9/307",
    journal = "Class. Quant. Grav.",
    volume = "18",
    pages = "1703--1714",
    year = "2001"
}

@book{carr2012applications,
  title={Applications of centre manifold theory},
  author={Carr, Jack},
  volume={35},
  year={2012},
  publisher={Springer Science \& Business Media}
}

@book{lefschetz1957,
  title={Differential Equations: Geometric Theory},
  author={Lefschetz, Solomon},
  year={1957},
  publisher={Interscience Publishers},
  address={New York}
}

@book{lynch2007,
  title={Dynamical Systems with Applications using MATLAB},
  author={Lynch, Stephen},
  year={2007},
  publisher={Springer},
  address={New York}
}

@article{wainwright1989,
    author = "Wainwright, J. and Hsu, L.",
    title = "{A dynamical systems approach to Bianchi cosmologies: Orthogonal models of class A}",
    doi = "10.1088/0264-9381/6/10/011",
    journal = "Class. Quant. Grav.",
    volume = "6",
    pages = "1409--1431",
    year = "1989"
}

@article{Rave-Franco:2021yvu,
    author = "Rave-Franco, Geovanny A. and Escamilla-Rivera, Celia and Levi Said, Jackson",
    title = "{Dynamical complexity of the teleparallel gravity cosmology}",
    eprint = "2101.06347",
    archivePrefix = "arXiv",
    primaryClass = "gr-qc",
    doi = "10.1103/PhysRevD.103.084017",
    journal = "Phys. Rev. D",
    volume = "103",
    number = "8",
    pages = "084017",
    year = "2021"
}

@article{Kadam:2022lgq,
    author = "Kadam, S. A. and Mishra, B. and Said Levi, Jackson",
    title = "{Teleparallel scalar-tensor gravity through cosmological dynamical systems}",
    eprint = "2205.04231",
    archivePrefix = "arXiv",
    primaryClass = "gr-qc",
    doi = "10.1140/epjc/s10052-022-10648-8",
    journal = "Eur. Phys. J. C",
    volume = "82",
    number = "8",
    pages = "680",
    year = "2022"
}

@article{Khyllep:2022spx,
    author = "Khyllep, Wompherdeiki and others",
    title = "{Cosmology in f(Q) gravity: A unified dynamical systems analysis of the background and perturbations}",
    eprint = "2207.02610",
    archivePrefix = "arXiv",
    primaryClass = "gr-qc",
    doi = "10.1103/PhysRevD.107.044022",
    journal = "Phys. Rev. D",
    volume = "107",
    number = "4",
    pages = "044022",
    year = "2023"
}

@article{Zonunmawia:2018xvf,
    author = {Zonunmawia, Hmar and others},
    title = "{Cosmological dynamics of brane gravity: A global dynamical system perspective}",
    eprint = "1810.03816",
    archivePrefix = "arXiv",
    primaryClass = "gr-qc",
    doi = "10.1103/PhysRevD.98.083532",
    journal = "Phys. Rev. D",
    volume = "98",
    number = "8",
    pages = "083532",
    year = "2018"
}

@article{Vishwakarma:2024bnd,
    author = "Vishwakarma, Pooja and Shah, Parth and Bamba, Kazuharu",
    title = "{Autonomous systems and attractor behaviors in non-metricity gravity: stability analysis and cosmic acceleration}",
    eprint = "2401.05455",
    archivePrefix = "arXiv",
    primaryClass = "gr-qc",
    reportNumber = "APS/123-QED, FU-PCG-122",
    doi = "10.1140/epjc/s10052-025-14040-0",
    journal = "Eur. Phys. J. C",
    volume = "85",
    number = "3",
    pages = "322",
    year = "2025"
}

@article{Vishwakarma:2024qvw,
    author = "Vishwakarma, Pooja and Shah, Parth",
    title = "{Qualitative behaviour of higher-curvature gravity with boundary terms i.e the f(Q) gravity models by dynamical system analysis}",
    eprint = "2402.07951",
    archivePrefix = "arXiv",
    primaryClass = "gr-qc",
    doi = "10.1140/epjc/s10052-024-12523-0",
    journal = "Eur. Phys. J. C",
    volume = "84",
    number = "2",
    pages = "159",
    year = "2024"
}

@article{ligo2017gravitational,
    author = "Abbott, B. P. and others",
    collaboration = "LIGO Scientific, Virgo",
    title = "{GW170817: Observation of Gravitational Waves from a Binary Neutron Star Inspiral}",
    eprint = "1710.05832",
    archivePrefix = "arXiv",
    primaryClass = "gr-qc",
    reportNumber = "LIGO-P170817",
    doi = "10.1103/PhysRevLett.119.161101",
    journal = "Phys. Rev. Lett.",
    volume = "119",
    number = "16",
    pages = "161101",
    year = "2017"
}

@article{Nojiri:2006be,
    author = "Nojiri, Shin'ichi and Odintsov, Sergei D.",
    editor = "Apostolopoulos, P. and Bona, Carles and Carot, J. and Mas, Ll. and Sintes, A. M. and Stela, J.",
    title = "{Modified gravity and its reconstruction from the universe expansion history}",
    eprint = "hep-th/0611071",
    archivePrefix = "arXiv",
    doi = "10.1088/1742-6596/66/1/012005",
    journal = "J. Phys. Conf. Ser.",
    volume = "66",
    pages = "012005",
    year = "2007"
}

@article{DeFelice:2010aj,
    author = "De Felice, Antonio and Tsujikawa, Shinji",
    title = "{f(R) theories}",
    eprint = "1002.4928",
    archivePrefix = "arXiv",
    primaryClass = "gr-qc",
    doi = "10.12942/lrr-2010-3",
    journal = "Living Rev. Rel.",
    volume = "13",
    pages = "3",
    year = "2010"
}

@article{Cai:2015emx,
    author = "Cai, Yi-Fu and others",
    title = "{f(T) teleparallel gravity and cosmology}",
    eprint = "1511.07586",
    archivePrefix = "arXiv",
    primaryClass = "gr-qc",
    doi = "10.1088/0034-4885/79/10/106901",
    journal = "Rept. Prog. Phys.",
    volume = "79",
    number = "10",
    pages = "106901",
    year = "2016"
}

@article{Bahamonde:2015zma,
    author = {Bahamonde, Sebastian and B{\"o}hmer, Christian G. and Wright, Matthew},
    title = "{Modified teleparallel theories of gravity}",
    eprint = "1508.05120",
    archivePrefix = "arXiv",
    primaryClass = "gr-qc",
    doi = "10.1103/PhysRevD.92.104042",
    journal = "Phys. Rev. D",
    volume = "92",
    number = "10",
    pages = "104042",
    year = "2015"
}

@article{BeltranJimenez:2018vdo,
    author = "Beltr{\'a}n Jim{\'e}nez, Jose and Heisenberg, Lavinia and Koivisto, Tomi S.",
    title = "{Teleparallel Palatini theories}",
    eprint = "1803.10185",
    archivePrefix = "arXiv",
    primaryClass = "gr-qc",
    reportNumber = "NORDITA-2018-023, IFT-UAM/CSIC-18-035, IFT-UAM-CSIC-18-035",
    doi = "10.1088/1475-7516/2018/08/039",
    journal = "Journal of Cosmology and Astroparticle Physics ",
    volume = "08",
    pages = "039",
    year = "2018"
}

@article{Hehl:1994ue,
    author = "Hehl, Friedrich W. and others",
    title = "{Metric affine gauge theory of gravity: Field equations, Noether identities, world spinors, and breaking of dilation invariance}",
    eprint = "gr-qc/9402012",
    archivePrefix = "arXiv",
    reportNumber = "TAUP-N192-94, TAUP-192-94",
    doi = "10.1016/0370-1573(94)00111-F",
    journal = "Phys. Rept.",
    volume = "258",
    pages = "1--171",
    year = "1995"
}

@book{yano2020theory,
  title={The theory of Lie derivatives and its applications},
  author={Yano, Kentaro},
  year={2020},
  publisher={Courier Dover Publications}
}

@article{Hehl:1976kj,
    author = "Hehl, F. W. and others",
    title = "{General Relativity with Spin and Torsion: Foundations and Prospects}",
    doi = "10.1103/RevModPhys.48.393",
    journal = "Rev. Mod. Phys.",
    volume = "48",
    pages = "393--416",
    year = "1976"
}

@article{Haghani:2012bt,
    author = "Haghani, Zahra and others",
    title = "{Weyl-Cartan-Weitzenboeck gravity as a generalization of teleparallel gravity}",
    eprint = "1202.1879",
    archivePrefix = "arXiv",
    primaryClass = "gr-qc",
    doi = "10.1088/1475-7516/2012/10/061",
    journal = "Journal of Cosmology and Astroparticle Physics ",
    volume = "10",
    pages = "061",
    year = "2012"
}

@article{Lehmkuhl:2014qhu,
    author = "Lehmkuhl, Dennis",
    title = "{Why Einstein did not believe that general relativity geometrizes gravity}",
    doi = "10.1016/j.shpsb.2013.08.002",
    journal = "Stud. Hist. Phil. Sci. B",
    volume = "46",
    pages = "316--326",
    year = "2014"
}

@article{Heisenberg:2018vsk,
    author = "Heisenberg, Lavinia",
    title = "{A systematic approach to generalisations of General Relativity and their cosmological implications}",
    eprint = "1807.01725",
    archivePrefix = "arXiv",
    primaryClass = "gr-qc",
    doi = "10.1016/j.physrep.2018.11.006",
    journal = "Phys. Rept.",
    volume = "796",
    pages = "1--113",
    year = "2019"
}

@article{BeltranJimenez:2017doy,
    author = "Beltran Jimenez, Jose and others",
    title = "{Born{\textendash}Infeld inspired modifications of gravity}",
    eprint = "1704.03351",
    archivePrefix = "arXiv",
    primaryClass = "gr-qc",
    reportNumber = "IFIC-17-23",
    doi = "10.1016/j.physrep.2017.11.001",
    journal = "Phys. Rept.",
    volume = "727",
    pages = "1--129",
    year = "2018"
}

@article{BeltranJimenez:2020sih,
    author = "Beltr{\'a}n Jim{\'e}nez, Jose and Heisenberg, Lavinia and Koivisto, Tomi",
    title = "{The coupling of matter and spacetime geometry}",
    eprint = "2004.04606",
    archivePrefix = "arXiv",
    primaryClass = "hep-th",
    doi = "10.1088/1361-6382/aba31b",
    journal = "Class. Quant. Grav.",
    volume = "37",
    number = "19",
    pages = "195013",
    year = "2020"
}

@article{Nester:1998mp,
    author = "Nester, James M. and Yo, Hwei-Jang",
    title = "{Symmetric teleparallel general relativity}",
    eprint = "gr-qc/9809049",
    archivePrefix = "arXiv",
    reportNumber = "NCU-CCS-980904",
    journal = "Chin. J. Phys.",
    volume = "37",
    pages = "113",
    year = "1999"
}

@article{DAmbrosio:2020nev,
    author = "D'Ambrosio, Fabio and Garg, Mudit and Heisenberg, Lavinia",
    title = "{Non-linear extension of non-metricity scalar for MOND}",
    eprint = "2004.00888",
    archivePrefix = "arXiv",
    primaryClass = "gr-qc",
    doi = "10.1016/j.physletb.2020.135970",
    journal = "Phys. Lett. B",
    volume = "811",
    pages = "135970",
    year = "2020"
}

@article{Bajardi:2020fxh,
    author = "Bajardi, Francesco and Vernieri, Daniele and Capozziello, Salvatore",
    title = "{Bouncing Cosmology in f(Q) Symmetric Teleparallel Gravity}",
    eprint = "2011.01248",
    archivePrefix = "arXiv",
    primaryClass = "gr-qc",
    doi = "10.1140/epjp/s13360-020-00918-3",
    journal = "Eur. Phys. J. Plus",
    volume = "135",
    number = "11",
    pages = "912",
    year = "2020"
}

@article{Atayde:2021pgb,
    author = "Atayde, Lu{\'\i}s and Frusciante, Noemi",
    title = "{Can $f(Q)$ gravity challenge $\Lambda$CDM?}",
    eprint = "2108.10832",
    archivePrefix = "arXiv",
    primaryClass = "astro-ph.CO",
    doi = "10.1103/PhysRevD.104.064052",
    journal = "Phys. Rev. D",
    volume = "104",
    number = "6",
    pages = "064052",
    year = "2021"
}

@article{Capozziello:2022wgl,
    author = "Capozziello, Salvatore and D'Agostino, Rocco",
    title = "{Model-independent reconstruction of f(Q) non-metric gravity}",
    eprint = "2204.01015",
    archivePrefix = "arXiv",
    primaryClass = "gr-qc",
    doi = "10.1016/j.physletb.2022.137229",
    journal = "Phys. Lett. B",
    volume = "832",
    pages = "137229",
    year = "2022"
}

@article{BeltranJimenez:2019odq,
    author = "Beltr{\'a}n Jim{\'e}nez, Jose and others",
    title = "{General teleparallel quadratic gravity}",
    eprint = "1909.09045",
    archivePrefix = "arXiv",
    primaryClass = "gr-qc",
    doi = "10.1016/j.physletb.2020.135422",
    journal = "Phys. Lett. B",
    volume = "805",
    pages = "135422",
    year = "2020"
}

@article{Heisenberg:2022mbo,
    author = "Heisenberg, Lavinia and Hohmann, Manuel and Kuhn, Simon",
    title = "{Homogeneous and isotropic cosmology in general teleparallel gravity}",
    eprint = "2212.14324",
    archivePrefix = "arXiv",
    primaryClass = "gr-qc",
    doi = "10.1140/epjc/s10052-023-11462-6",
    journal = "Eur. Phys. J. C",
    volume = "83",
    number = "4",
    pages = "315",
    year = "2023"
}

@article{Hohmann:2022mlc,
    author = "Hohmann, Manuel",
    title = "{Teleparallel Gravity}",
    eprint = "2207.06438",
    archivePrefix = "arXiv",
    primaryClass = "gr-qc",
    doi = "10.1007/978-3-031-31520-6_4",
    journal = "Lect. Notes Phys.",
    volume = "1017",
    pages = "145--198",
    year = "2023"
}

@article{BeltranJimenez:2019bnx,
    author = "Beltr{\'a}n Jim{\'e}nez, Jose and Heisenberg, Lavinia and Koivisto, Tomi S.",
    title = "{The canonical frame of purified gravity}",
    eprint = "1903.12072",
    archivePrefix = "arXiv",
    primaryClass = "gr-qc",
    reportNumber = "NORDITA 2019-030",
    doi = "10.1142/S0218271819440127",
    journal = "Int. J. Mod. Phys. D",
    volume = "28",
    number = "14",
    pages = "1944012",
    year = "2019"
}

@article{Koivisto:2022nar,
    author = "Koivisto, Tomi",
    title = "{Energy in the relativistic theory of gravity}",
    eprint = "2202.11522",
    archivePrefix = "arXiv",
    primaryClass = "gr-qc",
    doi = "10.1142/S0219887822400072",
    journal = "Int. J. Geom. Meth. Mod. Phys.",
    volume = "19",
    number = "Supp01",
    pages = "2240007",
    year = "2022"
}

@article{BeltranJimenez:2021kpj,
    author = "Beltr{\'a}n Jim{\'e}nez, Jose and Koivisto, Tomi S.",
    title = "{Noether charges in the geometrical trinity of gravity}",
    eprint = "2111.04716",
    archivePrefix = "arXiv",
    primaryClass = "gr-qc",
    doi = "10.1103/PhysRevD.105.L021502",
    journal = "Phys. Rev. D",
    volume = "105",
    number = "2",
    pages = "L021502",
    year = "2022"
}

@article{Heisenberg:2022nvs,
    author = "Heisenberg, Lavinia and Kuhn, Simon and Walleghem, Laurens",
    title = "{Wald{\textquoteright}s entropy in Coincident General Relativity}",
    eprint = "2203.13914",
    archivePrefix = "arXiv",
    primaryClass = "gr-qc",
    doi = "10.1088/1361-6382/ac987d",
    journal = "Class. Quant. Grav.",
    volume = "39",
    number = "23",
    pages = "235002",
    year = "2022"
}

@article{Blixt:2018znp,
    author = "Blixt, Daniel and Hohmann, Manuel and Pfeifer, Christian",
    title = "{Hamiltonian and primary constraints of new general relativity}",
    eprint = "1811.11137",
    archivePrefix = "arXiv",
    primaryClass = "gr-qc",
    doi = "10.1103/PhysRevD.99.084025",
    journal = "Phys. Rev. D",
    volume = "99",
    number = "8",
    pages = "084025",
    year = "2019"
}

@inproceedings{Blixt:2019ene,
    author = "Blixt, Daniel and others",
    title = "{Hamiltonian analysis in new general relativity}",
    booktitle = "{15th Marcel Grossmann Meeting on Recent Developments in Theoretical and Experimental General Relativity, Astrophysics, and Relativistic Field Theories}",
    eprint = "1905.11919",
    archivePrefix = "arXiv",
    primaryClass = "gr-qc",
    doi = "10.1142/9789811258251_0038",
    month = "5",
    year = "2019"
}

@article{Blixt:2020ekl,
    author = "Blixt, Daniel and others",
    title = "{Review of the Hamiltonian analysis in teleparallel gravity}",
    eprint = "2012.09180",
    archivePrefix = "arXiv",
    primaryClass = "gr-qc",
    doi = "10.1142/S0219887821300051",
    journal = "Int. J. Geom. Meth. Mod. Phys.",
    volume = "18",
    number = "supp01",
    pages = "2130005",
    year = "2021"
}

@article{Dambrosio:2020wbi,
    author = "D'ambrosio, Fabio and Heisenberg, Lavinia",
    title = "{Classification of primary constraints of quadratic non-metricity theories of gravity}",
    eprint = "2007.05064",
    archivePrefix = "arXiv",
    primaryClass = "gr-qc",
    doi = "10.1007/JHEP02(2021)170",
    journal = "JHEP",
    volume = "02",
    pages = "170",
    year = "2021"
}

@article{DAmbrosio:2023asf,
    author = "D'Ambrosio, Fabio and Heisenberg, Lavinia and Zentarra, Stefan",
    title = "{Hamiltonian Analysis of f(Q)$f(\mathbb {Q})$ Gravity and the Failure of the Dirac{\textendash}Bergmann Algorithm for Teleparallel Theories of Gravity}",
    eprint = "2308.02250",
    archivePrefix = "arXiv",
    primaryClass = "gr-qc",
    doi = "10.1002/prop.202300185",
    journal = "Fortsch. Phys.",
    volume = "71",
    number = "12",
    pages = "2300185",
    year = "2023"
}

@article{Koivisto:2018loq,
    author = "Koivisto, Tomi and Tsimperis, Georgios",
    title = "{The spectrum of teleparallel gravity}",
    eprint = "1810.11847",
    archivePrefix = "arXiv",
    primaryClass = "gr-qc",
    reportNumber = "NORDITA-2018-99",
    doi = "10.3390/universe5030080",
    month = "10",
    year = "2018"
}

@article{Buchdahl:1970ynr,
    author = "Buchdahl, Hans A.",
    title = "{Non-linear Lagrangians and cosmological theory}",
    journal = "Mon. Not. Roy. Astron. Soc.",
    volume = "150",
    pages = "1",
    year = "1970"
}

@article{Sotiriou:2008rp,
    author = "Sotiriou, Thomas P. and Faraoni, Valerio",
    title = "{f(R) Theories Of Gravity}",
    eprint = "0805.1726",
    archivePrefix = "arXiv",
    primaryClass = "gr-qc",
    doi = "10.1103/RevModPhys.82.451",
    journal = "Rev. Mod. Phys.",
    volume = "82",
    pages = "451--497",
    year = "2010"
}

@article{Ferraro:2018tpu,
    author = "Ferraro, Rafael and Guzm{\'a}n, Mar{\'\i}a Jos{\'e}",
    title = "{Hamiltonian formalism for f(T) gravity}",
    eprint = "1802.02130",
    archivePrefix = "arXiv",
    primaryClass = "gr-qc",
    doi = "10.1103/PhysRevD.97.104028",
    journal = "Phys. Rev. D",
    volume = "97",
    number = "10",
    pages = "104028",
    year = "2018"
}

@article{Li:2011rn,
    author = "Li, Miao and Miao, Rong-Xin and Miao, Yan-Gang",
    title = "{Degrees of freedom of $f(T)$ gravity}",
    eprint = "1105.5934",
    archivePrefix = "arXiv",
    primaryClass = "hep-th",
    reportNumber = "USTC-ICTS-11-06",
    doi = "10.1007/JHEP07(2011)108",
    journal = " Journal of High Energy Physics ",
    volume = "07",
    pages = "108",
    year = "2011"
}

@article{Blagojevic:2020dyq,
    author = "Blagojevi{\'c}, Milutin and Nester, James M.",
    title = "{Local symmetries and physical degrees of freedom in $f(T)$ gravity: a Dirac Hamiltonian constraint analysis}",
    eprint = "2006.15303",
    archivePrefix = "arXiv",
    primaryClass = "gr-qc",
    doi = "10.1103/PhysRevD.102.064025",
    journal = "Phys. Rev. D",
    volume = "102",
    number = "6",
    pages = "064025",
    year = "2020"
}

@article{DAmbrosio:2021zpm,
    author = "D'Ambrosio, Fabio and others",
    title = "{Black holes in f(Q) gravity}",
    eprint = "2109.03174",
    archivePrefix = "arXiv",
    primaryClass = "gr-qc",
    doi = "10.1103/PhysRevD.105.024042",
    journal = "Phys. Rev. D",
    volume = "105",
    number = "2",
    pages = "024042",
    year = "2022"
}

@article{DAmbrosio:2021pnd,
    author = "D'Ambrosio, Fabio and Heisenberg, Lavinia and Kuhn, Simon",
    title = "{Revisiting cosmologies in teleparallelism}",
    eprint = "2109.04209",
    archivePrefix = "arXiv",
    primaryClass = "gr-qc",
    doi = "10.1088/1361-6382/ac3f99",
    journal = "Class. Quant. Grav.",
    volume = "39",
    number = "2",
    pages = "025013",
    year = "2022"
}

@article{Khyllep:2021wjd,
    author = "Khyllep, Wompherdeiki and others",
    title = "{Background evolution and growth of structures in interacting dark energy scenarios through dynamical system analysis}",
    eprint = "2111.01268",
    archivePrefix = "arXiv",
    primaryClass = "gr-qc",
    doi = "10.1103/PhysRevD.105.043511",
    journal = "Phys. Rev. D",
    volume = "105",
    number = "4",
    pages = "043511",
    year = "2022"
}

@article{Anagnostopoulos:2022gej,
    author = "Anagnostopoulos, Fotios K. and others",
    title = "{New models and big bang nucleosynthesis constraints in f(Q) gravity}",
    eprint = "2205.11445",
    archivePrefix = "arXiv",
    primaryClass = "gr-qc",
    doi = "10.1140/epjc/s10052-023-11190-x",
    journal = "Eur. Phys. J. C",
    volume = "83",
    number = "1",
    pages = "58",
    year = "2023"
}

@article{Vishwakarma:2023brw,
    author = "Vishwakarma, Pooja and Shah, Parth",
    title = "{Stability analysis of f(Q) gravity models using dynamical systems}",
    eprint = "2401.09004",
    archivePrefix = "arXiv",
    primaryClass = "gr-qc",
    doi = "10.1142/S0218271823500712",
    journal = "Int. J. Mod. Phys. D",
    volume = "32",
    number = "11",
    pages = "2350071",
    year = "2023"
}

@article{Gannouji:2008wt,
    author = "Gannouji, R. and Moraes, B. and Polarski, D.",
    title = "{The growth of matter perturbations in f(R) models}",
    eprint = "0809.3374",
    archivePrefix = "arXiv",
    primaryClass = "astro-ph",
    doi = "10.1088/1475-7516/2009/02/034",
    journal = "Journal of Cosmology and Astroparticle Physics ",
    volume = "02",
    pages = "034",
    year = "2009"
}

@article{Goncalves:2024sem,
    author = "Gon{\c{c}}alves, Tiago B. and Atayde, Lu{\'\i}s and Frusciante, Noemi",
    title = "{Cosmological study of a symmetric teleparallel gravity model}",
    eprint = "2404.01742",
    archivePrefix = "arXiv",
    primaryClass = "gr-qc",
    doi = "10.1103/PhysRevD.109.084003",
    journal = "Phys. Rev. D",
    volume = "109",
    number = "8",
    pages = "084003",
    year = "2024"
}

@article{Planck:2018fzr,
    author = "Aghanim, N. and others",
    collaboration = "Planck",
    title = "{Planck 2018 results. XII. Galactic astrophysics using polarized dust emission}",
    eprint = "1807.06212",
    archivePrefix = "arXiv",
    primaryClass = "astro-ph.GA",
    doi = "10.1051/0004-6361/201833885",
    journal = "Astron. Astrophys.",
    volume = "641",
    pages = "A12",
    year = "2020"
}

@article{Camarena:2019moy,
    author = "Camarena, David and Marra, Valerio",
    title = "{Local determination of the Hubble constant and the deceleration parameter}",
    eprint = "1906.11814",
    archivePrefix = "arXiv",
    primaryClass = "astro-ph.CO",
    doi = "10.1103/PhysRevResearch.2.013028",
    journal = "Phys. Rev. Res.",
    volume = "2",
    number = "1",
    pages = "013028",
    year = "2020"
}

@article{Bernstein:1988ad,
    author = "Bernstein, Jeremy and Brown, Lowell S. and Feinberg, G.",
    title = "{COSMOLOGICAL HELIUM PRODUCTION SIMPLIFIED}",
    reportNumber = "DOE/ER/40423-02-P8",
    doi = "10.1103/RevModPhys.61.25",
    journal = "Rev. Mod. Phys.",
    volume = "61",
    pages = "25",
    year = "1989"
}

@book{Kolb:1990vq,
    author = "Kolb, Edward W. and Turner, Michael S.",
    title = "{The Early Universe}",
    reportNumber = "FERMILAB-BOOK-1990-01",
    doi = "10.1201/9780429492860",
    isbn = "978-0-429-49286-0, 978-0-201-62674-2",
    publisher = "Taylor and Francis",
    volume = "69",
    month = "5",
    year = "2019"
}

@article{Olive:1999ij,
    author = "Olive, Keith A. and Steigman, Gary and Walker, Terry P.",
    title = "{Primordial nucleosynthesis: Theory and observations}",
    eprint = "astro-ph/9905320",
    archivePrefix = "arXiv",
    reportNumber = "UMN-TH-1802-99, TPI-MINN-99-29",
    doi = "10.1016/S0370-1573(00)00031-4",
    journal = "Phys. Rept.",
    volume = "333",
    pages = "389--407",
    year = "2000"
}

@article{Cyburt:2015mya,
    author = "Cyburt, Richard H. and others",
    title = "{Big Bang Nucleosynthesis: 2015}",
    eprint = "1505.01076",
    archivePrefix = "arXiv",
    primaryClass = "astro-ph.CO",
    reportNumber = "UMN-TH-3432-15, FTPI-MINN-15-19",
    doi = "10.1103/RevModPhys.88.015004",
    journal = "Rev. Mod. Phys.",
    volume = "88",
    pages = "015004",
    year = "2016"
}

@article{Asimakis:2021yct,
    author = "Asimakis, Petros and others",
    title = "{Big bang nucleosynthesis constraints on higher-order modified gravities}",
    eprint = "2112.10863",
    archivePrefix = "arXiv",
    primaryClass = "gr-qc",
    reportNumber = "KCL-PH-TH/2021-48",
    doi = "10.1103/PhysRevD.105.084010",
    journal = "Phys. Rev. D",
    volume = "105",
    number = "8",
    pages = "084010",
    year = "2022"
}

@article{Barrow:2020kug,
    author = "Barrow, John D. and Basilakos, Spyros and Saridakis, Emmanuel N.",
    title = "{Big Bang Nucleosynthesis constraints on Barrow entropy}",
    eprint = "2010.00986",
    archivePrefix = "arXiv",
    primaryClass = "gr-qc",
    doi = "10.1016/j.physletb.2021.136134",
    journal = "Phys. Lett. B",
    volume = "815",
    pages = "136134",
    year = "2021"
}

@article{Torres:1997sn,
    author = "Torres, Diego F. and Vucetich, Hector and Plastino, A.",
    title = "{Early universe test of nonextensive statistics}",
    eprint = "astro-ph/9705068",
    archivePrefix = "arXiv",
    doi = "10.1103/PhysRevLett.79.1588",
    journal = "Phys. Rev. Lett.",
    volume = "79",
    pages = "1588--1590",
    year = "1997",
    note = "[Erratum: Phys.Rev.Lett. 80, 3889 (1998)]"
}

@article{Lambiase:2005kb,
    author = "Lambiase, Gaetano",
    title = "{Lorentz invariance breakdown and constraints from big-bang nucleosynthesis}",
    eprint = "astro-ph/0510386",
    archivePrefix = "arXiv",
    doi = "10.1103/PhysRevD.72.087702",
    journal = "Phys. Rev. D",
    volume = "72",
    pages = "087702",
    year = "2005"
}

@article{Lambiase:2011zz,
    author = "Lambiase, G.",
    title = "{Dark matter relic abundance and big bang nucleosynthesis in Horava's gravity}",
    doi = "10.1103/PhysRevD.83.107501",
    journal = "Phys. Rev. D",
    volume = "83",
    pages = "107501",
    year = "2011"
}

@article{Capozziello:2017bxm,
    author = "Capozziello, S. and Lambiase, G. and Saridakis, E. N.",
    title = "{Constraining f(T) teleparallel gravity by Big Bang Nucleosynthesis}",
    eprint = "1702.07952",
    archivePrefix = "arXiv",
    primaryClass = "astro-ph.CO",
    doi = "10.1140/epjc/s10052-017-5143-8",
    journal = "Eur. Phys. J. C",
    volume = "77",
    number = "9",
    pages = "576",
    year = "2017"
}

@article{Coc:2003ce,
    author = "Coc, Alain and others",
    title = "{Updated Big Bang nucleosynthesis confronted to WMAP observations and to the abundance of light elements}",
    eprint = "astro-ph/0309480",
    archivePrefix = "arXiv",
    doi = "10.1086/380121",
    journal = "Astrophys. J.",
    volume = "600",
    pages = "544--552",
    year = "2004"
}

@article{Olive:1996zu,
    author = "Olive, Keith A. and Skillman, Evan and Steigman, Gary",
    title = "{The Primordial abundance of He-4: An Update}",
    eprint = "astro-ph/9611166",
    archivePrefix = "arXiv",
    reportNumber = "UMN-TH-1514-96, OSU-TA-24-96",
    doi = "10.1086/304281",
    journal = "Astrophys. J.",
    volume = "483",
    pages = "788",
    year = "1997"
}

@article{Izotov:1998mj,
    author = "Izotov, Yuri I. and Thuan, Trinh X.",
    title = "{The Primordial abundance of 4-He revisited}",
    doi = "10.1086/305698",
    journal = "Astrophys. J.",
    volume = "500",
    pages = "188",
    year = "1998"
}

@article{Fields:1998gv,
    author = "Fields, Brian D. and Olive, Keith A.",
    title = "{On the evolution of helium in blue compact galaxies}",
    eprint = "astro-ph/9803297",
    archivePrefix = "arXiv",
    reportNumber = "UMN-TH-1625-98",
    doi = "10.1086/306248",
    journal = "Astrophys. J.",
    volume = "506",
    pages = "177",
    year = "1998"
}

@article{Izotov:1999wa,
    author = "Izotov, Y. I. and others",
    title = "{Helium abundance in the most metal-deficient blue compact galaxies: I zw 18 and sbs 0335-052}",
    eprint = "astro-ph/9907228",
    archivePrefix = "arXiv",
    doi = "10.1086/308119",
    journal = "Astrophys. J.",
    volume = "527",
    pages = "757--777",
    year = "1999"
}

@article{Kirkman:2003uv,
    author = "Kirkman, David and others",
    title = "{The Cosmological baryon density from the deuterium to hydrogen ratio towards QSO absorption systems: D/H towards Q1243+3047}",
    eprint = "astro-ph/0302006",
    archivePrefix = "arXiv",
    doi = "10.1086/378152",
    journal = "Astrophys. J. Suppl.",
    volume = "149",
    pages = "1",
    year = "2003"
}

@article{Izotov:2003xn,
    author = "Izotov, Y. I. and Thuan, T. X.",
    title = "{Systematic effects and a new determination of the primordial abundance of He-4 and dY/dZ from observations of blue compact galaxies}",
    eprint = "astro-ph/0310421",
    archivePrefix = "arXiv",
    doi = "10.1086/380830",
    journal = "Astrophys. J.",
    volume = "602",
    pages = "200--230",
    year = "2004"
}

@article{Bamba:2010wb,
    author = "Bamba, Kazuharu and others",
    title = "{Equation of state for dark energy in $f(T)$ gravity}",
    eprint = "1011.0508",
    archivePrefix = "arXiv",
    primaryClass = "astro-ph.CO",
    doi = "10.1088/1475-7516/2011/01/021",
    journal = "Journal of Cosmology and Astroparticle Physics ",
    volume = "01",
    pages = "021",
    year = "2011"
}

@article{Wu:2010av,
    author = "Wu, Puxun and Yu, Hong Wei",
    title = "{$f(T)$ models with phantom divide line crossing}",
    eprint = "1008.3669",
    archivePrefix = "arXiv",
    primaryClass = "gr-qc",
    doi = "10.1140/epjc/s10052-011-1552-2",
    journal = "Eur. Phys. J. C",
    volume = "71",
    pages = "1552",
    year = "2011"
}

@article{deRitis:1990ba,
    author = "de Ritis, R. and others",
    title = "{New approach to find exact solutions for cosmological models with a scalar field}",
    doi = "10.1103/PhysRevD.42.1091",
    journal = "Phys. Rev. D",
    volume = "42",
    pages = "1091--1097",
    year = "1990"
}

@article{Demianski:1992tu,
    author = "Demianski, M. and others",
    title = "{Scalar fields and anisotropy in cosmological models}",
    doi = "10.1103/PhysRevD.46.1391",
    journal = "Phys. Rev. D",
    volume = "46",
    pages = "1391--1398",
    year = "1992"
}

@article{Capozziello:1993vr,
    author = "Capozziello, S. and de Ritis, R.",
    title = "{Relation between the potential and nonminimal coupling in inflationary cosmology}",
    doi = "10.1016/0375-9601(93)90364-6",
    journal = "Phys. Lett. A",
    volume = "177",
    pages = "1--7",
    year = "1993"
}

@article{Bahamonde:2017ize,
    author = "Bahamonde, Sebastian and others",
    title = "{Dynamical systems applied to cosmology: dark energy and modified gravity}",
    eprint = "1712.03107",
    archivePrefix = "arXiv",
    primaryClass = "gr-qc",
    doi = "10.1016/j.physrep.2018.09.001",
    journal = "Phys. Rept.",
    volume = "775-777",
    pages = "1--122",
    year = "2018"
}

@article{Gonzalez-Espinoza:2020jss,
    author = "Gonzalez-Espinoza, Manuel and Otalora, Giovanni",
    title = "{Cosmological dynamics of dark energy in scalar-torsion $f(T,\phi )$ gravity}",
    eprint = "2011.08377",
    archivePrefix = "arXiv",
    primaryClass = "gr-qc",
    doi = "10.1140/epjc/s10052-021-09270-x",
    journal = "Eur. Phys. J. C",
    volume = "81",
    number = "5",
    pages = "480",
    year = "2021"
}

@article{Gonzalez-Espinoza:2021qnv,
    author = "Gonzalez-Espinoza, Manuel and others",
    title = "{Reconstructing inflation in scalar-torsion $f(T,\phi )$ gravity}",
    eprint = "2106.06145",
    archivePrefix = "arXiv",
    primaryClass = "gr-qc",
    doi = "10.1140/epjc/s10052-021-09542-6",
    journal = "Eur. Phys. J. C",
    volume = "81",
    number = "8",
    pages = "731",
    year = "2021"
}

@article{Gonzalez-Espinoza:2020azh,
    author = "Gonzalez-Espinoza, Manuel and Otalora, Giovanni",
    title = "{Generating primordial fluctuations from modified teleparallel gravity with local Lorentz-symmetry breaking}",
    eprint = "2005.03753",
    archivePrefix = "arXiv",
    primaryClass = "gr-qc",
    doi = "10.1016/j.physletb.2020.135696",
    journal = "Phys. Lett. B",
    volume = "809",
    pages = "135696",
    year = "2020"
}

@article{Capozziello:1999xr,
    author = "Capozziello, S. and Lambiase, G.",
    title = "{Selection rules in minisuperspace quantum cosmology}",
    eprint = "gr-qc/9912083",
    archivePrefix = "arXiv",
    doi = "10.1023/A:1001967102409",
    journal = "Gen. Rel. Grav.",
    volume = "32",
    pages = "673--696",
    year = "2000"
}

@article{Hohmann:2018rwf,
    author = {Hohmann, Manuel and J{\"a}rv, Laur and Ualikhanova, Ulbossyn},
    title = "{Covariant formulation of scalar-torsion gravity}",
    eprint = "1801.05786",
    archivePrefix = "arXiv",
    primaryClass = "gr-qc",
    doi = "10.1103/PhysRevD.97.104011",
    journal = "Phys. Rev. D",
    volume = "97",
    number = "10",
    pages = "104011",
    year = "2018"
}

@book{Faraoni:2004pi,
    author = "Faraoni, Valerio",
    title = "{Cosmology in scalar tensor gravity}",
    doi = "10.1007/978-1-4020-1989-0",
    isbn = "978-1-4020-1988-3",
    year = "2004"
}

@article{Kucukakca:2013mya,
    author = "Kucukakca, Yusuf",
    title = "{Scalar tensor teleparallel dark gravity via Noether symmetry}",
    eprint = "1404.7315",
    archivePrefix = "arXiv",
    primaryClass = "gr-qc",
    doi = "10.1140/epjc/s10052-013-2327-8",
    journal = "Eur. Phys. J. C",
    volume = "73",
    number = "2",
    pages = "2327",
    year = "2013"
}

@article{Rubano:2003et,
    author = "Rubano, C. and others",
    title = "{Exponential potentials for tracker fields}",
    eprint = "astro-ph/0311537",
    archivePrefix = "arXiv",
    doi = "10.1103/PhysRevD.69.103510",
    journal = "Phys. Rev. D",
    volume = "69",
    pages = "103510",
    year = "2004"
}

@article{Denzel:2020zuq,
    author = "Denzel, Philipp and others",
    title = "{The Hubble constant from eight time-delay galaxy lenses}",
    eprint = "2007.14398",
    archivePrefix = "arXiv",
    primaryClass = "astro-ph.CO",
    doi = "10.1093/mnras/staa3603",
    journal = "Mon. Not. Roy. Astron. Soc.",
    volume = "501",
    number = "1",
    pages = "784--801",
    year = "2021"
}

@article{Vishwakarma:2025yph,
    author = "Vishwakarma, Pooja and others",
    title = "{Noether symmetry and exact solutions in $f(Q,\phi)$ gravity with scalar field dynamics}",
    doi = "10.1142/s0219887825502615",
    journal = " Int.J.Geom.Meth.Mod.Phys.",
    month = "7",
    year = "2025"
}

@article{Vishwakarma:2026zmm,
    author = "Vishwakarma, Pooja and Kumar, Ravi and Shah, Parth",
    title = "{Autonomous systems and the dynamics of f(Q,{\ensuremath{\phi}}) gravity: Stability and cosmic acceleration}",
    doi = "10.1142/s0219887826501069",
    journal = " Int.J.Geom.Meth.Mod.Phys.",
    month = "1",
    year = "2026"
}

@article{Duchaniya:2022fmc,
    author = "Duchaniya, L. K. and others",
    title = "{Dynamical systems analysis in $f(T,\phi )$ gravity}",
    eprint = "2209.03414",
    archivePrefix = "arXiv",
    primaryClass = "gr-qc",
    doi = "10.1140/epjc/s10052-022-11155-6",
    journal = "Eur. Phys. J. C",
    volume = "83",
    number = "1",
    pages = "27",
    year = "2023"
}

@article{DESI:2025euz,
    author = "Bahr-Kalus, B. and others",
    title = "{Model-independent measurement of the matter-radiation equality scale in DESI 2024}",
    eprint = "2505.16153",
    archivePrefix = "arXiv",
    primaryClass = "astro-ph.CO",
    reportNumber = "FERMILAB-PUB-25-0369-PPD",
    doi = "10.1103/yqm1-ybbv",
    journal = "Phys. Rev. D",
    volume = "112",
    number = "6",
    pages = "063553",
    year = "2025"
}

@article{Zhang:2011qp,
    author = "Zhang, Yi and others",
    title = "{Notes on $f(T)$ Theories}",
    eprint = "1103.0719",
    archivePrefix = "arXiv",
    primaryClass = "astro-ph.CO",
    doi = "10.1088/1475-7516/2011/07/015",
    journal = "Journal of Cosmology and Astroparticle Physics ",
    volume = "07",
    pages = "015",
    year = "2011"
}

@article{Heisenberg:2023lru,
    author = "Heisenberg, Lavinia",
    title = "{Review on f(Q) gravity}",
    eprint = "2309.15958",
    archivePrefix = "arXiv",
    primaryClass = "gr-qc",
    doi = "10.1016/j.physrep.2024.02.001",
    journal = "Phys. Rept.",
    volume = "1066",
    pages = "1--78",
    year = "2024"
}

@article{Duchaniya:2022hiy,
    author = "Duchaniya, L. K. and Mishra, B. and Levi Said, Jackson",
    title = "{Noether symmetry approach in scalar-torsion $f(T,\phi )$ gravity}",
    eprint = "2210.11944",
    archivePrefix = "arXiv",
    primaryClass = "gr-qc",
    doi = "10.1140/epjc/s10052-023-11792-5",
    journal = "Eur. Phys. J. C",
    volume = "83",
    number = "7",
    pages = "613",
    year = "2023"
}
\newpage 
\begin{spacing}{1.5}
\chapter*{\centering \large\textbf{LIST OF PUBLICATIONS}}
\addcontentsline{toc}{section}{\bf ~~~~~~~~~~~LIST OF PUBLICATIONS}
\noindent
\begin{enumerate}

\harvarditem{P.~Vishwakarma \harvardand\ P.~Shah}{2023}{vishwakarma}
P.~Vishwakarma \harvardand\ P.~Shah  \harvardyearleft 2023\harvardyearright
, `Stability analysis of $f(Q)$ gravity models using dynamical systems', {\em Int. J. Mod. Phys. D} {\bf 32}(11),~2350071.

\harvarditem{P.~Vishwakarma \harvardand\ P.~Shah}{2024}{vishwakarma}
P.~Vishwakarma \harvardand\ P.~Shah  \harvardyearleft 2024\harvardyearright
, `Qualitative behaviour of higher-curvature gravity with boundary terms i.e the $f(Q)$ gravity models by dynamical system analysis', {\em Eur. Phys. J. C } {\bf 84}(2),~159.

\harvarditem{P.~Vishwakarma, P.~Shah \harvardand\ K.~Bamba}{2025}{vishwakarma}
P.~Vishwakarma, P.~Shah \harvardand\ K.~Bamba (2025), `Autonomous systems and attractor behaviors in non-metricity gravity: stability analysis and cosmic acceleration', {\em Eur. Phys. J. C} {\bf 85},~322.

\harvarditem{P.~Vishwakarma, R.~K.~Aswathy, V.~K.~Chaurasiya \harvardand\ P.~Shah}{2025}{vishwakarma}
P.~Vishwakarma, R.~K.~Aswathy, V.~K.~Chaurasiya \harvardand\ P.~Shah (2025), `Noether symmetry and exact solutions in $f(Q,\phi)$ gravity with scalar field dynamics', {\em  Int. J. Geom. Meth. Mod. Phys.},~2550261, {\em DOI: 10.1142/s0219887825502615}.

\harvarditem{P.~Vishwakarma, R.~Kumar \harvardand\ P.~Shah}{2026}{vishwakarma}
P.~Vishwakarma, R.~Kumar \harvardand\ P.~Shah (2026), `Autonomous systems and the dynamics of f(Q,{\ensuremath{\phi}}) gravity: Stability and cosmic acceleration', {\em  Int. J. Geom. Meth. Mod. Phys.},~2650106, {\em DOI:10.1142/s0219887826501069}.

\end{enumerate}

\end{spacing}

\end{document}